\documentclass[12pt]{report}
\usepackage[centertags]{amsmath}
\usepackage{amsfonts}
\usepackage{amssymb}
\usepackage{amsthm}
\usepackage{newlfont}
\usepackage{graphicx}
\usepackage[scriptsize,bf]{caption} %\usepackage[small,hang,bf]{caption}
\usepackage{floatflt}
\usepackage[perpage,symbol*]{footmisc}
\usepackage{color}
\usepackage[left=3cm,top=2.5cm,right=3cm,bottom=2.5cm,nohead]{geometry}
\usepackage{setspace}
\usepackage{amsmath}
\usepackage{graphics}

\newcommand{\A}[1]{\boldsymbol{#1}}
\newcommand{\pd}[2]{\frac{\partial #1}{\partial #2}}
\newcommand{\HALF}{\frac{1}{2}}
\begin{document}

\begin{titlepage}

\begin{center}

\large
\mbox{\textsc{\textbf{The Hebrew University of Jerusalem}}} \\

\vspace{7mm}

\mbox{\textsc{\textbf{The Faculty of Mathematics and Natural Sciences}}} \\

\vspace{4mm}

\mbox{\textsc{\textbf{The Racah Institute of Physics}}} \\

\vspace{15mm}

\Large
\textbf{Construction of a Multidimensional\\} \vspace{4mm}
\textbf{Parallel Adaptive Mesh Refinement\\} \vspace{4mm}
\textbf{Special Relativistic Hydrodynamics Code\\} \vspace{4mm}
\textbf{for Astrophysical Applications}

\vspace{17mm}

\large
M.Sc. Thesis in Natural Sciences

\vspace{6mm}

by \\

\vspace{5mm}

\large
\textbf{Ygal Klein} \\

\vspace{10mm}

\large
The work was carried out under the supervision of \\
\vspace{5mm}
Prof. Ami Glasner \\
Prof. Eli Waxman \\

\vspace{7mm}

August 2010

\end{center}

\end{titlepage}

%---------------------------------------
\newpage
\thispagestyle{empty}

\begin{center}

\textbf{Acknowledgements}

\vspace{10mm}

\end{center}

I would like to express my gratitude to Prof. Ami Glasner and Prof. Eli Waxman for their guidance and advice throughout the work on this thesis.

\vspace{5mm}

I would like to warmly thank Yoni Elbaz, for his very close help and guidance.

\vspace{5mm}

The following Ph.D.\ students at WIS were of great help: Nahliel Wygoda, Doron Kushnir, Nir Sapir and Eli Krailser.
\newpage

%---------------------------------------
\begin{abstract}
We have developed a new computer code, RELDAFNA, to solve the
conservative equations of special relativistic hydrodynamics (SRHD)
using adaptive mesh refinement (AMR) on parallel computers. We have
implemented a characteristic-wise, finite volume Godunov scheme
using the full characteristic decomposition of the SRHD equations,
to achieve second and third order accuracy in space (both PLM and
PPM reconstruction). For time integration, we use the method of
directional splitting with symmetrization, which is second order
accurate in time. We have also implemented second and third order
Runge-Kutta time integration scheme for comparison. The
implementation of AMR is based on the Fully Threaded Tree algorithm.
Parallelization is based on mesh decomposition using a Hilbert or
Morton space-filling curves, and utilizes work-load balance between
the processors. RELDAFNA is modular and includes the capability to
easily swap hydrodynamics solvers, reconstruction methods and
physics modules. In addition to the hydrodynamics solvers we have
implemented approximate Riemann solvers along with an exact Riemann
solver. We examine the ability of RELDAFNA to accurately simulate
special relativistic flows efficiently in number of processors,
computer memory and over all integration time. We show that a wide
variety of test problems can be solved as accurately as solved by
higher order programs, such as RAM, GENESIS, or PLUTO, but with a less number of
variables kept in memory and computer calculations than most schemes, an advantage
which is crucial for 3D high resolution simulations to be of
practical use for scientific research in computational astrophysics.
RELDAFNA has been tested in one, two and three dimensions and in
Cartesian, cylindrical and spherical geometries. We present the
ability of RELDAFNA to assist with the understanding of open
questions in high energy astrophysics which involve relativistic
flows.

\end{abstract}

%---------------------------------------

\tableofcontents

%---------------------------------------

\chapter{Introduction} \label{intro}

Nowadays, it is quite positive that many astrophysical phenomena involve flows with velocities very close to the speed of light. One unresolved phenomena is the origin of $\gamma$-ray bursts (GRBs) and their afterglow. The common feature of the various models for the evolution of GRBs, is the high Lorentz factor of a fraction of the matter in the fireball, which gets as high as a hundred and more \cite{piran05,waxman03}. The various models also base their interpretation on the beaming of the ejecta producing GRBs and their afterglows into jets. Detailed investigation of these interpretations and current afterglow observations, require high-resolution multidimensional simulations of relativistic shocks, jets and flows \cite{zhang06,zhang09}.

The equations governing the flow of matter with velocities close to the speed of light with no gravitational or magnetic field, the Special Relativistic Hydrodynamics Equations (SRHD hereafter), are subjected to a numerical solution in various techniques and approaches. At early years of numerical and computational investigation of these equations (the 70s), the use of artificial viscosity (AV) methods in Eulerian grids to catch shocks or contact discontinuities was common \cite{wilson72}. As high-order essentially non-oscillatory (ENO) shock capturing schemes \cite{harten87} began to appear, and successfully solve hyperbolic systems such as the Eulerian gas dynamics equations, ENO-based methods have been implemented for SRHD \cite{dolezal95,donat98,del-zanna02}, and it seems they have managed to overcome the numerical difficulties which arise when using AV methods to describe ultra-relativistic flows. As astrophysical applications became more and more demanding, multidimensional SRHD
codes were developed \cite{mignone05b}, some utilizing massive parallelism, e.g. GENESIS \cite{aloy99} and PLUTO \cite{mignone07}. Recently, the use of adaptive mesh refinement (AMR hereafter) has been implemented to solve the SRHD equations. Fairly new examples are RAM \cite{zhang06}, AMRVAC \cite{meliani07} and the methods proposed in \cite{wang08}, which use a characteristic-wise, finite difference, weighted essentially non-oscillatory (WENO) methods, and achieves fifth-order accuracy in space, with a third-fourth- and fifth- order total variation diminishing (TVD) Runge-Kutta scheme for the time integration.

The numerical solution of the SRHD equations for the aid of understanding and studying the astrophysical phenomena modeled with relativistic flows is the aim of this thesis work. This would be achieved by a description of a new code, RELDAFNA. RELDAFNA is a multidimensional parallel adaptive mesh refinement code for the numerical solution of the SRHD Equations using a Godunov method on an Eulerian grid.

RELDAFNA is based on the so-called reconstruct-solve-average (RSA) strategy. First, volume averages are reconstructed inside each computational cell. Then, an approximate Riemann solver is called at each interface of the mesh to resolve the discontinuity between the left and right states. Finally, the conserved quantities are advanced in time. Adding new physics modules to the code is quite straightforward since the code is modular. RELDAFNA provides different spatial reconstructions and different time integration methods, so a user can pick the suitable algorithm to solve a specific problem.

RELDAFNA uses AMR to concentrate the computational work near regions of interest in the flow, which are changing their location as time advances. The computational refinement and derefinement of the mesh is implemented using the FTT algorithm \cite{khokhlov98} with minimum user interface needed. RELDAFNA is composed entirely of Fortran90 modules and is completely parallelized, including the AMR. The sweep of the mesh is being done using space-filling curves, and parallelization is being implemented using the MPI library solely, and no other external libraries are called by RELDAFNA. During a calculation, the mesh is being continuously divided between the participating processors, in order to balance the work load between the processors at any given time and mesh size.

The thesis is organized as follows.

Chapter~\ref{Backg} provides the relevant physical background. In particular, section~\ref{applications} contains a brief description of some phenomena in astrophysics which are believed to contain special relativistic flow. Section~\ref{hydrodynamics} contains a review of basic concepts in Newtonian hydrodynamics, and section~\ref{eqns} presents the special relativistic hydrodynamics theory and equations, along with its unique features in comparison with Newtonian hydrodynamics.

Chapter~\ref{framework} describes the numerical and computational framework with which RELDAFNA was built. In section~\ref{num_methods} the various numerical methods relevant for the solution of the SRHD equations are presented. In section~\ref{logic} the basic concepts of RELDAFNA are explained, and its different techniques for the solution of the SRHD equations are listed. In sections~\ref{AMR}-\ref{parallel} the techniques of AMR and parallelization are shortly presented.

Chapter~\ref{schemes} is the major part of this work. It includes an extensively detailed explanation of the calculations being performed in RELDAFNA in order to solve the SRHD equations using different numerical methods. First, in section~\ref{initialization} the preparation of every timestep is described. In section~\ref{reconstruct} the different spatial reconstruction methods are presented (PCM, PLM and PPM). In section~\ref{solve} the different time integration schemes used are presented (Runge-Kutta, Characteristic Tracing and the approximate Riemann solvers).

%In chapter~\ref{code-tests} we present numerical tests for RELDAFNA in both one and two dimensions. In section~\ref{1d_tests} we demonstrate the ability of RELDAFNA to solve analytically solved problems, and in section~\ref{sec:rie2d} we show the solution of RELDAFNA to an extensively studied two dimensional problem in the literature.

In chapter~\ref{computational-astrophysics} we present calculations of designed problems for astrophysical applications in one and two dimensions, to emphasize the ability of RELDAFNA to contribute to research of up to date questions regarding high energy astrophysical phenomena.

Finally, chapter~\ref{summary} summarizes the current treatise.

\chapter{Background} \label{Backg}

\section{Astrophysical Applications} \label{applications}

In the past half of a century, a number of high energy astrophysical
phenomena were discovered observationally. Famous examples are
Active Galactic Nuclei (AGN's), Gamma Ray Bursts (GRBs's) and Cosmic
Rays (CR's). As more and more observational data was collected over
the years, different conclusions were made about the nature of such
energetic and cataclysmic events. These conclusions became to be
constraints on theoretical models trying to describe the various
ingredients involved in each observed phenomena and its apparent
different stages. Today, almost in every high energy astrophysical
event, the standard theoretical model supporting any of its stages
(if present), involves the acceleration of a portion of the matter
participating in the dynamics, to relativistic velocities ranging
from $\left|\A{v}\right| \approx 0.85c$ with a Lorentz factor of
$\Gamma \sim 2$, to $\left|\A{v}\right| \approx 0.99999c$ with a
Lorentz factor of $\Gamma \sim few \times 10^2$. Of course, after
the presentation of a theoretical model stating that SRHD is
concluded in the dynamics leading to, participating at or following
the high energetic event, one must turn to solving these equations
and that is done most efficiently numerically on computers. The
solution of these equations in a designed geometry and background
conditions, leads to the extraction of quantitative results regarding
the model proposed, as well as qualitative insight for the physics
of the SRHD, which some times lead to a deeper understanding of the
astrophysical event in general.

The scope of phenomena involving relativistic flows within their models range beyond the cover of this thesis. In the following section we will review one such phenomena, $\gamma$~-Ray Bursts, and point out the conclusions made from observed data that led to the role played by SRHD in the theoretical models explaining these events. We refer the reader to some other phenomena explained with relativistic flow such as the investigation of radio sources in AGN's \cite{begelman84,blandford74,blandford77,rees66} and relativistic shock breakout \cite{campana06,waxman07}.

\subsection{Gamma Ray Bursts}\label{GRBs}

In 1967, a short, intense flash of $\gamma$ radiation
was first detected. Since then, on an average, a few $(\sim 1-2)$
bursts are detected everyday. This phenomena was given the name
$\gamma$-Ray Burst. $\gamma$-Ray Bursts (GRBs) are short and intense
pulses of soft $\gamma$-rays. The bursts last from a fraction of a
second to several hundred seconds. GRBs arrive from cosmological
distances from random directions in the sky. The overall observed
fluxes range from $10^{-4}$ergs/cm$^2$ to $10^{-7}$ergs/cm$^2$
(detectable minimum). This corresponds to isotropic luminosity of
$10^{51}-10^{52}$ergs/sec, making GRBs the most luminous objects in
the sky. However, we know today that most GRBs are narrowly beamed
and the corresponding energies are "only" around $10^{51}$ergs,
making them comparable to Supernova in the total energy release.
Indeed, a connection between a GRB and a Supernova has been observed \cite{campana06}.
The GRBs are followed by an afterglow - lower energy, long lasting
emission in the X-ray, optical and radio. The radio afterglow was
observed in some cases several years after the burst. Within the
host galaxies there is evidence that (long duration) GRBs arise
within star forming regions, and there is evidence that they follow
the star formation rate \cite{piran05}.

Every model trying to explain this phenomenon is trying to accomplish three tasks. First, the search for a model of the mechanism for the GRB prompt emission. Evidently, this mechanism involves particles with very high energy, so the second task is to find the engine that drives particles to such high energies. The third task is the search for the explanation of the spectrum observed in the afterglow and its time dependence.

The first task is, as for today, accomplished by the well-accepted fireball model \cite{rees92}. In the fireball model, the expansion energy of a relativistic fireball is converted to radiation via interaction with external medium to the fireball (for example, the ISM or a stellar wind), or by internal shocks within the fireball. A summarized description of the fireball model following \cite{waxman03,piran99,piran05} will be given in the following section.

\subsubsection{The Fireball Model}
\label{fireball}

We consider the release of the energy observed in a GRB, $E \sim 10^{51}$~ergs, in a time scale of a typical long GRB $T \sim 10$'s of seconds, from a source whose length scale is of the order of a light travel time of the variability time of a GRB $\sim 1$~ms, i.e. $r_0\sim10^7$~cm. The release of a large amount of energy in such a compact volume, results in an expansion of the matter in the volume, and an explosive spherical shock wave to run into the surrounding medium \cite{zeldovich66,blandford76}. The shock wave can be either Newtonian ($\Gamma_s \lesssim 1.5$), mildly relativistic ($\Gamma_s \lesssim 100$) or ultra-relativistic ($\Gamma_s \gtrsim 100$). Starting at early years of the study of GRBs, the kinematics regime of the spherical shock wave, and the spatial and temporal dimensions considered, raised the compactness problem \cite{Ruderman75}. GRBs show a non thermal spectrum with a significant high energy tail. If the fireball kinematics is Newtonian, we must constrain its
dimensions to be $\lesssim 10^7$~cm. From this constraint one can calculate the optical depth of the source. The number density of photons at the source $n_\gamma$ is approximately given by
\begin{equation}
L_\gamma = 4\pi r_0^2 cn_\gamma\epsilon \, ,
\end{equation}
\noindent where $\epsilon\simeq1$~MeV is a characteristic photon energy in GRB's radiation. Now, two $\gamma$-rays can annihilate and produce e$^+$e$^-$ pairs, if the energy in their CM frame is larger than $2 m_e c^2$. The optical depth for pair creation is:
\begin{equation}
\tau_{\gamma\gamma}\sim r_0 n_\gamma\sigma_T\sim\frac{\sigma_TL_\gamma}
{4\pi r_0 c\epsilon}\sim10^{15},
\end{equation}
\noindent where $\sigma_T$ is the Thomson cross section. This is, of course, inconsistent with the non-thermal spectrum observed for a GRB. As mentioned above, the analysis which led to a high optical depth took an assumption, that the fireball is Newtonian. Removing that assumption, that is assuming for example that the emitting matter is moving relativistically towards the observer, one would have to take into account some relativistic corrections in the above analysis. One correction refers to the energy of the photons. If the emitting matter is moving relativistically with a lorentz factor $\Gamma$, the observed photons energy at the source frame is lower by a factor $\Gamma$. Lowering the energy of photons in the source compared to that in the observer frame, leads to a conclusion that most photons would be below the pair production threshold, thus lowering the optical depth. Another correction refers to the dimensions of the source. The dimension of a relativistically moving source is multiplied by a factor of $\Gamma^2$. This effect directly modifies the density estimate by a factor $\Gamma^{-4}$, and thus it is also lowering the optical depth. These two corrections lead one to estimate $\Gamma \approx few \times 100$ to obtain an optically thin source. So, we conclude that the fireball is expanding with relativistic velocities, and as its volume increases, the temperature decreases and thus the rate of annihilation of pairs increases and the photons are ready to escape from the fireball. In order to prevent the baryonic matter in the fireball to increase the optical depth via Thomson scattering, and reconvert the radiation energy back to the relativistic acceleration and by that block again the photons from escaping, it was proposed that the burst we observe is a result of a dissipation process of the ultra-relativistic fireball, which occurs at a radius $\gtrsim$ the Thomson radius. It has been shown that in order to explain the variability time scales of GRB's, a suitable dissipation process is internal shocks with in the fireball. The internal shocks are produced when a faster shell catches up with a slower one, in the following way. The relativistic expansion has two phases: radiation dominated phase and a matter dominated phase. At the beginning of the expansion, the shock wave is radiation dominated and the entropy is provided by the photons which obey the EOS $p = e/3\propto T_\gamma^4$ where $p$, $e$ and $T_\gamma$ are the pressure, energy density and radiation temperature, respectively. Conservation of entropy implies
\begin{equation}
r^2 \Gamma(r) r_0 T_\gamma^3(r) = const ,
\label{eq:entropy}
\end{equation}
\noindent and conservation of energy implies
\begin{equation}
r^2 \Gamma(r) r_0 \Gamma(r) T_\gamma^4(r) = const\,.
\label{eq:energy}
\end{equation}
\noindent where $\Gamma(r)$ is the shell Lorentz factor. From these two equations we have in the radiation dominated phase $\Gamma(r)\propto r$. When the fireball reaches a radius where the kinetic energy of the baryons is comparable to the fireball energy, $\Gamma M c^2 \sim E$, the acceleration becomes matter dominated, and the matter flies with a roughly constant Lorentz factor, $\Gamma \sim E / Mc^2$, a ratio that determines the evolution of the fireball as will be explained ahead.

The transition between the phases depends on the ratio between the explosion energy and the rest mass energy of the baryonic matter. If the baryonic matter mass is too low, $M<10^{-12}{\mbox{~M$_\odot$}}E_{52}^{1/2}R_{i7}^{1/2}$, the acceleration is only radiation dominated, the temperature (and thus pair opacity) decreases with expansion, and most energy escapes as radiation. If we increase the baryonic mass a little, $10^{-12}{\mbox{~M$_\odot$}}E_{52}^{1/2}R_{i7}^{1/2}<M<2\times10^{-7}{\mbox{~M$_\odot$}}E_{52}^{2/3}R_{i7}^{2/3}$, ($R_{i7}$ is the initial size of the fireball in units of $10^7$~cm and $E_{52}$ is the energy in units of $10^{52}$~erg) the baryonic matter is not negligible and its effect is by the free electrons associated with it, which determine the opacity, slowing its reduction with temperature reduction, but still the fireball remains radiation dominated and no matter dominated phase exists before most of the energy escapes as radiation. Increasing the baryonic mass to a significant fraction of the fireball energy, $2\times10^{-7}{\mbox{~M$_\odot$}}E_{52}^{2/3}R_{i7}^{2/3}<M<5\times10^{-3}{\mbox{~M$_\odot$}}E_{52}$, we get a matter dominated phase before the energy can escape as radiation, and we have a conversion of the explosion energy to kinetic energy of the relativistically flying baryonic matter. This scenario is a reasonable fireball situation for producing the observed GRB radiation, since it will not result in a thermal spectrum. Increasing the baryonic mass a little too high, $5\times10^{-3}{\mbox{~M$_\odot$}}E_{52}<M$, we make the radiation energy lower than the rest mass energy and the acceleration is only matter dominated and remains Newtonian from the start.

Since the third scenario presented above is the one resulting in a non thermal spectrum, we can extract a range for the constant Lorentz factor of the flying matter in the matter dominated phase. As explained above, this Lorentz factor is $\sim E / Mc^2$, i.e. the ratio of the rest mass energy to the explosion energy, since it is the Lorentz factor when the two energies becomes comparable. Demanding the baryonic mass to be in the third bin, resulting in the third scenario, leads one to demand the Lorentz factor to be in the range $10^2\lesssim\Gamma\lesssim10^3$.

In the radiation dominated phase, the fluid is relativistic and its speed of sound is $\sim c$. We would consider an expansion of a relativistic shell of radius $r$ to be significant after a time scale $\sim \frac{r}{c}$ (since the expansion velocity is very close to $c$) in our frame (the observer frame), which is $\frac{r}{\Gamma c}$ in the shell frame. This significance expansion time allows sound waves to travel $c\frac{r}{\Gamma c}$ in the shell frame which is $\frac{r}{\Gamma^2}=\frac{r_0}{r}r_0$ (remember that $\Gamma(r)\propto r$ and $\Gamma(r_0)=1$) in the observer frame, and we can see that in early stages of the expansion, regions separated by more than $r_0$ could not influence each other. So, if we take the values considered, a fireball of thickness $cT$ (which for the typical values considered is $\gg r_0$) would develop, and its expansion would result in a collection of non interacting subshells of thickness $\sim r_0$. Since no interaction between the subshells occur, each would reach its own final Lorentz factor. Define the velocities difference between the subshells by $\Delta v$, and we get that two shells emitted with a delay of $\frac{r_0}{c}<\Delta t < T$ would collide after a time $t_c\sim c\Delta t/\Delta v$ where its radius is $r\approx2\Gamma^2c\Delta t \sim 10^{13}$~cm. Table~\ref{t:radii} summarizes the different length scales and is taken from \cite{piran99}.

%\begin{center}
\begin{table*}[h]
\begin{center}
\begin {tabular}{|c|c|c|c|}

  \hline
  $R_i$ & Initial Radius & $c \delta t$  & $ \approx 10^7-10^8$cm  \\
  $R_\eta$&Matter dominates &$R_i  E / M c^2$ & $ \approx 10^9$cm \\
  $R_{pair}$& Optically thin to pairs & $
   [ (3 E /4 \pi R_i^3 a)^{1/4}/T_p ] R_i $ &
  $ \approx 10^{10}$cm \\
  $R_e$& Optically thin & $ ({\sigma_T E/  4 \pi m_p c^2 E / M c^2} )^{1/2}$
  &$\approx 10^{13}$cm \\
  $R_\delta$&Internal collisions &$\delta  \Gamma^2$ &
  $\approx 10^{12}-10^{14}$cm\\
  $R_\gamma$ & External Newtonian Shocks& $l \Gamma^{-2/3}$&
  $\approx 10^{16}$cm \\
  $R_\Delta$ & External Relativistic shocks&$ l^{3/4} \Delta^{1/4}$ &
  $\approx 10^{16}$cm \\
  $l$ or $L$ &Non relativistic external shock  &$l$ $^{(a)}$ or  $l\Gamma^{-1/3}$
  $^{(b)}$ & $\approx 10^{17}-10^{18}$cm \\
  l & Sedov Length  &$l=(3E/4 \pi n_{ism} m_p c^2)^{1/3}$ & $\approx 10^{18}$cm\\
  \hline

\end{tabular}

\end{center}
\caption{Critical Radii (a) - adiabatic fireball; (b) - radiative fireball  \protect\label{t:radii}}
\end{table*}
%\end{center}

This variability between the subshells, resulting in collisions, would produce internal shocks that will convert the kinetic energy back to internal energy (this is the dissipation of the relativistic motion). This reconverted internal energy is radiated as $\gamma$-rays by synchrotron and inverse Compton emission of the electrons. This $\gamma$-ray radiation is the prompt emission of the GRB that we observe. We can conclude that $\gamma$-ray observations indicate that, regardless of the nature of the underlying sources, GRBs are produced by the dissipation of the kinetic energy of a relativistically expanding fireball \cite{waxman03}. The fireball evolution is a multi-length scale problem, with stages characterized by lengths spreading over orders of magnitude.
Along with multi length scales, during the expansion the hydrodynamics turn from ultra relativistic to mildly relativistic, and then to Newtonian kinematics regime. An example of the evolution of the Lorentz factor with time and the different regimes involved is given in Fig.~\ref{figs_from_KPS99} from a numerical simulation presented in \cite{kobayashi99}:

\begin{figure*}[h]
\centering
\includegraphics[scale=0.4]{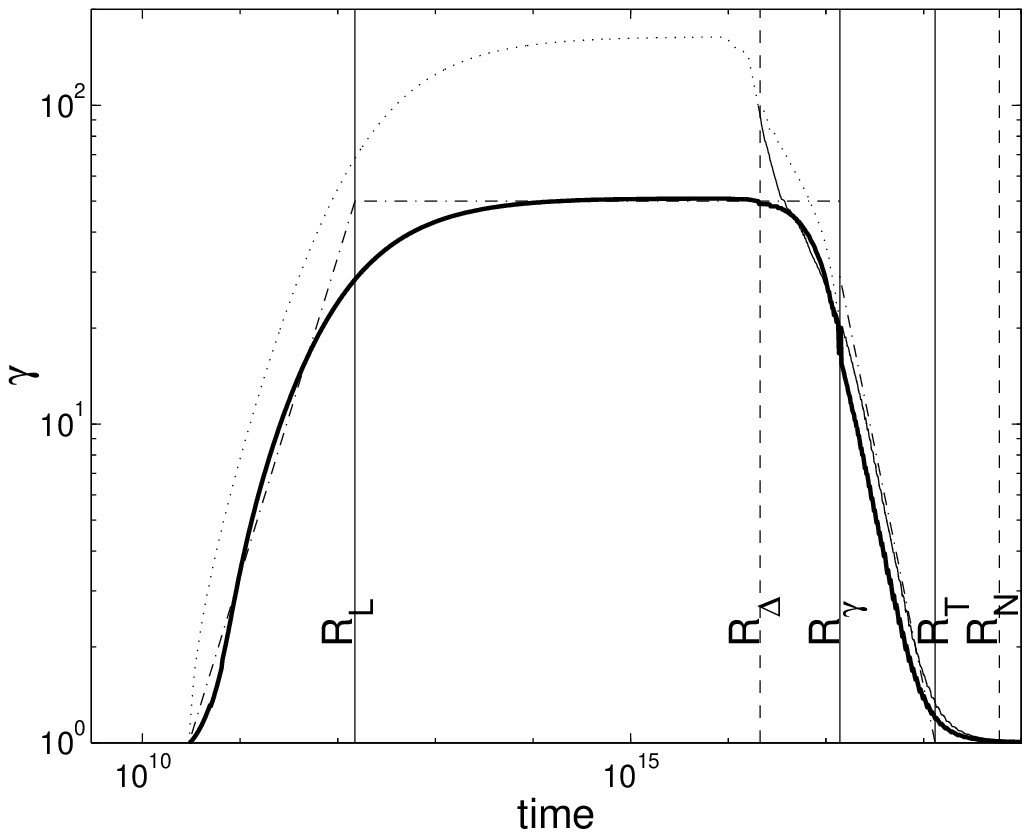}
\includegraphics[scale=0.4]{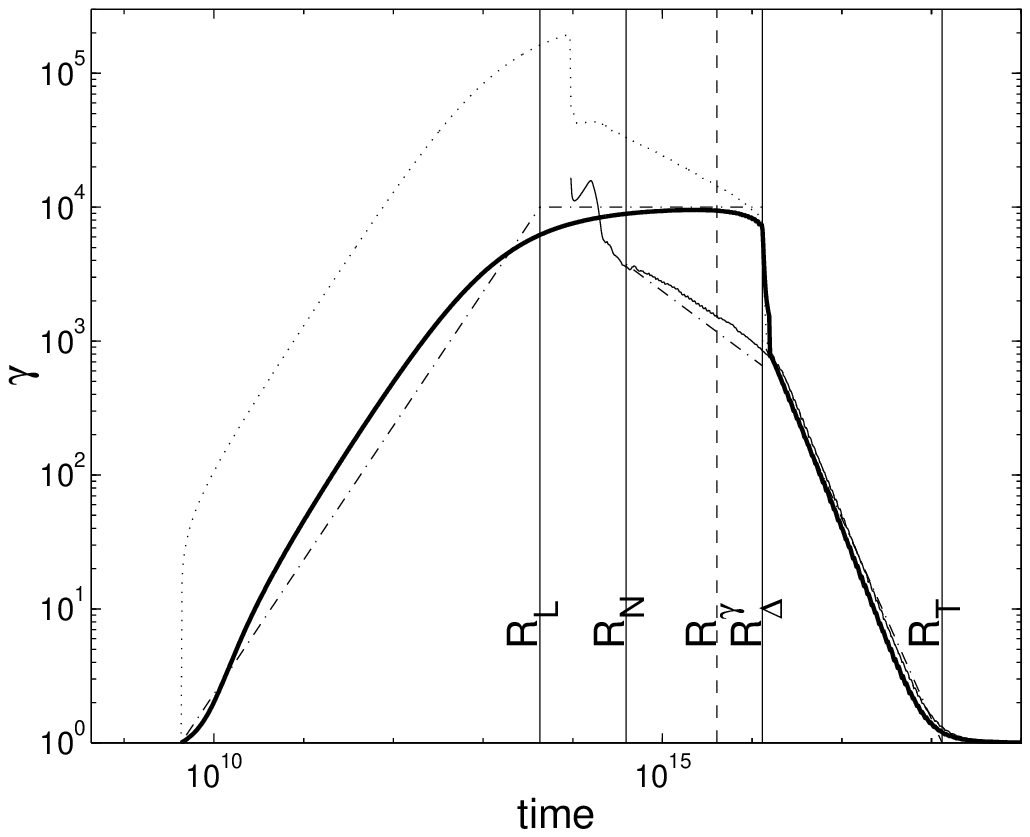}
\caption{ Fireball evolution  from its initial
formation at rest to the final Newtonian Sedov solution. The energy
extraction is due to the interaction with the ISM  via a
relativistic forward shock and a Newtonian reverse shock (right) or a relativistic reverse shock (left). Shown are the average value of the Lorentz
factor (thick solid line), the value at the forward shock (thin solid
line), the maximal value (dotted line) and an analytic estimate
(dashed dotted line).}
\label{figs_from_KPS99}
\end{figure*}

\subsubsection{Quest for The Progenitor} \label{progenitor}

The second task mentioned above, is not accomplished, as for today. Different models were suggested for the engine driving the energies to the particles radiating the burst, from black hole accretion, Neutron Star formation, Neutron Star death, binary Neutron Star-Neutron Star merger to rotation of black holes \cite{piran05}. One example for a model which demands extensive numerical simulations is the Collapsar model \cite{woosley93}. The model suggests that an iron core of a rotating star of mass $\sim 30 {\mbox{~M$_\odot$}}$ collapses to a black hole. The black hole would be surrounded by an accretion disk feeding it. Along the rotational axis, where there is less rotational support, matter would start to fly out, forming jets. In some cases the jets penetrate the stellar envelope and produce the GRB. Extensive numerical simulations have been performed to test this model, for example \cite{zhang03,aloy00}. These simulations try to describe the propagation of collimated jets through collapsing stars and their breakout from the stellar surface or wind. It has been found that even not very (initially) collimated jets, with half opening angle of $20^\circ$, get highly collimated with their advance in the star, to a half opening angle of $5^\circ$. The morphology of relativistic jets has also been extensively studied numerically \cite{marti97}, and it turns out that during its evolution, the jet becomes very hot. The high internal energy of the jet is converted to kinetic energy at the surface, with matter flowing at $\Gamma\sim 150$ along the axis. Instabilities in the surface of the beam lead to mixing with the stellar material which is being penetrated. This mixing is the cause of variations in the mass loading along the jet, which lead to variations in $\Gamma$, variations responsible for the internal shocks in the fireball dissipating later the kinetic energy back to internal energy radiating the $\gamma$-rays (see section~\ref{fireball}).

The mechanism of relativistic acceleration of matter in blast waves in exploding stars has also been investigated \cite{tan01}. Since one demands ultra relativistic ejecta at the stellar surface, one wonders what should be the nature of the blast wave at the center and along its propagation in the star. It has been found that even trans-relativistic blast waves can channel a fraction of the explosion energy to relativistic ejecta at the star surface since stars have a steeply declining density profile. In a sufficiently steep declining density profile of a star, the blast wave first decelerates since it sweeps more and more mass, but as it propagates more and more it starts to accelerate down the declining density profile. These acceleration and deceleration mechanisms compete, and one can impose conditions on the star density profile or the explosion energy, so that there will be relativistic acceleration of a fraction of the mass before the shock reaches the stellar surface. There is additional acceleration after the stellar surface has been shocked, since matter with very high pressure continues to accelerate the gas at the surface which is ejecting from the star because it is slower than the shock front. This study, which also involves transitions from Newtonian hydrodynamics to mildly relativistic flow has been modeled semi-analytically, and some estimates for the connection between the shock velocity and the final expansion velocity were made \cite{tan01}.

\subsubsection{Reproducing the Afterglow}
\label{afterglow}

As mentioned, the spectrum and evolution of the afterglow of GRBs
play a dominant role in the demands from any model trying to
explain GRBs. It is accepted that the afterglow is a result of the external shocks driven by the fireball into the surrounding medium. It starts when most of the fireball energy has been transferred to the shocked medium, at a radius $\sim 10^{16}$~cm. The external shock in the ISM, keeps heating new gas and producing relativistic electrons that produce the afterglow radiation we observe \cite{waxman97}. As described in \cite{granot02}, the synchrotron emission of the electrons is strongly dependent on the hydrodynamics of the electrons through their Lorentz factor and the electrons energy distribution. Analytical models presume that the hydrodynamics of the shockwave is well described by the Blandfrod-Mckee self-similar solution \cite{blandford76}. The energy distribution of the electrons is assumed to have a power law shape of their Lorentz factor $N(\Gamma)=\Gamma^{-p}$ immediately behind the shock. The temporal decay of the afterglow is fitted with combined power laws of time for each frequency, i.e. a search is made for the spectral indexes $\alpha$ and $\beta$ in Eq.~\ref{power-law-afterglow}
\begin{equation}\label{power-law-afterglow}
    F_\nu \propto t^{-\alpha} \nu^{-\beta} \,.
\end{equation}
%In Fig.~\ref{power_law_dist_afterglow} from \cite{granot02}, all the powerlaw segments that appear in the afterglow theory are shown.
%\begin{figure*}
%\centering
%\includegraphics[scale=0.7]{figures-for-article/afterglow-GS02.eps}
%\caption[]{ The different possible broad band synchrotron spectra from a
%  relativistic blast wave, that accelerates the electrons to a power
%  law distribution of energies. The thin solid line shows the
%  asymptotic power law segments (PLSs), and their points of
%  intersection, where the break frequencies, $\nu_b$, and the
%  corresponding flux densities, $F_{\nu_b,ext}$, are defined.
%  The different Power law segments (PLSs) are labeled A through H, while the different
%  break frequencies are labeled 1 through 11. The temporal scalings of
%  the PLSs and the break frequencies, for an ISM ($k=0$, i.e. constant density profile) or stellar
%  wind ($k=2$, quadratically declining density profile) environment, are indicated by the arrows.  The thick
%  solid line shows the spectrum we calculated in \cite{granot02}. The different spectra are labeled 1 through
%  5, from top to bottom. The relevant spectrum is determined by the
%  ordering of the break frequencies.}
%\label{power_law_dist_afterglow}
%\end{figure*}
The afterglow light curve can distinguish between spherical and jetted outflow \cite{sari99}. A jetted flow is being defined by an opening angle $\theta$. At the beginning, before the jet decelerates, we have $\Gamma\gg\theta^{-1}$. The motion in this phase is practically spherical, i.e. the jet behaves as a patch of an expanding sphere and its evolution is not affected by its own non spherical geometry. However, when $\Gamma \sim \theta ^{-1}$, sideways expansion becomes significant and hence the hydrodynamics change. Since the radiation emitted is strongly dependent on the flow conditions, one would expect the light curve to change as a result of that and the fact that with $\Gamma \sim \theta ^{-1}$ relativistic beaming becomes less effective, as in general special relativistic effects becomes less effective as the jet decelerates and approaches its spherical Newtonian phase \cite{piran01}. One could see one or two breaks in the light curve depending on the speed of the sideways expansion. Again, we have a transition from relativistic motion to Newtonian flow, a transition which has no analytical solution. Most of the energy in the dynamics of the ISM as the shock heats it is concentrated in a shell of width $\Delta R_s \sim 10^{15}$~cm. The radius at which the Newtonian phase begins is given by the Sedov-Taylor length scale \cite{zeldovich66},
\begin{equation}
  l_{ST}=  \left(\frac{E}{(4\pi/3)\rho_0c^2}\right)^{1/3} \simeq
  7 \times 10^{17} \textrm{cm} \sim 700\times\Delta R_s\,,
  \label{l-ST}
\end{equation}
for a typical explosion of $E\simeq 2 \times 10^{51}$~erg and $\rho_0=1.67 \times 10^{-24}$~g cm$^{-3}$. Detailed multi-dimensional numerical simulations of jets heating the ISM with different profile densities, must be performed in order to learn the hydrodynamic evolution of the sideways expansion, and to extract predicted light curves of the afterglow \cite{zhang09}.

\section{Newtonian Hydrodynamics} \label{hydrodynamics}

The motion of matter through space in the
continuum limit, flow, is a physics discipline being thoroughly
studied and analyzed, both theoretically and experimentally, for
hundreds of years, and in the last half of a century also
computationally and numerically. Everyday life phenomena, such as
the sound we all hear and sea waves, are explained in the study of
this theory, hydrodynamics. The physical and structural design of
machines and weapons, is also strongly connected with the need to
describe motion of information through matter and space, with
physical phenomena involved in their design such as shock waves,
detonation and deflagration waves and solids stress and strength. In
addition, the flow of matter in outer space has a prominent and
dominant role in a wide spectrum of physical processes that occur in
stars, star clusters and galaxies, and they are a crucial ingredient
in analyzing high energy astrophysical processes.

The basic theoretical analysis of hydrodynamics in the Newtonian
regime, $\left|\A{v}\right| << c$, is based on the evolution in time
and spatial dependence of five physical variables: the velocity of the
flowing matter (three components), the pressure and the mass density of
the matter. The dependence and connections between these five
variables, is being described by three conservation laws along with
a function describing the thermodynamic state of the matter flowing,
an equation of state, $f(p,\rho,e)=0$ where $p,\, \rho,$ and $e$
are the pressure, mass density and internal energy density (per mass), respectively.

The three conservation laws are based on the three following
principles. First, in the absence of a source for an addition of
matter to a physical system, or a sink for the loss of matter from a
physical system, there is no reason that with its flow through space
the overall amount of mass should change, i.e. conservation of mass
throughout the fluid dynamics within a closed physical system.
Second, according to the second Newton's law, in the absence of
external forces acting on a physical system the overall linear
momentum must be conserved. This rule is being physically
interpreted via the fact that the change in the velocity of a mass
element must only be a result of non-zero pressure spatial gradient
in the matter flowing. Finally, the third conservation rule is the
energy conservation principle, being represented by the fact that in
the absence of external work being done on the flowing matter, any
change in internal energy of a mass element must only be the result
of a compression work on the mass element by the matter surrounding
it inside the system (in the absence of heat flux), and this
compression work is being formed in changing the mass element
volume. Neglecting viscosity and heat transfer leads to a conclusion
that the flow is adiabatic. These three conservation laws are being
mathematically posed by the following three differential equations:
\begin{equation}
\begin{array}{rcl}
(mass \; \; conservation) \;\; \pd{\rho}{t} + \A{\nabla} \cdot (\rho\A{v}) & = & 0 , \\
 \noalign{\medskip}
(momentum \; \; conservation) \;\; \pd{(\rho v_i)}{t} + \A{\nabla} \cdot (\rho v_i \A{v}) & = & -\pd{p}{x_i} , \, i \in \{1...dim.\} , \\
 \noalign{\medskip}
(energy \; \; conservation) \;\; \pd{E}{t} + \A{\nabla} \cdot (\A{v}(E+p)) & = & 0 , \\
 \noalign{\medskip}
\end{array}
\label{newtonian_conservation}
\end{equation}
where $E=\rho e + \frac{1}{2} \rho \A{v}^2$ is the total energy per
unit volume.

\section{Equations of Special Relativistic Hydrodynamics} \label{eqns}

As was shown in section~\ref{applications}, a large variety of processes in astrophysics are
modeled and explained with the use of flow of matter with velocities
close to the speed of light, relativistic velocities. When such relativistic velocities are present in the flow,
and in the absence of strong gravitational fields one must
reconsider the equations (\ref{newtonian_conservation}), since with
such velocities the Newtonian kinematics equations must be replaced
with the special relativistic laws of motion. The conservation laws
to be derived should consider relativistic variables of mass,
momentum and energy. From this point on we shall work with units where the speed of light
$c$ equals unity. The speed of sound will be denoted $c_s$.

A derivation of the conservation laws for special relativistic
hydrodynamics is given in textbooks such as \cite{landau87,weinberg72}, and in the following we briefly summarize it. Let us define a four-vector to describe points in space-time and a
four vector to describe velocity in space time,
\begin{equation}
 x^\mu= \left( \begin{array}{c}
 t \\
 \noalign{\medskip}
 x \\
 \noalign{\medskip}
 y \\
 \noalign{\medskip}
 z \\
 \noalign{\medskip}
\end{array}\right)\,, \;\;
 u^\mu= \left( \begin{array}{c}
 \Gamma \\
 \noalign{\medskip}
 \Gamma v_x \\
 \noalign{\medskip}
 \Gamma v_y \\
 \noalign{\medskip}
 \Gamma v_z \\
 \noalign{\medskip}
\end{array}\right)\,,
\end{equation}
where $\A{v}$ is the three velocity and $\Gamma=\frac{1}{\sqrt{1-\A{v}^2}}$ is the Lorentz factor. Define the energy-momentum tensor:
\begin{equation}
T^{\mu\nu}=
 \left( \begin{array}{cc}
 energy \;\; density & energy \;\; flux\\
 \noalign{\medskip}
 energy \;\; flux & momentm \;\; flux \;\; tensor
\end{array}\right)\,.
\end{equation}
A perfect fluid is defined by the existence of a velocity function
defined at each point in space, such that if an observer followed
that velocity function the fluid would seem isotropic to him.
This definition means that for a perfect fluid one can write
the energy-momentum tensor in the rest frame by
\begin{equation}
T^{\mu\nu}=
 \left( \begin{array}{cccc}
 e & 0 & 0 & 0\\
 \noalign{\medskip}
 0 & p & 0 & 0\\
 \noalign{\medskip}
 0 & 0 & p & 0\\
 \noalign{\medskip}
 0 & 0 & 0 & p\\
 \noalign{\medskip}
\end{array}\right) \;\;
 u^\mu= \left( \begin{array}{c}
 1 \\
 \noalign{\medskip}
 0 \\
 \noalign{\medskip}
 0 \\
 \noalign{\medskip}
 0 \\
 \noalign{\medskip}
\end{array}\right)\,,
\end{equation}
(i.e. isotropic means a diagonal tensor). Denoting the internal energy density (per mass) by $\epsilon$ we get
that the relativistic internal energy $e$ is a sum of the mass
density $\rho$ which contributes the rest mass energy $mc^2$ and
$\rho \epsilon$, so we can write the energy-momentum tensor in the rest frame
\begin{equation}
T^{\mu\nu} = \rho \left( 1+\epsilon+\frac{p}{\rho} \right) u^\mu
u^\nu + pg^{\mu\nu} = \rho h u^\mu u^\nu + pg^{\mu\nu}\,,
\end{equation}
where $\rho$ is the mass density in the rest frame, $h\equiv 1 + \epsilon + p/\rho$ is the relativistic specific enthalpy,
$p$ is the pressure and $g^{\mu\nu}$ is the inverse spacetime metric.

The equations dominating the flow of an inviscid fluid at the
relativistic regime of velocity, without a gravitational or magnetic
field, the special relativistic hydrodynamics (SRHD) equations, are
similar in concept to the ones dominating a Newtonian flow. The
(rest) mass, momentum and energy (energy-momentum tensor) must not
be created without a source, and must not be reduced without a sink.
The energy-momentum conservation is a principle which must be right
in the relativistic regime just as it is true in the Newtonian
regime. The conservation of (rest) mass is an assumption of number
of particles conservation, that we add in order to define the set of
SRHD equations to be solved by RELDAFNA and discussed in this
thesis. For this assumption we define a four-vector to describe the
particle number density,
\begin{equation}
 n^\mu=nu^\mu= \left( \begin{array}{c}
 n\Gamma \\
 \noalign{\medskip}
 n\Gamma v_x \\
 \noalign{\medskip}
 n\Gamma v_y \\
 \noalign{\medskip}
 n\Gamma v_z \\
 \noalign{\medskip}
\end{array}\right)\,,
\end{equation}
where $n$ is a scalar denoting the proper number density of the
particles. This conservation of rest mass and energy-momentum can be summarized
by the following two relations:
\begin{equation}
(n^\mu)_{;\mu}=0 \Rightarrow m(n u^\mu)_{;\mu}=0 \Rightarrow (\rho u^\mu)_{;\mu}=0
\label{continuity}
\end{equation}
and
\begin{equation}
(T^{\mu\nu})_{;\nu}=0\,. \label{stress-energy}
\end{equation}
\\
Denote the rest mass, momentum and energy densities by $D$, $S^i \; (i \in \{1...dim.\})$ and $\tau$, respectively.
Equations (\ref{continuity})-(\ref{stress-energy}) are expanded to the following set of partial differential equations:
\begin{equation}
\begin{array}{rcl}
(rest \; \; mass  \; \; conservation) \;\; \pd{D}{t} + \A{\nabla} \cdot (D\A{v}) & = & 0 , \\
 \noalign{\medskip}
(momentum \; \; conservation) \;\; \pd{S^i}{t}+ \A{\nabla} \cdot (S^i\A{v}) & = & -\pd{p}{x_i} , i \in \{1...dim.\} , \\
 \noalign{\medskip}
(energy \; \; conservation) \;\; \pd{\tau}{t} + \A{\nabla} \cdot (\A{S}-D\A{v}) & = & 0 .\\
 \noalign{\medskip}
\end{array}
\label{SRHD_conservation}
\end{equation}
The conserved quantities are all measured in the laboratory frame, and their connection to the fluid frame hydrodynamic variables (primitive variables, hereafter) is given by
\begin{equation}
\begin{array}{rcl}
D & = & \rho\Gamma , \\
\noalign{\medskip}
S^i & = & \rho h \Gamma^2 v^i\,, \; \; i\in\{1...dim.\} , \\
\noalign{\medskip}
\tau & = & \rho h \Gamma^2 -p-\rho\Gamma . \\
\noalign{\medskip}
\end{array}
\label{conserved}
\end{equation}
One can verify that in the Newtonian regime ($\left|\A{v}\right|<<1, h\simeq1$), the conserved variables in Equations~\ref{conserved} reach the Newtonian conserved variables appearing in Equations~\ref{newtonian_conservation}, $\rho$, $\rho \A{v}$ and $\rho E$, and that Equations~\ref{SRHD_conservation} reach Equations~\ref{newtonian_conservation}.

For convenience in future reference we define the vector of conserved quantities
\begin{equation}
 \A{U} = \left(D, S^1,... S^{dim.}, \tau\right)^{T}\, ,
 \label{U}
\end{equation}
and the matrix of matched fluxes
\begin{equation}
 \A{F}^j = \left(Dv^j, S^1v^j+p\delta^j_{\ 1},...,S^{dim.}v^{j}+p\delta^{j}_{\ dim.}, S^{j}-Dv^j\right)^{T} \, .
  \label{F}
\end{equation}

\subsection{Unique Features of SRHD} \label{features}

The SRHD equations possess some features which are different from the Newtonian hydrodynamic theory, and these differences usually make their numerical solution much more complicated.

\begin{itemize}

\item A major difference is the lack of maximum compression ratio through shock waves in SRHD. As was already mentioned, the conserved variables in Newtonian hydrodynamics are $\rho$, $\rho \A{v}$ and $\rho E$. These variables must also be conserved over a shock front, which means that their fluxes given in Equations~\ref{newtonian_conservation} are equal from both sides of the shock front. This constraint leads to the well known \textit{Rankine-Hugoniot} jump conditions, which induce algebraic connections between the hydrodynamic variables in the shocked and unshocked states. Using an ideal gas equation of state $p(\rho,e)=(\gamma-1) \rho e$, one can explicitly write the ratios of any thermodynamic variable between the two sides of the shock front, for example the compression ratio over the shock is given by \cite{zeldovich66} to be
    \begin{equation}
    \frac{\rho_1}{\rho_0}=\frac{(\gamma+1)p_1+(\gamma-1)p_0}{(\gamma-1)p_1+(\gamma+1)p_0}\,,
    \end{equation}
    where a subscript $1$ refers to the shocked medium and a subscript $0$ refers to the unshocked medium.
    One can see that the limiting density ratio across a strong shock wave is
    \begin{equation}
    \frac{\rho_1}{\rho_0} \xrightarrow[p_1\gg p_0]{} \frac{\gamma+1}{\gamma-1}\,,
    \end{equation}
    a limit depending only on the thermodynamic properties of the shocked medium. In SRHD however, the conserved variables are given by $\rho \Gamma$, $\rho h \Gamma^2 \A{v}$ and $\rho h \Gamma^2 -p -\rho \Gamma$. The conservation rules still demand that over a shock front the fluxes given in Equations~\ref{SRHD_conservation} would remain unchanged, and from that one can deduce jump conditions for shock waves in SRHD. The limiting density ratio across a strong shock is extracted from these relations and takes the form \cite{blandford76}
    \begin{equation}
    \frac{\rho_1}{\rho_0} \xrightarrow[\frac{p_1}{n_1}\gg \frac{p_0}{n_0}]{} \frac{\gamma+1}{\gamma-1} + \frac{\gamma (\Gamma_1-1)}{\gamma-1}\,,
    \end{equation}
    a limit depending not only on $\gamma$ but also on a hydrodynamic variable $\Gamma_1$, which is the Lorentz factor of the shocked medium as measured in the unshocked medium frame. From this relation we see that the compression ratio in SRHD does not have a maximum, and it can be arbitrarily big as we increase $\Gamma_1$. This means that capturing a shock in SRHD numerically is much more complicated than in Newtonian hydrodynamics numerical methods. This feature would be tested in appendix~\ref{1d_tests}.

\item Another crucial difference between the two regimes comes directly from the limiting value of propagation speed in SRHD, the lack of minimum thickness of dense shells. In both theories, discontinuities between two separate thermodynamic states (the so-called Riemann problem) can evolve to a combination of rarefaction wave, contact discontinuity and a shock wave (not necessarily all of them together). When a contact discontinuity and a strong shock evolve, we have in the Newtonian regime \cite{zeldovich66}
    \begin{equation}
    \rho_0 u_0=\rho_1 u_1 \Rightarrow \frac{u_0}{u_1}\xrightarrow[p_1\gg p_0]{} \frac{\gamma+1}{\gamma-1}\,,
    \end{equation}
    where $u_0$ and $u_1$ are the velocities of the unshocked and shocked medium in the shock rest frame, respectively. In the unshocked medium rest frame we assign the shock velocity the term $u_s$ and the contact surface velocity the term $u_c$. From these definitions we have $u_s=-u_0$ and $u_c=u_1-u_0$, resulting with the ratio between the propagation velocity of the contact discontinuity and the shock wave velocity
    \begin{equation}
    \frac{u_s}{u_c}=\frac{u_0}{u_0-u_1}=\frac{1}{1-\frac{u_1}{u_0}}\xrightarrow[p_1\gg p_0]{} \frac{1}{1-\frac{\gamma-1}{\gamma+1}}=\frac{\gamma+1}{2} \,,
    \end{equation}
    and again we have a limiting value that depends only on the properties of the matter being shocked, and more importantly this ratio, for a given $\gamma$, is always \textit{"safely"} $>1$. The velocities ratio determines the thickness of the zone between the contact discontinuity and the shock front, a zone which is very dense since it has been shocked. If we have a limiting value to the velocities ratio, the contact discontinuity and the shock front would not be too close and the zone between them, which is a very important structure in the problem, and might be the one with minimum length scale, would be large enough so that the numerical resolution needed for a calculation would be attainable computationally. In SRHD however, the velocities ratio is much lower and with appropriate conditions it can reach very close to unity. The relation between the shock front velocity and the contact discontinuity velocity is \cite{blandford76}
    \begin{equation}
    u_s \xrightarrow[\frac{p_1}{n_1}\gg \frac{p_0}{n_0}]{} \sqrt{1-\frac{\gamma(2-\gamma)(\Gamma_1-1)+2}{(\Gamma_1+1)(\gamma(\Gamma_1-1)+1)^2}}\,,
    \end{equation}
    and in Fig.~\ref{fig:vel_ratio} we can see the velocities ratio $\frac{u_s}{u_c}$ as a function of $u_c$ (for $\gamma = \frac{4}{3}$), and we see that $\frac{u_s}{u_c}\xrightarrow[u_c \ll 1]{} \frac{\frac{4}{3}+1}{2} \approx 1.166$.

    \begin{figure}[h]
    \centering
    \includegraphics[scale=0.5, trim= 80mm 7mm 80mm 0]{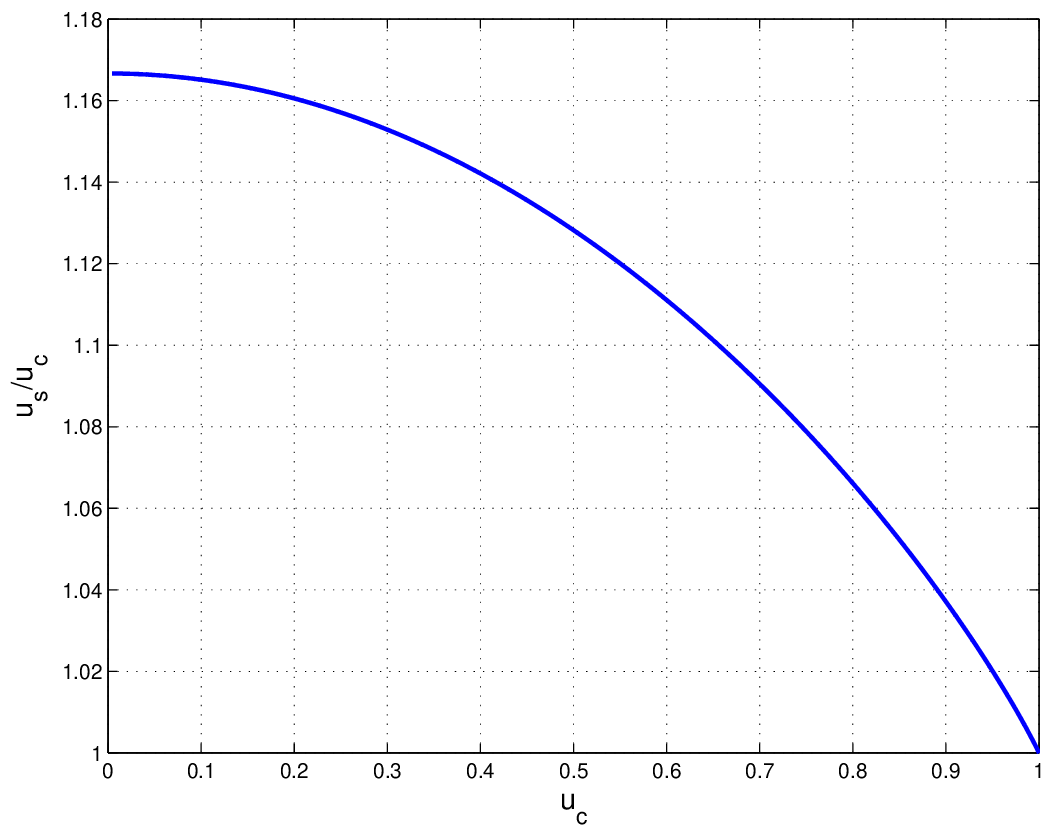}
    \caption[]{The ratio between shock velocity and contact surface velocity as a function of contact surface velocity (for the case $\gamma= \frac{4}{3}$).\label{fig:vel_ratio}}
    \end{figure}

    This feature, the lack of minimum width for shocked medium, is the major cause for the crucial need of AMR techniques to be used in numerical solution of the SRHD equations. As mentioned, these shocked medium are very important structures in the problem, and the fact that the density jump over the front is arbitrarily large and their width is arbitrarily small means that one has to concentrate resolution within them.
    As mentioned in section~\ref{afterglow}, these thin structures may travel distances which are orders of magnitude larger than their own length scale, which means that the space needed to be solved is very large compared to that length scale. Describing the space with equivolume cells in the size needed for that structure, is not possible even in some one dimensional problems, not to mention multidimensional simulations, and one has to be able to concentrate resolution in interesting locations in space, and these are constantly changing and moving as time advances.

\item A difference rising in multidimensional dynamics is the coupling of different components of the velocity between the Equations~\ref{SRHD_conservation} via the Lorentz factor $\Gamma$ and the enthalpy $h$. This difference makes multidimensional SRHD a difficult task for numerical methods, in a manner not found in schemes for Newtonian hydrodynamics. For example, consider a one dimensional Riemann problems (i.e. the evolution of a discontinuity between two adjacent thermodynamic states), with a transverse velocity component. In Newtonian hydrodynamics this problem is the same as if the transverse velocity is zero, since the transverse momentum is just advected through the discontinuity without any disturbance. In SRHD, however, all the velocity components are time dependent (although the problem is still one dimensional, since the spatial dependence is only through the normal direction to the discontinuity). Examples of such problems and the difficulties they induce on the numerical schemes are shown in appendix~\ref{1d_tests}.

\item One last difference is a numerical difference. When one examines the conserved variables in each theory, and their connection to the set of primitive variables used to describe the hydrodynamic and thermodynamic state of a fluid, i.e. $\rho$, $\A{v}$ and $p$, one immediately sees that while in the Newtonian theory the algebraic connection between the two sets is bidirectionally straightforward (for example $\A{v}=\frac{(\rho \A{v})}{\rho}$), in the SRHD theory one can write Equations~\ref{conserved} and calculate the conserved variables from the primitive ones, but the opposite calculation is non trivial and numerical manipulations (mostly root finding techniques) must be used. The use of such a transformation is presented in section~\ref{initialization}.
\end{itemize}

All of these features make the numerical solution of the SRHD equations a complicated task even one dimensionally. The different techniques for solving Equations~\ref{SRHD_conservation}, together with an equation of state $f(p,\rho,e)=0$, initial and boundary conditions, form the study of special relativistic hydrodynamics.

\chapter{Numerical and Computational Framework} \label{framework}

All the different stages presented in section~\ref{GRBs} included problems involving different kinematics regime ($\Rightarrow$ one needs numerical simulations), and also multi-length scale such as presented in Table~\ref{t:radii} or Equation~\ref{l-ST} ($\Rightarrow$ one needs very high resolution over a huge numerical space). Adding to that the unique features present in special relativistic flows (especially the lack of maximum density ratio and lack of minimum width of structures, see section~\ref{features}), one faces a difficult task when approaching such cataclysmic astrophysical phenomena. The complex hydrodynamic processes lead us to a conclusion that one must treat the problem numerically, with a technique to overcome the unreachable computational resolution being imposed on a numerical scheme using a constant resolution over the whole space. One has to be able to concentrate resolution via an AMR module.

For that goal, we have developed RELDAFNA, providing a numerical treatment of complex multidimensional special relativistic flows with multi-length scales. Indeed, in chapter~\ref{computational-astrophysics} we give a number of examples for the ability of RELDAFNA to aid the understanding of the processes presented in section~\ref{GRBs}.

In the following chapter, we will present the numerical methods used to treat the SRHD equations, and we will shortly address some computational aspects which are decoupled from the numerical treatment of the SRHD equations within RELDAFNA, but they still play a crucial role in its ability to provide qualitative answers to questions in computational astrophysics such as the ones presented in section~\ref{applications}.

\section{Numerical Methods} \label{num_methods}

The first attempt to solve the SRHD equations numerically is presented in \cite{wilson72}. It relies on artificial viscosity techniques widely used in algorithms for the solution of the Newtonian hydrodynamics equations. As was shown in \cite{centrella84}, the artificial viscosity techniques are not stable in problems where the Lorentz factor is of the order of 2 or more. There were several attempts to modify the artificial viscosity techniques to achieve stability for ultra relativistic flows, but only some one dimensional advances were made, and these techniques were left only for problems that involve matter flowing with a Lorentz factor $\lesssim$ 2. The attempts to develop a numerical scheme to deal with ultra relativistic flows, led to trying some
non-artificial viscosity approaches. SPH (Smooth Particle Hydrodynamics) and FCT (Flux Corrected Transport) algorithms were developed, but their accuracy was similar to the artificial viscosity schemes already used. For a more detailed review of the development of numerical methods for the solution of the SRHD equations we refer the reader to \cite{marti03}.

The development of high resolution shock capturing (HRSC) methods was the kind of breakthrough the SRHD equations were waiting for. HRSC schemes are written in conservation form, a form the SRHD equations acquire. These schemes advance zone averaged state vectors in time, using numerical fluxes evaluated at zone interfaces. The numerical fluxes are calculated using the solution of a Riemann problem between the two sides of an interface of two adjacent computational zones. The zone averaged state vectors are interpolated by monotonic polynomials, and the order of the polynomial interpolation sets the spatial order of the scheme. A major part of HRSC methods is the Riemann solver used. Using an analytical Riemann solver is a time consuming computation, and usually an
approximate solver gives the same results using considerably less computer time.

\begin{figure*}[h]
\centering
\includegraphics[scale=0.6]{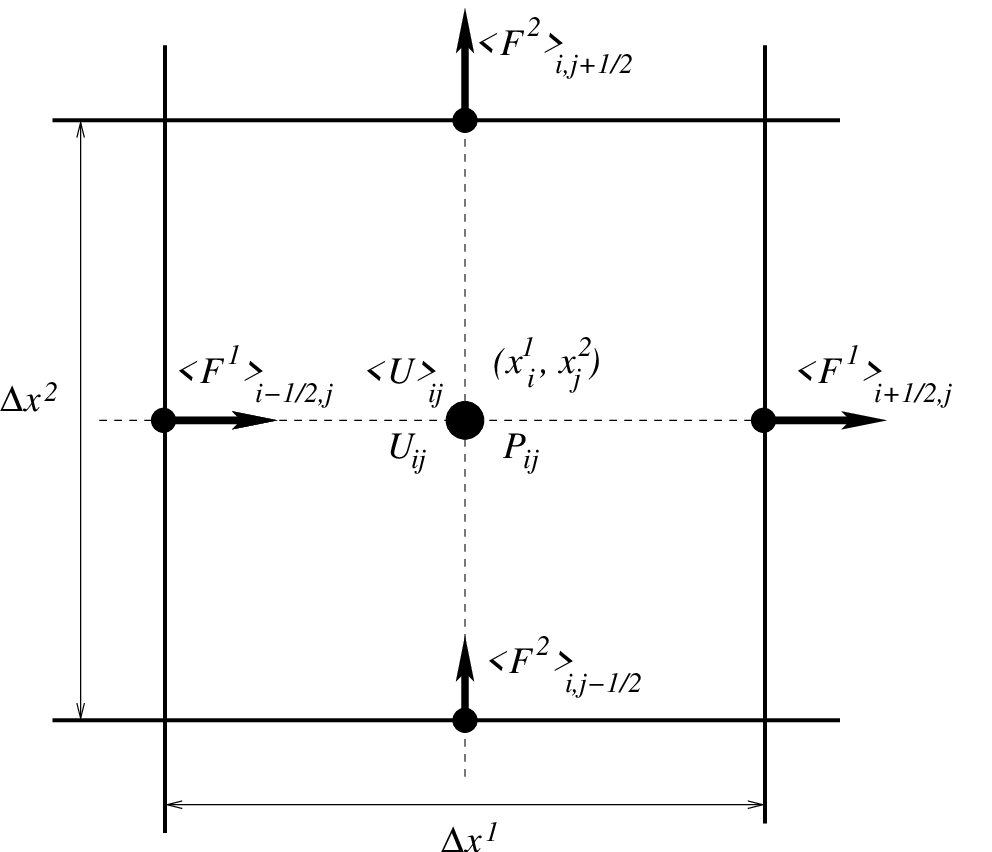}
\caption{Location of quantities within the grid cell $\Delta_{ij}$. Thick solid lines show the grid of cell interfaces and dashed lines show the grid of cell centers. The cell center is indicated by the large dot where the primitive $P_{ij}$, point conserved $U_{ij}$, and average conserved $\langle{U}\rangle_{ij}$ quantities are located. The fluxes are located at the centers of grid cell interfaces shown by smaller dots.\protect\label{fig:2d-cell}}
\end{figure*}

Since RELDAFNA is based on a Godunov-type method, we will give a slightly wider background on such methods. Let us consider a conservation law
\begin{equation}
  \frac{\partial{\mathbf{U}}}{\partial{t}} + \sum_{j=1}^{3}
  \frac{\partial{\mathbf{F}^j}}{\partial{x^j}} = 0\,. \label{general-conservation}
\end{equation}
We use a set of discrete points $\{(x_i,y_j,z_k,t^n)\}^{n=1,N}_{i=1,I;j=1,J;k=1,K}$ and transform Equation~\ref{general-conservation} to a finite difference form defined as
\begin{equation}
\begin{array}{rcl}
x_i & = & (i - 1/2) \Delta x, \,\,\,\, i=1,2,\ldots I  \, , \\
\noalign{\medskip}
y_j & = & (j - 1/2) \Delta y, \,\,\,\, j=1,2,\ldots J  \, , \\
\noalign{\medskip}
z_k & = & (k - 1/2) \Delta z, \,\,\,\, k=1,2,\ldots K  \, , \\
\noalign{\medskip}
\label{spatial-descretization}
\end{array}
\end{equation}
and
\begin{equation}
t^n = n \Delta t, \,\,\,\, n=0,1,2,\ldots N\, ,
\end{equation}
where $\Delta t$ and $\Delta x$, $\Delta y$, $\Delta z$ are the time step and the zone sizes,
respectively. This discretization allows one to obtain
approximations to the solution at the new time, $U_{i,j,k}^{n+1}$, from
the approximations in previous time steps, in the following finite difference form,
\begin{eqnarray}
\nonumber    U_{i,j,k}^{n+1} = U_{i,j,k}^n
    - \Delta t^n \left( \frac{\mathbf{F}^1_{i+1/2,j,k} -
      \mathbf{F}^1_{i-1/2,j,k}}{\Delta x}
    + \frac{\mathbf{F}^2_{i,j+1/2,k} -
      \mathbf{F}^2_{i,j-1/2,k}}{\Delta y} \right. \\
    \left. + \frac{\mathbf{F}^3_{i,j,k+1/2} - \mathbf{F}^3_{i,j,k-1/2}}{\Delta z}\right)\,.
    \label{finite-difference}
\end{eqnarray}
The fluxes are evaluated at cell interfaces while the conserved quantities are located at cell centers as depicted in the 2D Fig.~\ref{fig:2d-cell} from \cite{tchekhovskoy07}.

Equation~\ref{finite-difference} is the final step of every numerical scheme for the solution of conservation laws, and it just emphasizes the fact that the evolution in time of a set of grid cell averages of conserved quantities obeying Equation~\ref{general-conservation}, is done by considering binary interactions between adjacent cells, since the equations are spatially local. The main difference between the schemes is the method used to calculate the fluxes at cell interfaces. Some schemes \cite{zhang06} deal only with the fluxes functions in order to obtain the cell interface flux, and do not treat the conserved variables at all until the final step of Equation~\ref{finite-difference}, i.e. they look for the solution of an equation of the sort \begin{equation}
 \mathbf{F}_{i+1/2} =
  F(\mathbf{F}_{i-r},...,\mathbf{F}_{i+s}),
  \label{flux-reconstruction}
\end{equation}
where the stencil $(i-r,i-r+1,...,i+s)$ depends on the choice of the reconstruction scheme. In Godunov methods the interactions between adjacent cells are usually approximated as one-dimensional. The solution is based on the dynamic evolution of two different hydrodynamic and thermodynamic states separated by a diaphragm being removed in no time, the so-called Riemann problem. The simplest form of the Godunov method is to calculate the decay of the discontinuity in the hydrodynamic variables between neighboring cells, through the solution of the Riemann problem between the two cells averaged conserved variables. In accordance with Equation~\ref{flux-reconstruction} one can write this simple form,
\begin{equation}
 \mathbf{F}_{i+1/2} = R(\mathbf{U}_{i+1/2}^{-}, \mathbf{U}_{i+1/2}^{+}),
\end{equation}
where $-$ and $+$ denote the left and right side of the interface $i+1/2$, respectively. The function $R$ is the Riemann solver (exact or approximate).

In order to achieve higher accuracy than the simple form, one attempts to build a smoother spatial function for the conserved variables at the time step, and also to update the conserved variables within the timestep. The different techniques for doing these two tasks are the characteristics of different numerical schemes based on the Godunov method. Building a higher order polynomial spatial profile for a quantity makes the Riemann problem much more complicated since it is not a discontinuity between just two constant states any more, but an effective diaphragm in the middle of a smooth profile with a discontinuity inside it. This could even lead to a non linearity of the characteristics which will separate between the different states being produced once the diaphragm is removed \cite{tchekhovskoy07}. Even so, most numerical schemes (including the ones used in RELDAFNA) use the Riemann solvers of two constant states, since their accuracy is sufficient considering the computational cost of general Riemann solvers.

\begin{figure*}[h]
\centering
\includegraphics[scale=0.35]{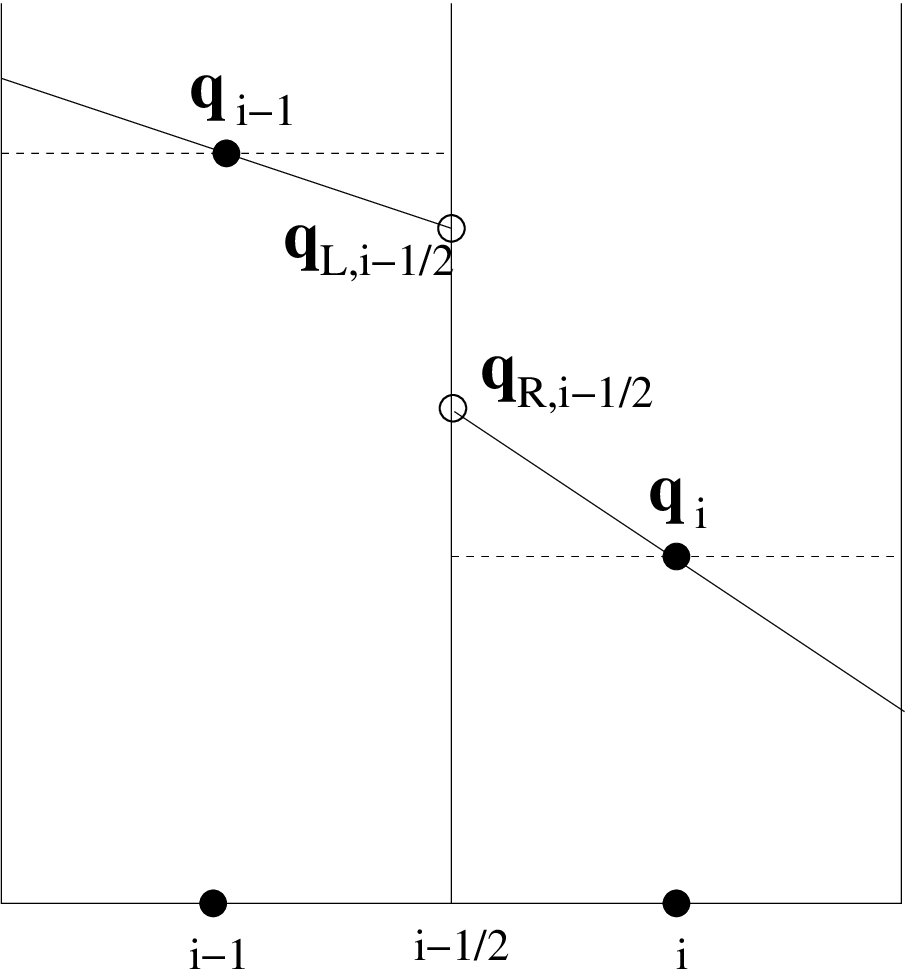}
\includegraphics[scale=0.35,trim= 0 80mm 90mm 0]{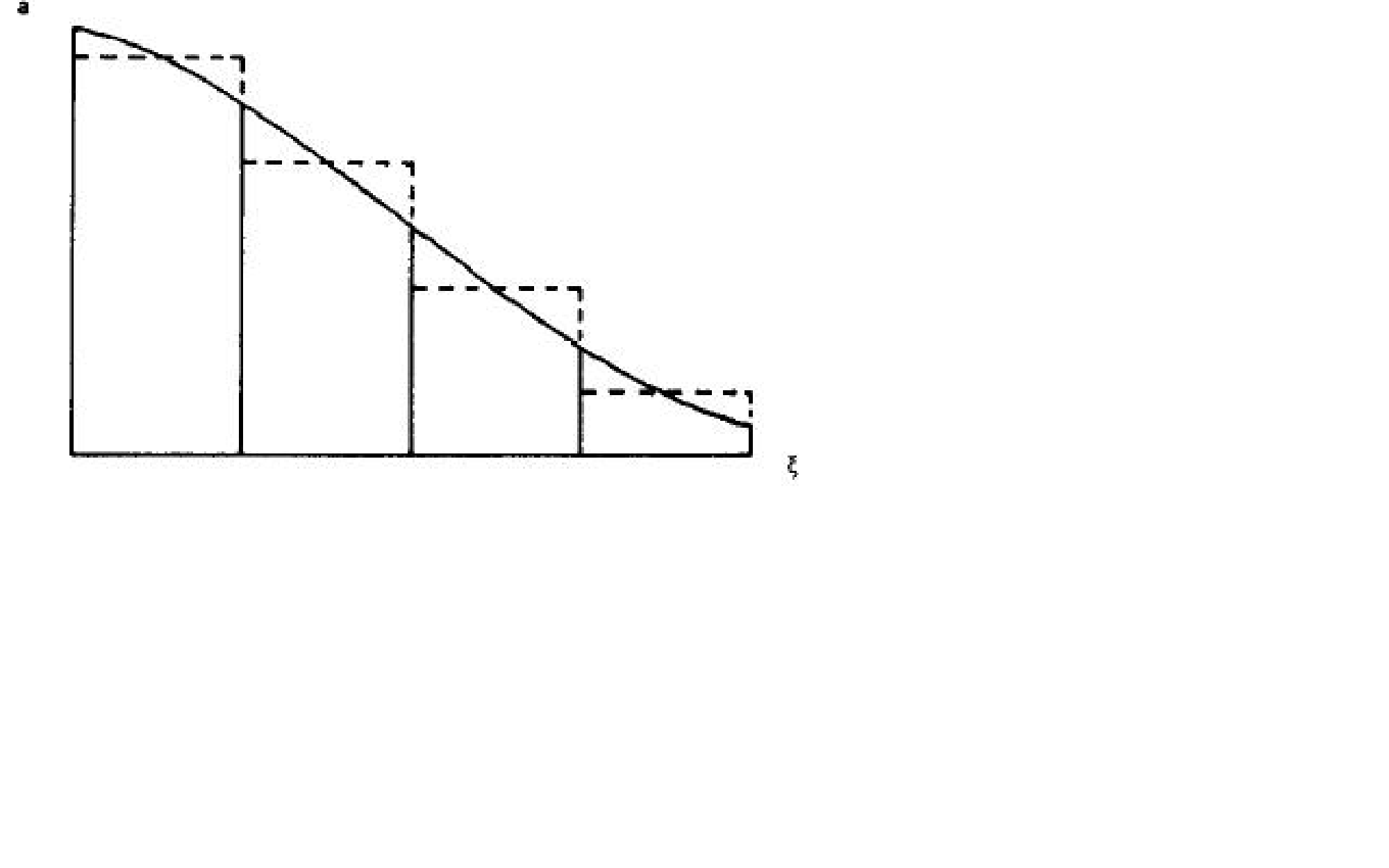}
\caption{Polynomial spatial reconstruction profiles, PLM (left, from \cite{stone08}) and PPM (right, from \cite{collela84})\protect\label{fig:PLM-PPM}}
\end{figure*}

The spatial reconstructions are usually made by imposing a polynomial profile for the variables, although other profiles have also been implemented \cite{lucas04}. The frequently used orders are linear (PLM) \cite{miller96} and parabolic (PPM) \cite{collela84,miller02}. In Fig.~\ref{fig:PLM-PPM} we show a schematic representation of a polynomial profile.

The temporal update is based on an upwind characteristic tracing method \cite{collela85}. It is based on a linearization of the PDEs~\ref{general-conservation}
\begin{equation}
  \frac{\partial{\mathbf{u}}}{\partial{t}} + \mathbf{A}\frac{\partial{\mathbf{u}}}{\partial{x^j}} = 0\,, \label{quasilinear}
\end{equation}
where the matrix $\mathbf{A}$ is the Jacobian of the fluxes
\begin{equation}
\mathbf{A} =
\frac{\partial{\mathbf{F}(\mathbf{u})}}{\partial{\mathbf{u}}}\,.
\end{equation}
The exact solution of Equation~\ref{quasilinear} is given by \cite{miller96}
\begin{equation}
u^{n+1/2}_{i\pm 1/2} = u^n_i+\frac{1}{2}(\pm \Delta x I -\Delta t A)\pd{u^n}{x} \, ,
\end{equation}
but upwind methods such as Godunov take into consideration only the characteristics which are propagating in the direction being calculated, i.e. when evaluating the right interface of a cell only the positive eigenvalues of $\mathbf{A}$ are being calculated while the negative ones are set to zero.

One can summarize the Godunov scheme in a RSA flow diagram:

\begin{itemize}
\item \textbf{Reconstruct}: Building the spatial profile of the variables using a smooth function, usually a polynomial.

\item \textbf{Solve}: Updating temporally the cell interface variables and solve the Riemann problems imposed between the two sides of the interface between adjacent cells.

\item \textbf{Average}: Using the Riemann solution obtaining fluxes at cell interfaces, and calculating the new conserved variables via Equation~\ref{general-conservation}.
\end{itemize}

\begin{figure*}[h]
\centering
\includegraphics[scale=0.85]{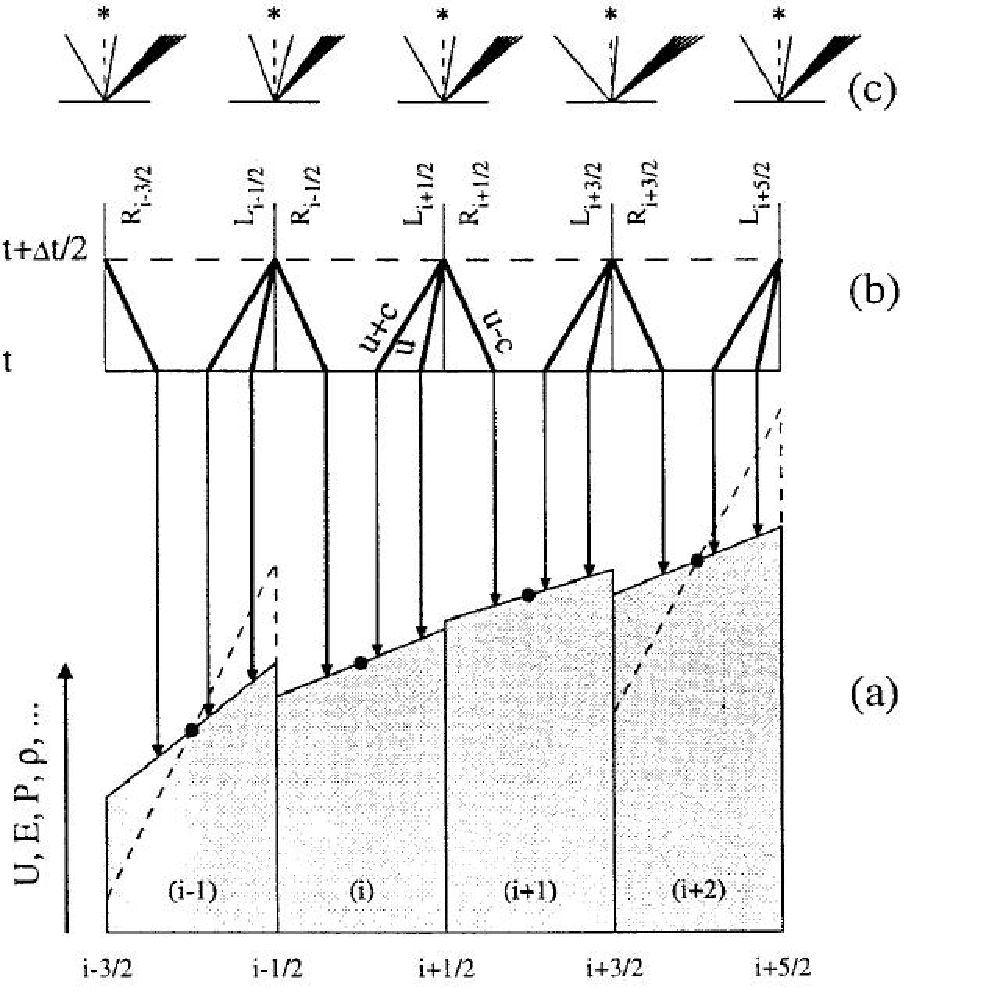}
\caption{Schematic representation of a high-order Godunov method. (a) Cell centered variables (filled dots) are used to construct a central difference approximation to the slope. The resultant distributions (dashed volumes) may be limited to satisfy certain monotonicity constraints (shaded volumes). (b) Upwind characteristic tracing is used to deduce the time-centered cell-interfaces states that feed the Riemann problem. (c) The middle state of the Riemann problem determines the time-centered fluxes used in Equation~\ref{general-conservation}.\protect\label{fig:RSA}}
\end{figure*}

Fig.~\ref{fig:RSA} (taken from \cite{miller96}) depicts the RSA flow.

\section{Logic} \label{logic}

RELDAFNA was developed in a thoughtful manner so that a physicist who did not participate in its development could use it to approach relativistic flow problems at his/her interest. In order to do that, we have completely separated the numerical solution modules from computational ones such as the building of the computational domain, its decomposition to different processors, the connections and information transfer between the processors during a computation and the Adaptive Mesh Refinement. This decoupling we have made, led us to a somewhat non-conventional language for the implementation of the numerical algorithms. RELDAFNA is entirely composed of grid based methods, so naturally one would tend to attach each cell a discrete address with a certain order of the cells, and addresses connecting the address of one cell to the address of its neighbors, as in Equation~\ref{spatial-descretization}. However, since RELDAFNA utilizes AMR and general mesh decomposition parallelism, as will be shortly presented ahead, the grid based methods implemented within it are dealing only with discrete and isolated cells logically. The meaning of that is, since the methods we use are RSA (Reconstruct-Solve-Average) methods, we make calculations only on cells within their own information, or calculate fluxes which are crossing a single interface between two neighboring cells in a defined direction. The fact that two cells sharing an interface, through which the flux is being transferred, could have different sizes and could also be on two different processors, makes the old language of indexed cells a bit cumbersome. Therefore, in our way of describing the calculation we refer to isolated cells and interfaces only. A cell does not have a defined address, and it only knows its nearest neighbors in the different directions of the calculations, one in the positive side of the direction and one in the negative side (sometimes referred to as left $\&$ right, up $\&$ down and inside $\&$ outside). The code is written in such a way that the physical and numerical calculations are decoupled from the AMR and parallel layers. The implementation of the computational and numerical techniques in RELDAFNA is robust, and includes within it a treatment of $1$, $2$ and $3$~dimensional problems, in Cartesian, cylindrical or spherical geometries. The outcome of that is that a user can first design a simple $1$~dimensional calculation for first insights on his/her problem, and then turn it easily to a $2$ or $3$~dimensional calculation for a more detailed investigation. In that spirit, we are describing the physical and numerical steps used in RELDAFNA, so that there would be minimal disturbances from the fact that cells might have different sizes and could be held by different processors.

\section{Adaptive Mesh Refinement} \label{AMR}

As was highlighted throughout section~\ref{applications}, the relevant questions in high energy astrophysics that involve SRHD are multi length scale problem, and are very hard to solve using an Eulerian static mesh describing the whole space needed for the calculation (see for example Table~\ref{t:radii}). RELDAFNA solves the SRHD equations on a structured mesh, i.e. a mesh of rectangular shaped cells. In order to increase spatial resolution in zones that require high resolution at a given time, it uses Adaptive Mesh Refinement via the Fully Threaded Tree algorithm \cite{khokhlov98}, in a way presented ahead.

\begin{figure*}[h]
\centering
\includegraphics[scale=0.45,trim= 50mm 0 50mm 0]{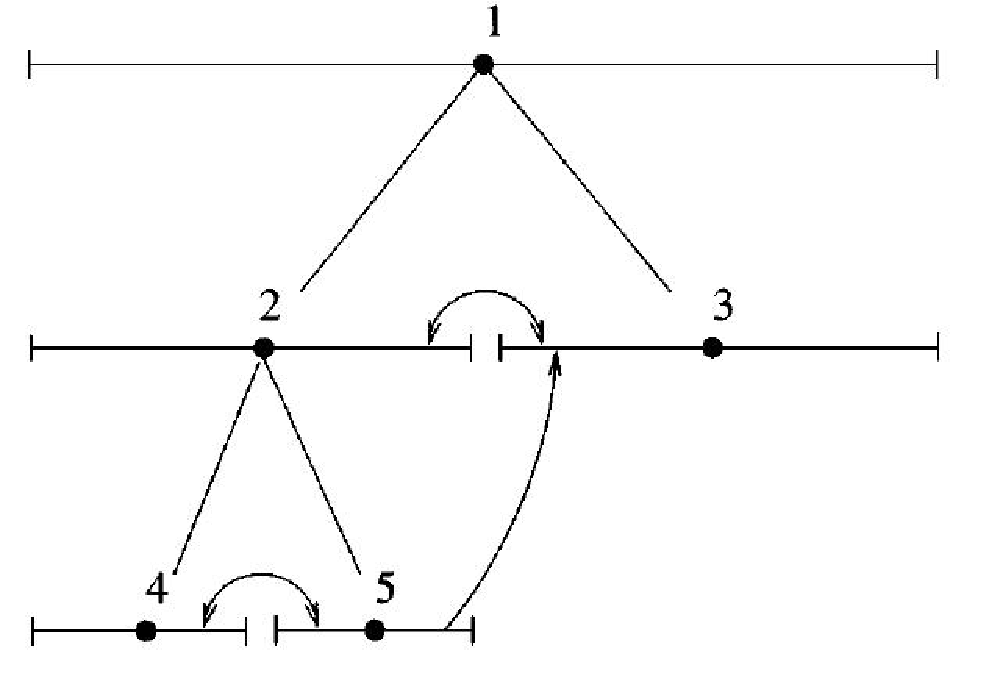}
\caption{Relationship between cells for a one-dimensional, binary, fully threaded tree. Each cell is represented by a horizontal barred line. The length of a line corresponds to the geometrical size of a cell. Cell 1 is the root of the tree, representing the entire computational domain. Cells 1 and 2 are split cells. Cells 3, 4 and 5 are leaves. Pointers to children and parents are indicated by  straight lines without arrows. Pointers to neighbors are indicated by arrows.\protect\label{fig:1d-tree}}
\end{figure*}

The FTT approach is to treat each cell individually for refinement or derefinement at any given time, instead of patching grids of different resolutions one on the other. The result of that approach is a mesh of arbitrary shape, although the cells still have rectangular shape. When one treats every cell separately, all there has to be done when needed, is to split the rectangular shaped cell into $2^{dimension}$ equally sized rectangular shaped cells covering it. These cells will be treated from now on as individual cells themselves, and when high resolution is not needed any more at their location, they will be erased from the cells list and their ancestor will become the highest resolution cell in that area. Of course, this procedure can be done a number of times in each location depending on the need, and that would result in a tree of cells, each level of the tree is doubled in resolution than the one level above it. The crucial feature of a threaded tree, which saves a great amount of time searching it for neighboring cells, is the fact that every cell uses the same amount of memory allocation. "Parent" cells use memory to point the location of their "children" cells, but "leave" cells that are not refined use that memory to point to shortcuts on the tree for an efficient search up and down the tree. It has been shown that in order to prevent shocks to run from a refined mesh, one must keep the resolution ratio between two adjacent cells $\leq 2$, and also to prepare buffer cells in advance and ahead of shocks. A schematic figure of a one dimensional tree where the whole domain is being refined several times is presented in Fig.~\ref{fig:1d-tree} from \cite{khokhlov98}. The shortcuts mentioned and seen in Fig.~\ref{fig:1d-tree} are based on the fact that when a cell is split, its children are created simultaneously and therefore can be stored in neighboring places in the memory so that only one pointer (from the ancestor to the "children") is needed to find all the children. Moreover, since all the children (siblings) are also neighbors in the memory, neighbor--neighbor relations between siblings are known automatically. Also, a neighbor of a cell is either his "sibling", i.e. they share a parent, his "uncle" or "cousin", thus, a child's neighbor that belongs to a different parent can be determined without search if parents keep pointers to their own neighbors (see for example the arrow from cell no.~$5$ in Fig.~\ref{fig:1d-tree}).

\begin{figure*}[h]
\centering
\includegraphics[scale=0.4]{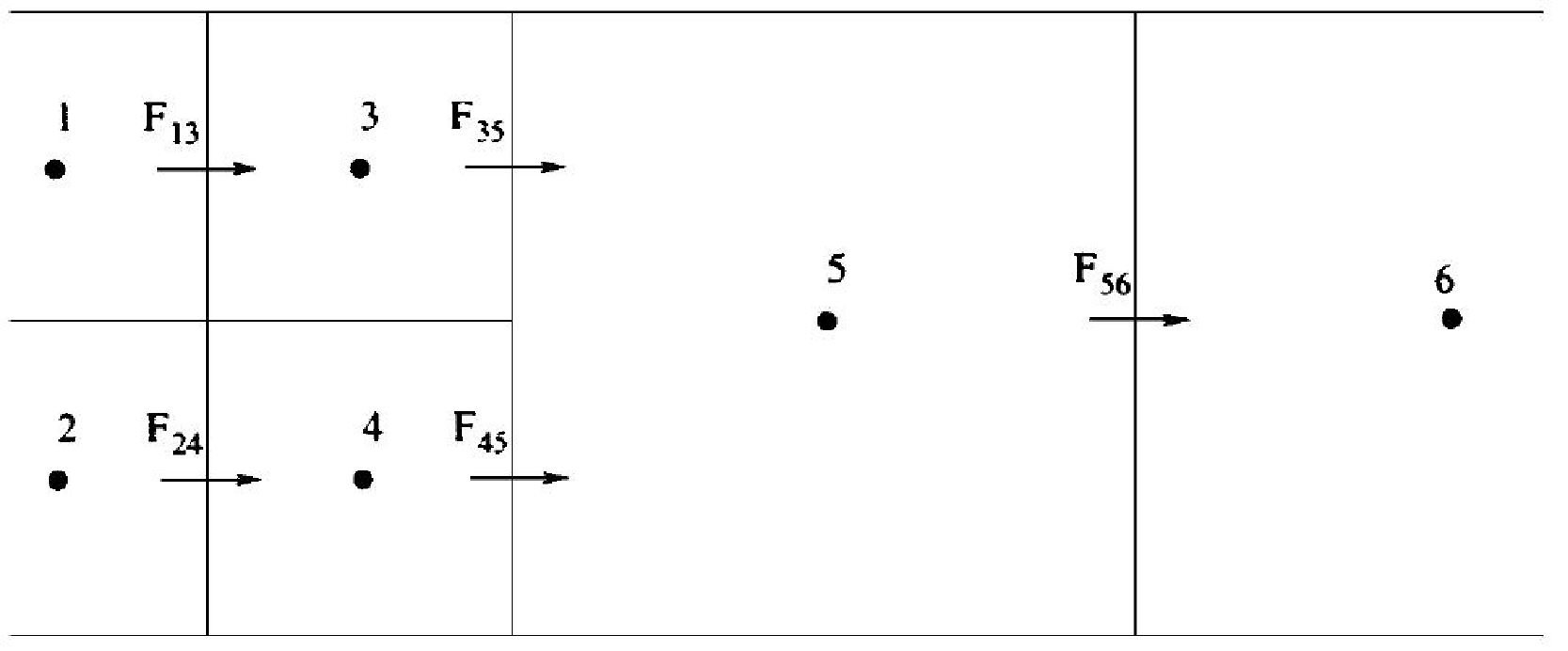}
\caption{An illustration of flux evaluation at two different levels of the tree. Fluxes at interfaces between fine cells (1--3 and 2--4), and fluxes between fine and coarse cells (3--5 and 4--5) are evaluated twice as often as fluxes between coarse cells (5--6). Four fluxes from the fine side (3--5 and 4--5 computed at two different times), and one flux (5--6) from the coarse side contribute to changes of the state vector of cell 5 during a coarse time step.\protect\label{fig:2d-fluxes}}
\end{figure*}

The simple topology of a one dimensional space, with which cells have defined neighbors, i.e. the cells to their right and left, regardless of their size, is not so simple in a two dimensional space. In Fig.~\ref{fig:2d-fluxes} (from \cite{khokhlov98}) we demonstrate the interfaces between cells with different sizes in a two dimensional space, interfaces which transform fluxes from one cell to another. These connections lead one to keep physical state information $\mathbf{U}$ in both split and unsplit cells. Information kept in split cells is needed to evaluate the fluxes $\mathbf{F}$ across interfaces between leaves of different size (see Fig.~\ref{fig:2d-fluxes}), and to make decisions about refinement and derefinement. The physical state vectors for split cells are updated by averaging over children, when required.

\begin{figure*}[h]
\centering
\includegraphics[scale=0.4]{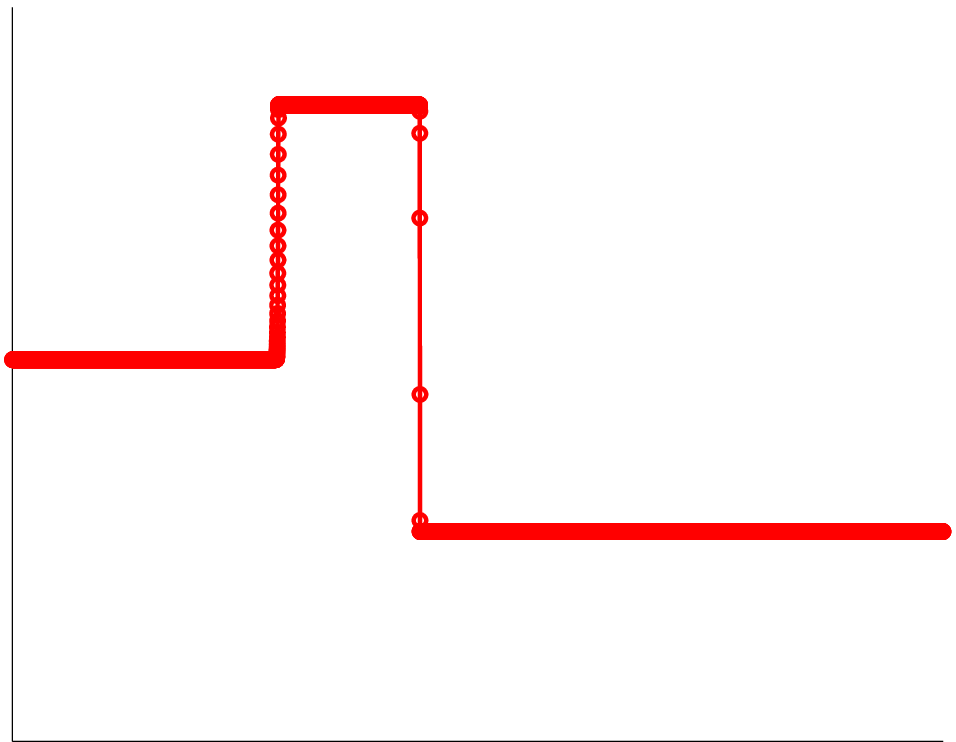}
\includegraphics[scale=0.4]{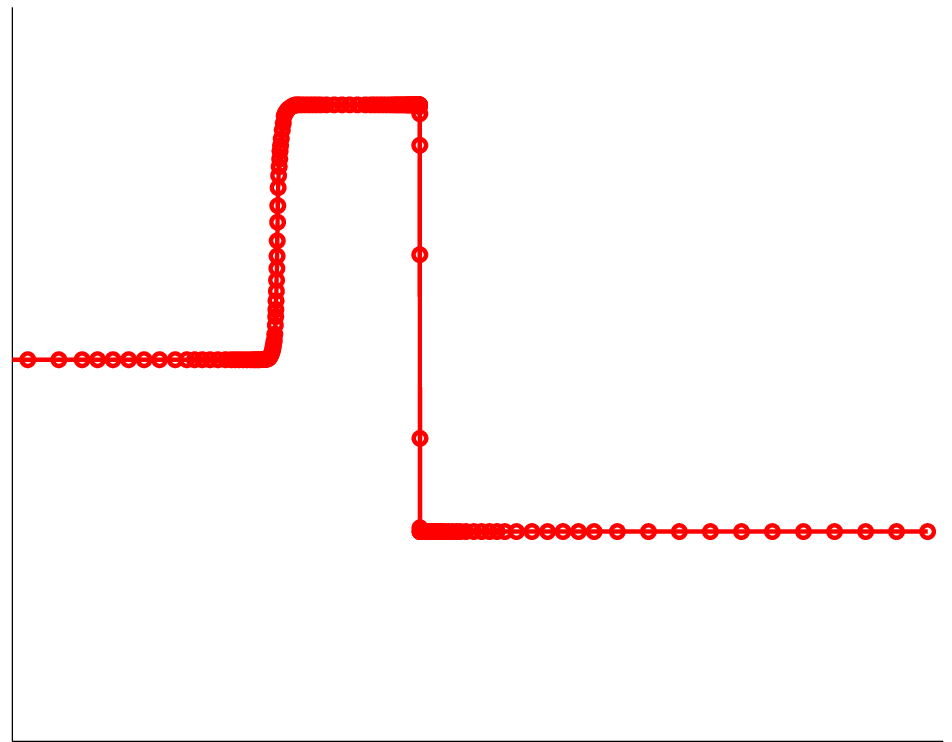}
\caption{A profile from a one-dimensional calculation using a constant grid (\textit{left}) and an AMR grid (\textit{right}).\protect\label{fig:amr-grids-1d}}
\end{figure*}

In Fig.~\ref{fig:amr-grids-1d} we show a spatial profile of some hydrodynamic variable from a one dimensional calculation with and without AMR. At the right and left sides of the profile, where the variable does not exhibit a high gradient, one can see that the calculation using AMR is sampling the area with a significantly lower number of cells. At the discontinuities, however, the number of cells used is the same since the AMR resolution is equivalent to the constant grid. One can see that the number of cells used increases gradually at the zero gradient zones which are close to the discontinuities.

\begin{figure*}[h]
\centering
\includegraphics[scale=0.4,trim=0 -20mm 0 0]{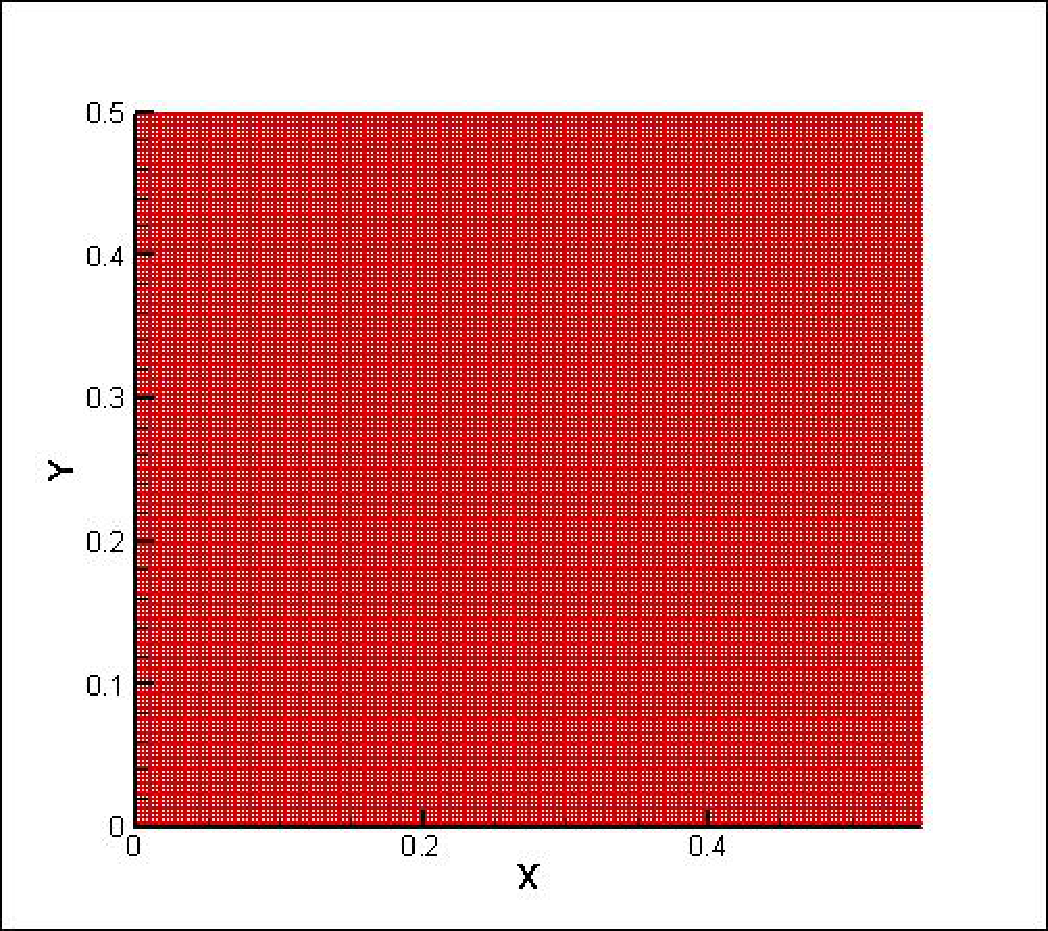}
\includegraphics[scale=0.4,trim=0 -20mm 0 0]{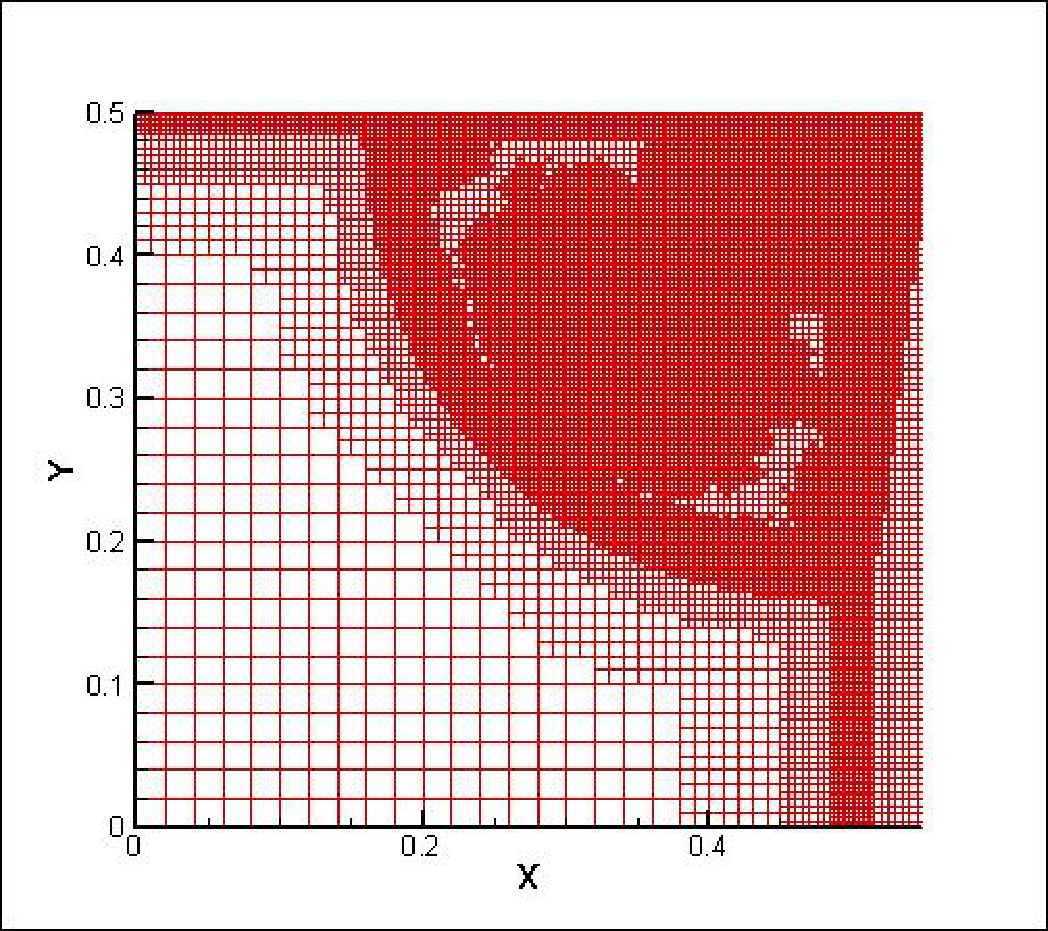}
\caption{A patch of the mesh from a two-dimensional calculation  using a constant grid (\textit{left}) and an AMR grid (\textit{right}).\protect\label{fig:amr-grids-2d}}
\end{figure*}

In Fig.~\ref{fig:amr-grids-2d} we show a patch from the grid of a two dimensional calculation with and without AMR. At the left bottom area of the patch, where unperturbed matter is flowing, one can see that in the calculation using AMR the mesh is coarse there. The bow shaped layer of highly refined cells is a shock front advancing towards the point $(0,0)$, and one can see the buffer cells ahead of it signaling the unperturbed areas the high refinement coming. The high refinement at the lines $x=0.5$ and $y=0.5$ is a result of the fact that we are showing a patch from a two-dimensional Riemann problem presented in appendix~\ref{sec:rie2d}, and there is a strong gradient at these areas (see also Fig.~\ref{fig:2d-parallel}).

Now we turn to describing the criterion for refinement and derefinement. Generally, one would like to refine cells which are close to strong gradients of the hydrodynamic variables, signaling the advance of a shock wave or a contact discontinuity that should be resolved with high resolution. Consider a hydrodynamic variable $q$ and a given cell that would be denoted by a superscript $i$. We go over all directions in the calculation and look for a steep gradient in the variable $q$, between the cell being treated and its neighbors in that directions. The effective gradient in a direction $k \, \left(k\in\{1\ldots dimension \}\right)$ is being calculated as,
\begin{equation}
dq_k=max\left(\frac{\left|q^r-q^i\right|}{min\left(\left|q^r\right|,\left|q^i\right|\right)},
              \frac{\left|q^l-q^i\right|}{min\left(\left|q^l\right|,\left|q^i\right|\right)}\right)
\end{equation}
where the superscripts $r$ and $l$ denote the right and left neighbors of the cell $i$ in direction $k$, respectively. If $\sum_{k=1}^{k=dim.}dq_k<d_{deref}$ for all hydrodynamic variables participating in the criterion the cell is marked for derefinement, and if $\sum_{k=1}^{k=dim.}dq_k>d_{ref}$ for one hydrodynamic variable participating in the criterion the cell is marked for refinement. The values $d_{deref}$ and $d_{ref}$ are predefined values tuned by the user to achieve an optimized calculation in both accuracy and time (and memory) consumption. As mentioned above, cells could be marked for refinement if they are close to a shock wave even if they do not obey the criterion, and that is also determined by the user, the size of the buffer layers coating the refined cells in shock waves and contacts. In RELDAFNA, the variables being calculated in the criterion are the density, pressure and Lorentz factor.

\section{Parallel Implementation} \label{parallel}

Multidimensional computations for high energy astrophysics require a very high number of cells allocated in memory $\left(\gtrsim 10^8\right)$, even when using AMR with a high maximum level of refinement. Therefore, the need to use parallel computers in order to store this amount of memory and to speed up the calculation is unavoidable. In order to that the MPI library is used for parallel commands.

\begin{figure*}[h]
\centering
\includegraphics[scale=0.4]{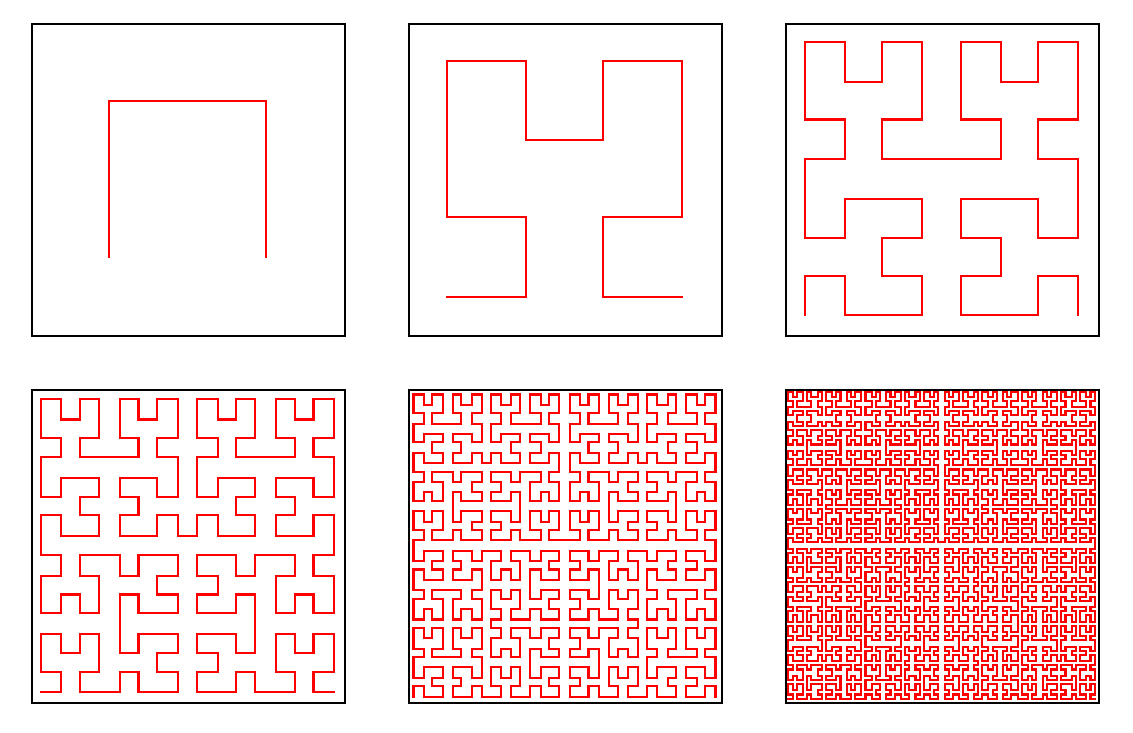}
\caption{A 2D Hilbert space filling curve. \protect\label{fig:hilbert}}
\end{figure*}

The mesh is being decomposed to the number of processors participating in the calculation. Every processor keeps its own patch of the grid and also overlaying ghost cells, which represent cells kept in another processor, and they are filled with information from neighboring processors. In order to minimize the connections between different processors, a space-filling curve (such as Hilbert, as shown in Fig.~\ref{fig:hilbert}) is used to mark the cells and order them in an effective one dimensional array \cite{aftosmis04}. This array is being cut to spread the cells between the participating processors, so that the decentralization of the cells between the processors would be optimized.

\begin{figure*}[h]
\centering
\includegraphics[scale=0.5,angle=270,trim= 0mm 100mm 0mm 100mm]{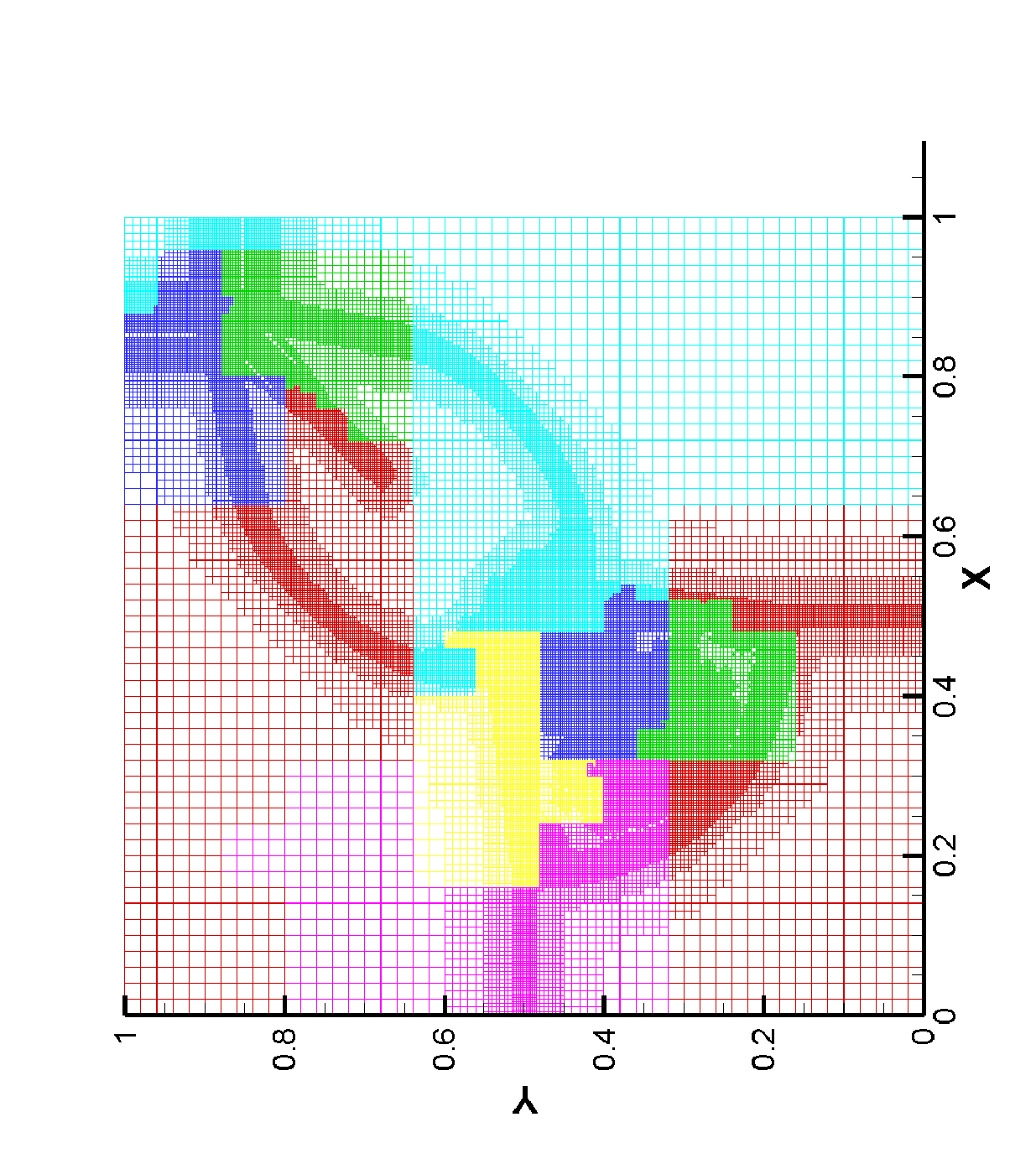}
\caption{A $2$~dimensional space shared by $10$~processors, each of which is colored differently. \protect\label{fig:2d-parallel}}
\end{figure*}

The decomposition of a $2$~dimensional space between $10$~processors is shown in Fig.~\ref{fig:2d-parallel}.

Since RELDAFNA uses AMR, the number of cells at a specific location in space can change dramatically during a computation, as shock waves and contact discontinuities coast through space. As a result of that, even if at the beginning of a calculation the division of space to the processors was equal, the work load on the processors can get out of balance as refinement and derefinement occur. In order to over come that and keep the load balance between the processors at a maximum, we use the space filling curve again and again during the computation, and redistribute the cells between the processors whenever they get out of load balance. The fact that the cells can be ordered in a one dimensional array, using the space filling curve, allows one to just move the boundaries between the processors inside that array, and thus achieving load balance with minimum transfer of cells between processors in that procedure. It has been found that parallelizing an AMR computation without a load balance procedure over and over again, does not achieve the desired speed up one would want to get out of the parallel cluster of processors.

\section{The User Interface} \label{user}

We list the numerous constants a user should feed the computation using RELDAFNA. This list is not complete but it includes the most necessary values:
\begin{itemize}

\item The spatial boundaries of the space being calculated, and the resolution in each direction for the base mesh.

\item The geometry of the space (Cartesian, cylindrical or spherical).

\item The initial conditions of the hydrodynamic variables in space.

\item The maximum level of refinement.

\item The values $d_{ref}$ and $d_{deref}$ for the density, pressure and Lorentz factor.

\item The number of processors participating in the calculation.

\item The Numerical Scheme to solve the problem computationally. The schemes will be presented in Chapter~\ref{schemes}.

\end{itemize}

\chapter{Numerical Schemes \& Code Description} \label{schemes}

RELDAFNA is a numerical code for the numerical solution of the SRHD Equations. It is based on the Godunov method. Two time integration schemes are implemented within it (RK and CT), along with two spatial reconstruction methods (PLM and PPM). The Riemann problems are solved with an accurate (iterative) solver or with one of two approximate solvers (HLL and HLLC). The spatial domain is divided into cells, and the variables are kept within cells as averages over the whole cell volume. At the beginning of a calculation the space is being filled with a density, velocity and energy distribution as given by the user to solve his/her specific problem, and the average values within cells are being advanced with time throughout the simulation. The implementation of the various techniques is explained with great detail in the following chapter, with the description of the flow of one time step. This flow is being iterated again and again until the simulation reaches its final time given by the user.

\begin{itemize}
\item \textbf{Thermodynamics}: At the beginning of a time step the conserved variables $D$, $\A{S}$ and $\tau$, are known. From the conserved ones, the primitive variables $\rho$, $\A{v}$ and $p$ are calculated,  using the EOS.

\item \textbf{AMR}: Once every two time steps, the need to refine or derefine cells is checked over the whole space, according to the criterions presented in section~\ref{AMR}. After that, the thermodynamic variables of \textit{'parent'} cells are updated according to the information kept by their \textit{'children'} cells.

\item \textbf{Load Balance}: Once every two time steps, the need to transfer cells between processors is examined, in order to load the balance and optimize the parallelization as explained in section~\ref{parallel}.

\item \textbf{Time Step Adjustment}: The size of the time advancing in the specific cycle is being calculated.
\end{itemize}

All the above are considered initialization of the time step and will be presented in section~\ref{initialization}.

The following steps produce the hydrodynamic solution for the advance in time of the conserved variables.

\begin{itemize}
\item \textbf{Reconstruction}: A smooth spatial profile for the primitive variables is built, as will be presented in section~\ref{reconstruct}.

\item \textbf{Solve}: The interface values of the primitive variables are advanced half of a time step, and the Riemann problems imposed on the interfaces are solved to evaluate the fluxes, as will be presented in section~\ref{solve}.

\item \textbf{Average}: The cell centered conserved variables are advanced in time using the fluxes over the cell's interface and predictor solutions (in RK\# methods). This step will be presented in section~\ref{average}.
\end{itemize}

\section{Timestep Initialization} \label{initialization}

\subsection{Thermodynamics} \label{thermodynamics}

RELDAFNA uses the interface values of the primitive variables in order to calculate the fluxes.
Therefore, one has to calculate the cell centered primitive variables, which when put into Equations~\ref{conserved} will result with the conserved variables that are present at the cell center. As was pointed out in section~\ref{features}, this procedure is not straightforward since the algebraic relation is not bidirectional. The variable searched is the pressure which on one hand is given by the EOS $f(p,\rho,e)=0$, and on the other hand can be calculated from the conserved variables and the other primitive variables. Following a scheme similar to the one proposed in \cite{aloy99,marti96}, from Equations~\ref{conserved} one can evaluate the Lorentz factor
\begin{equation}
\Gamma = \frac{1}{\sqrt{1-\left( \A{S} / \left(\tau+D+p\right)\right)^2}} \,,
\end{equation}
and then the density and internal energy density
\begin{equation}
\begin{array}{rcl}
\rho & = & \frac{D}{\Gamma} \,, \\
 \noalign{\medskip}
e & = & \frac{\tau+D\left(1-\Gamma\right)+p\left(1-\Gamma^2\right)}{D\Gamma}\,.\\
 \noalign{\medskip}
 \end{array}
 \label{effective_rho_e}
\end{equation}
Now all one has to do is to find the root of the EOS $f(p,\rho,e)=0$, when one uses $\rho$ and $e$ from Equations~\ref{effective_rho_e}. The specific implementation in RELDAFNA is for an ideal gas equation of state, and it is given in great detail in appendix~\ref{Newton-Raphson}. Using a different equation of state in RELDAFNA is straightforward since all one has to do is to find the root of the EOS using its explicit form.

\subsection{Time Step Adjustment} \label{dtcalc}

The average values of the physical variables are being calculated at discrete points of the calculation time. The difference between the discrete points is changing throughout the simulation according to the hydrodynamic profiles at the current point of time. The size of the time step is being determined so that information would not cross a cell in one time step (the CFL condition). For each cell, define
\begin{equation}
x_l=min\left(\Delta x_i\right) \,, u_l=max\left(\left|u_i\right|\right)\,,\, i\in \{1...dim.\}\,,
\end{equation}
and the timestep would be calculated as
\begin{equation}
dt = CFL \cdot min_{all
cells}\left(\frac{x_l}{\frac{u_l+c_s}{1+u_lc_s}}\right)\,,
\end{equation}
where CFL is a positive constant $<1$. Actually, one should use the values in Equations~\ref{eq:eigenvalues} for the CFL condition, but as also noted in section~\ref{riemann_solvers} this condition is sufficient (in our calculations $CFL\le0.5$) and cheaper computationally. A user can easily swap the condition to include Equations~\ref{eq:eigenvalues} if needed.

\section{Reconstruct} \label{reconstruct}

As was mentioned in section~\ref{num_methods}, the hydrodynamic solution in Godunov methods is being calculated through the RSA flow. In the following sections, each stage of the Reconstruct-Solve-Average would be presented, starting with \textbf{Reconstruct} in the following section.

The first stage of the hydrodynamic solver is to build a smooth profile for the primitive variables. The profiles are usually polynomials within the cells, with the order of the polynomial setting the order of spatial accuracy.\\
An obvious and straightforward reconstruction is the piecewise constant method (PCM), a zero-order polynomial ($f\left(x\right)=constant$) within a cell, which approximates the spatial reconstruction inside a cell as uniform, resulting in assigning the side of an interface the value kept for the cell at that side. PCM is available in RELDAFNA but never used, since it is poorly accurate in resolving structures of an approximated solution, in either 1D simple demonstration problems or astrophysical applications.

The next order for a polynomial reconstruction is a linear profile within a cell, and that will be presented in detail ahead. We have also implemented yet a higher order polynomial, that is a parabolic reconstruction (PPM),  which is due to \cite{collela84}. PPM is similar in concept to PLM, though it involves complicated calculations. The detailed implementation of the spatial reconstruction using PPM is given in appendix~\ref{spatial-PPM}.

\subsection{Piecewise Linear Method} \label{spatial-PLM}

\begin{figure*}[h!]
\centering
\includegraphics[scale=0.5, trim=0 65mm 0 0]{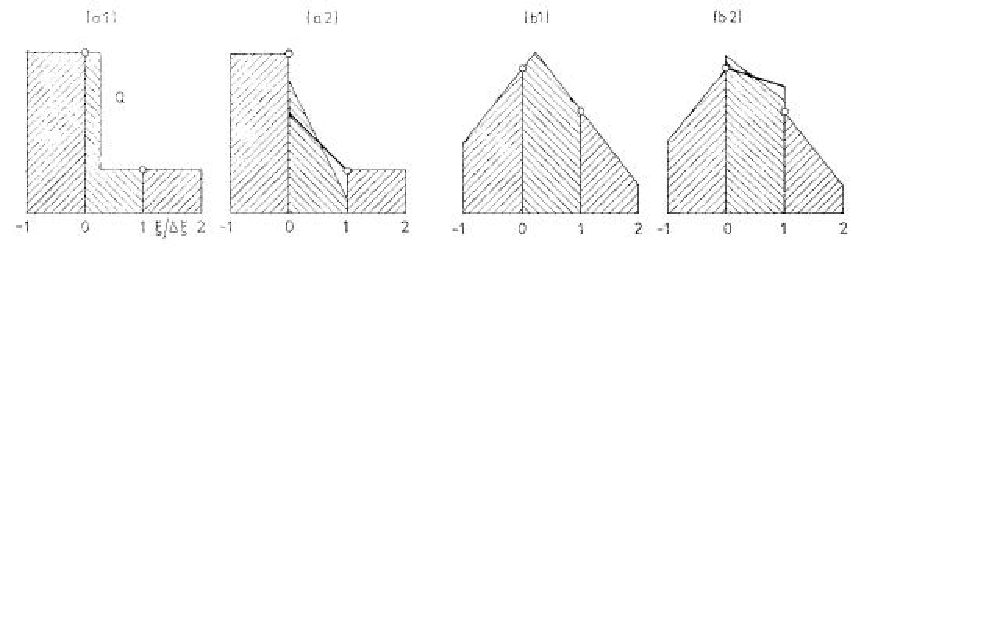}
\caption{The linear profile and the monotonicity condition. (a1) The initial values represent a shock inside the cell $[\xi_0,\xi_1]$. (a2) The linear profile and its monotonization (heavy line). The monotonicity condition is that the profile inside the cell must not take values outside the range defined by the actual values at $\xi_0$ and $\xi_1$ (circles). (b1) The initial values have a peak inside the cell $[\xi_0,\xi_1]$. (b2) The peaked actual profile is replaced by a monotonized linear one (heavy line). The highest value allowed to appear in the cells is the actual value at $\xi_0$.  \protect\label{fig:monotinization-VL97}}
\end{figure*}

In the Piecewise Linear Method (PLM) reconstruction method, the derivatives of the hydrodynamic variables with respect to the generalized volume coordinates $\xi$ at each direction \cite{miller96} must be known. The derivative is being calculated in two iterations, and each iteration depends only on information kept by the nearest neighboring cells. In this stage, \textbf{cells} are being treated. Throughout all the description, a variable of the cell being treated, $q$, will be denoted as bare with no superscript or subscript. At any given direction, denote the physical quantities of the cell adjacent to the cell treated in the positive direction by a superscript $'P'$, and the physical quantities of the cell adjacent to the cell treated in the negative direction by a superscript $'N'$. The $1^{st}$~order derivative is being calculated using the \textit{generalized minmod} function. This function gives a linear profile of distinct average values, preserving the integral averages over cells without presenting any additional extremum as shown in Fig.~\ref{fig:monotinization-VL97} from \cite{vanleer97}.
The first iteration for calculating the derivative is implemented in the following way. Define
\begin{equation}
F^1=\frac{q-q^{N}}{d\xi}\,, \quad F^2=\frac{q^{P}-q}{d\xi}\,.
\end{equation}
If $F^1 \times F^2 <0$ the derivative is zero. Otherwise, define
\begin{equation}
\begin{array} {rcl}
DR_{LR2} & = & \frac{d\xi^{N}+2d\xi+d\xi^{P}}{2}\,, \\
 \noalign{\medskip}
F^3 & = & \frac{q^{P}-q^{N}}{DR_{LR2}}\,, \\
 \noalign{\medskip}
\end{array}
\end{equation}
and the derivative is temporarily calculated as
\begin{equation}
\pd{q}{\xi}=\frac{F^1+F^2}
{\left|F^1+F^2\right|}min\left(\theta\left|F^1\right|,\theta\left|F^2\right|,\left|F^3\right|\right)\,,
\label{1st-app.-for-derivative}
\end{equation}
where $\theta \in [1,2]$ is a constant value tuned by the user. After the first approximation for all the cells, we make another iteration for a better approximation of the derivative, define
\begin{equation}
\begin{array}{rcl}
DR_{LR4} & = & \frac{d\xi^{N}+4d\xi+d\xi^{P}}{4}\,, \\
 \noalign{\medskip}
F^3 & = & \frac{q^P-\frac{d\xi^P}{4}\pd{q^P}{\xi}-q^N-\frac{d\xi^N}{4}\pd{q^N}{\xi}}{DR_{LR4}}\,, \\
 \noalign{\medskip}
 \end{array}
\label{2nd-def.-of-F3}
\end{equation}
and the new and more exact approximation for the derivative in the cell is given again by Equation~\ref{1st-app.-for-derivative}, with a new definition of $F^3$ given by Equation~\ref{2nd-def.-of-F3}.

In RELDAFNA we calculate the derivatives of the following variables: $\rho$, $\A{v}$, $p$ and $\Gamma$.

Every cell has two interfaces in a specific direction, on its negative and positive sides. With the derivatives one can calculate the \textbf{interface} values induced by the cell, $q\pm\frac{1}{2}\left(\pd{q}{\xi}\right) d\xi$.

\section{Solve} \label{solve}

After a static smooth spatial profile for the hydrodynamic variables is built, RELDAFNA turns to temporal advance of the interface values, either by Runge-Kutta iterations (section~\ref{Runge-Kutta}) or by Characteristic-Tracing method (section~\ref{Characteristic Tracing}).

\subsection{Runge-Kutta Scheme} \label{Runge-Kutta}

One method for the "Solve" part of the RSA strategy is the Runge-Kutta method. This method does not advance in time the interface values built in section~\ref{reconstruct}, but instead it iterates the timestep a fixed number of times, and then interpolates the whole time step using these iterations as will be presented in section~\ref{average}.\\
In this stage, \textbf{interfaces} are being treated. During each RK iteration the interface values at its two sides are calculated, and the Riemann problem on the interface is induced by these values. Each interface has two cells sharing it. Denote the negative and positive sides of the interface by subscripts $1$ and $2$, respectively. At the end of section~\ref{reconstruct} the following variables: $\rho_1$, $\A{v}_1$, $p_1$, $\Gamma_1$, $\rho_2$, $\A{v}_2$, $p_2$ and $\Gamma_2$ were calculated. From them, the whole thermodynamic state at the two sides of the interface is calculated via the EOS. In RELDAFNA, an ideal gas EOS is used and the interface values are, for $i \in \{1,2\}$
\begin{equation}
\begin{array}{rcl}
\left(\rho e\right)_i & = & \frac{p_i}{\gamma-1} \,,\\
 \noalign{\medskip}
h_i & = & 1+\frac{\gamma p_i}{\left(\gamma -1 \right)\rho_i} \,,\\
 \noalign{\medskip}
c_{s_i} & = & \sqrt{\frac{\gamma p_i}{h_i \rho_i}} \,.\\
 \noalign{\medskip}
 \end{array}
 \label{thermodynamic-variables}
\end{equation}
For the velocity and Lorentz factor one has to give special care, since unphysical values could be reconstructed by the derivatives of the velocity (i.e. $\left|\A{v_i}\right|\ge1$ which in SRHD is unphysical). Therefore, one demands consistency between the interface values of the Lorentz factor and the velocity, so that $\left|\A{v_i}\right|<1$ and $\Gamma_i\ge1$. If $\left|\A{v}_i\right| < 1$ for $i \in \{1,2\}$, the interface values of the Lorentz factor are set to be
\begin{equation}
    \Gamma_i=\frac{1}{\sqrt{1-\left|\A{v}_i\right|^2}}\,.
    \label{Lorentz-factor}
\end{equation}
On the other hand, if for any $i \in \{1,2\}$ one has $\left|\A{v}_i\right| \ge 1$ we change the velocity interpolation according to the Lorentz factor interpolation. In order to do that, one has to verify that $\Gamma_i\ge1$. If for any $i \in \{1,2\}$ one has $\Gamma_i<1$ which is unphysical, a piecewise constant interpolation for the Lorentz factor is set $\Gamma_i=\Gamma$ (side $1$ is given the value of the cell in the negative direction to the interface and similarly for side $2$). The velocity at the interface is calculated from the Lorentz factor by
\begin{equation}
\A{v}_i = \sqrt{\frac{\Gamma_i^2-1}{\Gamma_i \left|\A{v}_i\right|}} \, \A{v}_i\,,
\label{velocity_correction}
\end{equation}
and one can verify that now the interface values are consistent, i.e. Equation~\ref{Lorentz-factor} holds. In RELDAFNA a user can change the maximum limit for $\left|\A{v}_i\right|$ to switch from the direct interpolation of the velocity to an interpolation from the Lorentz factor (Of course, this limit must be $<1$).

The interface values at its two sides impose a Riemann problem on the interface, whose solution methods are presented in section~\ref{riemann_solvers}. The outcome of the Riemann problem are the fluxes of the conserved variables which appear again in section~\ref{average} for the iteration final solution.

\subsection{Characteristic Tracing Scheme} \label{Characteristic Tracing}

As an alternative to the Runge-Kutta integration scheme, an operator splitting scheme using characteristic tracing is also implemented in RELDAFNA. In this scheme, one does not have to iterate the whole RSA flow a few times per timestep (i.e. reconstruction and Riemann solution twice or three times per timestep), but instead one advances the static interface values constructed in section~\ref{reconstruct} to a half timestep, and uses that time advanced interface values to impose the Riemann problem on the interface. In this scheme one treats the problem as one dimensional, and uses Strang splitting where the direction being calculated is changing in an alternating order (i.e. from $1$ to the dimension of the problem on odd timesteps and from the dimension to $1$ on even timesteps). In this stage, \textbf{interfaces} are being treated. In the direction being treated, the calculations explained ahead are being preformed over all interfaces.

Since the treatment of the problem is one dimensional, one can utilize the fact that the SRHD
equations in one direction dependence can be rewritten in quasi-linear form \cite{mignone05b}:
\begin{equation}
 \pd{\A{V}}{t} + \frac{1}{h}\A{A}\left(\A{V}\right)\cdot\pd{\A{V}}{x} =
 \A{S}\left(\A{V}\right)\,.
 \label{quasi-linear-from}
\end{equation}
where
\begin{equation}
 \A{V} = \left(\rho, v^1,... v^{dim.}, p\right)^{T}\, ,
\end{equation}
(beware and do not confuse $\A{V}$ with $\A{U}$ which is defined in Equation~\ref{U}), $\A{A}\left(\A{V}\right) = \pd{\mathbf{F}}{\mathbf{U}}$ is the Jacobian and $\A{S}\left(\A{V}\right)$ is a geometric source term. The term $dim.$ is the dimension of the problem being computed. The dimensions of $\A{V}$ and $\A{A}$ depend on the dimension of the problem. As mentioned in section~\ref{logic}, the multidimensional implementation in RELDAFNA is modular and is not affected by the fact that it should treat different problems in different dimensions. However, we think that the most convenient way to present the calculations being made by RELDAFNA, is to give an example of a treatment of the $1^{st}$ direction of a $3$~-dimensional problem. The generalization to other directions or to lower dimensions is straightforward.\\
The Jacobian has the following eigenvalues \cite{donat98}
\begin{eqnarray}\label{eq:eigenvalues}
\nonumber \lambda^1 & = & \frac{u(1-c_s^2)-c_s^2\sqrt{(1-\left|\A{v}\right|^2)(1-u^2-(\left|\A{v}\right|^2-u^2)c_s^2)}}{1-\left|\A{v}\right|^2c_s^2} ,\\
 \lambda^{2,3,4} & = & u , \\ \nonumber
 \lambda^{5} & = & \frac{u(1-c_s^2)+c_s^2\sqrt{(1-\left|\A{v}\right|^2)(1-u^2-(\left|\A{v}\right|^2-u^2)c_s^2)}}{1-\left|\A{v}\right|^2c_s^2} ,
\end{eqnarray}
where $u$ is the velocity component of the $1^{st}$ direction. The left and right eigenstates matrices are given by
%R =
% \left( \begin{array}{ccccc}
% -\frac{\rho \Gamma^2}{c} & \eta \Gamma & -\frac{v_2\left(\Gamma \eta u - c_s\right)}{1-u^2} & -\frac{v_3\left(\Gamma \eta u - c_s\right)}{1-u^2} & -\rho h \Gamma^2c_s\\
% \noalign{\medskip}
% 1 & 0 & 0 & 0 & 0\\
% \noalign{\medskip}
% 0 & 0 & 1 & 0 & 0\\
% \noalign{\medskip}
% 0 & 0 & 0 & 1 & 0\\
% \noalign{\medskip}
% \frac{\rho \Gamma^2}{c} & \eta \Gamma & -\frac{v_2\left(\Gamma \eta u + c_s\right)}{1-u^2} &  -\frac{v_3\left(\Gamma \eta u + c_s\right)}{1-u^2}  & \rho h \Gamma^2c_s\\
% \noalign{\medskip}
%\end{array}\right) \;\;
\begin{eqnarray} \label{eq:eigenstates}
R & = &
 \left( \begin{array}{ccccc}
 -\frac{\rho \Gamma^2}{c_s} & 1 & 0 & 0 &  \frac{\rho \Gamma^2}{c_s} \\
 \noalign{\medskip}
 \eta \Gamma & 0 & 0 & 0 & \eta \Gamma \\
 \noalign{\medskip}
 -\frac{v_2\left(\Gamma \eta u - c_s\right)}{1-u^2} & 0 & 1 & 0 & -\frac{v_2\left(\Gamma \eta u + c_s\right)}{1-u^2} \\
 \noalign{\medskip}
 -\frac{v_3\left(\Gamma \eta u - c_s\right)}{1-u^2} & 0 & 0 & 1 & -\frac{v_3\left(\Gamma \eta u + c_s\right)}{1-u^2} \\
 \noalign{\medskip}
 -\rho h \Gamma^2c_s & 0 & 0 & 0 & -\rho h \Gamma^2c_s\\
 \noalign{\medskip}
\end{array}\right) \\ \nonumber
L & = &
 \left( \begin{array}{ccccc}
 0 & 1 & 0 & 0 & 0  \\
 \noalign{\medskip}
 \frac{1}{2\eta\Gamma} & 0 & \frac{uv_2}{1-u^2} & \frac{uv_3}{1-u^2} & \frac{1}{2\eta\Gamma} \\
 \noalign{\medskip}
 0 & 0 & 1 & 0 & 0 \\
 \noalign{\medskip}
 0 & 0 & 0 & 1 & 0 \\
 \noalign{\medskip}
 -\frac{1}{2h\rho\Gamma^2c_s} & -\frac{1}{h c_s^2} & \frac{v_2}{\Gamma^2\rho h \left(1-u^2\right)} & \frac{v_3}{\Gamma^2\rho h \left(1-u^2\right)} & \frac{1}{2h\rho\Gamma^2c_s} \\
 \noalign{\medskip}
\end{array}\right) \,,
\end{eqnarray}
%
%\begin{equation}
% \begin{array}{rcl}
% r^1(1)=-\frac{\rho\Gamma^2}{c} \, & , & \quad r^1(1+dir.)=\eta\Gamma ,\\
% \noalign{\medskip}
% r^1(\stackrel{dir.^1 \neq dir.}{1+dir.^1})=\frac{v_{dir.^1}(\Gamma \eta u-c)}{1-u^2} \, & , & \quad r^1(2+dim.)=-\rho h\Gamma^2c ,\\
% \noalign{\medskip}
% r^{1+dir.}(1)=1 \, & , & \quad r^{\stackrel{dir.^1 \neq dir.}{1+dir.^1}}(\stackrel{dir.^1 \neq dir.}{1+dir.^1})=1 ,\\
% \noalign{\medskip}
% r^{2+dim.}(1)=\frac{\rho\Gamma^2}{c} \, & , & \quad r^{2+dim.}(1+dir.)=\eta\Gamma ,\\
% \noalign{\medskip}
% r^{2+dim.}(\stackrel{dir.^1 \neq dir.}{1+dir.^1})=-\frac{v_{dir.^1}(\Gamma \eta u+c)}{1-u^2} \, & , & \quad r^{2+dim.}(2+dim.)=\rho h\Gamma^2c ,\\
% \noalign{\medskip}
% l^1(1+dir.)=\frac{1}{2\eta\Gamma} \, & , & \quad l^1(2+dim.)=-\frac{1}{2h\rho\Gamma^2c} ,\\
% \noalign{\medskip}
% l^{1+dir.}(1)=1 \, & , & \quad l^{1+dir.}(2+dim.)=-\frac{1}{hc^2} ,\\
% \noalign{\medskip}
% l^{\stackrel{dir.^1 \neq dir.}{1+dir.^1}}(1+dir.)=\frac{uv_{dir.^1}}{1-u^2} \, & , & \quad l^{\stackrel{dir.^1 \neq dir.}{1+dir.^1}}(\stackrel{dir.^1 \neq dir.}{1+dir.^1})=1 ,\\
% \noalign{\medskip}
% l^{\stackrel{dir.^1 \neq dir.}{1+dir.^1}}(2+dim.) & = & \frac{v_{dir.^1}}{\Gamma^2\rho h(1-u^2)} , \\
% \noalign{\medskip}
% l^{2+dim.}(1+dir.)=\frac{1}{2\eta\Gamma} \, & , & \quad l^{2+dim.}(2+dim.)=\frac{1}{2\rho h\Gamma^2c}
% ,\\
% \noalign{\medskip}
% \end{array}
%\end{equation}
where
\begin{equation}
\eta=\sqrt{1-u^2-c_s^2(\left|\A{v}\right|^2-u^2)}\,.
\end{equation}
After laying out the quasi-linear form, RELDAFNA turns to advance the static interface values from section~\ref{reconstruct} half of a time step, using the characteristics information influencing the interface from the cells sharing it. This step is different for PLM and PPM spatial reconstructions (Similar to descriptions given in e.g, \cite{miller96,miller02}). The predictor step for the PPM spatial reconstruction is given in appendix~\ref{CT-PPM}. The predictor step for the PLM reconstruction is presented in detail ahead.\\

\subsubsection{Predictor step for PLM spatial reconstruction} \label{CT-PLM}

Following the notation of section~\ref{Runge-Kutta}, at the beginning of this stage the following values are known: $\rho_1$, $\A{v}_1$, $p_1$, $\Gamma_1$, $\rho_2$, $\A{v}_2$, $p_2$ and $\Gamma_2$. These variables are used to construct the whole static interface state the same way as in section~\ref{Runge-Kutta}, i.e. an execution of Equations~\ref{thermodynamic-variables}-\ref{velocity_correction}.

We start the description of this stage by considering the half time step advance of the interface values on its negative side. As a result, the bare values will denote the values of the \textbf{cell} on the negative side of the interface. The eigenvalues of the Jacobian (Equations~\ref{eq:eigenvalues}) are calculated using the cell values. The geometric values of the interface are
\begin{equation}
\begin{array}{rcl}
R_1 & = & R+\frac{dr}{2} \,, \\
 \noalign{\medskip}
\xi_1 & = & \xi+\frac{d\xi}{2}.\\
 \noalign{\medskip}
\end{array}
\label{places-interface}
\end{equation}
$\rho_1$, $\A{v}_1$, $p_1$, $h_1$, $c_{s_1}$ and $\Gamma_1$ are used to calculate the left and right eigenstates matrices (Equations~\ref{eq:eigenstates}). The first time dependent approximation of the interface values is calculated using the positive eigenvalues. For every $\lambda^i>0\,, i\in\{1 \ldots 5 \}$ define
\begin{equation}
\begin{array}{rcl}
\xi^\lambda & = & \xi(R_1-\lambda \Delta t) \, , \\
 \noalign{\medskip}
d\xi^\lambda & = & \frac{\xi^\lambda+\xi_1}{2}-\xi\,.\\
 \noalign{\medskip}
\end{array}
\end{equation}
With the eigenvalues and the eigenstates, RELDAFNA turns to calculate the half timestep advanced values of $\A{V}$, that will be fed to the Riemann problem as the negative side values,
\begin{equation}
\A{V}_1^\frac{\Delta t}{2}=\A{V}+d\xi^{\lambda^5}\left(\pd{\A{V}}{\xi}\right)+\sum_{i=1}^{4}
\frac{max\left(\lambda^i,0\right)}{\lambda^i}\left(d\xi^{\lambda^i}-d\xi^{\lambda^5}\right)
\left(L^{:i}\bullet\left(\pd{\A{V}}{\xi}\right)\right)R^{:i}
\,,
\end{equation}
where the notation $L^{:i}$ refers to the $i^{th}$ column of the matrix $L$ (which is a one dimensional vector).

When problems with non-Cartesian geometries are treated, geometrical factors are added to $\A{V}_1^\frac{\Delta t}{2}$. In RELDAFNA, $1$~dimensional spherical and $1$ and $2$~dimensional cylindrical geometries are available. That is, when the radial direction is dealt, the following geometrical factors are defined
\begin{equation}
\begin{array}{rcl}
Cylindrical: \;\; G & = & \frac{1}{R} \,, \\
 \noalign{\medskip}
Spherical: \;\; G & = & \frac{2}{R} \,. \\
 \noalign{\medskip}
\end{array}
\end{equation}
The interface values on its negative side at a half time step are given by
\begin{equation}
\begin{array}{rcl}
\rho_1^{\frac{\Delta t}{2}} & = & \A{V}_1^\frac{dt}{2}\left(1\right)-\frac{\rho^1\times G\times u^1dt}{2\Delta}\,,\\
 \noalign{\medskip}
\A{v}_1^{\frac{\Delta t}{2}} & = & \A{V}_1^\frac{dt}{2}\left(2\ldots 4\right)+\frac{\A{v}^1\times G\times u^1dt\left(c_s^1\right)^2}{2\Delta\left(\Gamma^1\right)^2}\,,\\
 \noalign{\medskip}
p_1^{\frac{\Delta t}{2}} & = &
\A{V}_1^\frac{dt}{2}\left(5\right)-\frac{\rho^1 h^1 \left(c_s^1\right) ^2 \times G\times u^1dt}{2\Delta}\,,\\
 \noalign{\medskip}
 \label{ro-v-p-calc-negative}
\end{array}
\end{equation}
where $\Delta = 1-\left|\A{v}_1\right|^2c_{s_1}^2$. The rest of the thermodynamic variables at the half timestep are calculated according to the EOS as in Equation~\ref{thermodynamic-variables}.

The half time step advance of the interface values on its positive side, is calculated in a similar way using the values of the \textbf{cell} on the positive side of the interface. As a result, the bare values will denote the values of the \textbf{cell} on the positive side of the interface. Since it is very similar to the procedure described by Equations~\ref{places-interface}-\ref{ro-v-p-calc-negative}, we will describe it shortly with the changes needed to be made to adjust it to the opposite direction. The eigenvalues of the Jacobian (Equations~\ref{eq:eigenvalues}) are calculated using the cell values. The geometric values of the interface are the same as before. The left and right eigenstates matrices (Equations~\ref{eq:eigenstates}) are calculated using the values on the \textbf{positive} side. The first time dependent approximation of the interface values is calculated using the \textbf{negative} eigenvalues. The values $\xi^\lambda$ and $d\xi^\lambda$ are calculated as before using the values of the cell in the \textbf{positive} side of the interface. With the eigenvalues and the eigenstates, RELDAFNA turns to calculate the half timestep advanced values of $\A{V}$, that will be fed to the Riemann problem as the \textbf{positive} side values,
\begin{equation}
\A{V}_2^\frac{\Delta t}{2}=\A{V}+d\xi^{\lambda^1}\left(\pd{\A{V}}{\xi}\right)+\sum_{i=2}^{5}
\frac{min\left(\lambda^i,0\right)}{\lambda^i}\left(d\xi^{\lambda^i}-d\xi^{\lambda^1}\right)
\left(L^{:i}\bullet\left(\pd{\A{V}}{\xi}\right)\right)R^{:i}
\,.
\end{equation}
The geometrical factors are calculated in a similar way, and the half timestep advanced interface values from the positive side of the interface are being calculated in the same way as Equations~\ref{ro-v-p-calc-negative}.

Similar to the static reconstruction, one can get unphysical interface values at half timestep for the Lorentz factor and the velocity components. If for all $i\in\{1,2\}$, $\left|\A{v}_i^{\frac{\Delta t}{2}}\right|<1$, the Lorentz factor is advanced by its definition,
\begin{equation}
\Gamma_i^{\frac{\Delta t}{2}}=\frac{1}{\sqrt{1-\left|\A{v}_i^{\frac{\Delta t}{2}}\right|^2}}.
\end{equation}
If on the other hand, for any $i\in\{1,2\}$, $\left|\A{v}_i^{\frac{\Delta t}{2}}\right|\ge1$, the Lorentz factors are advanced via their own interpolation and then impose a renormalization constraint on the velocities, in the following manner:\\
On the negative side the Lorentz factor is advanced using eigenvalue no.~$2$ (the one attached to the direction being treated, beware the superscript $2$ does not mean squared)
\begin{equation}
\Gamma_1^{\frac{\Delta t}{2}}=\Gamma +\frac{max\left(\lambda^2,0\right)}{\lambda^2}\left(\pd{\Gamma}{\xi}\right)d\xi^{\lambda^2}\,.
\end{equation}
For the positive side, the procedure is similar only using the values of the \textbf{positive} side of the interface,
\begin{equation}
\Gamma_2^{\frac{\Delta t}{2}}=\Gamma +\frac{min\left(\lambda^2,0\right)}{\lambda^2}\left(\pd{\Gamma}{\xi}\right)d\xi^{\lambda^2}\,.
\end{equation}
In cases where this interpolation fails to give a physically meaningful result, i.e. if for any $i\in\{1,2\}$, $\Gamma_i^{\frac{\Delta t}{2}}<1$, the static interpolation is not advanced in time, i.e.
\begin{equation}
\Gamma_i^{\frac{\Delta t}{2}}=\Gamma_i\,.
\label{simple-PLM-Gamma}
\end{equation}
After the half timestep advance of the interface Lorentz factors on its two sides, the half timestep advance of the velocity components are renormalized in the same manner as Equations~\ref{velocity_correction}.

\subsection{Riemann Solvers} \label{riemann_solvers}

\begin{figure*}[h]
\centering
\includegraphics[scale=0.7, trim= 120mm 0 170mm 0]{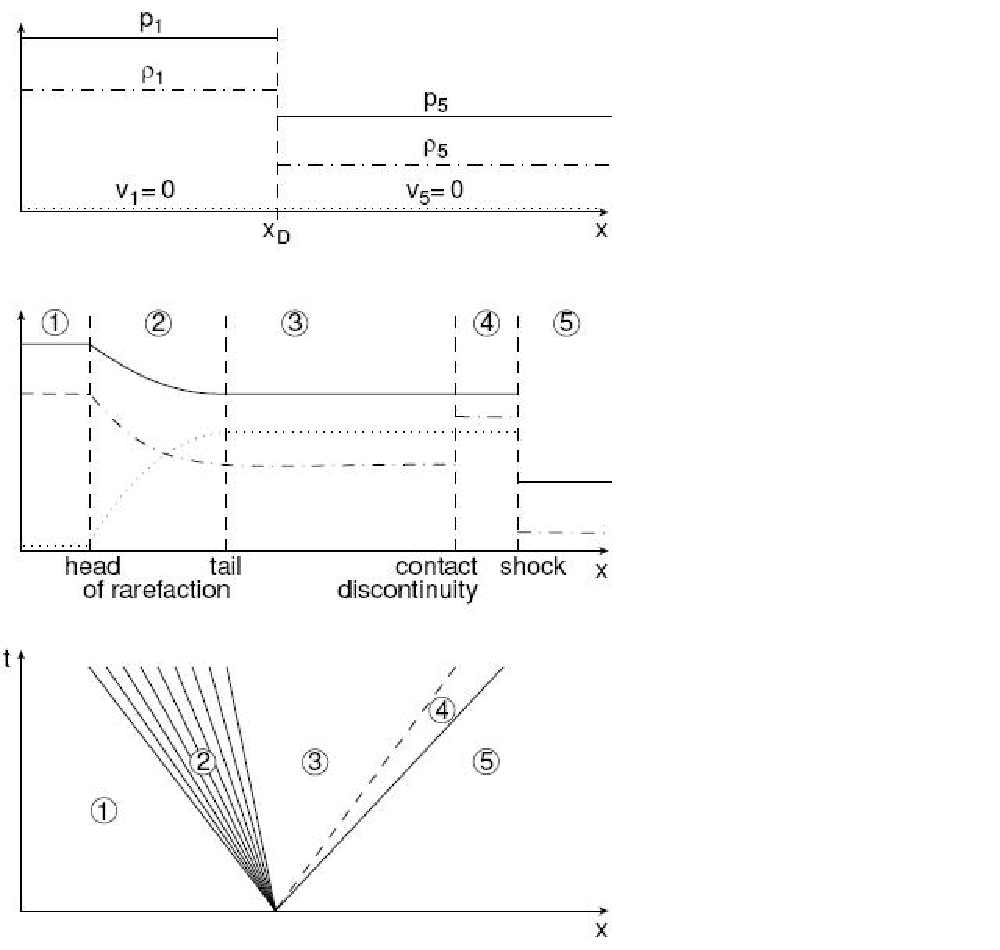}
\caption{Schematic solution of a Riemann problem.  The initial state at $t = 0$ (\textit{top}) consists of two constant states (1) and (5) with $p_1 > p_5$, $\rho_1 > \rho_5$ and $v_1 = v_2 = 0$ separated by a diaphragm at $x_D$. The evolution of the flow pattern once the diaphragm is removed (\textit{middle}) is illustrated in a  space--time diagram (\textit{bottom}) with a shock wave (solid line) and a contact discontinuity (dashed line) moving to the right. The bundle of solid lines represents a rarefaction wave propagating to the left.\protect\label{fig:riemannMM03}}
\end{figure*}

The Riemann problem consists of computing the breakup of a discontinuity, which initially separates two arbitrary constant states L (left) and R (right) in the gas (see Fig.~\ref{fig:riemannMM03} (from \cite{marti03}) with L $\equiv 1$ and R $\equiv 5$). Godunov methods use the solutions of such problems between adjacent numerical cells to advance the average cell values with time. We will briefy present the basic notions leading to the exact solution of the Riemann problem based on \cite{marti03}. We refer the reader to \cite{marti03,marti94,marti96,pons00} for a more comprehensive description of the exact solution of the Riemann problem.

The solution of this problem is self-similar, because it only depends on the two constant states defining the discontinuity $\A{V}_L$ and $\A{V}_R$, and on the ratio $\left(x-x_0\right)/\left(t-t_0\right)$, where $x_0$ and $t_0$ are the initial location of the discontinuity and the time of breakup, respectively. The discontinuity decays into two elementary nonlinear waves (shocks or rarefactions) which move in opposite directions towards the initial left and right states. Between these waves two new constant states $\A{V}_{L*}$ ($\equiv 3$ in Fig.~\ref{fig:riemannMM03}) and $\A{V}_{R*}$ ($\equiv 4$ in Fig.~\ref{fig:riemannMM03}) appear, which are separated from each other through a contact discontinuity moving with the fluid. Across the contact discontinuity the density exhibits a jump, whereas pressure and velocity are continuous.

The self-similar character of the flow through rarefaction waves and the \textit{Rankine-Hugoniot} conditions across shock fronts, result in algebraic connections between the intermediate states $\A{V}_{S*}$ ($S=L,R$) with the corresponding initial states $\A{V}_S$. They also allow one to express the fluid flow velocity in the intermediate states $v_{S*}$ as a function of the pressure $p_{S*}$ in these states. Adding these relations together with the demand that the pressure and normal velocity will be continuous across the contact, one gets the following two relations which are solved iteratively,
\begin{equation}
\begin{array}{rcl}
p_*=p_{L*}=p_{R*}\,,\\
 \noalign{\medskip}
v_{L*}(p_{*})=v_{R*}(p_{*}) \, .\\
 \noalign{\medskip}
\end{array}
\end{equation}
In RELDAFNA we have implemented an exact Riemann Solver for an ideal gas equation of state. The implementation is identical to the one presented in \cite{marti94,pons00,marti03,marti96}. The solution of the above equations result with an exact solution to the Riemann problem. Since the solution is usually obtained with iterative methods it is very expensive computationally even for an ideal gas equation of state. Therefore, approximate Riemann solvers were developed for a reasonable reduction of computation time.

Each of the Riemann solvers uses as input the interface values from both sides, which are computed as presented in sections~\ref{Runge-Kutta}-\ref{Characteristic Tracing}. In this stage \textbf{interfaces} are treated. Therefore, following the usual notation, The given values are $\rho_i$, $\left(\rho e\right)_i$, $\A{v}_i$, $p_i$, $c_{s_i}$ and $\Gamma_i$ where $i \in \{1,2\}$, either static when using Runge-Kutta method or half timestep advanced when using Characteristic Tracing method.

From these variables the conserved variables of the interface on its both sides are calculated, from Equations~\ref{conserved}, i.e. for $i\in\{1,2\}$
\begin{equation}
\begin{array}{rcl}
D_i & = & \rho_i \Gamma_i \,, \\
 \noalign{\medskip}
\A{S}_i & = & \left(\rho_i+\left(\rho e\right)_i+p_i\right) \left(\Gamma_i\right)^2 \A{v}_i \,, \\
 \noalign{\medskip}
\tau_i & = & \left(\rho_i+\left(\rho
e\right)_i+p_i\right) \left(\Gamma_i\right)^2 -p_i -D_i \,.\\
 \noalign{\medskip}
\end{array}
\end{equation}
We will give ahead a short description of the approximate Riemann solvers implemented in RELDAFNA (HLL and HLLC) (based on \cite{mignone05a,toro94}) and then present the detailed implementation. In general, the approximate solutions are not accurate as stand alone Riemann solvers. The Godunov method uses them only to calculate fluxes of average values across interfaces. The averaging of the variables over the whole cell volume allows one to use approximate solutions of the Riemann problem without presenting yet an additional source of error into the solution. In RELDAFNA it has been found that problems solved using the approximate solvers, result with solutions very similar to the results using the exact solver.

\subsubsection{HLL-scheme} \label{HLL}

Additional assumptions on the Riemann problem make it easier to solve. In \cite{harten83} the following assumptions were added to the Riemann problem:

\begin{itemize}

\item The only waves present are the left and right non-linear waves.

\item One can estimate the bounds for the propagation velocities of these non-linear waves from the initial conditions.

\end{itemize}

These assumptions wipe out the contact discontinuity and average the three states ($2,3,4$ in Fig.~\ref{fig:riemannMM03}) into a single constant state. This approximation reduces the number of states from $5$ (or $4$ since states $2$ and $3$ are actually one state) (see Fig.~\ref{fig:riemannMM03}) to $3$:
\begin{equation}\label{eq:hll_states}
  \A{U}(0,t) = \left\{\begin{array}{cll}
     \A{U}_L      & \; \textrm{if} \; & \lambda_L \ge 0 \,, \\ \noalign{\medskip}
     \A{U}^{hll}  & \; \textrm{if} \; & \lambda_L \le 0 \le \lambda_R \,, \\ \noalign{\medskip}
     \A{U}_R      & \; \textrm{if} \; & \lambda_R \le 0 \,, \\ \noalign{\medskip}
\end{array}\right.
\end{equation}
where the single state $\A{U}^{hll}$ is calculated using the fastest and slowest signal velocities $\lambda_L$ and $\lambda_R$ which are calculated from the initial states (see the second assumption):
\begin{equation}\label{eq:hll_state}
 \A{U}^{hll} = \frac{\lambda_R \A{U}_R - \lambda_L\A{U}_L +
                      \A{F}\left(\A{U}_L\right)- \A{F}\left(\A{U}_R\right)}{\lambda_R - \lambda_L} \,,
\end{equation}
where the notation $\A{F}\left(\A{U}\right)$ is based on Equations~\ref{U}-\ref{F}. Equation (\ref{eq:hll_state}) represents the integral average of the solution of the Riemann problem over the wave fan \cite{toro94}. This allows us to write the interface numerical flux as:
\begin{equation}\label{eq:hll_fluxes}
 \A{f} = \left\{\begin{array}{cll}
   \A{F}_L     & \;  \textrm{if} \;&   \lambda_L \ge 0 \,, \\ \noalign{\medskip}
   \A{F}^{hll} & \;  \textrm{if} \;&   \lambda_L \le 0 \le \lambda_R \,, \\ \noalign{\medskip}
   \A{F}_R     & \;  \textrm{if} \;&   \lambda_R \le 0 \,, \\ \noalign{\medskip}
\end{array}\right.
\end{equation}
where
\begin{equation}\label{eq:hll_flux}
 \A{F}^{hll} = \frac{\lambda_R\A{F}\left(\A{U}_L\right) - \lambda_L\A{F}\left(\A{U_R}\right) + \lambda_R\lambda_L \left(\A{U}_R - \A{U}_L\right)}
            {\lambda_R - \lambda_L}   \,.
\end{equation}
The maximum and minimum velocities of propagation are calculated as a relativistic summation of the velocity component of the direction being treated and the speed of sound at the interface from both sides,
\begin{equation}
\begin{array}{rcl}
a_1 & = & min(\frac{u_1-c{s_1}}{1-u_1c_{s_1}},\frac{u_2-c_{s_2}}{1-u_2c_{s_2}},0)\,,\\
 \noalign{\medskip}
a_2 & = &
max(\frac{u_1+c_{s_1}}{1+u_1c_{s_1}},\frac{u_2+c_{s_2}}{1+u_2c_{s_2}},0)\,.\\
 \noalign{\medskip}
\end{array}
\label{HLL-propagation}
\end{equation}
If desired, a more complicated (and expensive) calculation for the propagation velocities is available in RELDAFNA using the values in Equations~\ref{eq:eigenvalues}, which represent the maximum and minimum propagation velocities. It has been checked that the simple summation in Equation~\ref{HLL-propagation} is sufficient and one does not need to use the expensive calculation in the Riemann solver.
%Define
%\begin{equation}
%\begin{array}{rcl}
%\lambda_\pm^i & = & \frac{u^i\left(1-\left(c_s^i\right)^2\right) \pm c_s^i\sqrt{\left(1-\left|\A{v}^i\right|^2\right)\left(1-\left(u^i\right)^2-\left(\left|\A{v}i\right|^2
% -\left(u^i\right)^2\right)\right)\left(c_s^i\right)^2}}{1-\left|\A{v}^i\right|^2\left(c_s^i\right)^2}\,,\\
% \noalign{\medskip}
%\end{array}
%\end{equation}
%and the propagation velocities are calculated as
%\begin{equation}
%\begin{array}{rcl}
%a_1 & = & min\left(\lambda_-^N,\lambda_-^P\right)\,,\\
% \noalign{\medskip}
%a_2 & = & max\left(\lambda_+^N,\lambda_+^P\right)\,.\\
% \noalign{\medskip}
%\end{array}
%\end{equation}
The fluxes are computed using Equations~\ref{eq:hll_flux} to be
\begin{equation}
\begin{array}{rcl}
\mathbf{F}^* & = & \frac{a_2\mathbf{F}\left(\mathbf{U}_1\right)-a_1\mathbf{F}\left(\mathbf{U}_2\right)
+a_1a_2\left(\mathbf{U}_2-\mathbf{U}_1\right)}{a_2-a_1}\,,\\
 \noalign{\medskip}
\end{array}
\end{equation}
and the pressure at the * state is computed in a similar manner
\begin{equation}
p^*=\frac{a_2p_1-a_1p_2}{a_2-a_1}\,.
\end{equation}

\subsubsection{HLLC-scheme} \label{HLLC}

Although the HLL assumptions lead to a computationally inexpensive scheme, its inability to resolve contact discontinuities or tangential waves led \cite{toro94} to restore the contact surface into the assumptions made in the HLL scheme for Newtonian hydrodynamics. The relativistic version was proposed by \cite{mignone05a}.

\begin{figure*}[h]
\centering
\includegraphics[scale=0.5, trim= 80mm 7mm 80mm 0]{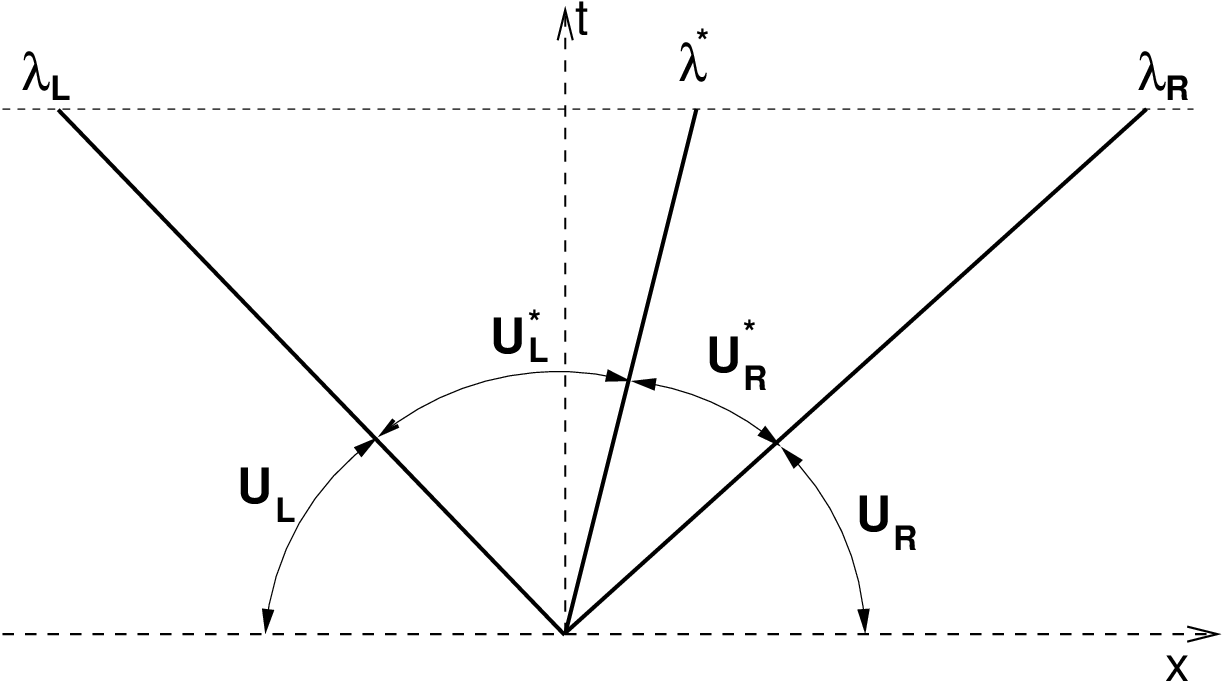}
\caption{Graphical representation of the Riemann fan in the $x-t$ plane.
          The two initial states $\A{U}_L$ and $\A{U}_R$ decay
          into two nonlinear waves (with speeds $\lambda_L$ and
          $\lambda_R$) and a linear contact wave with velocity
          $\lambda^*$. The resulting wave pattern divides the
          $x-t$ plane into four regions each defining
          a constant state: $\A{U}_L$, $\A{U}^*_L$, $\A{U}^*_R$ and $\A{U}_R$.\protect\label{fig:fanMB05}}
\end{figure*}

The HLLC scheme removes the assumption that the middle states can be averaged to one single state. This leads to a restoration of the full wave structure inside the Riemann fan by replacing the single averaged state defined by Equation~\ref{eq:hll_state} with two approximate states, $\A{U}^*_L$ and $\A{U}^*_R$ as seen in Fig.~\ref{fig:fanMB05}.

These two states are separated by a middle contact wave which is assumed to have constant speed $\lambda^*$, so that Equations~\ref{eq:hll_states} are now replaced by
\begin{equation}\label{eq:hllc_states}
  \A{U}(0,t) = \left\{\begin{array}{ccc}
   \A{U}_L    & \quad \textrm{if} & \; \lambda_L \ge 0    \,,              \\ \noalign{\medskip}
   \A{U}^*_L  & \quad \textrm{if} & \; \lambda_L \le 0 \le \lambda^* \,,\\ \noalign{\medskip}
   \A{U}^*_R  & \quad \textrm{if} & \; \lambda^* \le 0 \le \lambda_R \,,\\ \noalign{\medskip}
   \A{U}_R    & \quad \textrm{if} & \; \lambda_R \le 0              \,, \\ \noalign{\medskip}
\end{array}\right.
\end{equation}
and Equations~\ref{eq:hll_flux} are replaced by
\begin{equation}\label{eq:hllc_flux}
 \A{f} = \left\{\begin{array}{ccc}
   \A{F}\left(\A{U}_L\right)    & \quad \textrm{if} & \; \lambda_L \ge 0              \,, \\ \noalign{\medskip}
   \A{F}^*_L  & \quad \textrm{if} & \; \lambda_L \le 0 \le \lambda^*\,, \\ \noalign{\medskip}
   \A{F}^*_R  & \quad \textrm{if} & \; \lambda^* \le 0 \le \lambda_R\,, \\ \noalign{\medskip}
   \A{F}\left(\A{U}_R\right)    & \quad \textrm{if} & \; \lambda_R \le 0   \,.          \\ \noalign{\medskip}
\end{array}\right.
\end{equation}
The intermediate state fluxes $\A{F}^*_L$ and $\A{F}^*_R$ may be expressed in terms of $\A{U}^*_L$ and $\A{U}^*_R$ from the Rankine-Hugoniot jump conditions
\begin{equation}\label{eq:jump_1}
  \lambda \left(\A{U}^* - \A{U}\right) =
  \A{F}^* - \A{F} \,.
\end{equation}
If $\lambda_L$ and $\lambda_R$ are calculated using the second assumption of the HLL scheme, Equation~\ref{eq:jump_1} represent a system of $2n$ equations (where $n$ is the number of components of $\A{U}$) for the $4n + 1$ unknowns $\A{U}^*_L$,  $\A{U}^*_R$, $\A{F}^*_L$, $\A{F}^*_R$ and $\lambda^*$. The pressure and normal velocity must be continuous across the contact discontinuity, so the following must hold, $v^*_R = v^*_L$, $p^*_R = p^*_L$ and $\lambda^* = v_L^* = v_R^*$. The assumption made instead of the first assumption of HLL is that $\A{F}^* = \A{F}\left(\A{U}^*\right)$. Writing explicitly Equation~\ref{eq:jump_1} for the left or the right state along with the continuity constraints and the above assumption results in a quadratic equation for $\lambda^*$. From the propagation velocity of the contact discontinuity $p^*$ and $\A{U}^*$ are obtained.

As presented in section~\ref{HLL}, the maximum and minimum propagation velocities are calculated. For the velocity of the middle wave that the HLLC scheme adds to HLL, the quadratic equation whose coefficients are
\begin{equation}
\begin{array}{rcl}
A & = & a_1a_2\left(\tau_2+D_2-\tau_1-D_1\right)+a_2S_{1_1}-a_1S_{1_2}\,,\\
 \noalign{\medskip}
B & = & a_1\left(\tau_1+D_1\right)-S_{1_1}-a_2S_{1_1}\left(u_1-a_1\right)-p_1a_2-\\
 \noalign{\medskip}
 & - & a_2\left(\tau_2+D_2\right)+S_{1_2}+a_1S_{1_2}\left(u_2-a_2\right)+p_2a_1\,,\\
 \noalign{\medskip}
C & = & S_{1_1}\left(u_1-a_1\right)+p_1-S_{1_2}\left(u_2-a_2\right)-p_2\,,\\
 \noalign{\medskip}
\end{array}
\end{equation}
is solved. Remember that we are presenting a treatment of the $1^{st}$ direction, and that is the reason why $S_1$ appears in the last equation. The minus sign solution of the quadratic equation $Ax^2+Bx+C=0$, is the propagation velocity of the middle wave, $s_M$. The state vector for the middle state is calculated using Equations~\ref{eq:hllc_states} according to the direction of propagation as follows,\\
if $a_1\leq0$ and $s_M>0$ then
\begin{equation}
\begin{array}{rcl}
p^* & = & \frac{\left(\tau_1+D_1\right)a_1s_M+S_{1_1}\left(u_1-a_1-s_M\right)+p_1}{1-s_Ma_1}\,,\\
 \noalign{\medskip}
D^* & = & \frac{D_1\left(a_1-u_1\right)}{a_1-s_M}\,,\\
 \noalign{\medskip}
\A{S}^* & = & \frac{\A{S}_1\left(a_1-u_1\right)}{a_1-s_M}\,,\\
 \noalign{\medskip}
S_1^*& = & \frac{S_{1_1}\left(a_1-u_1\right)+p^*-p_1}{a_1-s_M}\,,\\
 \noalign{\medskip}
\tau^* & = & \frac{\tau_1\left(a_1-u_1\right)-p_1u_1+p^*s_M}{a_1-s_M}\,,\\
 \noalign{\medskip}
\end{array}
\end{equation}
if $a_2\geq0$ and $s_M \leq 0$ then
\begin{equation}
\begin{array}{rcl}
p^* & = & \frac{\left(\tau_2+D_2\right)a_2s_M+S_{1_2}\left(u_2-a_2-s_M\right)+p_2}{1-s_Ma_2}\,,\\
 \noalign{\medskip}
D^* & = & \frac{D_2\left(a_2-u_2\right)}{a_2-s_M}\,,\\
 \noalign{\medskip}
\A{S}^* & = & \frac{\A{S}_2\left(a_2-u_2\right)}{a_2-s_M}\,,\\
 \noalign{\medskip}
S_1^*& = & \frac{S_{1_2}\left(a_2-u_2\right)+p^*-p_2}{a_2-s_M}\,,\\
 \noalign{\medskip}
\tau^* & = & \frac{\tau_2\left(a_2-u_2\right)-p_2u_2+p^*s_M}{a_2-s_M}\,,\\
 \noalign{\medskip}
\end{array}
\end{equation}
if $a_1>0$ then
\begin{equation}
\begin{array}{rcl}
\mathbf{U}^* & = & \mathbf{U}_1\,,\\
 \noalign{\medskip}
p^* & = & p_1\,,\\
 \noalign{\medskip}
s_M & = & u_1\,,\\
 \noalign{\medskip}
\end{array}
\end{equation}
if $a_2<0$ then
\begin{equation}
\begin{array}{rcl}
\mathbf{U}^* & = & \mathbf{U}_2\,,\\
 \noalign{\medskip}
p^* & = & p_2\,,\\
 \noalign{\medskip}
s_M & = & u_2\,.\\
 \noalign{\medskip}
\end{array}
\end{equation}
The fluxes are finally calculated using the assumption made in the HLLC method
\begin{equation}
\begin{array}{rcl}
\mathbf{F}_D & = & D^*s_M\,,\\
 \noalign{\medskip}
\mathbf{F}_S & = & \A{S}^*s_M \,,\\
 \noalign{\medskip}
\mathbf{F}_\tau & = & S_1^*-D^*s_M\,.\\
 \noalign{\medskip}
\end{array}
\end{equation}

\section{Average} \label{average}

The average phase of the RSA flow presents the transfer of the fluxes over interfaces from one cell to its neighbors. In this stage \textbf{cells} are treated. Again, we are treating one direction only, and in that direction every cell has two interfaces. Following the notation of section~\ref{reconstruct}, denote by a superscript $N$ the interface in the \textbf{negative} side of a given cell, and by a superscript $P$ the interface in the \textbf{positive} side of a given cell. Fluxes are defined across interfaces, and average cell values are defined within cells. In the following equation, average cell values will appear with superscripts $n$ and $n+1$ denoting the most updated value before the current RSA flow and the updated value after the RSA flow, respectively. The averaging is being done with the aid of Equation~\ref{finite-difference},
\begin{equation}
\begin{array}{rcl}
D^{n+1} & = & D^n+\Delta t\left(F_D^N \frac{A^N}{V}-F_D^P \frac{A^P}{V}\right) \,,\\
 \noalign{\medskip}
S_i^{n+1} & = & S_i^n+\Delta t\left(F_{S_i}^N \frac{A^N}{V}-F_{S_i}^P \frac{A^P}{V}\right)
 +\delta_{1i}\left(p^{*^N}\frac{1}{dr}-p^{*^P}\frac{1}{dr}\right) \,, i\in\{1\ldots3\}\,,\\
 \noalign{\medskip}
\tau^{n+1} & = & \tau^n+\Delta t\left(F_\tau^N \frac{A^N}{V}-F_\tau^P \frac{A^P}{V}\right) \,,\\
 \noalign{\medskip}
 \label{time-advance}
\end{array}
\end{equation}
where $V$ is the volume of the cell, $A$ is the area of the interface the flux is crossing, $dr$ is the size of the cell in the direction being treated. Remember, of course, that we are demonstrating the treatment of the $1^{st}$ direction, and that is why $\delta_{1i}$ appears in the above equation. Since we use AMR in RELDAFNA, cells with different refinement levels can share an interface, and a big cell can have two different fluxes on one side, since that side is composed of two interfaces, and for the big cell the flux would be an average of these two fluxes (see Fig.~\ref{fig:2d-fluxes}). Therefore, the area of the interface has a superscript, and is always taken to be the area of half the interface for the big cell. Again, for the pressure spatial derivative we should consider the refinement level ratio, which also comes to compensate for the different spatial resolution between two neighboring cells. In RELDAFNA there are no neighboring cells differing in size by a factor $>2$, and so this ratio is $\frac{1}{2}$ for $2$~-dimensional calculations and $\frac{1}{4}$ for $3$~-dimensional calculations, of course only if the cells sharing the interface differ in size. In $1$~-dimensional calculations there is no change in the interface when it is shared by cells with different sizes. As a result, for multidimensional calculations, there is an additional factor if the cell sharing the $i$ interface ($i\in\{N,P\}$) is smaller than the calculated cell in Equation~\ref{time-advance}.

\begin{itemize}
\item \textbf{Characteristic Tracing:} Calculation of the fluxes over all the interfaces in all directions (in an alternating order), using one RSA flow for any dimension in one timestep and Equation~\ref{time-advance} to describe the time advance, since there is a predictor step before the Riemann solver was used.

\item \textbf{Runge-Kutta:} A number of iterations are executed every timestep. Each iteration involves one RSA flow for each direction. The number of iterations in one timestep is chosen by the user. In RELDAFNA $2^{nd}$ and $3^{rd}$~-order RK methods are available, but an extension to higher order methods is straightforward. The different iterations of the RSA flow in one timestep are averaged with different weights to advance the conserved variables the whole timestep.\\
    The input of the $1^{st}$ iteration is of course the values at the beginning of the timestep and the input of the $2^{nd}$ iteration is the state at the end of the $1^{st}$ iteration.

     \begin{itemize}
     \item \textbf{$2^{nd}$~-order RK:} The final state at the end of the timestep is given by an average
         \begin{equation}\label{RK2}
             \A{U}^{n+1}=\frac{1}{2}\left(\A{U}^n+\A{U}_2^n\right)\,,
         \end{equation}
         where $\A{U}_2^n$ is the outcome of the $2^{nd}$ iteration.

     \item \textbf{$3^{rd}$~-order RK:} The input of the $3^{rd}$ iteration is given by an average $\frac{3}{4}\A{U}^n+\frac{1}{4}\A{U}_2^n$, and the final state at the end of the timestep is given by an average
         \begin{equation}\label{RK3}
             \A{U}^{n+1}=\frac{1}{3}\A{U}^n+\frac{2}{3}\A{U}_3^n\,,
         \end{equation}
         where $\A{U}_3^n$ is the outcome of the $3^{rd}$ iteration.
     \end{itemize}
\end{itemize}

\chapter{Code Test \& Astrophysical Problems} \label{computational-astrophysics}

We have tested the ability of RELDAFNA to deal with a series of test problems. The aim of that is Verification $\&$ Validation. Any modern and advanced numerical code must reproduce analytical and previously obtained results. Moreover, it should be compared to other codes with a well defined metric to a known solution. In order to do that, some of the problems chosen to test the code have analytic solutions, e.g. the so-called one dimensional Riemann problems. Other problems do not have analytic solutions, but have been extensively studied by other authors, such as the two dimensional Riemann problem. RELDAFNA is composed with different modules. Different numerical schemes, combined with AMR and parallelism, dealing with multidimensional problems with a wide spectrum of hydrodynamical and thermodynamical states. The way to test all the features of RELDAFNA is by \textit{separation of variables}. One dimensional tests with an analytical solution at different kinematical regimes, each of which mimics astrophysical scenarios, are tested by all the numerical schemes at different constant resolutions and also with AMR. In all cases, our results are similar to published results. In this chapter we will compare six schemes of RELDAFNA (chapter~\ref{schemes}): the Characteristic-Tracing time integration with PPM, the Characteristic-Tracing time integration with PLM, and both $2^{nd}$ and $3^{rd}$~order Runge-Kutta time integration scheme with PPM and PLM spatial reconstruction, denoted by CT-PPM, CT-PLM, RK2-PPM, RK2-PLM, RK3-PPM and RK3-PLM, respectively. When PLM schemes were used, the value $\theta$ in the minmod (see Equation~\ref{1st-app.-for-derivative}) is set to be $1.5$, unless otherwise stated. A CFL number of 0.5 is used for these tests unless otherwise stated. The Riemann solver used by default is HLLC. In one dimensional problems, the difference between the solutions with different Riemann solvers is negligible with respect to the differences between the various integration schemes and spatial reconstruction methods. For the AMR calculations we use the following refinement criterions: $\rho_{ref}=p_{ref}=\Gamma_{ref}=0.1$ and $\rho_{deref}=p_{deref}=\Gamma_{deref}=0.03$ (for the definitions see section~\ref{AMR}).

\section{One-Dimensional Riemann Problem 1} \label{rie1d1} 

As was stated above, one dimensional Riemann problems have an
analytic solution. Indeed, the figures in this section include lines of the analytical solution, which simply represent an implementation of the analytical solution given by \cite{marti94,pons00}. In the following (and also in section~\ref{1d_tests}), the spatial domain is constructed of the line $0 \leq x \leq 1$. Putting
a subscript $L$ ($R$) refers to the line $0 \leq x \leq 0.5 \, (0.5<
x \leq 1)$. It is conventional to study the error between numerical
solutions and analytical solutions using $L_1$ norm defined as $L_1
= \Sigma_i|u_i-u(x^i)|\Delta x^i$, where $u_i$ is the numerical
solution, $u(x^i)$ is the analytical solution and $\Delta x^i$ is
the cell width. In several cases, the $L_1$ norm alone lacks the
ability to compare between two integration schemes, since it is very
sensitive to the location of structures and less sensitive to their
internal profiles, so we will also use the $L_0$ norm defined as
$L_0 =100 \times \frac{max(u_{num.})}{max(u_{ex.})}$ inside the
perturbed areas only.

The following problem is presented ahead. Other extensively tested problems are presented in appendix~\ref{code-tests}.
\begin{equation}
\begin{array}{rclrclrcl}
     \gamma & = & \frac{5}{3}\,, & & & & \\
     \noalign{\medskip}
     \textrm{Left:}\,p & = & 13.33\,,\rho & = & 10\,,\A{v} & = & 0\,,\\
     \noalign{\medskip}
     \textrm{Right:}\,p & = & 10^{-8}\,,\rho & = & 1\,,\A{v} & = & 0\,.\\
     \noalign{\medskip}
 \label{RAM-prob1}
\end{array}
\end{equation}

This problem is the most common test problem, tested by~\cite{del-zanna02,aloy99,donat98,lucas04,marti96,meliani07,mignone05a,cannizzo08,mignone05b,wang08,zhang06,tchekhovskoy07}. In this problem, the fluid is only mildly relativistic dynamically, but thermodynamically it shows the lack of maximum density ratio in relativistic dynamics (see section~\ref{features}).

\begin{figure*}[h!]
\centering
\includegraphics[scale=0.4,trim=0 -30mm 0 0]{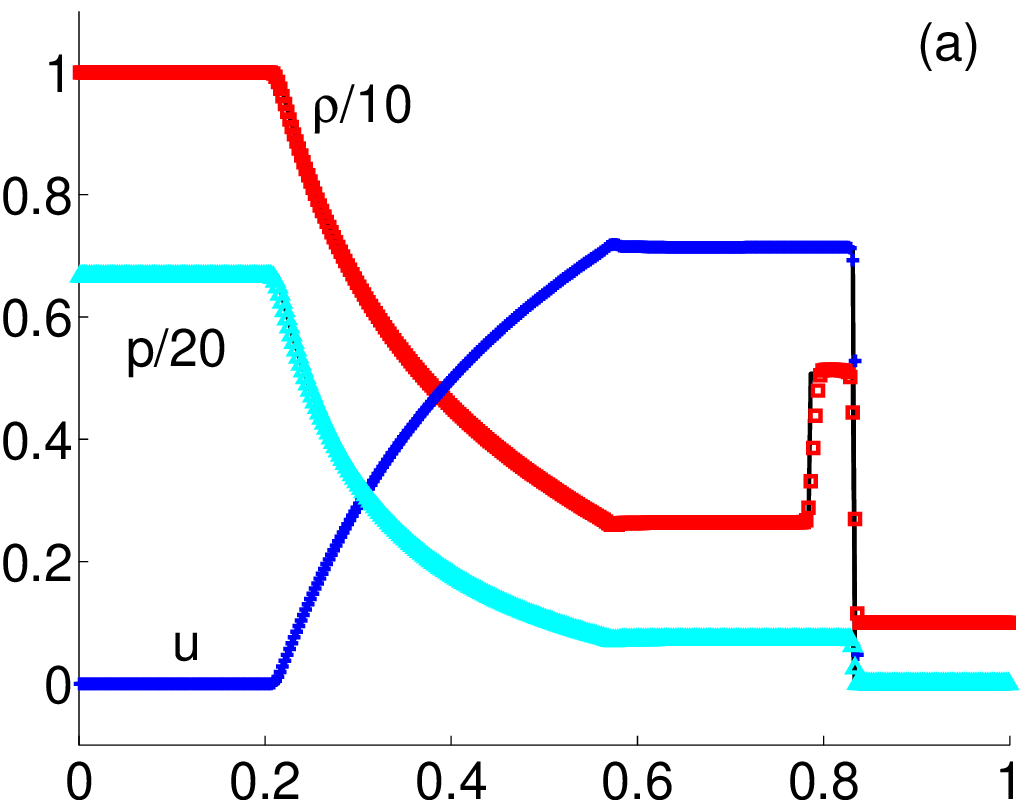}
\includegraphics[scale=0.4,trim=0 -30mm 0 0]{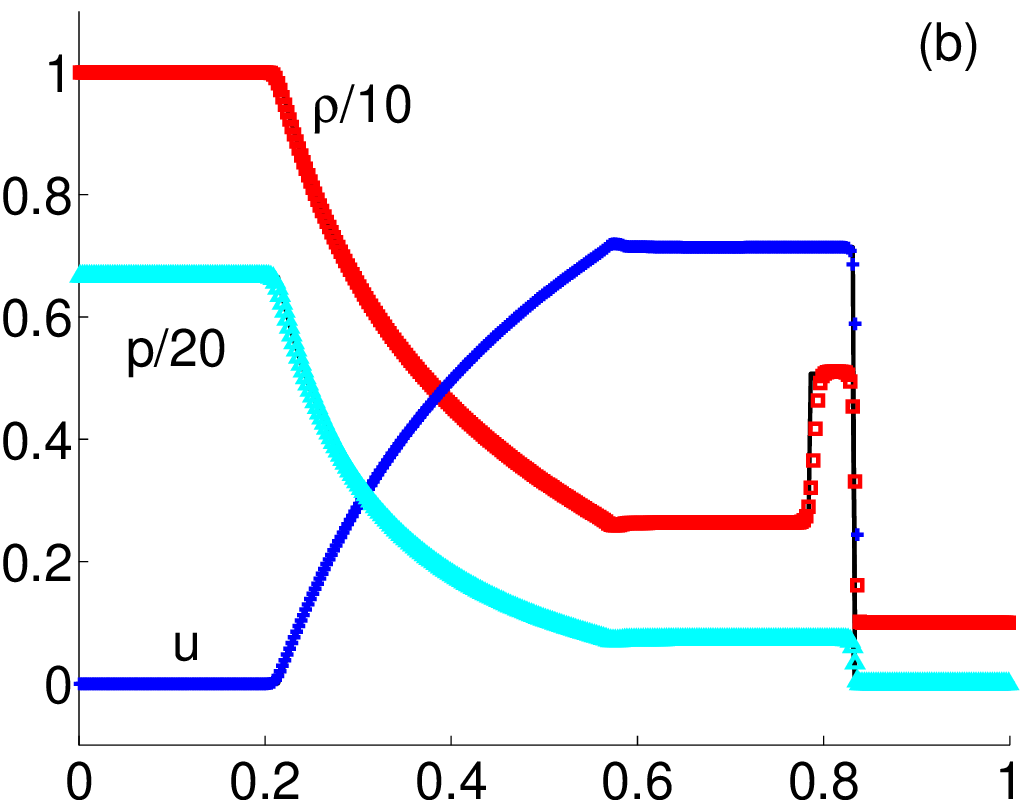}\\
\includegraphics[scale=0.4,trim=0 -30mm 0 0]{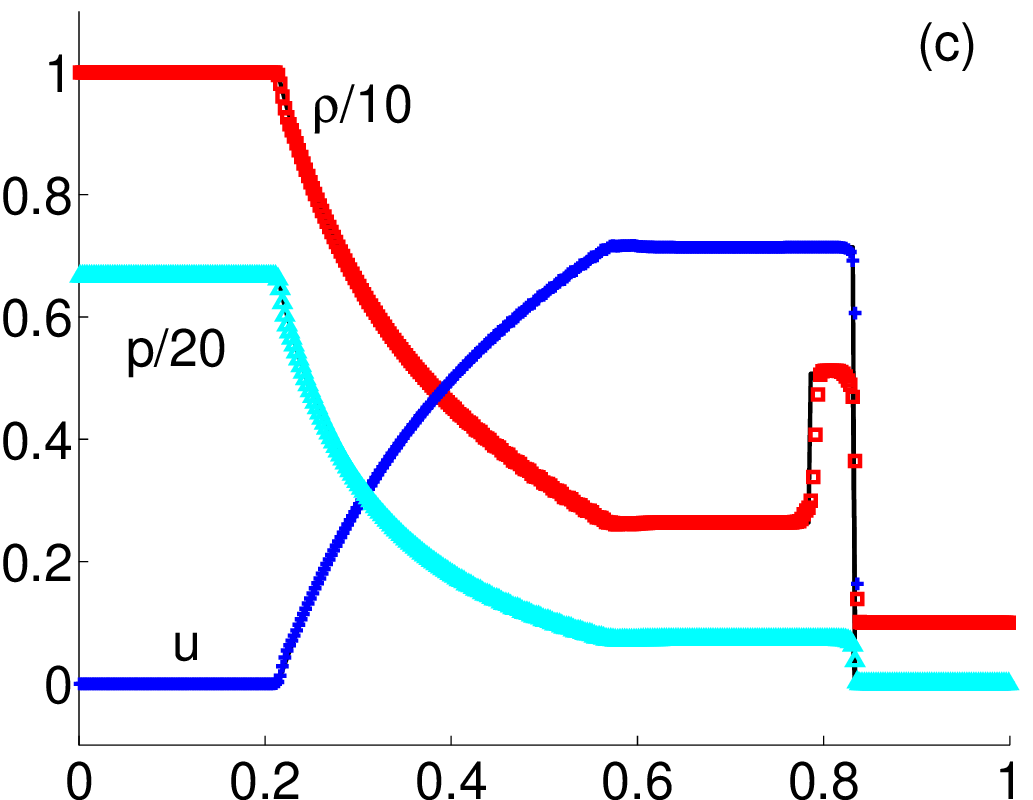}
\includegraphics[scale=0.4,trim=0 -30mm 0 0]{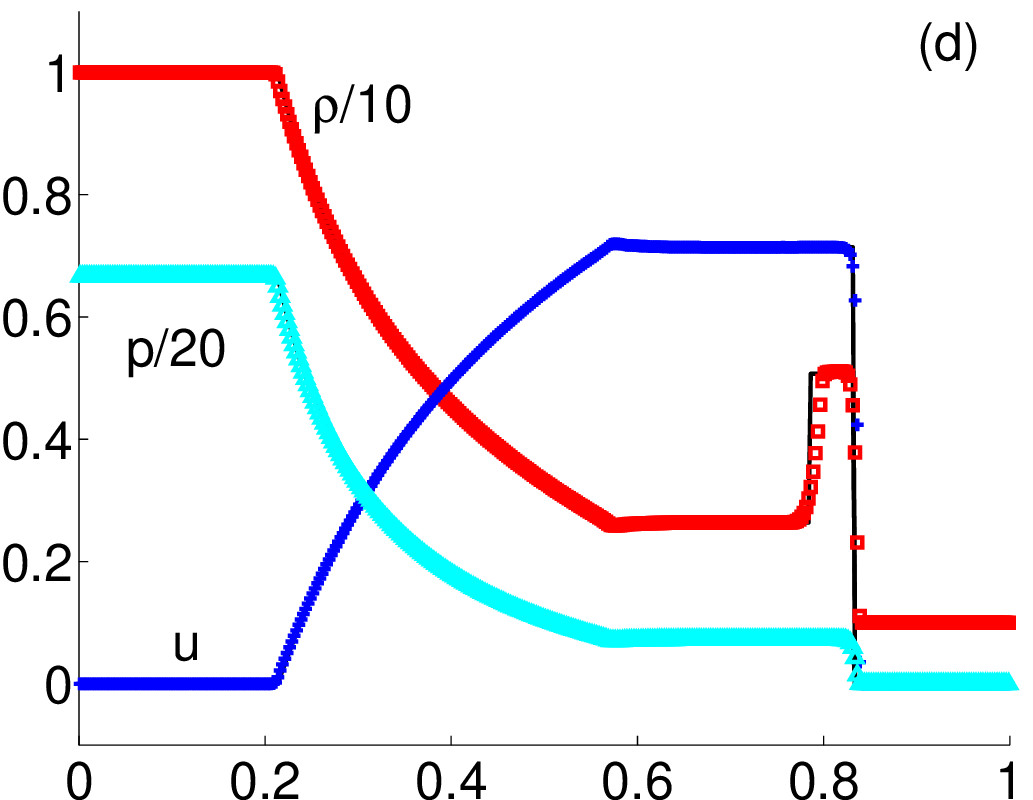}\\
\includegraphics[scale=0.4,trim=0 -30mm 0 0]{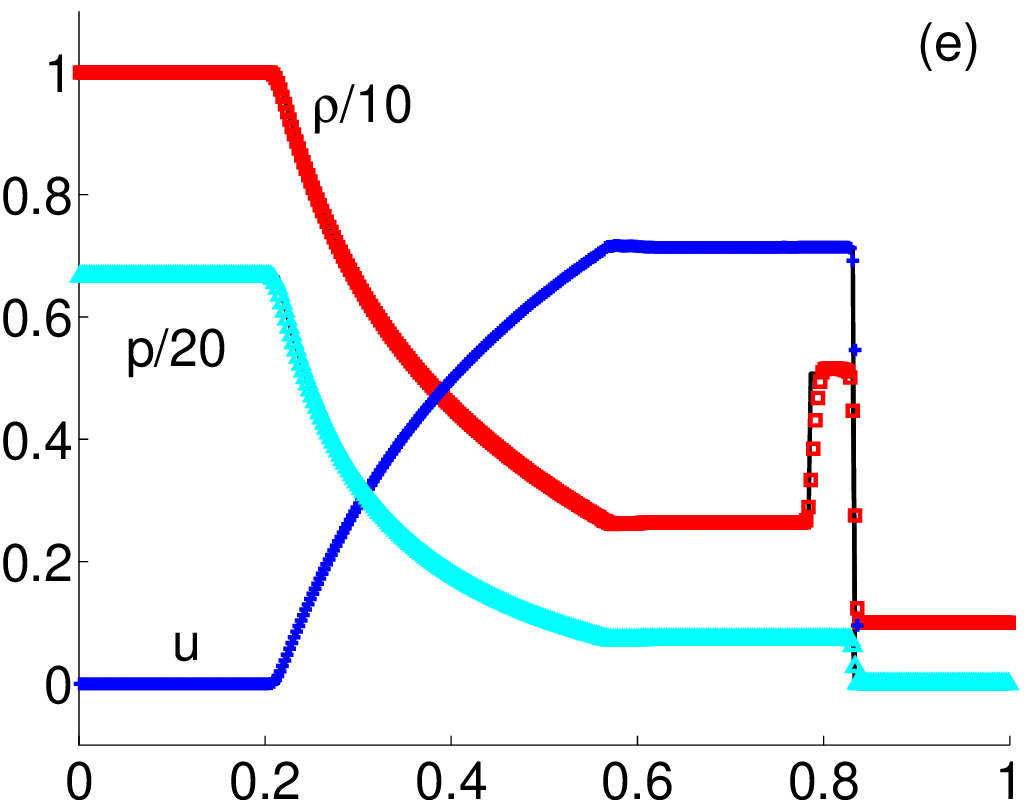}
\includegraphics[scale=0.4,trim=0 -30mm 0 0]{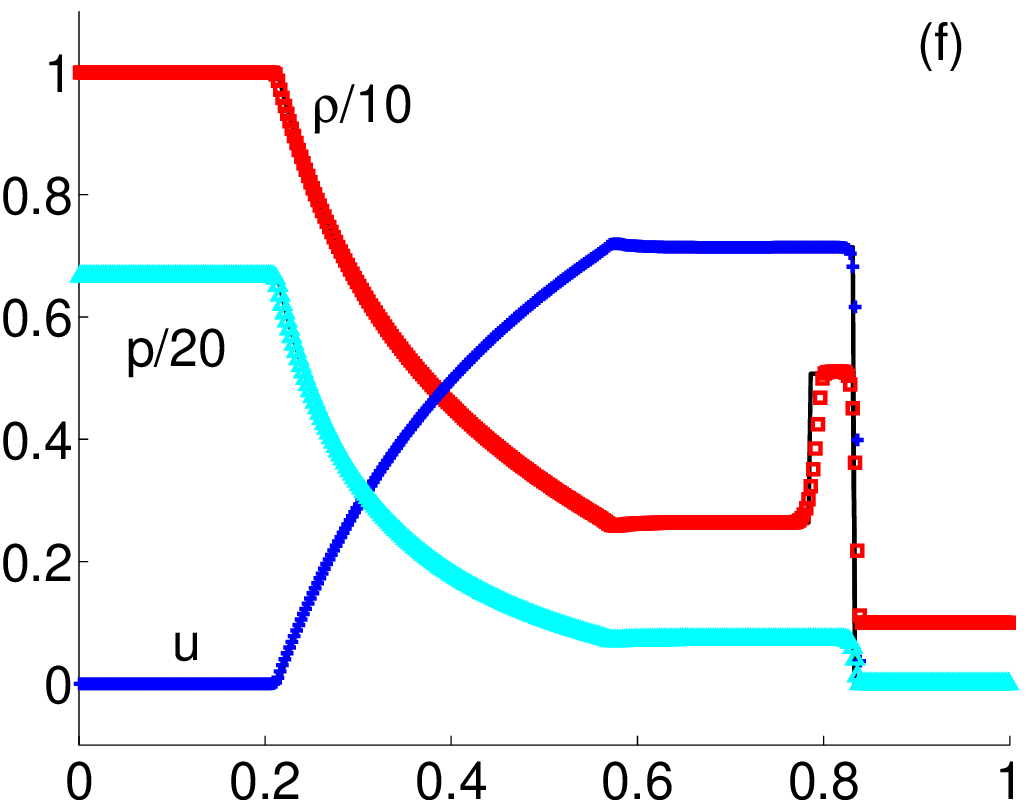}
\caption{One-dimensional Riemann problem 1 at $t=0.4$. Results for
 six schemes: (a) CT-PPM, (b) CT-PLM, (c) RK2-PPM, (d) RK2-PLM, (e) RK3-PPM and (f) RK3-PLM are shown.
 The computational grid consists of 400 zones. Numerical results are
 shown in symbols, whereas the exact solution is shown in solid
 lines. We show proper mass density (square), pressure (triangle)
 and velocity (plus sign)
\protect\label{fig:rie1d1}}
\end{figure*}

\begin{figure*}[h!]
\centering
\includegraphics[scale=0.4,trim=0 -30mm 0 0]{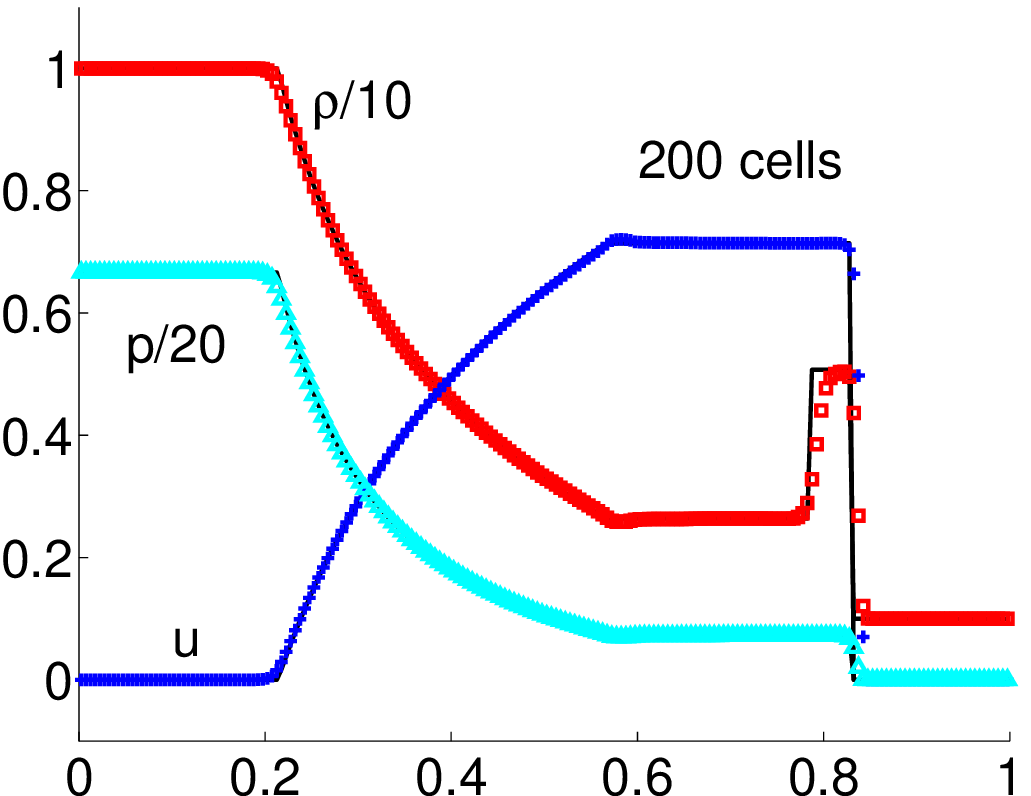}
\includegraphics[scale=0.4,trim=0 -30mm 0 0]{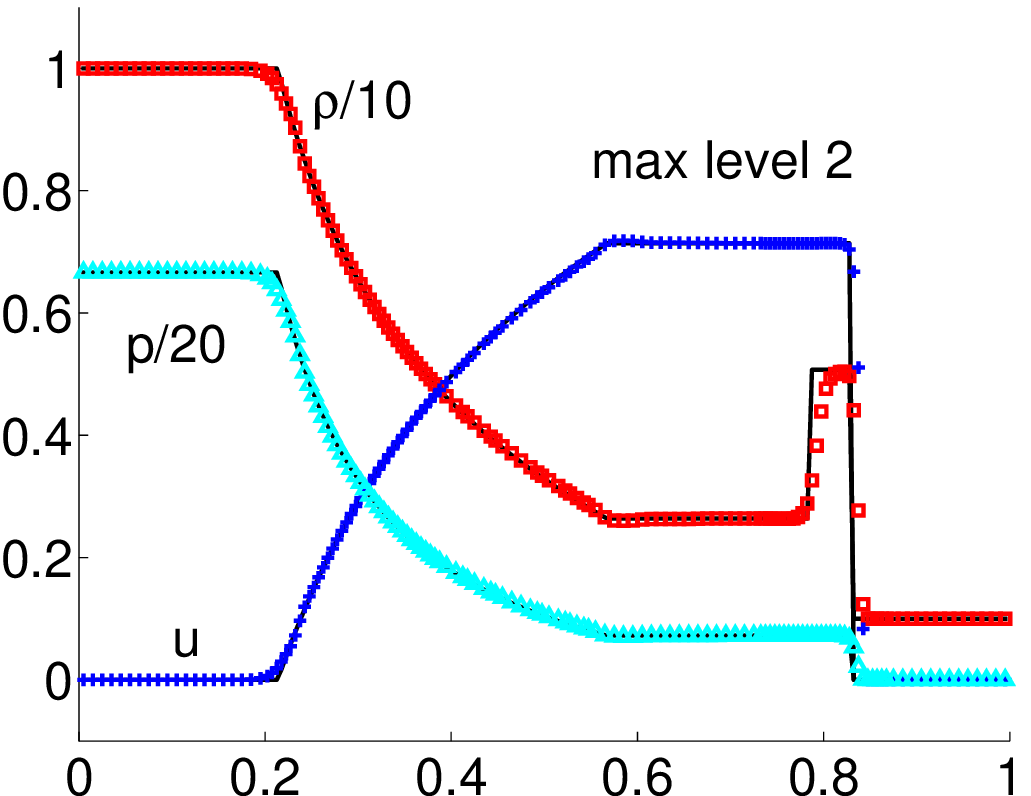}\\
\includegraphics[scale=0.4,trim=0 -30mm 0 0]{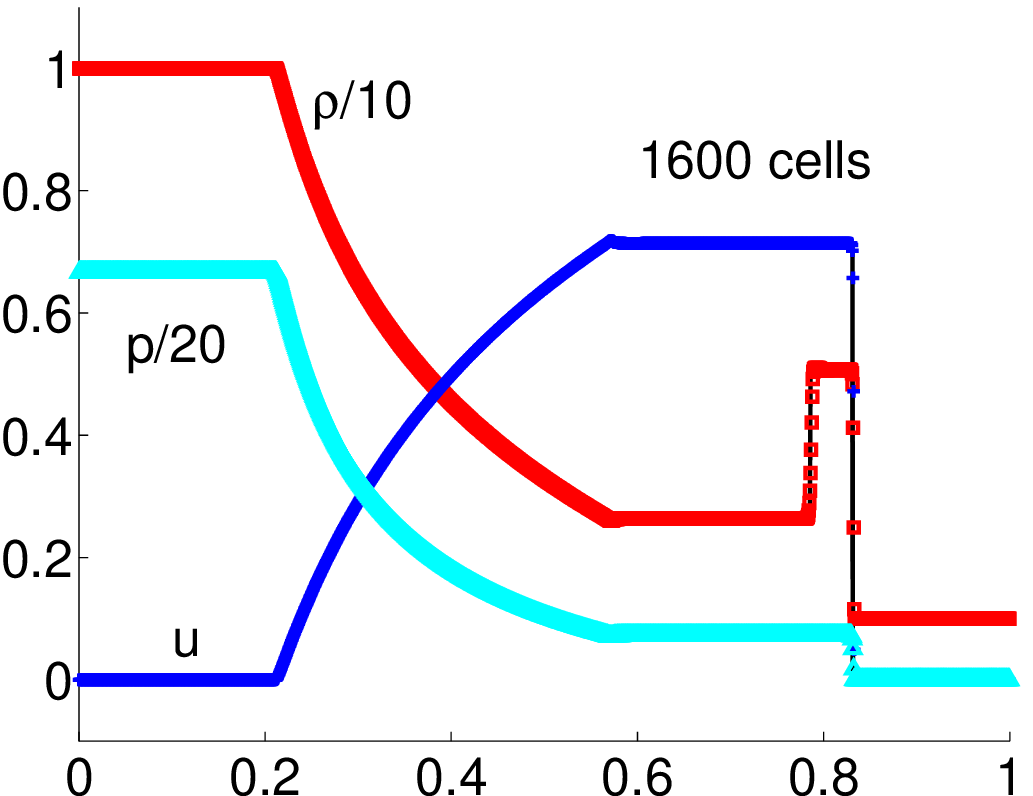}
\includegraphics[scale=0.4,trim=0 -30mm 0 0]{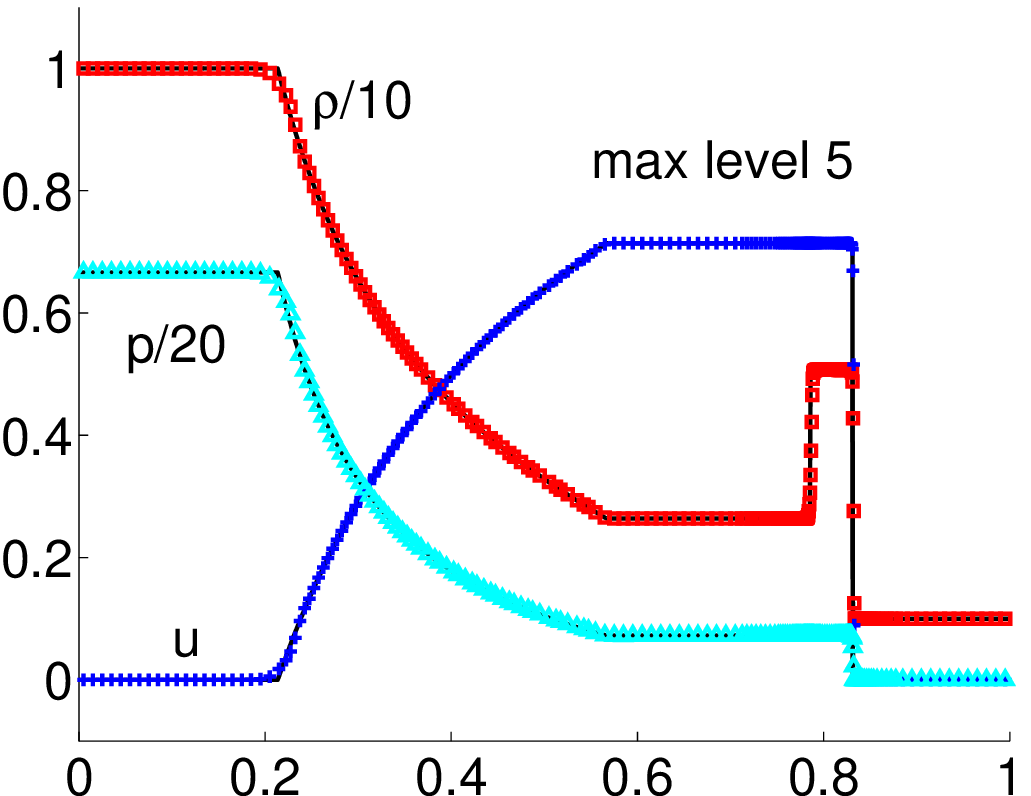}\\
\includegraphics[scale=0.4,trim=0 -30mm 0 0]{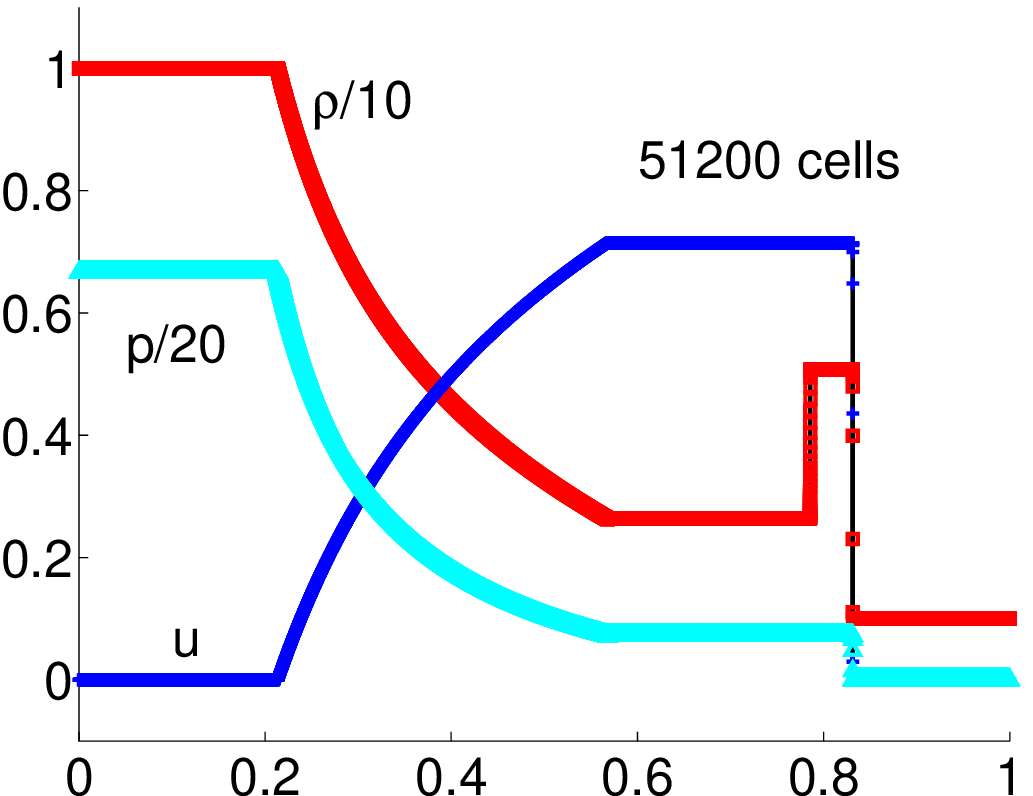}
\includegraphics[scale=0.4,trim=0 -30mm 0 0]{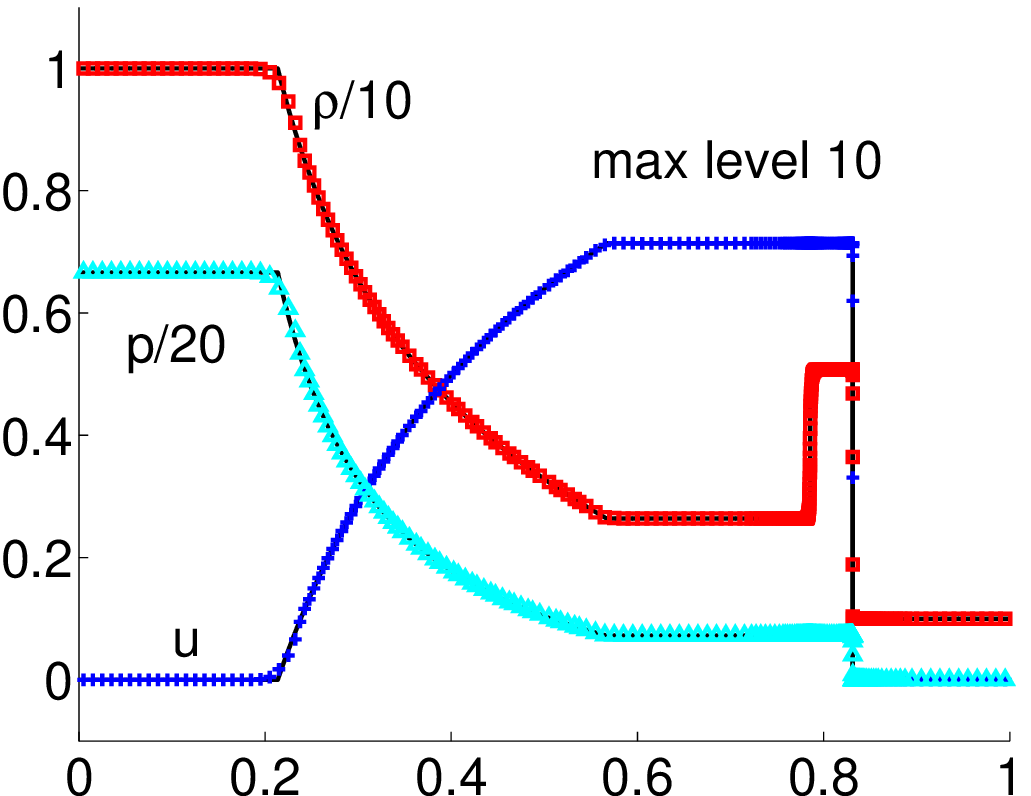}
\caption{One-dimensional Riemann problem 1 at $t=0.4$. Results for
AMR calculations using CT-PLM are shown.
At the right hand side are constant grid calculations,
and on the left hand side are equivalent resolution AMR calculations,
using a base mesh resolution of 100 zones. Numerical results are
shown in symbols, whereas the exact solution is shown in solid
lines. We show proper mass density (square), pressure (triangle)
and velocity (plus sign)
\protect\label{fig:rie1d1-amr}}
\end{figure*}

\begin{figure*}[h!]
\centering
\includegraphics[scale=0.7]{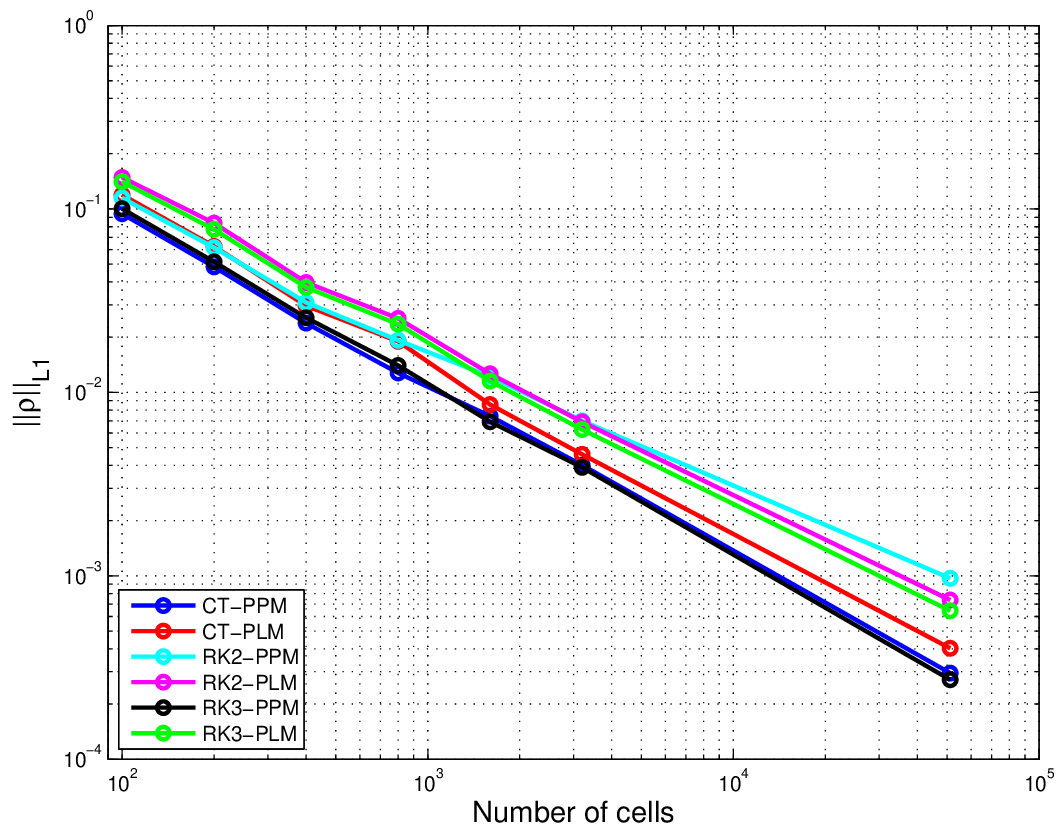}
\caption{$L_1$ errors of the density for the 1D Riemann Problem
 1. Seven different uniform grid resolutions (100,200,400,800,1600,3200 and 51200 cells) are
 shown at $t = 0.4$. Results for
 six schemes: (blue) CT-PPM, (red) CT-PLM, (cyan) RK2-PPM, (magenta) RK2-PLM, (black) RK3-PPM and (green) RK3-PLM are shown.
\protect\label{fig:conv-rie1d1}}
\end{figure*}

The results of the six schemes of RELDAFNA for this problem at $t=0.4$
are shown in Fig.~\ref{fig:rie1d1}. The initial discontinuity gives
rise to a transonic rarefaction wave propagating left, a shock wave
propagating right, and a contact discontinuity in between. This
problem is only mildly relativistic with a post-shock velocity
$0.72$ and shock velocity $0.83$. The $L_1$ norm errors of density
at $t=0.4$, for the six schemes with various grid resolutions, are
shown in Fig.~\ref{fig:conv-rie1d1} and Table.~\ref{tab:rie1d1}. The accuracy of our results is
comparable to that of \cite{zhang06} (see Table~\ref{tab:rie1d1}) and a little better than that
of \cite{wang08,tchekhovskoy07}. The order of the convergence rate is about $1$.
This is consistent with the fact that there are discontinuities in
the solution. We have also checked the ability of RELDAFNA to
calculate this problem using AMR. The base mesh resolution we used
is 100 cells (as the lowest resolution presented in
Fig.~\ref{fig:conv-rie1d1}), and the maximum level of AMR
corresponds to equivalent number of cells as some of the resolutions
presented in Fig.~\ref{fig:conv-rie1d1}, that is, maximum level of
$2$,$5$ and $10$. In a calculation with a maximum level of $10$, which
is equivalent to a constant grid of $51200$ cells, the maximum
number of cells used (during one cycle only) was $638$ cells, which
is only $1.25\%$. The results of a calculation using AMR with CT-PLM
scheme at $t=0.4$, is shown in Fig.~\ref{fig:rie1d1-amr}.

A look at Fig.~\ref{fig:rie1d1}, shows that the pressure and velocity profiles are similarly resolved by all six schemes, using a low number of cells. The shock wave is smeared over $\sim 3-4$ cells, and its location and jump condition is calculated in a satisfactory manner. The quantity which differs between the schemes is the density, since it is discontinuous over the contact discontinuity as well as the shock wave. The location of the contact discontinuity is found using $400$ cells with all schemes, but it is apparent that PPM schemes used less cells to capture the jump in the density (CT-PPM only $\sim 5$ cells). Using a selected spatial reconstruction method (either PPM or PLM), one would find that the CT time integration method is a better choice for this problem, since its flat density between the contact discontinuity and the shock wave seems uniform, and is closest to the analytic solution.
%Fig.~\ref{fig:conv-rie1d1} shows this separation between the spatial
%reconstruction methods in the low resolution regime ($\lesssim 5000$
%cells), where CT-PLM is closer to the PPM schemes relative to the
%RK$\#$-PLM schemes.
% The limiters used in the PPM reconstruction make
%the structure of the contact discontinuity (by that we mean the
%transition between the two values of proper density in the two sides
%of the discontinuity), more and more stair-like with resolution and
%one needs a better time integrator to sharpen that transition. As a
%result, RK-PPM becomes equivalent to the PLM schemes in high
%resolutions ($\gtrsim 1000$ cells), and a use of another iteration
%of Runge-Kutta, that is RK3-PPM, or the CT-PPM integrator leads to
%the lowest $L_1$ norm.
In Fig.~\ref{fig:rie1d1-amr} we can see that
the use of AMR does not affect the accuracy of the solution,
especially in zones where the effective resolution at a given time
is the same as the equivalent uniform grid. We could see that the
rarefaction wave is captured with less accuracy at its high tail,
but that is a result of the refinement criterions used for the
pressure jump. Tuning these criterions leads to a minor increase in
the number of cells used, and to a much closer agreement between the
uniform grid and AMR calculations, in all the hydrodynamic states
and structures apparent in this problem.

From the different tests we have made for RELDAFNA, we have concluded the following. For a detailed observation at these conclusions we refer the reader to section~\ref{1d_tests}.

\begin{itemize}

\item PPM schemes use less cells to capture the jumps.

\item CT time integration method achieves results closer to the analytical ones in dense shells and uses a lower number of cells to smear shocks.

\item \textbf{A comparison to spatially higher order schemes does not show any justifying improvements for the additional computational work. The norms of errors of analytically solved problems are similar to those obtained by other groups using higher order reconstruction methods.}

\item The use of AMR makes the low base mesh calculation very accurate.

\item Problems involving slowly moving structures produce oscillations.

\item Naturally, the location of thin structures is demanding a higher resolution than solving their initial structure.

\item When non-zero tangential velocity components are present, we observe a smearing of the sharp interfaces, using all schemes of RELDAFNA. This problem is well known in the literature \cite{zhang06,wang08} and is due to the fact the Lorenz factor in these cases is not continuous across contact discontinuities.
\end{itemize}

% tab 1
\begin{table}[hp!]
\scriptsize
 \centering
 \begin{tabular}{cccc}
$Scheme$ & $Number \; of \; cells$ & $L_1 \; Error$ & $Convergence \; Rate$\\
\hline
CT-PPM & 100 & 9.41e-2 & \\
 & 200 & 4.83e-2 & 0.96 \\
 & 400 & 2.39e-2 & 1.02 \\
 & 800 & 1.28e-2 & 0.9 \\
 & 1600 & 7.42e-3 & 0.78 \\
 & 3200 & 4.e-3 & 0.88 \\
 & 51200 & 2.95e-4 & \\
\hline
RK2-PPM & 100 & 1.147e-1 & \\
 & 200 & 6.15e-2 & 0.9 \\
 & 400 & 3.09e-2 & 0.99 \\
 & 800 & 1.92e-2 & 0.68 \\
 & 1600 & 1.22e-2 & 0.65 \\
 & 3200 & 7.00e-3 & 0.8 \\
 & 51200 & 9.7e-4 & \\
\hline
RK3-PPM & 100 & 1.00e-1 & \\
 & 200 & 5.14e-2 & 0.96 \\
 & 400 & 2.55e-2 & 1.01 \\
 & 800 & 1.40e-2 & 0.86 \\
 & 1600 & 6.92e-3 & 1.01 \\
 & 3200 & 3.90e-3 & 0.82 \\
 & 51200 & 2.72e-4 & \\
\hline
CT-PLM & 100 & 1.19e-1 & \\
 & 200 & 6.24e-2 & 0.93 \\
 & 400 & 2.97e-2 & 1.07 \\
 & 800 & 1.9e-2 & 0.64 \\
 & 1600 & 8.58e-3 & 1.14 \\
 & 3200 & 4.6e-3 & 0.9 \\
 & 51200 & 4.02e-4 & \\
\hline
RK2-PLM & 100 & 1.48e-1 & \\
 & 200 & 8.36e-2 & 0.83 \\
 & 400 & 3.97e-2 & 1.07 \\
 & 800 & 2.53e-2 & 0.65 \\
 & 1600 & 1.26e-2 & 1 \\
 & 3200 & 6.90e-3 & 0.87 \\
 & 51200 & 7.4e-4 & \\
\hline
RK3-PLM & 100 & 1.41e-1 & \\
 & 200 & 7.74e-2 & 0.86 \\
 & 400 & 3.74e-2 & 1.05 \\
 & 800 & 2.36e-2 & 0.66 \\
 & 1600 & 1.15e-2 & 1.03 \\
 & 3200 & 6.27e-3 & 0.88 \\
 & 51200 & 6.45e-4 & \\
\hline
F-WENO \cite{zhang06} & 100  & 1.31e-1 &      \\
       & 200  & 7.25e-2 & 0.85 \\
       & 400  & 3.32e-2 & 1.1  \\
       & 800  & 2.08e-2 & 0.67 \\
       & 1600 & 1.00e-2 & 1.1  \\
       & 3200 & 5.07e-3 & 0.98 \\
\hline
F-PLM \cite{zhang06} & 100  & 1.47e-1 &      \\
       & 200  & 8.50e-2 & 0.79  \\
       & 400  & 4.06e-2 & 1.1  \\
       & 800  & 2.33e-2 & 0.80 \\
       & 1600 & 1.22e-2 & 0.93 \\
       & 3200 & 7.48e-3 & 0.71 \\
\hline
U-PPM \cite{zhang06} & 100  & 1.27e-1 &      \\
       & 200  & 7.30e-2 & 0.80 \\
       & 400  & 3.47e-2 & 1.1  \\
       & 800  & 1.97e-2 & 0.82 \\
       & 1600 & 9.77e-3 & 1.0  \\
       & 3200 & 5.10e-3 & 0.94 \\
\hline
U-PLM \cite{zhang06} & 100  & 1.32e-1 &      \\
       & 200  & 8.57e-2 & 0.62 \\
       & 400  & 3.86e-2 & 1.2  \\
       & 800  & 2.27e-2 & 0.77 \\
       & 1600 & 1.15e-2 & 0.98 \\
       & 3200 & 6.48e-3 & 0.83
\end{tabular}
\caption{$L_1$ errors of the density for the 1D Riemann Problem
 1. Six schemes of RELDAFNA and four schemes of RAM \cite{zhang06} with various resolutions with uniform spacing are shown at $t = 0.4$.
}\label{tab:rie1d1}
\end{table}

\newpage

\section{Astrophysical Applications}

In chapter~\ref{applications}, a series of astrophysical problems involving special relativistic flows were presented. The different stages of the high energetic phenomena of GRBs were summarized, each of which with a highlight of the observations that led to a conclusion that SRHD is present. Among such flow patterns were relativistic jets crossing through an ambient medium with different hydrodynamic state and the mixing between them, and the interaction of collimated jets with the ISM regarding the external shocks produced in the ISM with connection to the observed afterglow. Another flow pattern regarding GRBs progenitors is the relativistic shock breakout from the surface of stars. All of these problems were the motivation of constructing RELDAFNA, and in the following chapter we present its ability to deal with such flow patterns.

\subsection{One Dimensional Trans-Relativistic Spherical Explosion}
\label{sec:tmm}

In \cite{tan01} there is a semi-analytical study of
trans-relativistic blastwaves propagating down spherical density profiles. The
motivation for that analysis is the option of trans-relativistic
blastwaves in stars to be progenitors for GRBs. The blast wave is launched into the stellar envelope by the collapse of the core during a supernova. The shock velocity decreases in most parts of the star as its energy is being deposited to an increasing amount of mass. If the shock is propagating down a sufficiently steep (declining) density profile, the velocity increases. The pressure is dominated by radiation. As a result, the moment the photons would be able to "escape" from the star, the shock would breakout of the star. This shock breakout moment may become before the shock accelerates to relativistic velocities in the acceleration down the declining density profile. This competition, between the acceleration of the shock to relativistic velocities and the breakout moment, depends on the energy of the explosion and the density profile of the star. The semi-analytical models proposed by \cite{tan01,gnatyk85}, predict the shock velocity as a function of this two parameters, the energy and density profile of a spherically symmetric star. Namely, they propose a multiplication of the two competing processes,
\begin{equation}
\label{nonrelshock}
\beta_s\left(r\right)=A \left(\frac{E_{in}}{m\left(r\right)} \right)^{1/2} \left(\frac{m\left(r\right)M_{ej}}{\rho\left(r\right) r^3} \right)^\alpha\,,
\end{equation}
where $E_{in}$ is the explosion energy in units of the rest mass energy of the material at the explosion, $m\left(r\right)$ is the mass of the shocked material in units of the mass of the material at the explosion, and $A$ and $\alpha$ are constants. This expression has been tested for two types of density profiles, and different amounts of energy deposited in the explosion. In \cite{tan01} the exact formula developed for the
blast wave velocity is being tested versus the formula proposed in \cite{gnatyk85}, and a one dimensional lagrangian code developed by the authors for that goal. This investigation leads to constraints on the explosion energy and the density profile for the shock to become relativistic before it breaks out of the star.

Here we test RELDAFNA ability to cope with spherical geometry and a nonhomogeneous media.
We will show RELDAFNA results of the shock wave velocity within the star versus the semi-analytic formulas given by \cite{tan01,gnatyk85}. Calculations including description of the ISM being shocked by the blastwave breaking out of the surface of star were also preformed, and the acceleration and deceleration phases of the shock wave in the ISM were analyzed.

\begin{figure*}[h]
\centering
\includegraphics[scale=0.4]{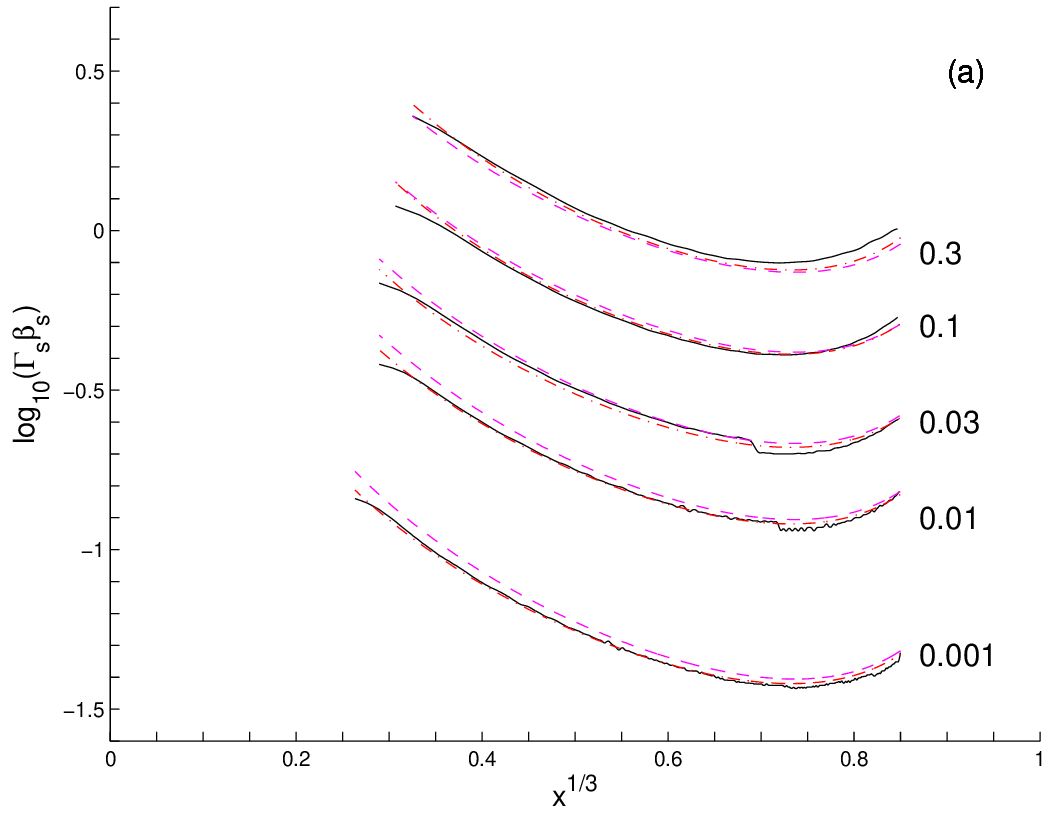}
\includegraphics[scale=0.4]{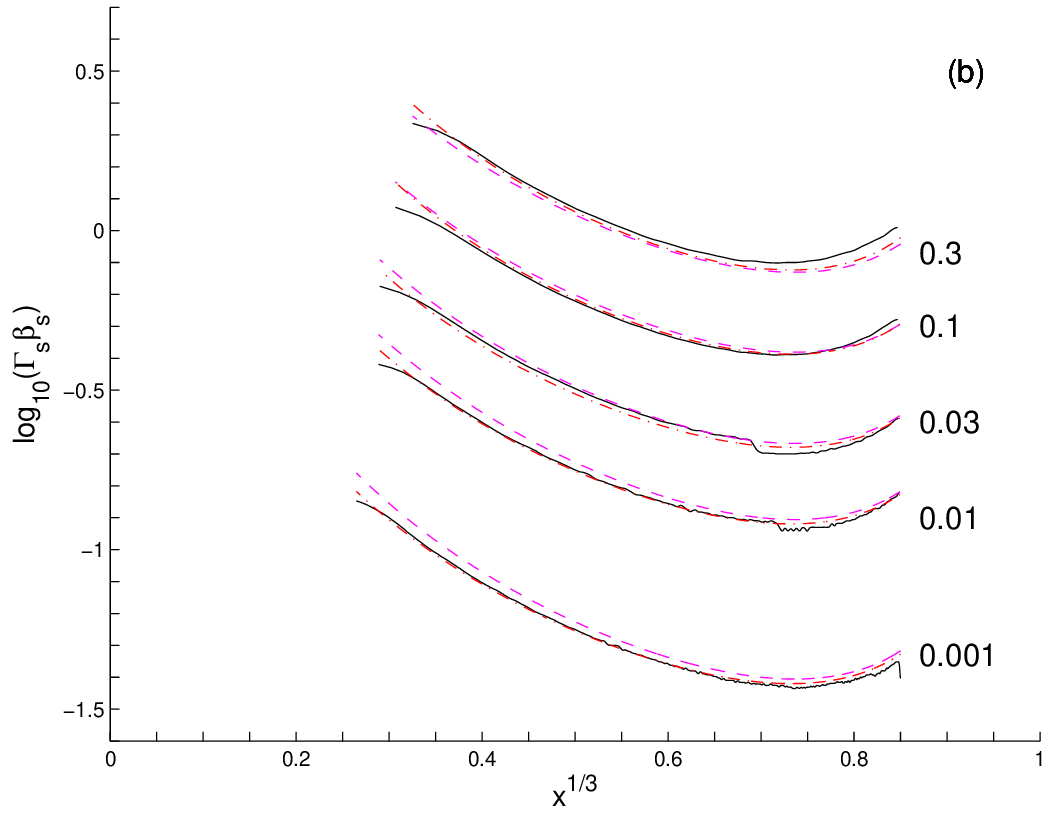}
\caption{Spherical shock propagation down a density profile
$\rho=\rho_h x^3$, where $x\equiv 1-r/R$, for different explosion
energies, $\tilde{E}_{in} \in \{0.001,0.01,0.03,0.1,0.3\}$.
Simulation results ({\it solid} lines) are compared to the analytic
prediction of \cite{tan01} (Eq. \ref{tmmformula1} and
\ref{tmmformula2}) ({\it red dash-dotted} lines), and to the
variation of the scaling predicted by \cite{gnatyk85} as expressed
by equation (\ref{gnatyk}) ({\it magenta dashed} lines). The
normalization constant, $A$, is 0.68 for $\tilde{E}_{in}\leq0.03$,
and 0.72, 0.74 for $\tilde{E}_{in}=0.1,0.3$, respectively. Results
for two schemes: (a) CT-PPM, (b) CT-PLM are shown.
\protect\label{fig:tmmsimple}}
\end{figure*}

\begin{figure*}[h]
\centering
\includegraphics[scale=0.4]{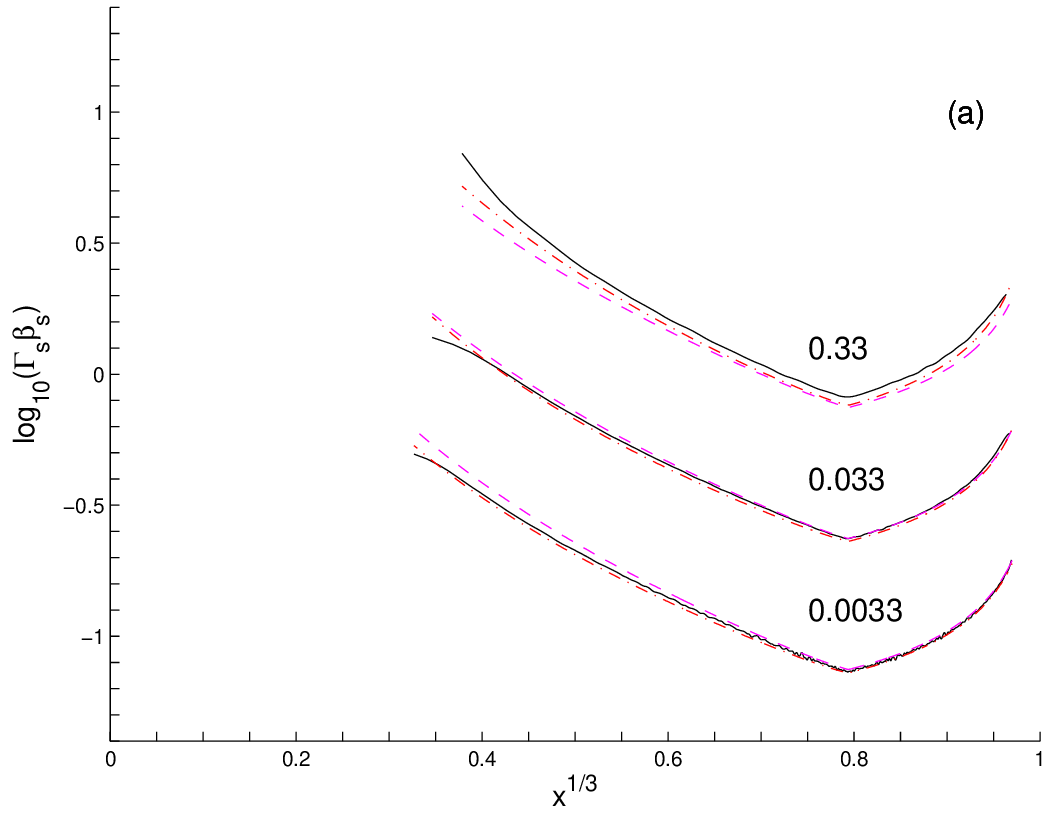}
\includegraphics[scale=0.4]{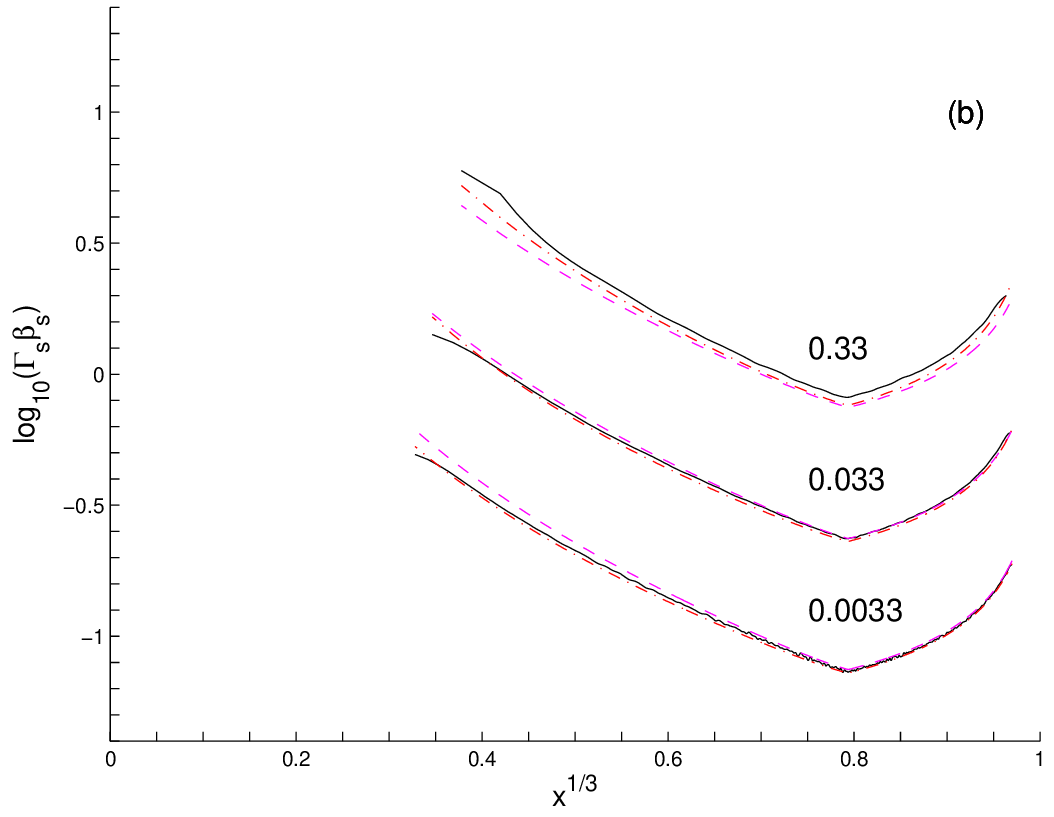}\\
\includegraphics[scale=0.4]{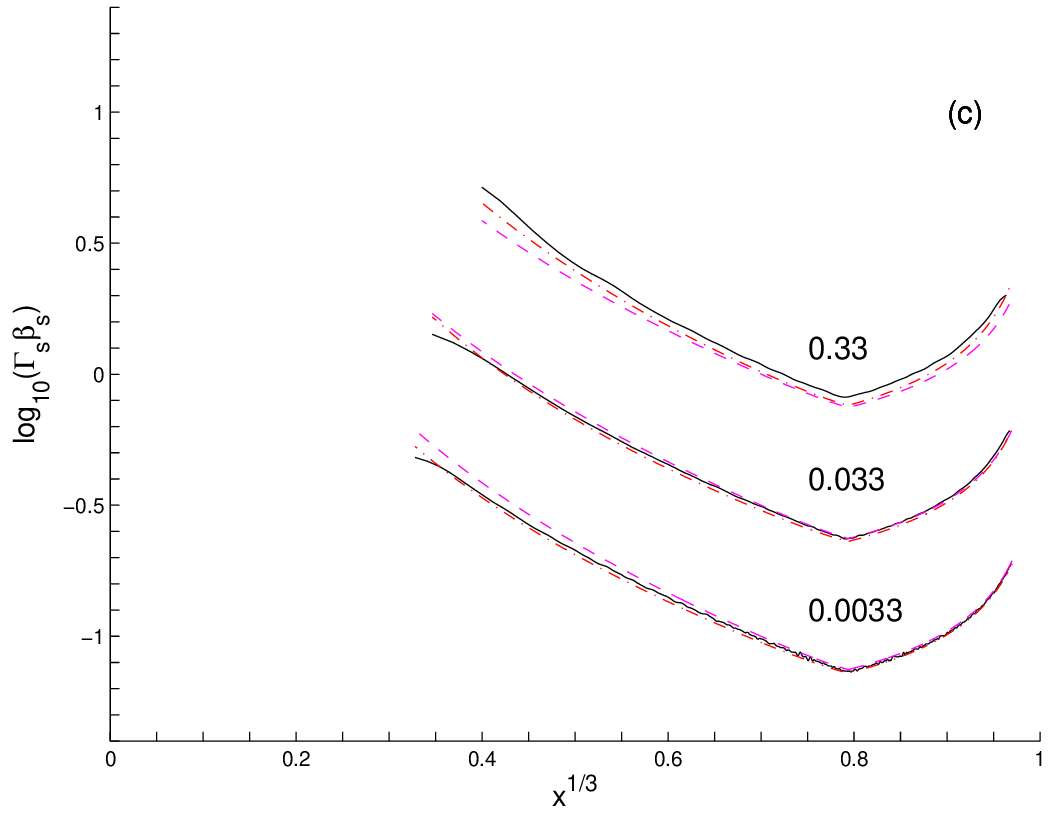}
\includegraphics[scale=0.4]{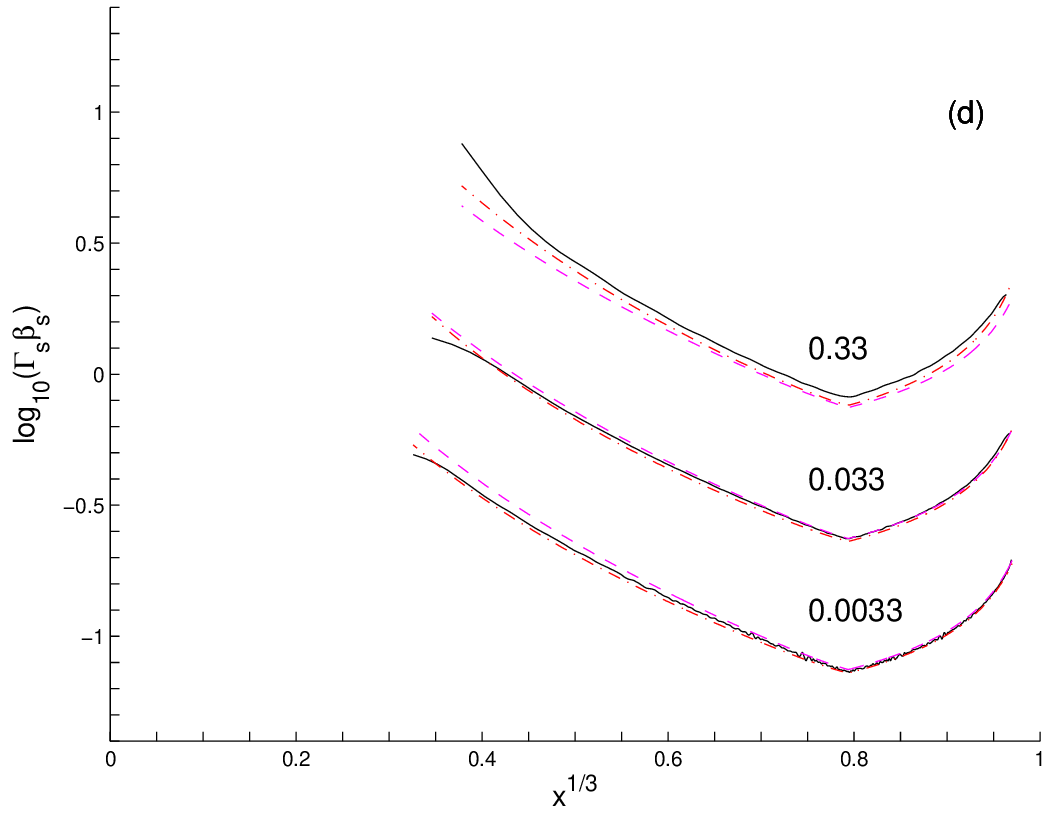}\\
\caption{Spherical shock propagation down a realistic stellar
density profile, given by Equation~\ref{rhostellar}, for different
explosion energies, $\tilde{E}_{in} \in \{0.0033,0.033,0.33\}$.
Simulation results ({\it solid} lines) are compared to the analytic
prediction of \cite{tan01} (Eq. \ref{tmmformula1} and
\ref{tmmformula2}) ({\it red dash-dotted} lines), and to the
variation of the scaling predicted by \cite{gnatyk85} as expressed
by equation (\ref{gnatyk}) ({\it magenta dashed} lines). The
normalization constant, $A$, is 0.736 for all energies. Results for
four schemes: (a) RK-PPM, (b) RK-PLM , (c) RK3-PLM, (d) RK3-PPM are
shown. \protect\label{fig:tmmstellar}}
\end{figure*}

Let there be a star with a spherical density profile to be defined
later $\rho(r)$, of an ideal gas with an adiabatic index
$\gamma=\frac{4}{3}$, and denote its radius by $R \; (0 \leq r \leq
R)$. Within a sphere of radius $r_{exp}=0.015R$ we inject internal
energy density in the following way:
\begin{equation}
e_{exp}=\frac{\tilde{E}_{in}*M(R)}{M(r_{exp})}
\end{equation}
where $M(r)$ is defined to be the mass of gas enclosed by a sphere of
radius $r$ and $\tilde{E}_{in}$ is a factor of order unity to change
from one case to another. The gas at $r \ge r_{exp}$ is given zero
internal energy density. We follow the definition of \cite{tan01}
and define a normalized mass coordinate with respect to the mass of
the explosion sphere
\begin{equation}
\tilde{m}(r)=\frac{M(r)}{M(r_{exp})}\,.
\end{equation}
With these definitions we write the formula given by \cite{gnatyk85}
\begin{equation}
(\Gamma_s \beta_s)(r) = A \left(\frac{e_{exp}}{\tilde{m}(r)} \right)^{1/2} \left(\frac{\tilde{m}(r) M(r_{exp})}{\rho(r) r^3} \right)^{\alpha_g}\,,
\label{gnatyk}
\end{equation}
and the formula given by \cite{tan01}
\begin{eqnarray}
(\Gamma_s \beta_s)(r) & = & p(r)(1+p(r)^2)^{0.12}, \label{tmmformula1}\\
p(r) & \equiv & A \left(\frac{e_{exp}}{\tilde{m}(r)} \right)^{1/2} \left(\frac{\tilde{m}(r) M(r_{exp})}{\rho(r) r^3} \right)^{\alpha_{nr}}\,.
\label{tmmformula2}
\end{eqnarray}
The two density profiles studied are a simplified model and a more realistic stellar profile. The simple model is given by
\begin{equation}
\rho(r)=\rho_h(x(r))^3\,,
\end{equation}
where
\begin{equation}
x(r) \equiv 1-\frac{r}{R}
\end{equation}
and $\rho_h=1$. The more realistic model is being composed by an inner power-law density distribution and an outer envelope given by the following:
\begin{equation}\label{rhostellar}
\rho(r) = \rho_h \times \left\{
\begin{array}{lc}
\left(\frac{R}{r_c}-1\right)^{3.85} \left(\frac{r}{r_c}\right)^{-1.9}, & r < r_c \\
\left(\frac{R}{r}-1\right)^{3.85}, & r_c <r<R.
\end{array}
\right.
\end{equation}
where we set the radius of the core to be $r_c=\frac{R}{2}$. In
Fig.~\ref{fig:tmmsimple} we show the results of two numerical
schemes dealing with the simplified density profile against the
semi-analytic formulas. In Fig.~\ref{fig:tmmstellar} we show the
results of the other four numerical schemes implemented in RELDAFNA
dealing with the more realistic stellar density profile against the
semi-analytic formulas. The calculations were made with a constant zoning, one dimensional mesh with spherical geometry using $6000$ cells describing the star $0\leq r\leq R_{star}$. This resolution is enough to obtain a converged velocity distribution, and calculations using AMR give similar results. In both density profiles we can clearly see that RELDAFNA is capable of resolving the blastwave velocity with
agreement to the semi-analytic formulas. In both density profile settings, we observe that in the outer layers of the star, the different schemes of RELDAFNA predict similar velocities at all energy depositions, but the shock velocity start to shift from the semi-analytical predictions. The shifts are different depending on the energy explosion. In the realistic stellar model, RELDAFNA predicts higher velocity than the analytical predictions. Since the numerical factors $A$ and $\alpha$ were determined using a numerical simulation \cite{tan01}, this only means that there is a slight difference between RELDAFNA and a one dimensional Lagrangian code used for the determination of the factors. The functional dependence of the shock velocity on the energy and density profile predicted by RELDAFNA agrees with the analytical predictions quite well.

\newpage

\subsection{Relativistic Jet In Two-Dimensional Cylindrical Geometry} \label{sec:jetC2}

The dynamics and morphology of relativistic jets is extremely
important in the analysis of GRBs afterglow structure
\cite{piran05,waxman03,zhang09}, and also in feeding the radio lobes
in powerful double radio sources \cite{blandford74,rees66}.
Therefore, the ability of RELDAFNA to resolve the structure and
dynamics of relativistic jets must be tested, for many astrophysical
questions rely critically on relativistic collimated flows.

The basic features of a jet were first analyzed for Newtonian hydrodynamics in \cite{norman82}. The morphology of the jet depend on a set of parameters such as the beam velocity, the densities ratios between the bean and ambient medium and the pressure ratio between them. These features include a beam, cocoon, Mach disk and bow shock. The beam is constantly feeding the medium and could be shocked by a reverse shock if the pressure gradient on the axis is negative (see for example appendix~\ref{rie1d3}). The layer separating beam and shocked ambient medium gas determines the velocity of the jet. If one increases the beam density the cocoon becomes less prominent. If the velocity of the jet decreases with its density fixed the cocoons turn into lobes of matter near the head of the jet. All of these features were observed also in the analysis of relativistic jets in \cite{marti97}. It was also found that relativistic jets are more stable and propagate more efficiently than Newtonian ones.

\begin{figure*}
\centering
\includegraphics[scale=0.35]{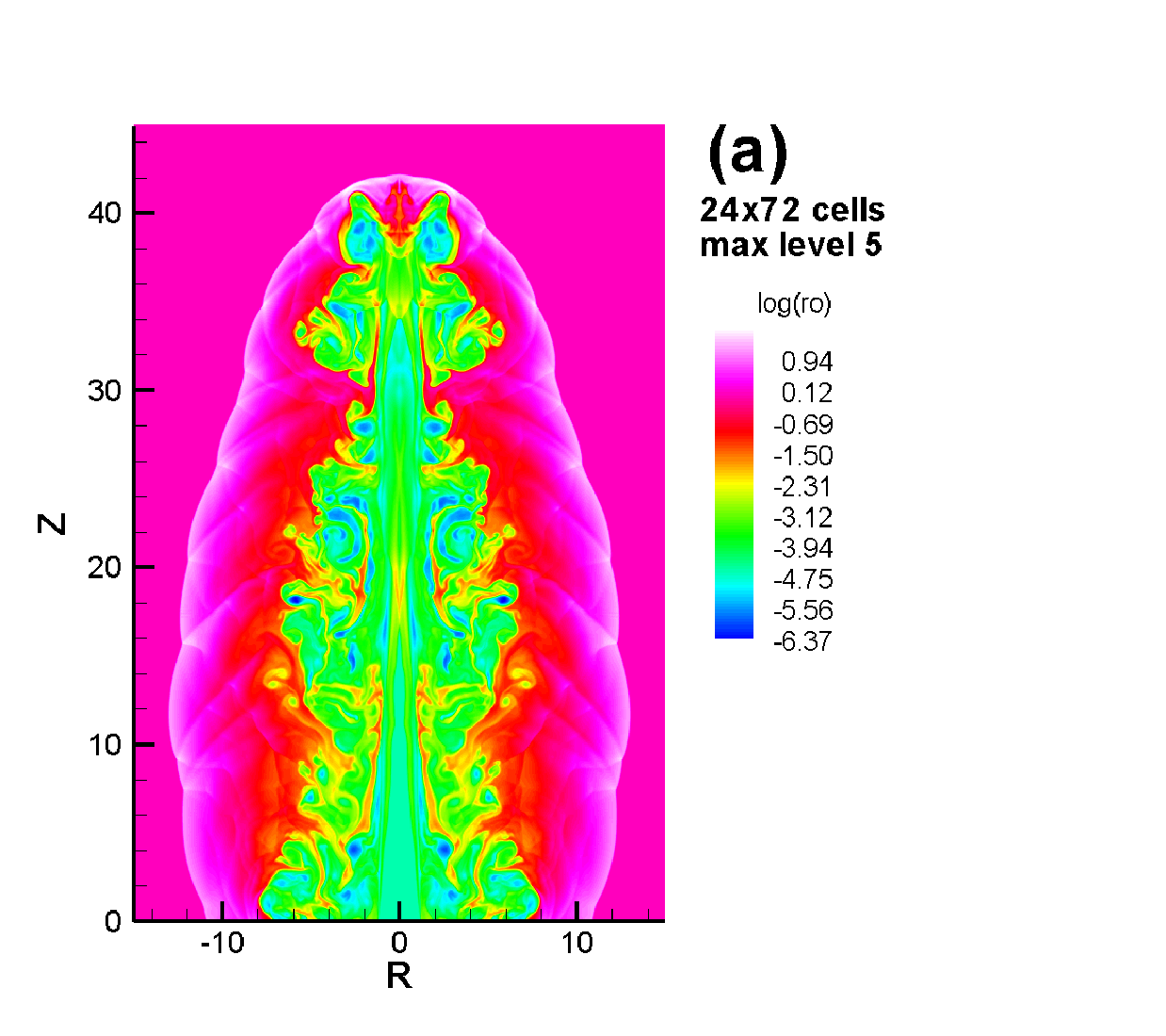}
\includegraphics[scale=0.35]{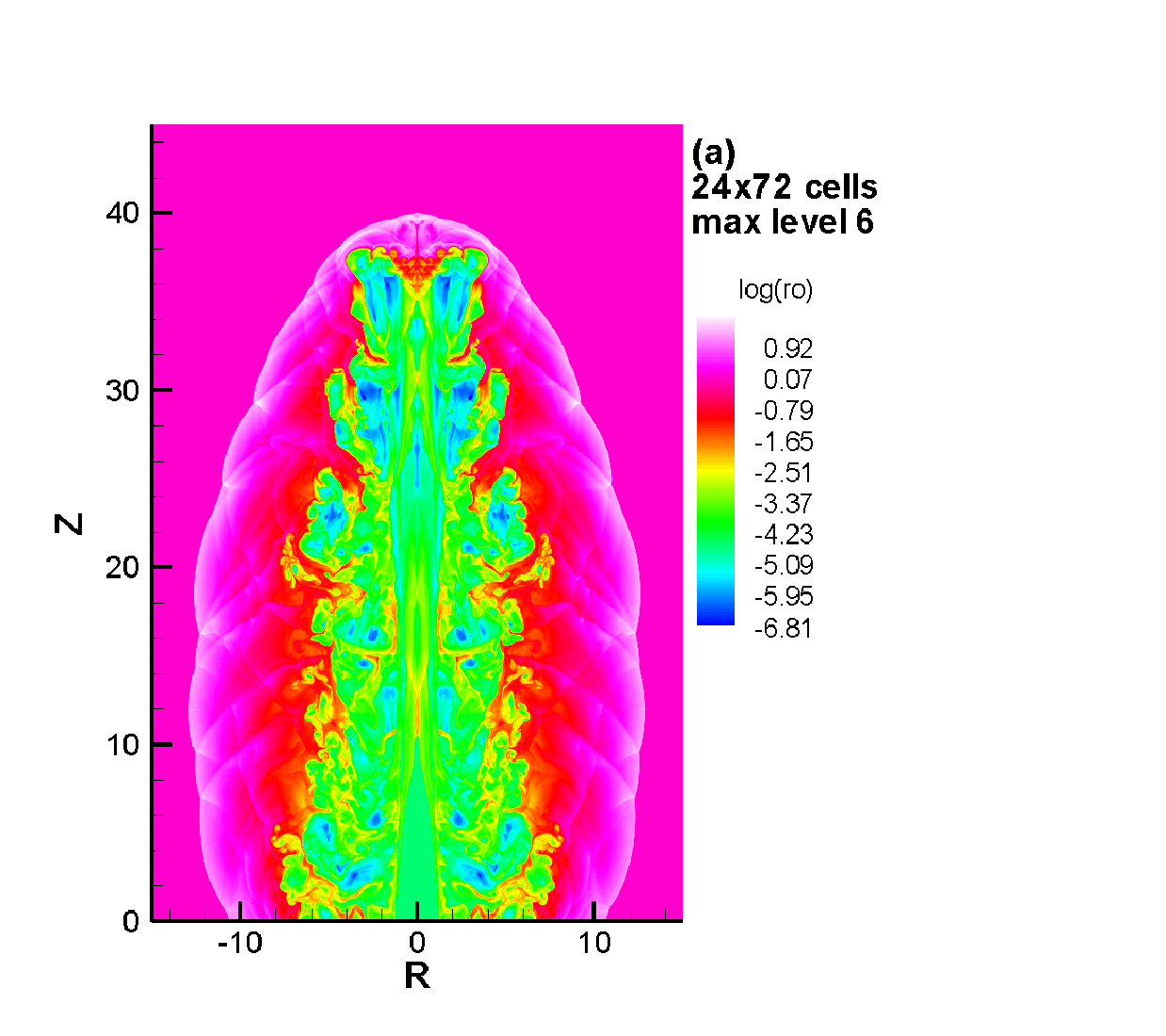}\\
\includegraphics[scale=0.35]{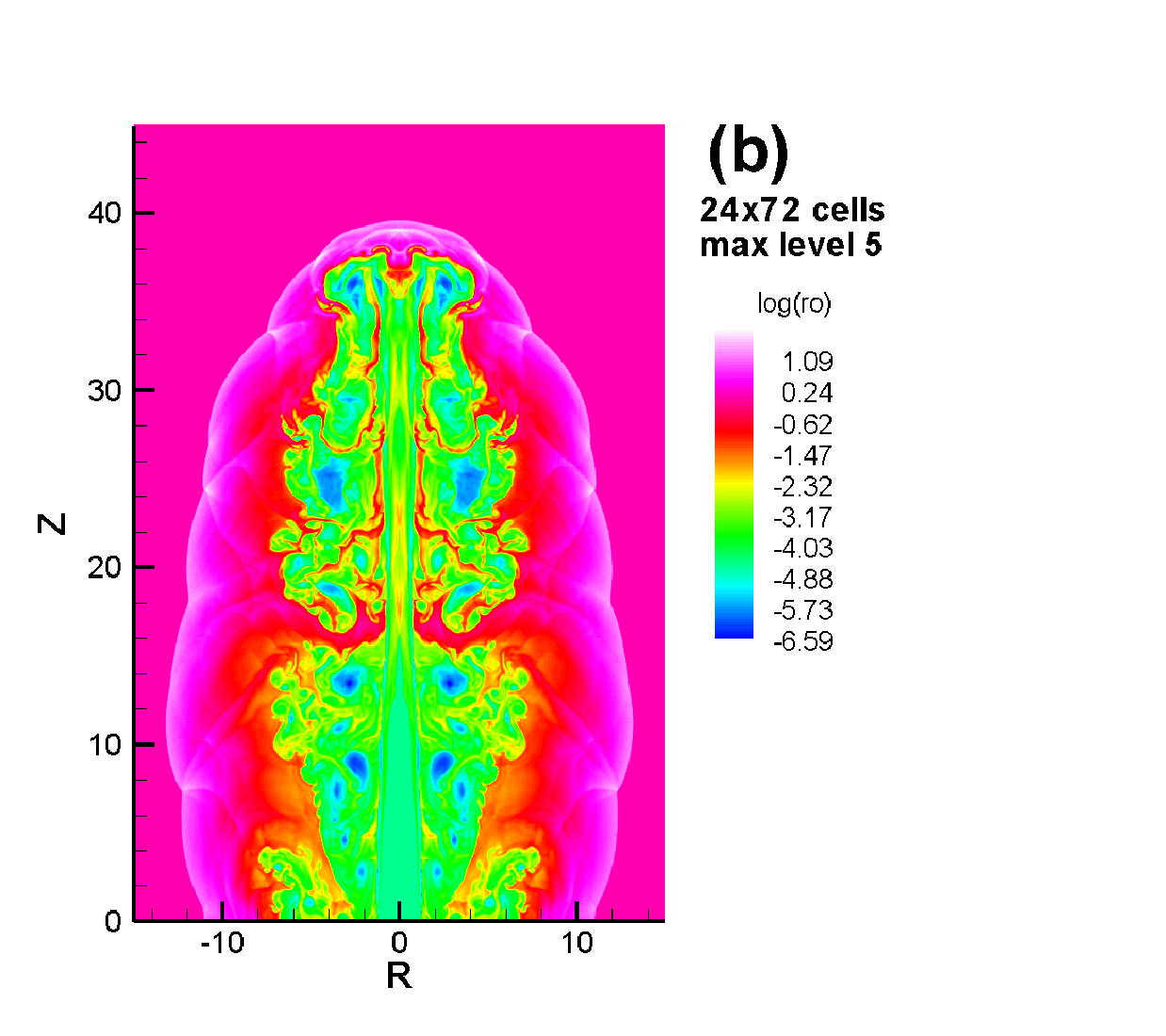}
\includegraphics[scale=0.35]{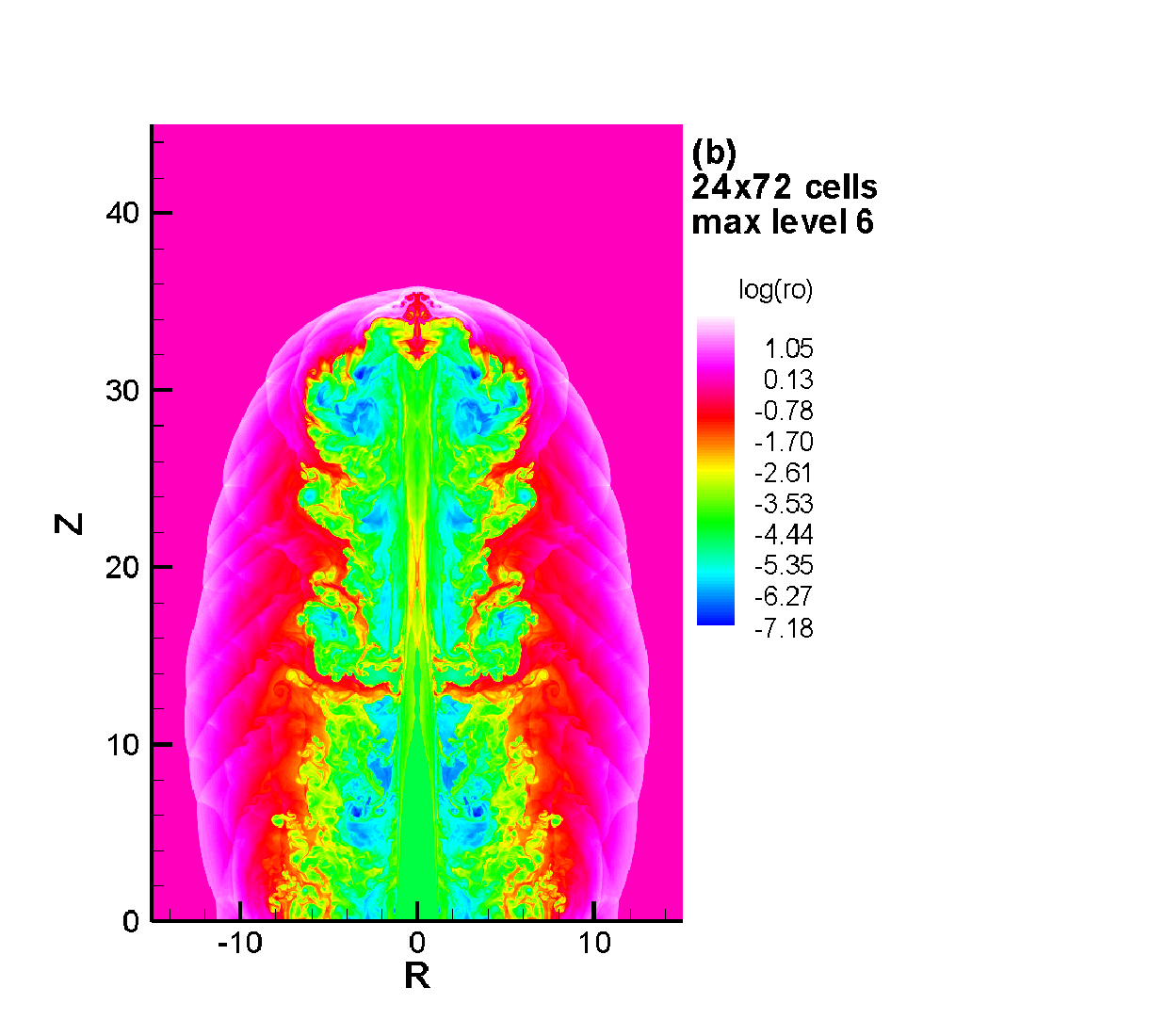}\\
\includegraphics[scale=0.35]{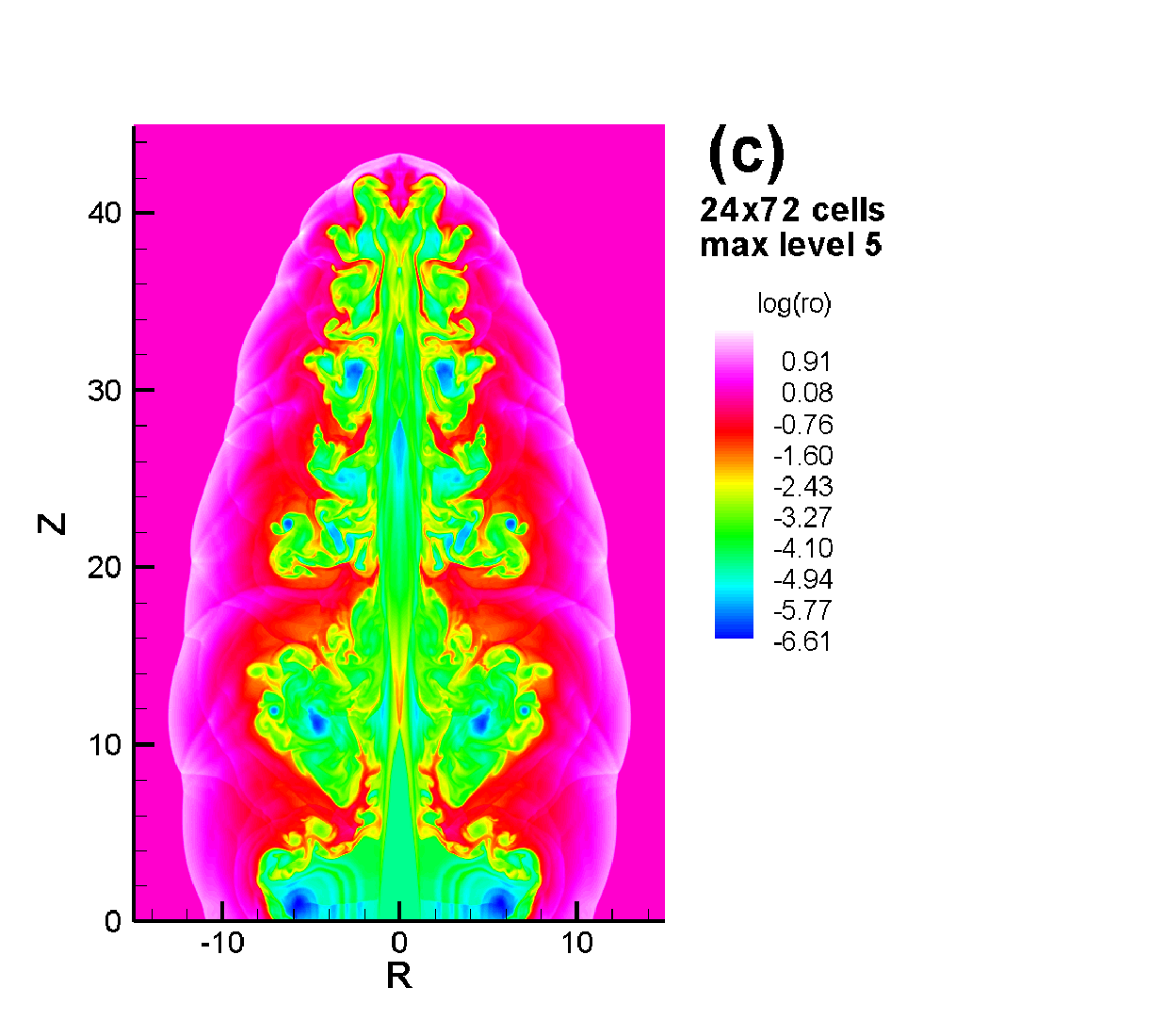}
\includegraphics[scale=0.35]{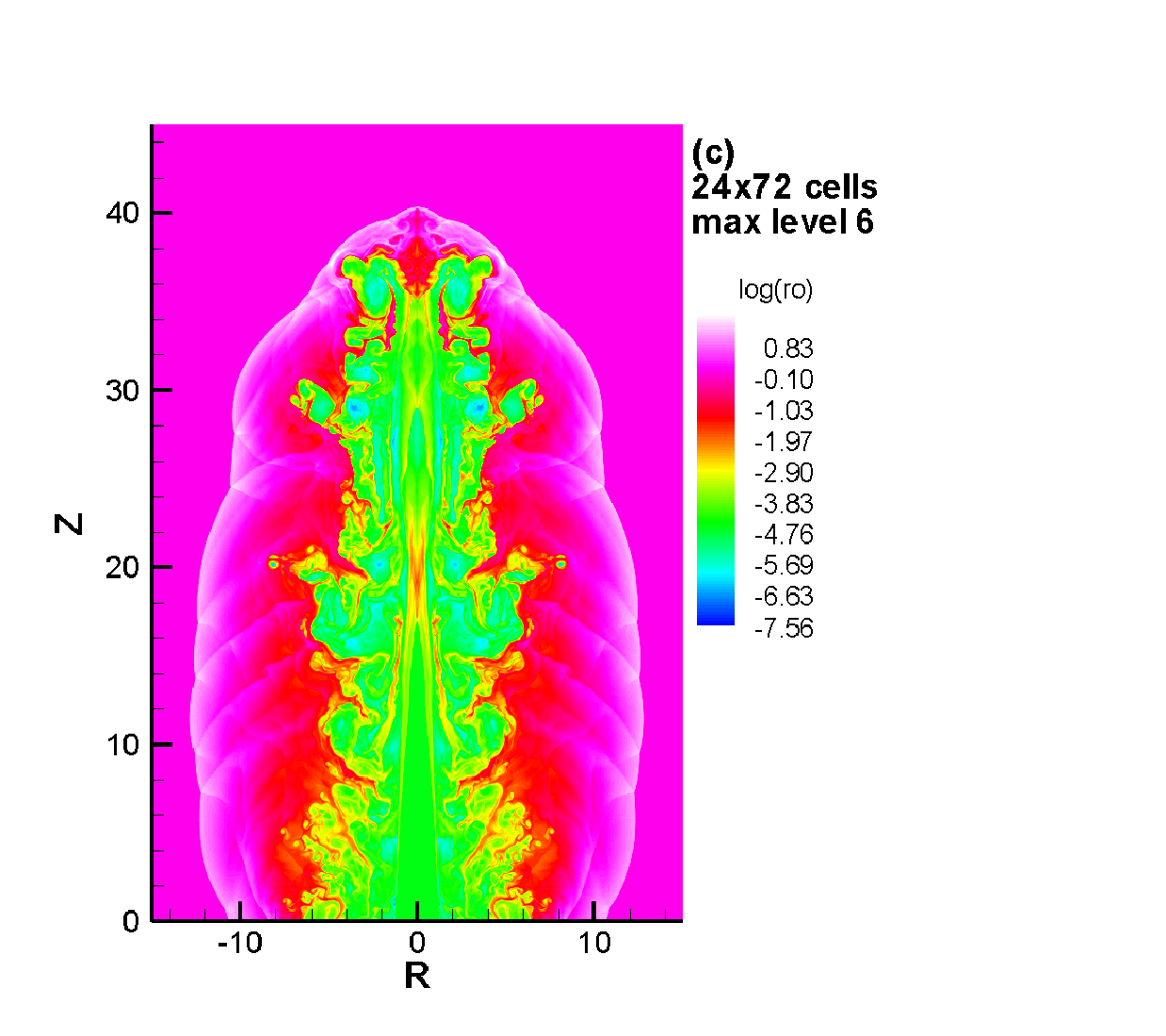}\\
\caption{Two dimensional cylindrical axis symmetric jet, C2 model,
at t=100 beam radii. Results of calculations using CT-PLM, RK2-PPM,
and RK2-PLM are shown in panels (a), (b), and (c), respectively. The
logarithm of proper density is plotted. The base mesh of the
calculations is $24 \times 72$ cells, and the maximum level of refinement is 5 (left) and 6 (right), with effective resolution of $384\times1152$ and $768\times2304$, respectively.\protect\label{fig:jet-C2}}
\end{figure*}

In the following test, we simulate the model C2 of \cite{marti97}
which was extensively tested numerically
\cite{marti97,zhang06,meliani07,wang08,tchekhovskoy07,aloy99}, with different solvers and different resolutions both with and without AMR. The computational region is a two-dimensional cylindrical box ($0 \le r
\le 15$, $0 \le z \le 45$) filled with an ideal gas with an
adiabatic index of $\gamma = 5/3$. The thermodynamical and
hydrodynamical state of the beam are, $v^z_b = 0.99$ , $v^r_b = 0$ ,
$\rho_b = 0.01$ and $p_b = 0.000170305$ and at $t=0$ it is injected
at the cylinder ($r \le 1$,$z \le 1$). Initially the computational
region is filled with a uniform medium with $v^z_m = 0$ , $v^r_m =
0$ ,$\rho_m = 1$ and $p_m = 0.000170305$. In the boundary ($r \le 1$
$z=0$) we force inflow boundary conditions with the beam
thermodynamical and hydrodynamical states. Outflow boundary
conditions with zero gradients of variables are used at all other
boundaries except the boundary ($r=0$) where reflective boundary
conditions are assumed (only half a cylinder is being calculated due
to the cylindrical symmetry). Our results of three schemes with
CFL=0.3 at $t = 100$ are shown in Fig.~\ref{fig:jet-C2}. In all the
calculations, the lowest level of the grid consists of $24 \times
72$ zones. The corresponding spatial resolution at this level is
$\sim 1.6$ zones per jet beam radius. Two maximum levels of refinement
$5$ and $6$ are shown. These correspond to equivalent resolutions of $\sim
26$ and $\sim 52$ zones per jet beam radius, respectively. The
model in this test is highly supersonic, and relativistic effects
from ultra-relativistic motion dominates those from internal energy
\cite{marti97}.

\begin{figure*}
\centering
\includegraphics[scale=0.5]{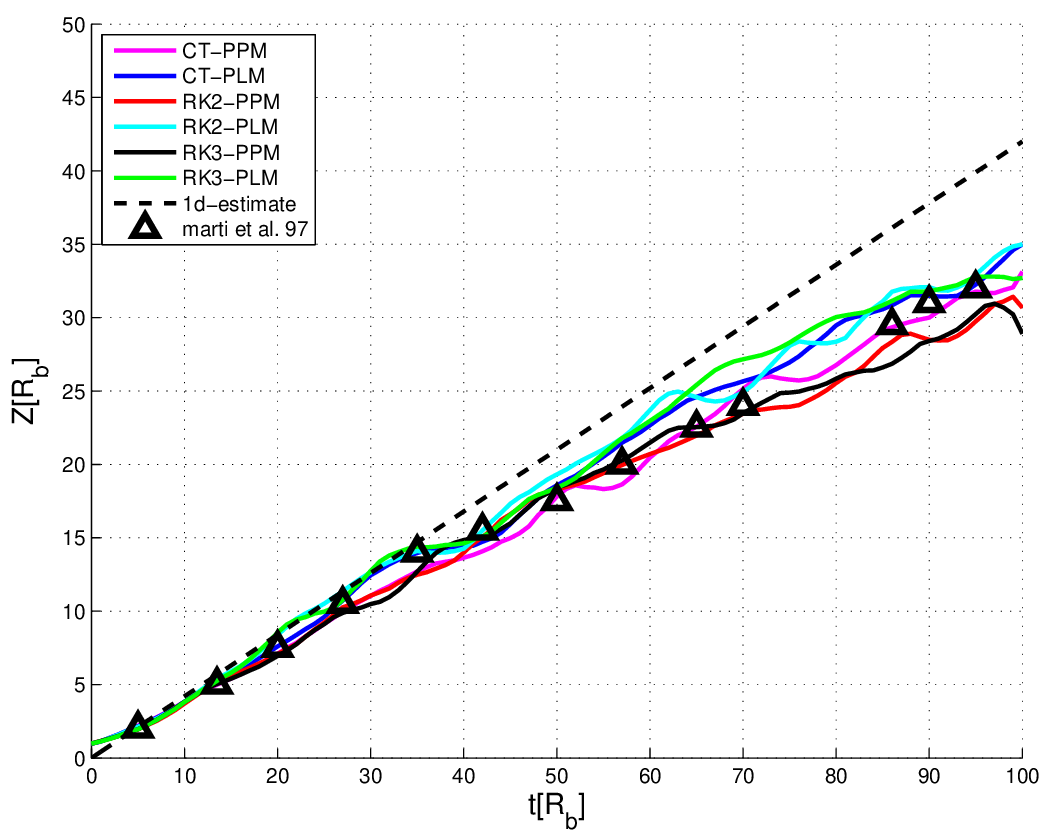}
\caption[]{The jet location as a function of time for the C2 model. Six schemes of RELDAFNA with maximum refinement level of $6$ are shown vs. a one dimensional estimate and data from Fig.~$6$ of \cite{marti97}.}
\label{fig:jet-velocity}
\end{figure*}

We can see in Fig.~\ref{fig:jet-C2} that the morphological
features of such a relativistic jet are observed in both resolutions. A bow shock is
formed due to the supersonic motion of the jet. The medium is
shocked by the bow shock. The jet beam is slowed down at the Mach
disk and feeds the cocoon. The shocked jet beam moves sideways and
then even backwards. The discontinuity between the shocked jet
material and shocked medium material admit Kelvin-Helmholtz
instabilities in the cocoon. As seen also by \cite{zhang06}, using higher resolution results in more mixing. The average speed of the jet head is given in Fig.~\ref{fig:jet-velocity} and is $\sim 35$, a little lower than the one-dimensional analytic constraint, $42$. A quantitative comparison to \cite{marti97} is also given, and we can see that the different numerical schemes agree with the value given by \cite{marti97}, which is an additional support for the ability of RELDAFNA to calculate complicated two dimensional flow patterns.

\begin{figure*}[h!]
\centering
\includegraphics[scale=0.53, trim= 0 0mm 0 0]{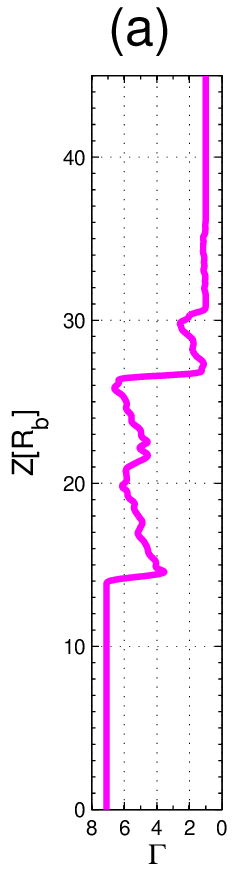}
\includegraphics[scale=0.53, trim= 0 0mm 0 0]{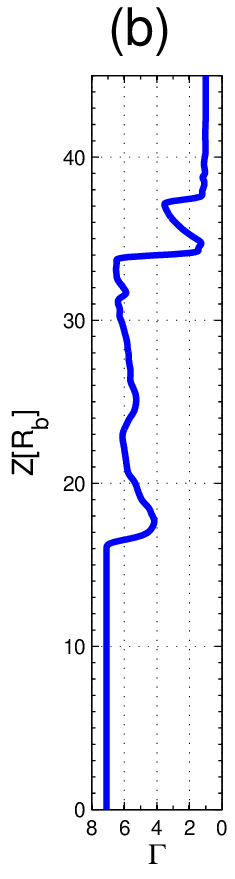}
\includegraphics[scale=0.53, trim= 0 0mm 0 0]{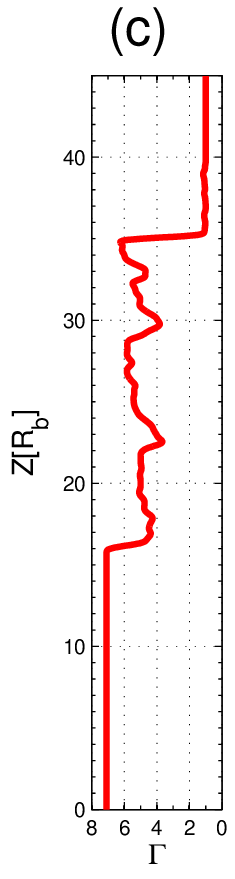}
\includegraphics[scale=0.53, trim= 0 0mm 0 0]{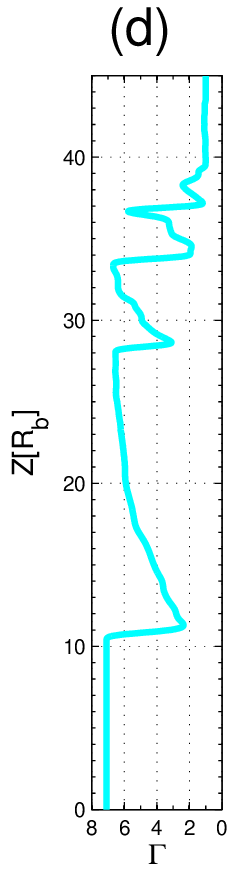}
\includegraphics[scale=0.53, trim= 0 0mm 0 0]{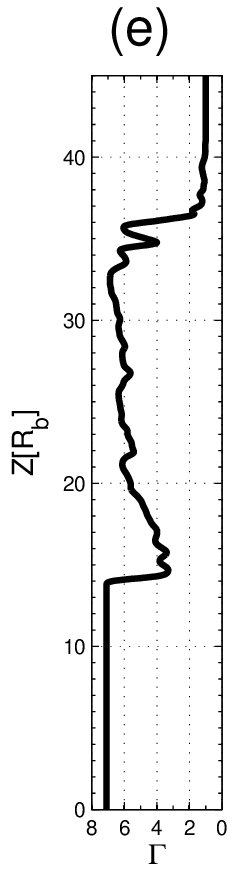}
\includegraphics[scale=0.53, trim= 0 0mm 0 0]{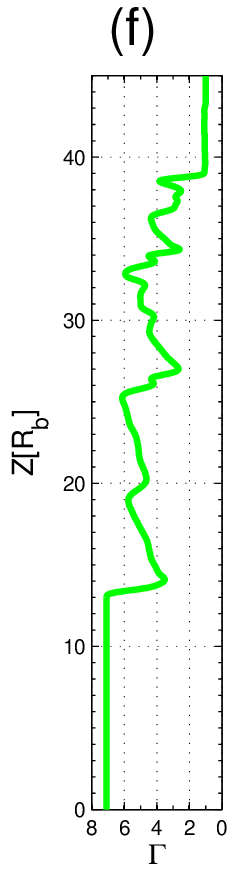}
\includegraphics[scale=0.72,angle=90]{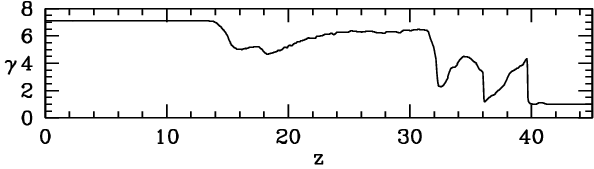}
\caption[]{The spatial profile of $\Gamma$ along the axis of the jet is presented. $6$ schemes of RELDAFNA are shown: (a) CT-PPM, (b) CT-PLM , (c) RK2-PPM, (d) RK2-PLM, (e) RK3-PPM, (f) RK3-PLM using a base mesh of $24 \times 72$~cells and maximum level of refinement of $5$, compared to Figure No.~$22$ from \cite{tchekhovskoy07} using equivalent constant resolution of $384\times1152$~cells.}
\label{fig:lorentz-on-axis}
\end{figure*}

A look at Fig.~\ref{fig:jet-C2} shows that different numerical schemes give different internal structures of the jet. This feature was noted by other groups too \cite{wang08,mignone05a}. Since no analytical results exist for this problem, one has to provide measures of analyzing the structure developed in this problem. Such measures must be compared between different schemes in order to obtain quantitative results that can bound observable values. One such measure is the profile of the Lorentz factor along the axis of the jet at a given time. In Fig.~\ref{fig:lorentz-on-axis} we compare the $6$ schemes of RELDAFNA with a calculation presented in \cite{tchekhovskoy07}. The time presented is $100 \times R_b$, and one can see again the spread between different schemes also seen in the jet location in Fig.~\ref{fig:jet-C2}. The maximum level of refinement is $5$ equivalent to the constant resolution used by \cite{tchekhovskoy07}, $384\times1152$~cells.

\begin{figure*}[h!]
\centering
\includegraphics[scale=0.47, trim= 0 -20mm 130mm 0]{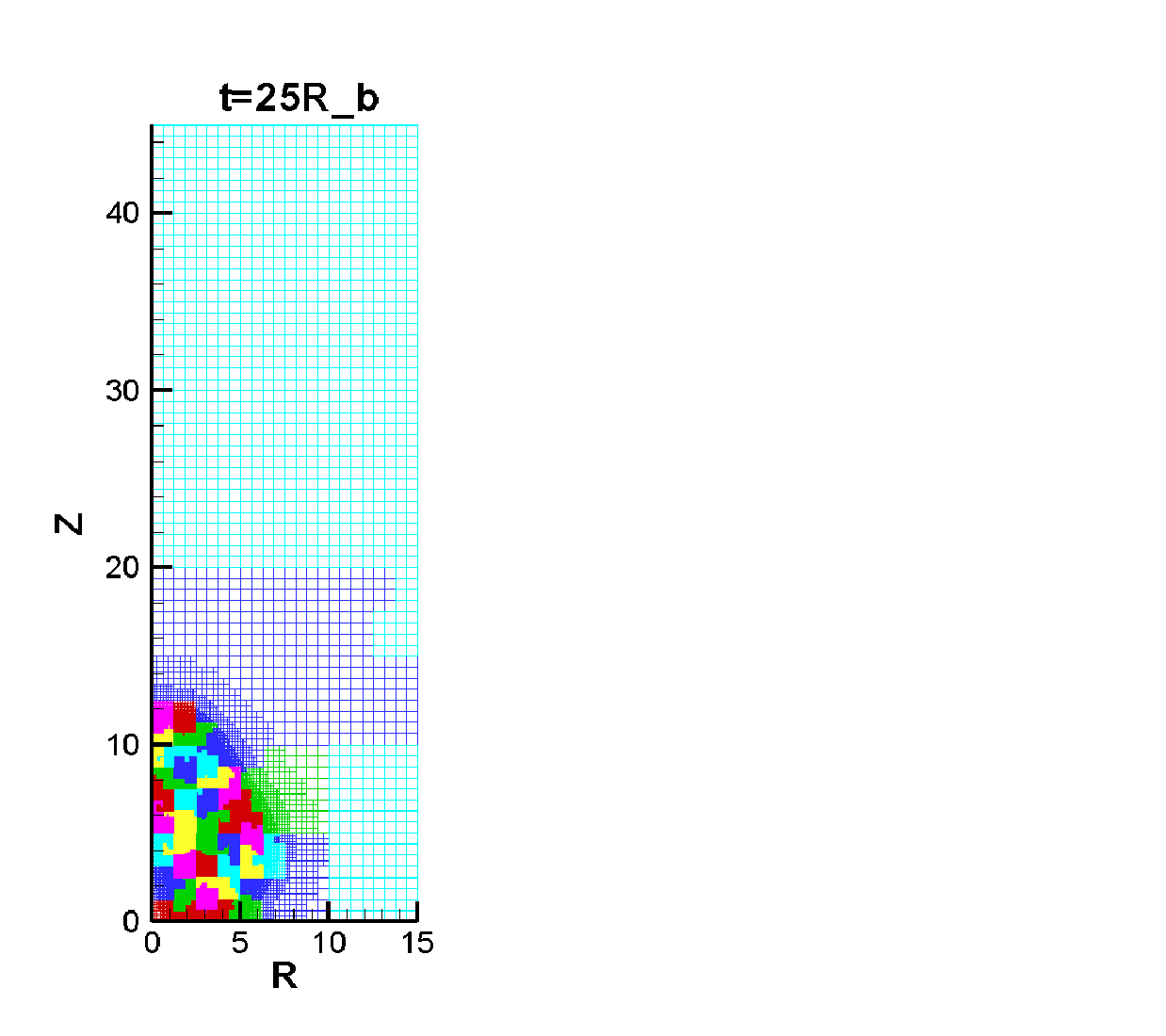}
\includegraphics[scale=0.47, trim= 0 -20mm 130mm 0]{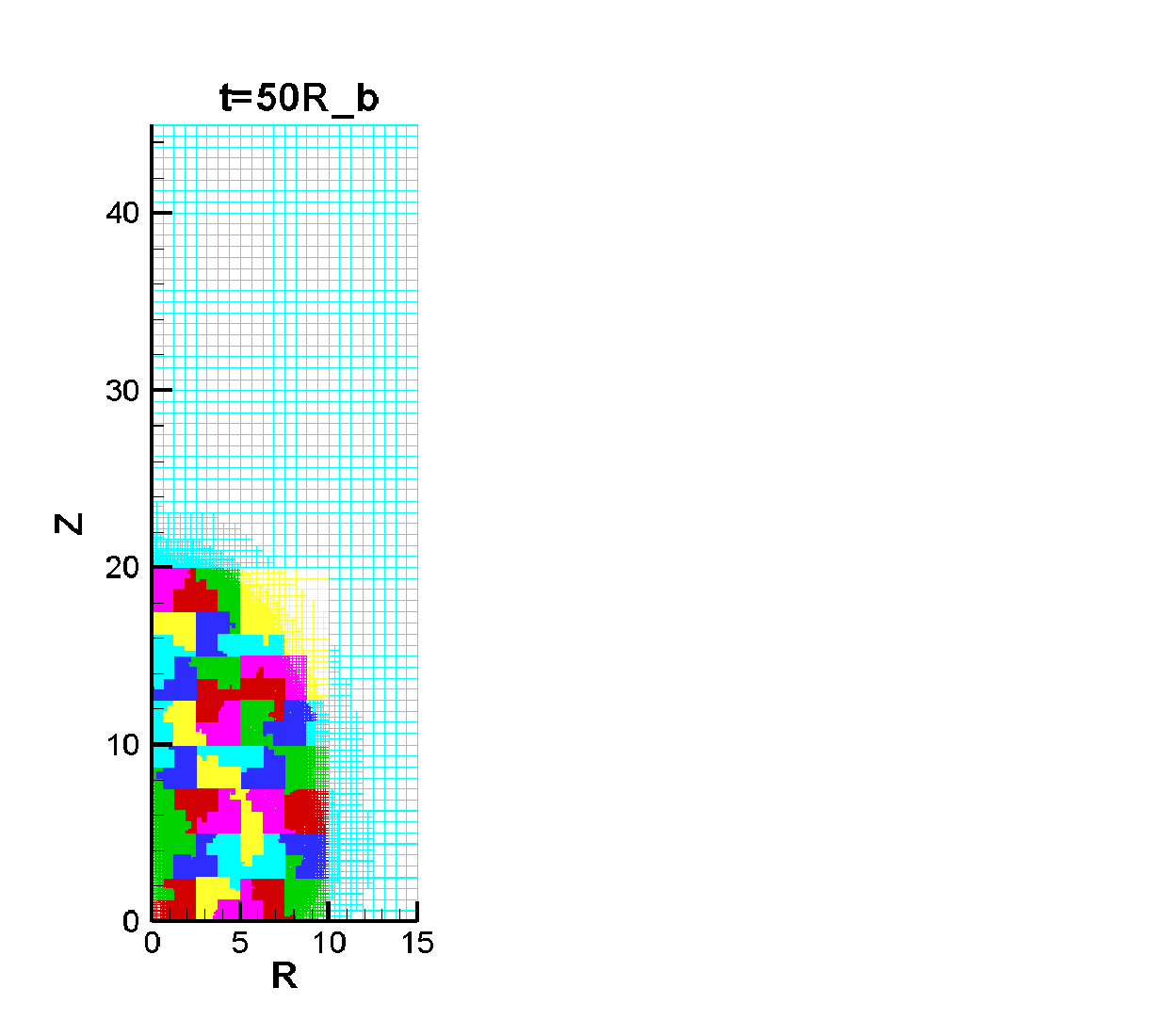}
\includegraphics[scale=0.47, trim= 0 -20mm 130mm 0]{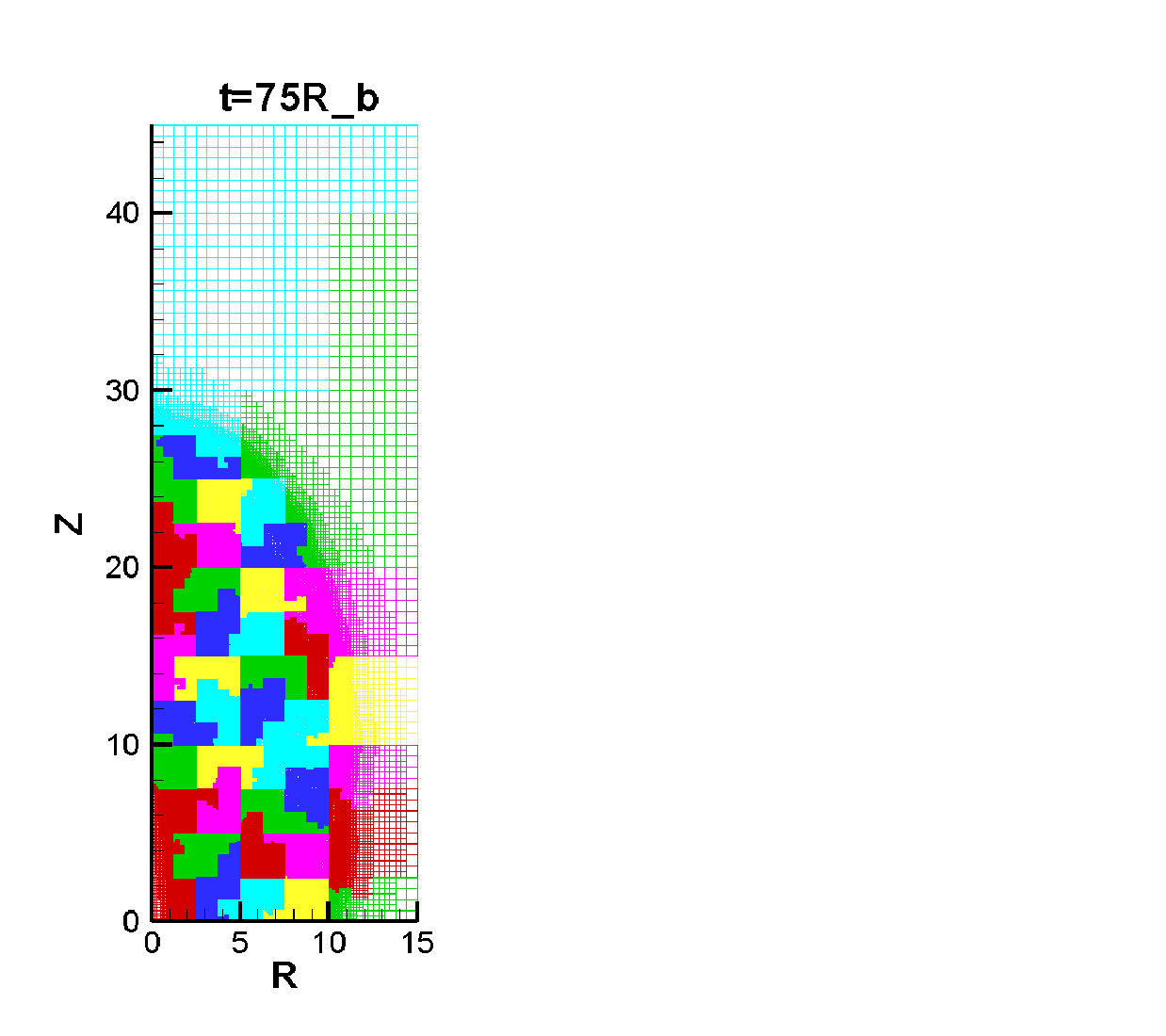}
\includegraphics[scale=0.47, trim= 0 -20mm 130mm 0]{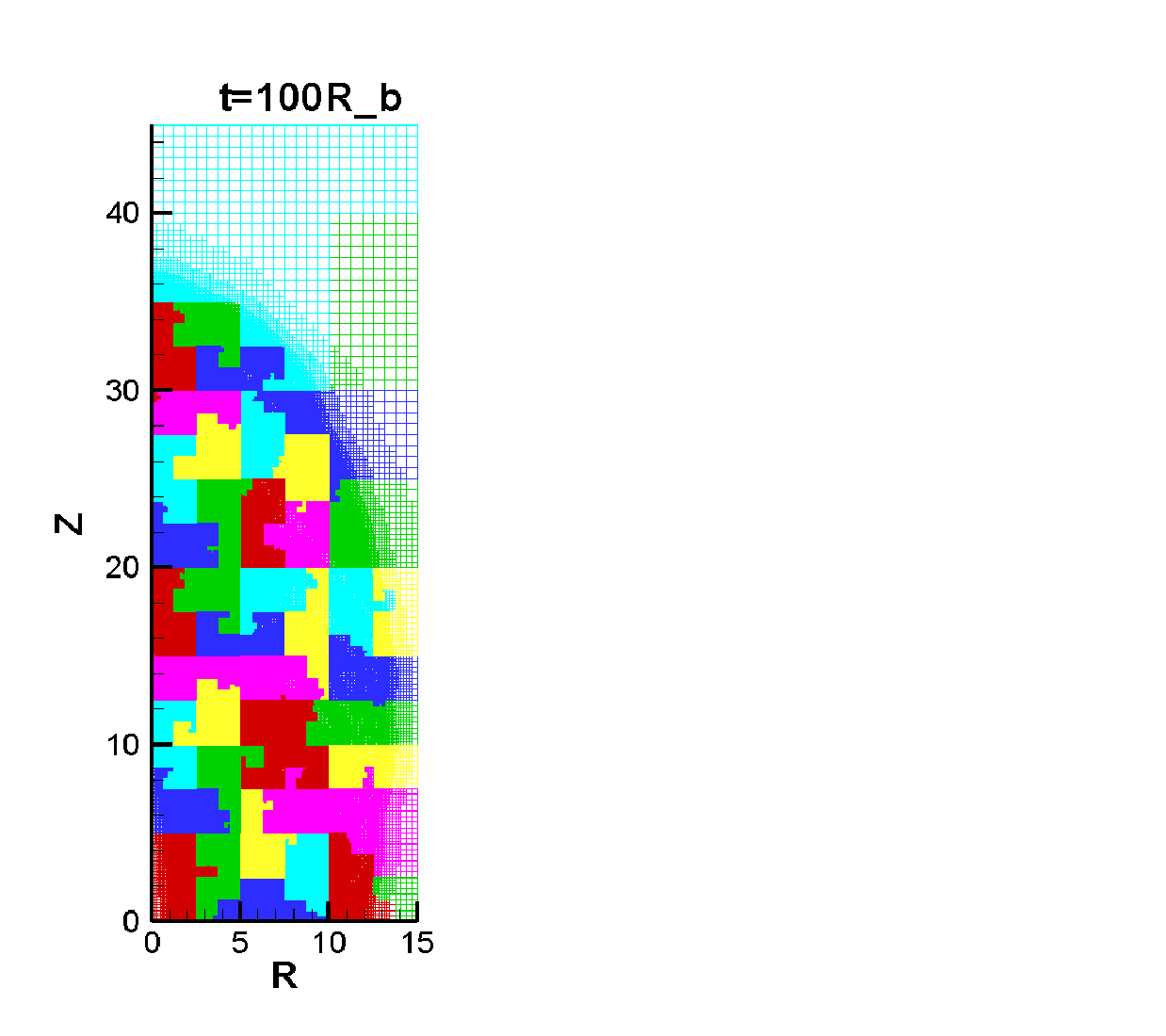}
\caption{Two dimensional cylindrical axis symmetric jet, C2 model. The mesh is shown in $4$ times, as the jet propagates to the ISM and the resolution is increasing through all the space. The base mesh of the
calculations is $24 \times 72$ cells, and the maximum level of refinement is $5$, with effective resolution of $384\times1152$. The different colors represent different processors exchanging cells to balance the work load.\protect\label{fig:jet-mesh-dynamics}}
\end{figure*}

From a numerical point of view, at the early stages of the flow, most of the space is filled with unperturbed ISM and one saves a lot of memory using AMR. At later times, most of the ambient medium is shocked and mixing occurs with in the jet, resulting in a resolution of the highest refinement level in most of the space, as can be seen in Fig.~\ref{fig:jet-mesh-dynamics}.

\begin{figure*}[h!]
\centering
\includegraphics[scale=0.43]{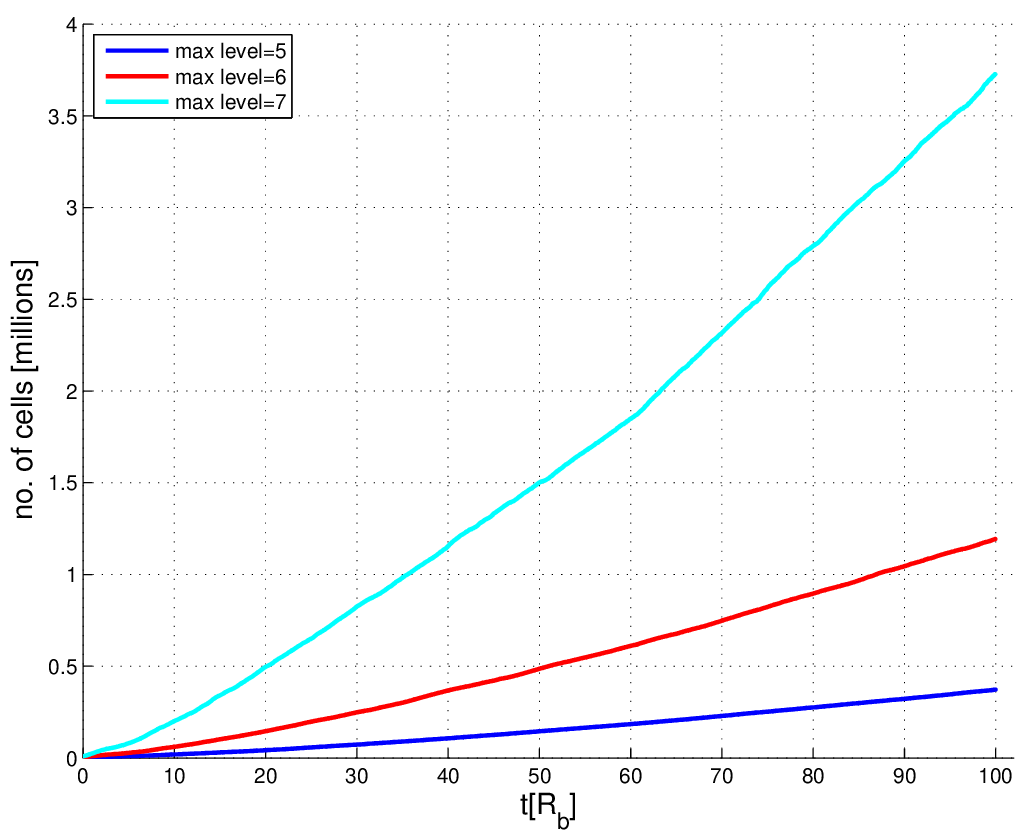}
\includegraphics[scale=0.43]{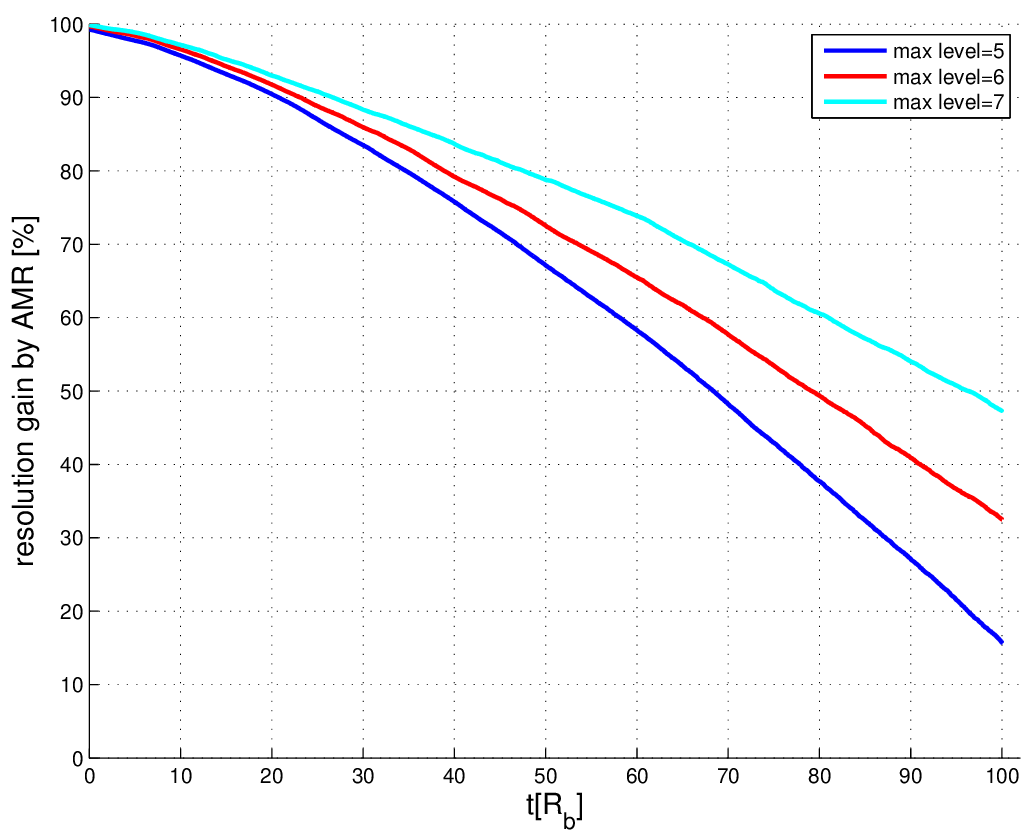}
\caption{Number of cells ({\it left}) and relative gain by the use of AMR ({\it right}) as a function of time, in calculations using CT-PLM of the C2 jet model.\protect\label{fig:number-of-cell-jet-C2}}
\end{figure*}

In Fig.~\ref{fig:number-of-cell-jet-C2}, we plot the number of cells used in different AMR calculation of the C2 model, as a function of time (and of course a function of the penetration depth of the jet into the ambient medium, see Fig.~\ref{fig:jet-velocity}). The relative gain from the use of AMR decreases as the jet penetrates the medium further and further. One can see, that when using a higher maximum level of refinement one saves more with respect to an equivalent constant zoning mesh, but uses a much higher number of cells overall. The calculation with a maximum refinement level of $5$ almost reaches a constant zoning grid at the end of the simulation.

\newpage

\subsection{Collimated Jets and External Shocks} \label{nahliel}

As mentioned in chapter~\ref{applications}, the afterglow of a GRB is produced during the slowdown of the event outflow by the surrounding media. The outflow is believed ultra-relativistic jets. As presented in section~\ref{afterglow}, the material in the jet expands like an angular patch of a spherical blast wave. Therefore, in its early relativistic stages it can be described by the solution of \cite{blandford76}. As the jet decelerates it becomes less and less relativistic, and expands sideways to become spherical and Newtonian. Therefore, at the late stages the GRB outflow obey the Sedov-von Neumann-Taylor non-relativistic self-similar solution.  However, the transition between these two phases is not solved analytically.

In this section, we present calculations made using RELDAFNA by a Ph.D.\ student at the Benoziyo Center for Astrophysics at WIS, Nahliel Wygoda. Following the work presented in \cite{zhang09}, the calculations describe the evolution of a GRB outflow during the afterglow phase. The spatial domain is constructed of the square $(r,z) \in [0,1.1\times10^{19}]\times[0,1.1\times10^{19}]$ with a cylindrical symmetry, filled with an ideal gas with an adiabatic index of $\gamma=\frac{4}{3}$. The base mesh is composed of $24\times24$ cells for the whole domain and the maximum level of refinement used is $16$. The density of the ISM is $1.67\times10^{-24}$~g cm$^{-3}$ and its internal energy density is set to $4.482\times10^{-13}$~erg cm$^{-3}$. Inside a cone defined by $\theta<0.2$~radians and a (spherical) radius of $R_s\le3.8\times10^{17}$~cm, the hydrodynamic variables are obeying the self-similar solution of \cite{blandford76} with $\Gamma=20$ just behind the shock. These variables account for a time $t=147$~days after the burst. After the calculation provides the density and energy distributions in the ISM, calculation of the synchrotron emission is made for comparison with afterglow models.

\begin{figure*}[h]
\centering
\includegraphics[scale=0.3,angle=270]{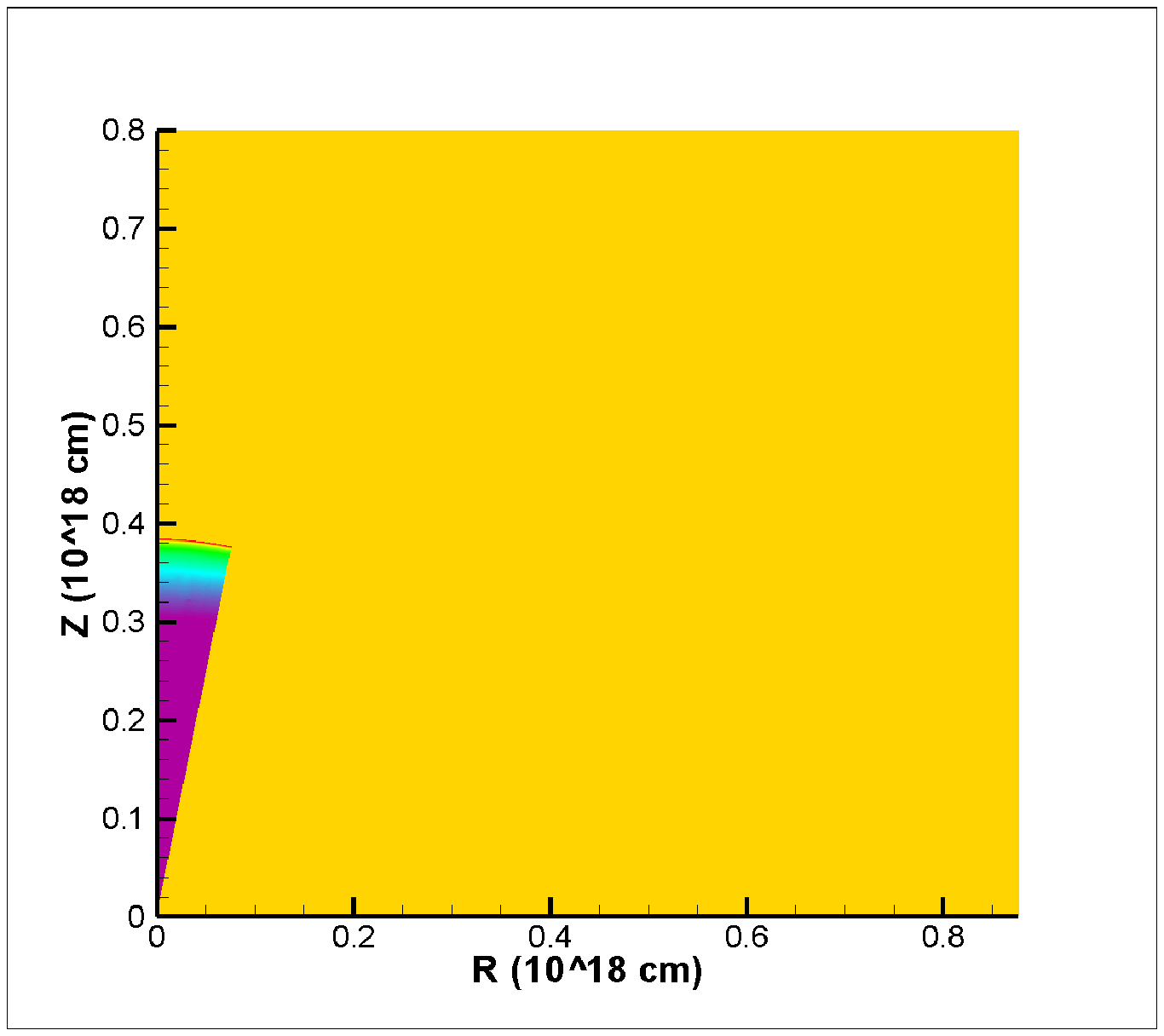}
\includegraphics[scale=0.3,angle=270]{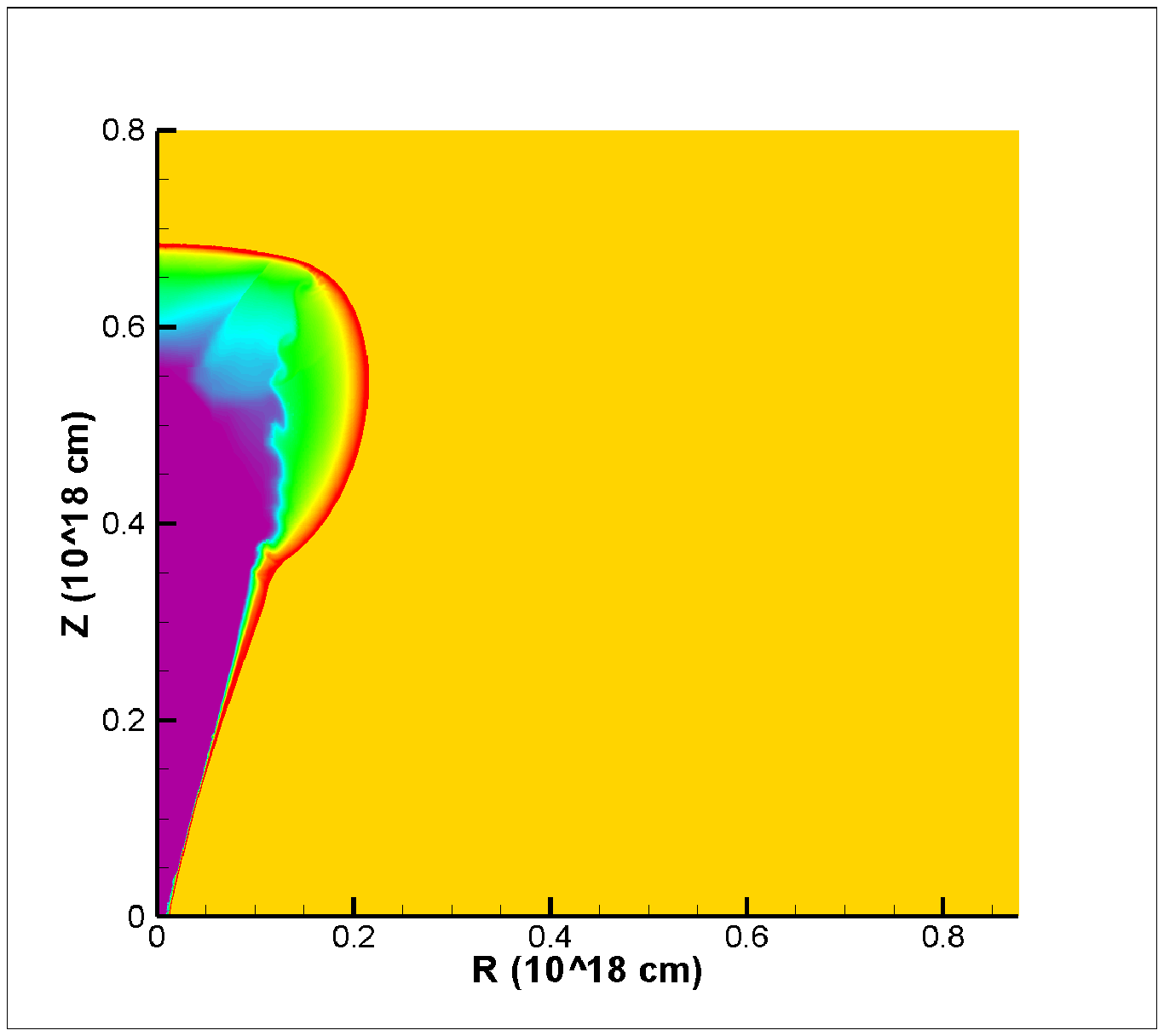}\\
\includegraphics[scale=0.3,angle=270]{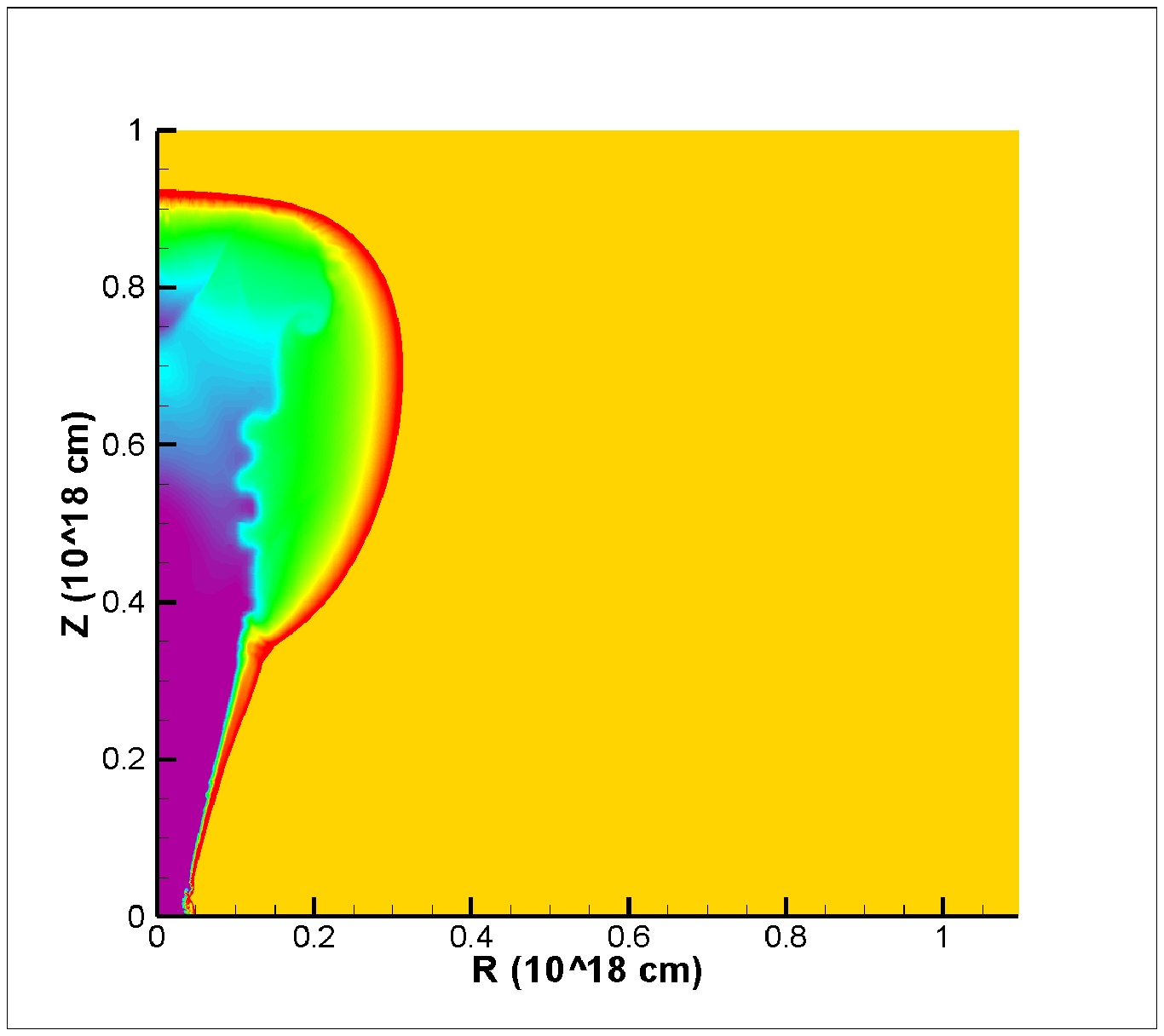}
\includegraphics[scale=0.3,angle=270]{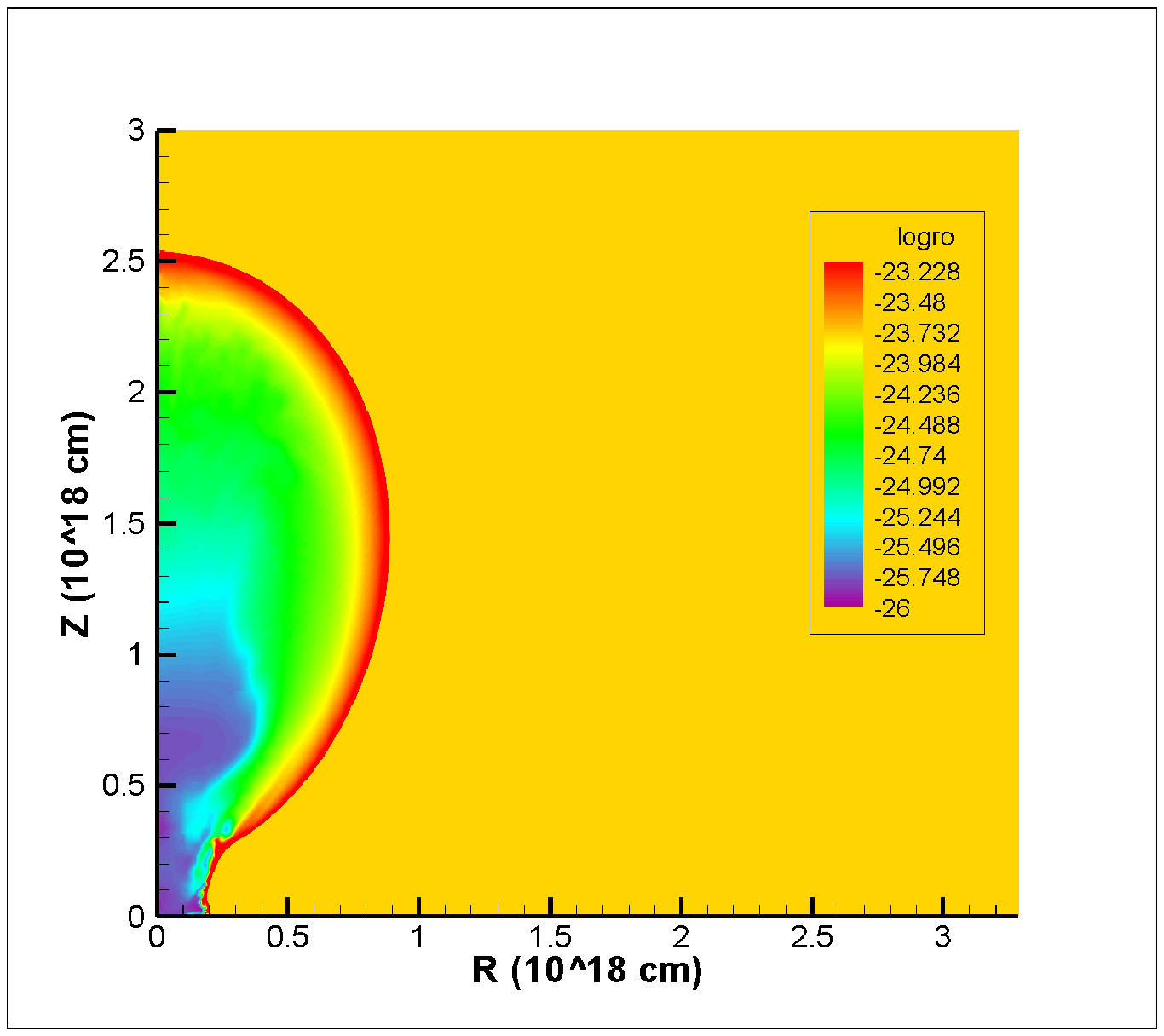}\\
\includegraphics[scale=0.3,angle=270]{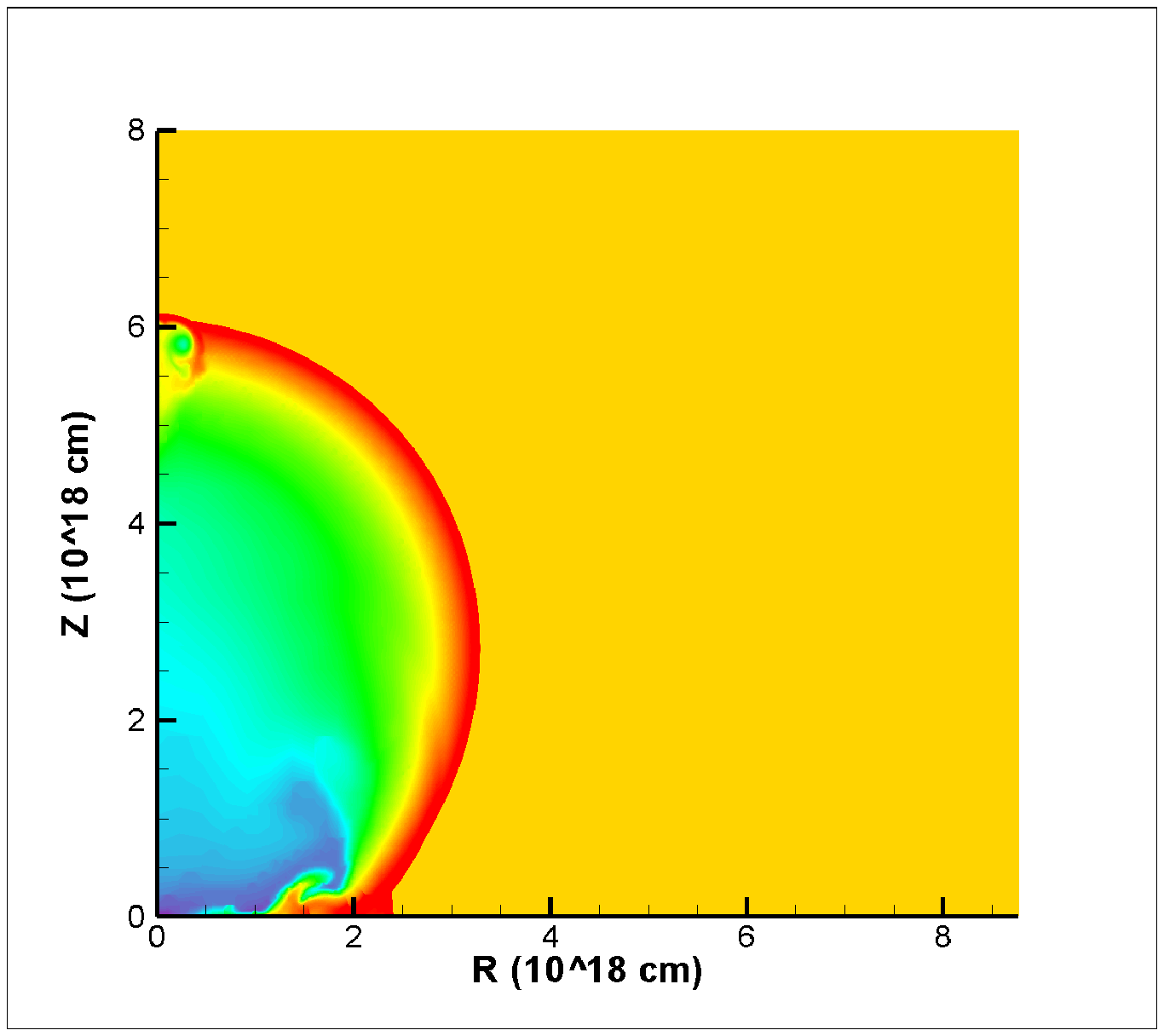}
\includegraphics[scale=0.3,angle=270]{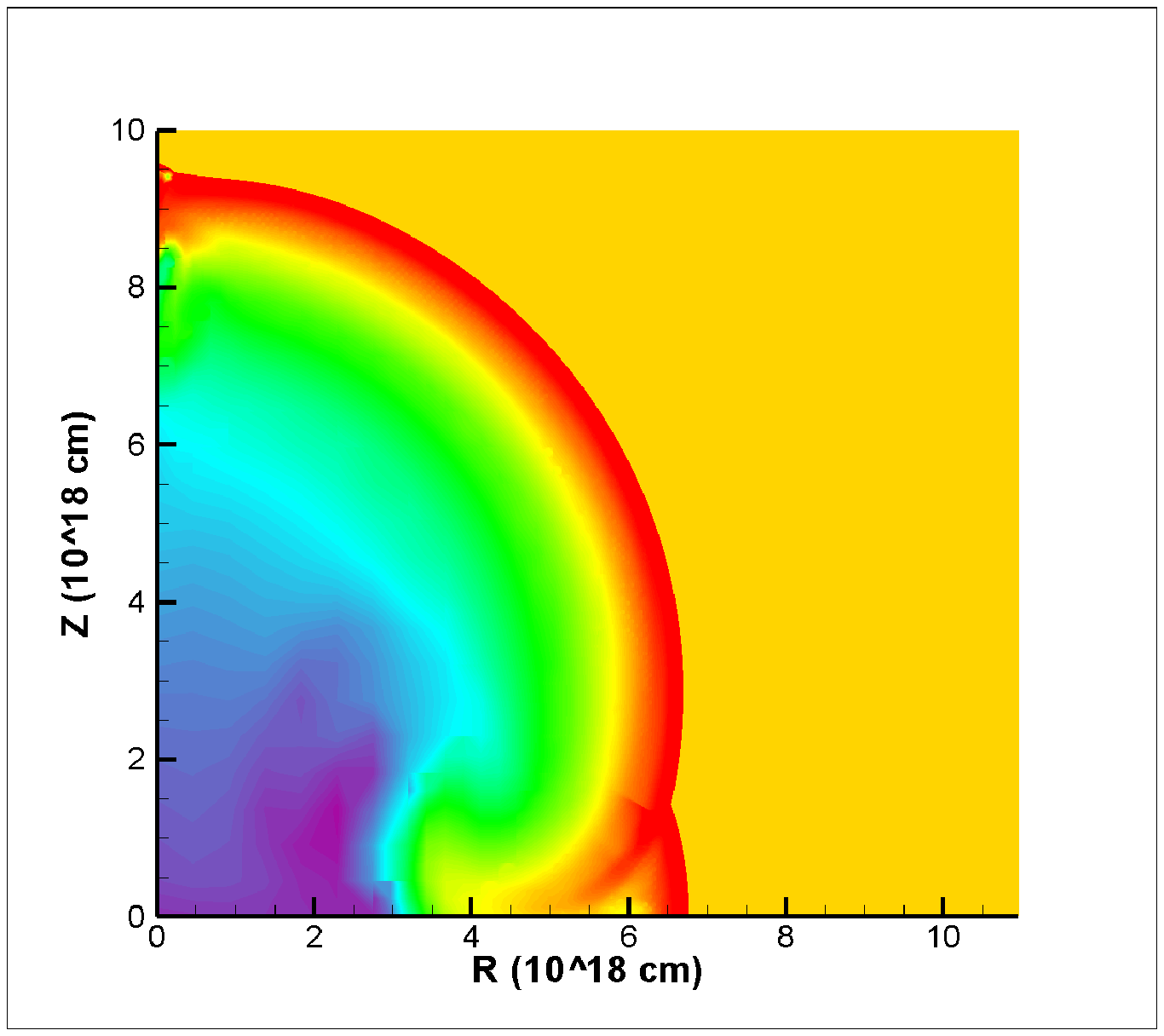}\\
\caption{Time evolution of the density in a logarithmic scale. Snapshots
  of the simulation are shown at $147$~days ({\it top left}), $264$~days, $356$~days, $1051$~days, $28.8$~years and $150$~years ({\it bottom right}) after the explosion.\protect\label{fig:nahliel}}
\end{figure*}

In Fig.~\ref{fig:nahliel} we present the time evolution of the GRB outflow into the ISM. The variable presented is the density in logarithmic scale, through a wide spectrum of time from $\sim100$~days to $\sim100$~years. One can see that the different subfigures of Fig.~\ref{fig:nahliel} have different length scales, noting again that the use of AMR is unavoidable. We see that at times $t\approx3$~years the outflow is still highly anisotropic. As seen also in section~\ref{sec:jetC2}, Kelvin-Helmholtz instability due to velocity shear produce mixing layers visible in early times $t\approx1$~year. The nearly vertical shock front near the equator seen at times $t\approx150$~years is a Mach stem, which is a result of the shock collision along the equator. The different features and results seen in Fig.~\ref{fig:nahliel} are very similar to the ones presented in \cite{zhang09} using the numerical code RAM \cite{zhang06}.

\chapter{Conclusions} \label{summary}

The work described in this thesis, contains the development of a new computer code for the solution of the Special Relativistic Hydrodynamics Equations. The need for a numerical solution of these equations is very wide in the field of high energy astrophysics, since most phenomena in that field involve the flow of matter in relativistic velocities.

The nature of such problems lead to multi length scales and multi-dimensional problems (collimated jets, for example), and therefore the need for an efficient numerical method, both in computer memory and computational effort is crucial in order to provide quantitative results, that should be confronted vs. observations of these high energetic phenomena.

Therefore, Adaptive Mesh Refinement and massive work-load parallelism must be implemented within the codes used, and that is kind of a computational aid for increasing the resolution and speed of calculations. More important, is the search for a method which minimizes the number of calculations made in a time step, but does not affect the accuracy of the numerical solution. In the code described in this thesis, RELDAFNA, we have implemented six schemes which are all accurate, compared to other published codes, but are all using a low number of variables and calculations during a time step, relative to higher order schemes. We have showed that there is no need to look for higher order methods when dealing with problems arising in astrophysics, and that for such applications one should look for more and more efficient codes using different tricks in the relatively low order schemes. The various techniques described, together with AMR and massive work-load balance parallelism, result in an active numerical code already being used to calculate designed problems, in the search for answers to various up to date high energy astrophysical problems.

We have tested RELDAFNA vs. analytical solutions of different one dimensional Riemann problem. Using these problems we studied the differences between the different methods within RELDAFNA. We have also presented the ability of RELDAFNA to deal with different problems regarding astrophysical applications, which do not have analytical solutions.

Today, RELDAFNA is already in use for astrophysical calculations. In the future, some of the following issues will be addressed through calculations made by RELDAFNA.

\begin{itemize}
  \item One dimensional spherical explosions were presented in this work. We plan to analyze the effects of deviations from spherical symmetry of the explosion due to non-spherical density profiles of the star or an off-center explosion. These effects may describe better the signal observed from shock breakouts.

  \item Propagation of relativistic jets through the ambient medium of the ISM were presented in this work. We plan to study their propagation through a denser medium \textit{deeper} within the star before the breakout. The mechanism of very high energy deposition would be analyzed through such calculations.

  \item The expansion of the GRB outflow through the ISM was presented as an application of RELDAFNA. Such calculations will be further improved to correctly interpret the afterglow observations.
\end{itemize}

Now that the code is in use for physical calculations, we also turn to additional numerical developments such as,
\begin{itemize}

 \item Different EOS models.

 \item Search for other efficient methods, starting at the search for an answer of the questions presented at the end of section~\ref{rie1d1}, i.e. the oscillations observed in slowly moving structures and the need of very high resolutions to resolve the thin structures developed when non-zero tangential velocities are present.

 \item Implementation of different geometries for the grid used, starting with a $2$~dimensional $r-\theta$ grid to enhance the conservation of spherical symmetry in problems where such symmetry exists.
     
 \item Addition of magnetic hydrodynamics module.
\end{itemize}

We hope that the work presented in this thesis would pay off in contributing results of advanced and future research, mainly in astrophysics and also in pure numerical investigation of the SRHD equations.

\addcontentsline{toc}{chapter}{Appendixes.}
\appendix

\chapter{The NR Procedure for the Variables Conversion} \label{Newton-Raphson}

Since primitive variables are needed in the reconstruction process, we need to convert conserved variables to
primitive variables. While conserved variables can be computed directly from primitive variables using Eq. (\ref{conserved}), the inverse operation is not straightforward.\\
As for now RELDAFNA deals with one kind of EOS, namely an ideal gas EOS which reads the form:
\begin{equation}
 p = (\gamma - 1) \rho \epsilon \, ,
\label{ideal-EOS}
\end{equation}
where $\gamma$ is the adiabatic index of the gas. We use a Newton-Raphson (NR) iteration to solve a nonlinear equation for the pressure, to recover primitive variables from conserved variables.\\
First, we fix unphysical result of the different integration schemes, meaning if in a cell the variable $D$ has been updated to a negative number, we say the following
\begin{equation}
p=\epsilon_p\,,\rho=\epsilon_\rho\,,
\rho\epsilon=\frac{\epsilon_p}{\gamma-1}\,, \A{v}=0\,, c=0\,,
\Gamma=1\,,
\end{equation}
where $\epsilon_p$ is a negligible pressure quantity, which in our calculation is usually set to be $10^{-16}$, and $\epsilon_{\rho}$ is also a negligible density quantity, which in our calculation is usually set to be $10^{-10}$. In the common case where the integration scheme has updated the conserved variables with physical quantities we turn to the NR operation. We define a minimum pressure available as
\begin{equation}
p_{min}=max(\left|S\right|-\tau-D,\epsilon_p)\,,
\end{equation}
and a first guess for the pressure as
\begin{equation}
p=\HALF(p_{min}+(\gamma-1)\tau)\, .
\end{equation}
We iterate the following procedure until a variable $err_p$ reduces to a negligible quantity (usually in our calculations this quantity is $10^{-8}$),
\begin{equation}
\begin{array}{rcl}
p & = & max(p,p_{min})\,,\\
 \noalign{\medskip}
\A{v} & = & \frac{\A{S}}{\tau+D+p}\,,\\
 \noalign{\medskip}
\Gamma & = & \frac{1}{\sqrt{1-\left|\A{v}\right|^2}}\,,\\
 \noalign{\medskip}
F & = & (\gamma-1)\frac{\tau+D(1-\Gamma)+p(1-\Gamma^2)}{\Gamma^2}-p\,,\\
 \noalign{\medskip}
\frac{dF}{dp} & = & (\gamma-1)\left|\A{v}\right|^2\frac{\tau+D(1-\Gamma)+p}{\tau+D+p}-1\,,\\
 \noalign{\medskip}
p^{new} & = & max(p^{old}-\frac{F}{\frac{dF}{dp}},p_{min})\,,\\
 \noalign{\medskip}
err_p & = &
\left|1-\frac{p^{new}}{p^{old}}\right|\,,\\
 \noalign{\medskip}
\end{array}
\end{equation}
after the pressure has been calculated iteratively we calculate the other primitive variables through Eq. (\ref{conserved})
\begin{equation}
\begin{array}{rcl}
\A{v} & = & \frac{\A{S}}{\tau+D+p}\,,\\
 \noalign{\medskip}
\Gamma & = & \frac{1}{\sqrt{1-\left|\A{v}\right|^2}}\,,\\
 \noalign{\medskip}
\rho & = & \frac{D}{\Gamma}\,,\\
 \noalign{\medskip}
c & = & \frac{\sqrt{p\gamma(\gamma-1)}}{p\gamma+\rho(\gamma-1)}\,,\\
 \noalign{\medskip}
\left(\rho\epsilon\right) & = & \frac{\tau+p+D}{\Gamma^2}-\rho-p\\
 \noalign{\medskip}
\end{array}
\end{equation}

\chapter{The Piecewise Parabolic Method Spatial Reconstruction} \label{spatial-PPM}

We now describe the Piecewise Parabolic Method (PPM) spatial construction leading to the values at cell's interfaces, a description mostly based on the idea given in \cite{collela84} and is also presented in \cite{marti96,mignone05b}.

As mentioned in section~\ref{reconstruct}, the spatial reconstruction is being calculated for a defined direction, either for Runge-Kutta time integration scheme, or for Characteristic-Tracing scheme. Therefore, the following calculations are made for all interfaces between cells in the treated direction. In this stage \textbf{interfaces} are being treated. Denote the cell in the negative direction of the interface with a subscript $1$, and the cell in the positive direction of the interface with a subscript $2$. Since the interpolation is $2^{nd}$ order in space we need information from two other neighboring cells of the interface in the treated direction. Denote the cell in the negative direction of cell $1$ with a subscript $m1$, and the cell in the positive direction of cell $2$ with a subscript $p1$. Define the following
\begin{equation}
\begin{array}{rcl}
B & = & \frac{d\xi_2(d\xi_2+d\xi_{p1})}{(d\xi_{m1}+d\xi_1+d\xi_2+d\xi_{p1})(d\xi_1+2d\xi_2)}\,,\\
 \noalign{\medskip}
C & = & \frac{d\xi_1(d\xi_{m1}+d\xi_1)}{(d\xi_{m1}+d\xi_1+d\xi_2+d\xi_{p1})(2d\xi_1+d\xi_2)}\,,\\
 \noalign{\medskip}
A & = & \frac{d\xi_1+2(Cd\xi_2-Bd\xi_1)}{d\xi_1+d\xi_2}\,,\\
 \noalign{\medskip}
\end{array}
\end{equation}
and the vector $\A{V}$ for each cell
\begin{equation}
 \A{V} = \left(\rho, v^1,... v^{dim.}, p, \Gamma\right)^{T}\, .
\end{equation}
Define
\begin{equation}
\begin{array}{rcl}
\delta \A{V}^1 & = & \left(\frac{\left(\A{V}_2-\A{V}_1\right)\left(2d\xi_{m1}+d\xi_1\right)}{d\xi_1+d\xi_2}
+\frac{\left(\A{V}_1-\A{V}_{m1}\right)\left(2d\xi_2+d\xi_1\right)}{d\xi_{m1}+d\xi_1}\right)
\frac{d\xi_1}{d\xi_{m1}+d\xi_1+d\xi_2}\,,\\
 \noalign{\medskip}
\delta \A{V}^2 & = &
\left(\frac{\left(\A{V}_{p1}-\A{V}_2\right)\left(2d\xi_1+d\xi_2\right)}{d\xi_{p1}+d\xi_2}+
\frac{\left(\A{V}_2-\A{V}_1\right)\left(2d\xi_{p1}+d\xi_2\right)}{d\xi_1+d\xi_2}\right)
\frac{d\xi_2}{d\xi_1+d\xi_2+d\xi_{p1}}\,.\\
 \noalign{\medskip}
\end{array}
\end{equation}
Define
\begin{equation}
 \bar{\delta \A{V}}_1 = \left\{\begin{array}{ccc}
 min(\left|\delta \A{V}_1\right|,2\left|\A{V}_1-\A{V}_{m1}\right|,2\left|\A{V}_1-\A{V}_2\right|)\frac{\left|\delta \A{V}_1\right|}{\delta \A{V}_1} & \quad \textrm{if} \; & \left(\A{V}_2-\A{V}_1\right)\left(\A{V}_1-\A{V}_{m1}\right)>0 \,, \\ \noalign{\medskip}
 0 & \quad \textrm{if} \; & \left(\A{V}_2-\A{V}_1\right)\left(\A{V}_1-\A{V}_{m1}\right) \leq 0 \,, \\ \noalign{\medskip}
\end{array}\right.
\end{equation}

\begin{equation}
 \bar{\delta \A{V}}_2 = \left\{\begin{array}{ccc}
 min(\left|\delta \A{V}_2\right|,2\left|\A{V}_2-\A{V}_1\right|,2\left|\A{V}_2-\A{V}_{p1}\right|)\frac{\left|\delta \A{V}_2\right|}{\delta \A{V}_2} & \quad \textrm{if} \; & \left(\A{V}_{p1}-\A{V}_2\right)\left(\A{V}_2-\A{V}_1\right)>0 \,, \\ \noalign{\medskip}
 0 & \quad \textrm{if} \; & \left(\A{V}_{p1}-\A{V}_2\right)\left(\A{V}_2-\A{V}_1\right) \leq 0 \,, \\ \noalign{\medskip}
\end{array}\right.
\end{equation}

The spatial reconstruction of the two cells sharing the interface is being calculated using the following equation
\begin{equation}
\A{V}_{R_1}=\A{V}_{L_2}=\A{V}_1+A(\A{V}_2-\A{V}_1)+B\bar{\delta \A{V}}_1-C\bar{\delta \A{V}}_2\,.
\label{VRVL}
\end{equation}
After Equation~\ref{VRVL} was calculated for all the interfaces, all the cells have two vectors $\A{V}_R$ and $\A{V}_L$ representing the spatial reconstruction of the variables $\rho$, $\A{v}$, $p$ and $\Gamma$ at the \textbf{positive} and \textbf{negative} sides of the cell, respectively. The vectors $\bar{\delta \A{V}}$ are also stored in memory. After the calculation of the vectors $\A{V}_R$ and $\A{V}_L$ for each cell, another vector is defined for each \textbf{cell},
\begin{equation}
\delta ^2 \rho_i = \frac{\frac{\rho_p-\rho_i}{d\xi_p+d\xi_i}-\frac{\rho_i-\rho_m}{d\xi_i+d\xi_m}}{d\xi_m+d\xi_i+d\xi_p}
\end{equation}
where a subscript $i$ refers to the cell being calculated, and subscripts $p,m$ refer to the cells neighboring cell $i$, in the positive and negative direction being treated, respectively. With the definition of $\delta ^2 \rho$ discontinuities in the reconstruction of the density profile can be detected, which is an unwanted feature of the PPM method. Define
\begin{equation}
 \tilde{\eta} = \left\{\begin{array}{ccc}
 \frac{\frac{\left(d\xi_i-d\xi_m\right)^3+\left(d\xi_p-d\xi_p\right)^3}{\rho_p-\rho_m}\left(\delta^2 \rho_m-\delta^2 \rho_p\right)}{d\xi_p-d\xi_m} & \quad \textrm{if} \; & \delta ^2 \rho_p \delta ^2 \rho_m < 0 \, , \left|\rho_p-\rho_m\right|-\epsilon_1min\left(\rho_p,\rho_m\right)>0 \,, \\ \noalign{\medskip}
 0 & \quad \textrm{otherwise} \,, \\ \noalign{\medskip}
\end{array}\right.
\end{equation}
and
\begin{equation}
 \eta = \left\{\begin{array}{ccc}
 max\left(0,min\left(\eta_1\left(\tilde{\eta}-\eta_2\right)\right),1\right) & \quad \textrm{if} \; & \gamma K_0 \frac{\rho_p-\rho_m}{min\left(\rho_p,\rho_m\right)} \ge \frac{p_p-p_m}{min\left(p_p,p_m\right)} \,, \\ \noalign{\medskip}
 0 & \quad \textrm{otherwise} \,, \\ \noalign{\medskip}
\end{array}\right.
\end{equation}
where $\gamma$ is the adiabatic index of the ideal gas, and $\eta_1,\eta_2,K_0,\epsilon_1$ are user defined parameters, to tune the discontinuities detection and smoothing. For our calculations we use the parameters proposed in \cite{collela84}, i.e $\eta_1=5$, $\eta_2=0.05$, $K_0=1$ and $\epsilon_1=0.1$. The density reconstruction is modified in the following way
\begin{equation}
\begin{array}{rcl}
\A{V}_{L_\rho}=\A{V}_{L_\rho}\left(1-\eta\right)+\left(\rho_m+\HALF\delta^m_m\right)\eta\,,\\
 \noalign{\medskip}
\A{V}_{R_\rho}=\A{V}_{R_\rho}\left(1-\eta\right)+\left(\rho_p-\HALF\delta^m_p\right)\eta\,.\\
 \noalign{\medskip}
\end{array}
\end{equation}
Finally, the reconstruction is monotonized. If $\left(\A{V}_R-\A{V}\right)\left(\A{V}-\A{V}_L\right)\le0$ set $\A{V}_R=\A{V}_L=\A{V}$. Define
\begin{equation}
\A{A}_1=\left(\A{V}_R-\A{V}\right)\left(\A{V}-\A{V}_L\right)\,,
\end{equation}
and
\begin{equation}
 \A{A}_2 = \left\{\begin{array}{ccc}
 \A{V}_L & \quad \textrm{if} \; & \A{A}_1=0 \,, \\ \noalign{\medskip}
 3\A{V}-2\A{V}_R & \quad \textrm{otherwise} \,, \\ \noalign{\medskip}
\end{array}\right.
\end{equation}

\begin{equation}
 \A{A}_3 = \left\{\begin{array}{ccc}
 \A{V}_R & \quad \textrm{if} \; & \A{A}_1=0 \,, \\ \noalign{\medskip}
 3\A{V}-2\A{V}_L & \quad \textrm{otherwise} \,. \\ \noalign{\medskip}
\end{array}\right.
\end{equation}

If $\left(\A{V}_R-\A{V}_L\right)\left(\A{V}_L-\A{A}_2\right)<0$ set $\A{V}_L=\A{A}_2$, and if $\left(\A{V}_R-\A{V}_L\right)\left(\A{A}_3-\A{V}_R\right)<0$ set $\A{V}_R=\A{A}_3$.

\chapter{The Characteristic Tracing PPM Predictor Step} \label{CT-PPM}

We now describe how the calculation of the half time advance of the interface values using a Piecewise Parabolic Method (PPM) for the reconstruction, a description mostly based on the ideas given in \cite{miller96,mignone05b}.

In this stage \textbf{interfaces} are being treated. Following the notation of Appendix~\ref{spatial-PPM} and section~\ref{reconstruct} define the following vectors for all interfaces in the treated direction
\begin{eqnarray}
\A{V}_P=\A{V}_{R_1}\, , \quad \A{V}_N=\A{V}_{L_2}\,,\\
 \noalign{\medskip}
\Delta \A{V}_i=\A{V}_{R_i}-\A{V}_{L_i}\, \, i \in \{1,2\}\,,\\
 \noalign{\medskip}
\A{V}^6_i=6(\A{V}_i-\HALF(\A{V}_{R_i}+\A{V}_{L_i}))\,  \, i\in\{1,2\}\,.
\end{eqnarray}
The eigenvalues of the Jacobian using the values of the cell in the negative side of the interface. As been noted in section~\ref{CT-PLM} the geometric values of the interface are
\begin{equation}
\begin{array}{rcl}
R_1 & = & R+\frac{dr}{2} \,, \\
 \noalign{\medskip}
\xi_1 & = & \xi+\frac{d\xi}{2}.\\
 \noalign{\medskip}
\end{array}
\end{equation}
For all the eigenvalues calculated for the cell in the negative side define
\begin{equation}
\begin{array}{rcl}
Y^\lambda & = & max\left(0,\lambda dt\right)\,,\\
 \noalign{\medskip}
\xi{1_\lambda} & = & \xi\left(R_1-Y_\lambda\right)\,,\\
 \noalign{\medskip}
X_\lambda & = & \frac{\xi_1-\xi_{1_\lambda}}{d\xi}\,,\\
 \noalign{\medskip}
\A{V}^\lambda & = & \A{V}_P-\HALF X_\lambda\left(\Delta
\A{V}_1-\left(1-\frac{2X}{3}\A{V}^6_1\right)\right)\,.\\
 \noalign{\medskip}
\end{array}
\label{X-Y-PPM}
\end{equation}
Since the time advance of the interface values is calculated time using the eigenstates of the Jacobian, consider a contraction of $\A{V}^\lambda$ to the components dealing with $\rho$, $\A{v}$ and $p$. This contraction will be named $\A{V}_M^\lambda$. This vector contains the approximation of the interface values for the primitive variables. These variables are used to construct the whole static interface state via Equations~\ref{thermodynamic-variables}-\ref{Lorentz-factor}. If the Lorentz factor interpolation is needed set a first approximate $\Gamma^1=\A{V}^5_6$ (Remember that we are presenting an example where one deals with a $3$~-dimensional problem). When this fails to give a physically meaningful result, i.e. a Lorentz factor $>1$, consider another approximation $\Gamma^1=\A{V}_{P_6}\,,$ and if this approximation also fails a constant approximation $\Gamma^1=\Gamma$ is used. After approximating the Lorentz factor at the interface, the approximated velocity components are renormalized so they would match the Lorentz factor as presented in Equation~\ref{velocity_correction}.\\
$\rho^1, \A{v}^1 p^1, h^1, c^1$ and $\Gamma^1$ are used to calculate the left and right eigenstates of the negative side of the interface (Equations~\ref{eq:eigenstates}). Define
\begin{equation}
\A{V}_1^{\frac{\Delta t}{2}}=\A{V}_M^5-\sum_{i=1}^{4}
\frac{max\left(\lambda^i,0\right)}{\lambda^i}\left(L^{:i}\bullet\left(\A{V}_M^5-\A{V}_M^i\right)\right)R^{:i}\,,
\end{equation}
where again the notation $L^{:i}$ refers to the $i^{th}$ column of the matrix $L$. The treatment of the geometrical factors is being done the same way as in the PLM case (section~\ref{CT-PLM}) following Equations (\ref{ro-v-p-calc-negative}), and the rest of the thermodynamic variables are calculated according to the EOS as in Equations (\ref{thermodynamic-variables}).\\
The calculation of the positive side is very similar to the calculation presented ahead. Therefore, we will briefly present the differences between them. The eigenvalues of the Jacobian are calculated using the values of the cell in the \textbf{positive} side of the interface. Denote the following
\begin{equation}
\begin{array}{rcl}
R_2 & = & R-\frac{dr}{2} \,, \\
 \noalign{\medskip}
\xi_2 & = & \xi-\frac{d\xi}{2}.\\
 \noalign{\medskip}
\end{array}
\end{equation}
For all the eigenvalues calculated for the cell in the \textbf{positive} side define
\begin{equation}
\begin{array}{rcl}
Y_\lambda & = & max\left(0,\A{-} \lambda dt\right)\,,\\
 \noalign{\medskip}
\xi_{2_\lambda} & = & \xi\left(R_2 \A{+} Y\right)\,,\\
 \noalign{\medskip}
X_\lambda & = & \frac{\xi_2-\xi_{2_\lambda}}{d\xi}\,,\\
 \noalign{\medskip}
\A{V}^\lambda & = & \A{V}_N \A{+} \HALF X_\lambda\left(\Delta
\A{V}_2+\left(1-\frac{2X}{3}V^6_2\right)\right)\,.\\
 \noalign{\medskip}
\end{array}
\end{equation}
$\A{V}^\lambda$ is contracted to the components $\rho$, $\A{v}$ and $p$, and the contraction is denoted $\A{V}_M^\lambda$. The interface state are built in the same way as the negative side of the interface. The eigenstates matrices are calculated using the values of the \textbf{positive} side of the interface. Define
\begin{equation}
\A{V}_2^{\frac{\Delta t}{2}}=\A{V}_M^1-\sum_{i=2}^{5}
\frac{min\left(\lambda^i,0\right)}{\lambda^i}\left(L^{:i}\bullet\left(\A{V}_M^1-\A{V}_M^i\right)\right)R^{:i}\,,
\end{equation}
The treatment of the geometrical factors and the rest of the thermodynamic variables is done the same way as the negative side.

Similar to the section~\ref{CT-PLM} unphysical interface values at half timestep for Lorentz factor and the velocity components might emerge. If the velocity norm calculated at half timestep at the two sides of the interface is $<1$ the Lorentz factor is interpolated by its definition, i.e. for $i \in\{1,2\}$
\begin{equation}
\Gamma_i^{\frac{\Delta t}{2}}=\frac{1}{\sqrt{1-\left|\A{v}_i^{\frac{\Delta t}{2}}\right|^2}}.
\end{equation}
If on the other hand, unphysical velocity norm has been calculated at any of the sides of the interface, i.e. $\left|\A{v}_i^{\frac{\Delta t}{2}}\right|\ge1$ the Lorentz factors are interpolated via their own spatial reconstruction and then the velocity components are renormalized according to the calculated Lorentz factors, in the following manner:

The first approximation is, for $i \in \{1,2\}$
\begin{equation}
\Gamma_i^{\frac{\Delta t}{2}}=\A{V}_\Gamma^2\,,
\end{equation}
i.e. the Lorentz factor component of the vector built for the $2^{nd}$ eigenvalue is assumed as the half time step advance (see Equations~\ref{X-Y-PPM} and remember that we are presenting the solution of the $1^{st}$ direction). In cases where this interpolation also fails to give a physically meaningful result, i.e. a Lorentz factor $\ge 1$, a simpler method is used such as, $\Gamma_1^{\frac{\Delta t}{2}}=\A{V}_{P_\Gamma}$ and $\Gamma_2^{\frac{\Delta t}{2}}=\A{V}_{N_\Gamma}$. When this approximation also fails the constant interpolation is used and the cell Lorentz factor is assigned to the interface. After interpolating the Lorentz factor at half timestep at the two sides of the interfaces, the calculation of the velocity components is corrected in the same manner as Equations~\ref{velocity_correction}.

\chapter{Rimann problems} \label{code-tests}

In chapter~\ref{computational-astrophysics} we have presented the solution of a one dimensional Riemann problem. Other kinematical regimes can be tested by different Riemann problems, both one dimensional and two dimensional. Such problems were tested versus every code developed for the solution of the SRHD equations, and one must verify that the code developed is, before all, consistent with analytical solutions, and previously obtained numerical results.

\section{One Dimensional Riemann Problems} \label{1d_tests}

The following one dimensional problems were tested in addition to the problem presented in section~\ref{rie1d1}.

\begin{itemize}

\item \textbf{One-Dimensional Riemann Problem 2}\\
\begin{equation}
\begin{array}{rclrclrcl}
     \gamma & = & \frac{5}{3}\, & &  &  &,\\
     \noalign{\medskip}
     \textrm{Left:}\,p & = & 1000\,,\rho & = & 1\,,\A{v} & = & 0\,,\\
     \noalign{\medskip}
     \textrm{Right:}\,p & = & 10^{-2}\,,\rho & = & 1\,,\A{v} & = & 0\,.\\
     \noalign{\medskip}
 \label{RAM-prob2}
\end{array}
\end{equation}
This problem is a much difficult test than the previous one. The fluid becomes both dynamically and thermodynamically relativistic resulting in a very thin and dense shell (see section~\ref{features}), which challenges the code in capturing its location and high density. This problem is also very common, tested by~\cite{del-zanna02,donat98,lucas04,marti96,meliani07,mignone05a,cannizzo08,mignone05b,wang08,zhang06,tchekhovskoy07}.

\vspace{45mm}

\item \textbf{One-Dimensional Riemann Problem 3}\\
\begin{equation}
\begin{array}{rclrclrcl}
     \gamma & = & \frac{4}{3}\,, & & & & \\
     \noalign{\medskip}
     \textrm{Left:}\,p & = & 1\,,\rho & = & 1\,,\A{v} & = & 0.9\,,\\
     \noalign{\medskip}
     \textrm{Right:}\,p & = & 10\,,\rho & = & 1\,,\A{v} & = & 0\,.\\
     \noalign{\medskip}
 \label{RAM-prob3}
\end{array}
\end{equation}
This problem mimics the interaction of a planar jet with the ISM. The negative pressure gradient gives rise to a strong reverse shock relevant to GRB calculations \cite{kobayashi99}. This problem was tested by~\cite{lucas04,mignone05a,cannizzo08,mignone05b,wang08,zhang06,tchekhovskoy07}.

\item \textbf{One-Dimensional Riemann Problem With Non-Zero Transverse
 Velocity: Problem 1}\\
\begin{equation}
\begin{array}{rclrclrclrcl}
     \gamma & = & \frac{5}{3}\,, & & & & & & \\
     \noalign{\medskip}
     \textrm{Left:}\,p & = & 1000\,,\rho & = & 1\,,\A{v} & = & 0 \,, & & \\
     \noalign{\medskip}
     \textrm{Right:}\,p & = & 10^{-2}\,,\rho & = & 1\,,v_x & = & 0 \,,v_y & = & 0.9\,.\\
     \noalign{\medskip}
 \label{RAM-prob4}
\end{array}
\end{equation}
This problem introduces tangential velocity components, which as opposed to Newtonian dynamics, are coupled to the dynamics through the Lorentz factor (see section~\ref{features}). When astrophysical jets heat the ISM shearing layers appear due to the difference in tangential velocity resulting in a mixing of ambient material with the jet. This sort of problems are the most difficult problems which still have analytical solutions and are therefore very important to test the ability of a numerical code. This problem was tested by~\cite{meliani07,cannizzo08,mignone05b,zhang06,tchekhovskoy07}.

\item \textbf{One-Dimensional Riemann Problem With Non-Zero Transverse
 Velocity: Problem 2}\\
\begin{equation}
\begin{array}{rclrclrclrcl}
     \gamma & = & \frac{5}{3}\,, & & & & & & \\
     \noalign{\medskip}
     \textrm{Left:}\,p & = & 1000\,,\rho & = & 1\,,v_x & = & 0 \,,v_y & = & 0.9\,,\\
     \noalign{\medskip}
     \textrm{Right:}\,p & = & 10^{-2}\,,\rho & = & 1\,,v_x & = & 0 \,,v_y & = & 0.9\,.\\
     \noalign{\medskip}
 \label{RAM-prob5}
\end{array}
\end{equation}
This problem introduces tangential velocity components in both sides of the discontinuity. In this problem extremely thin structures with a very high Lorentz factor evolve, resulting in the need of very high resolution (or equivalently AMR use) to catch the location of the different waves. This problem was tested by~\cite{meliani07,cannizzo08,mignone05b,wang08,zhang06,tchekhovskoy07}.
\end{itemize}

\subsection{One-Dimensional Riemann Problem 2} \label{sec:rie1d2}
 
\begin{figure*}
\centering
\includegraphics[scale=0.4]{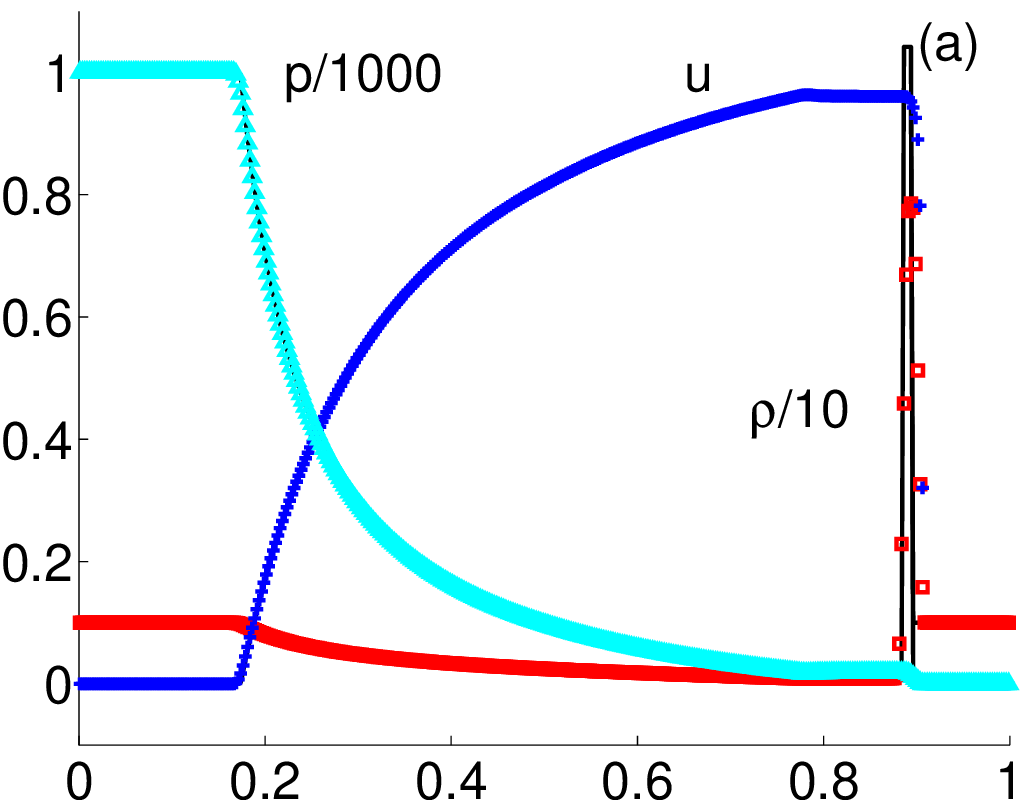}
\includegraphics[scale=0.4]{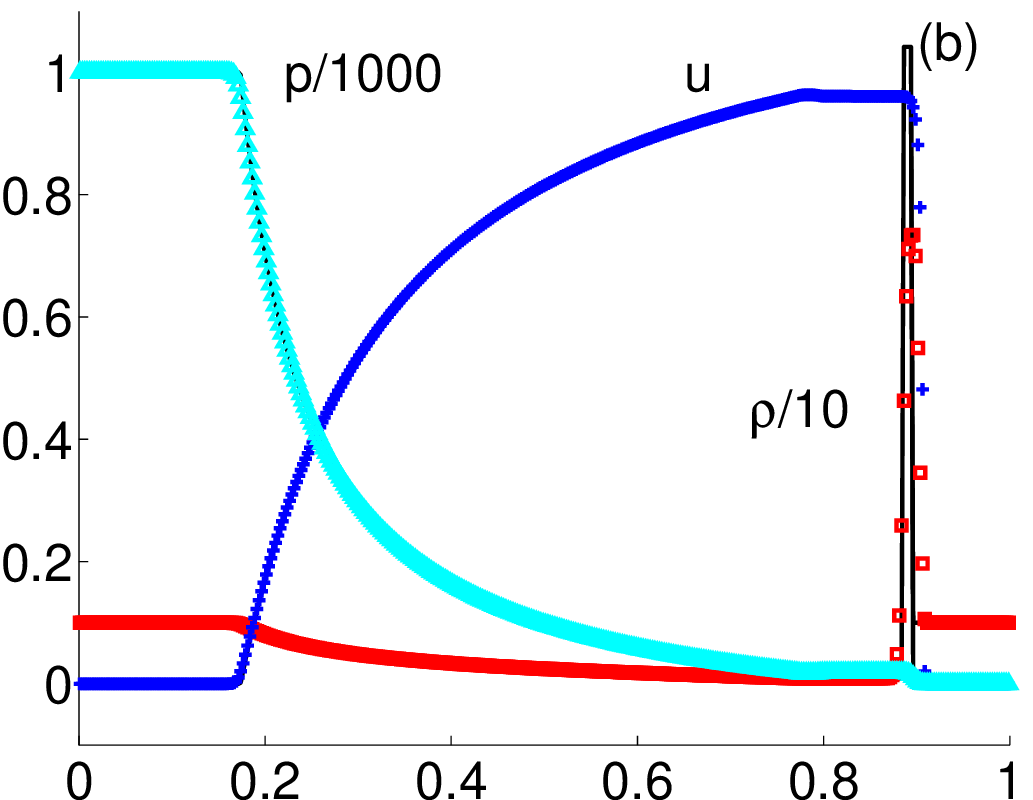}\\
\includegraphics[scale=0.4]{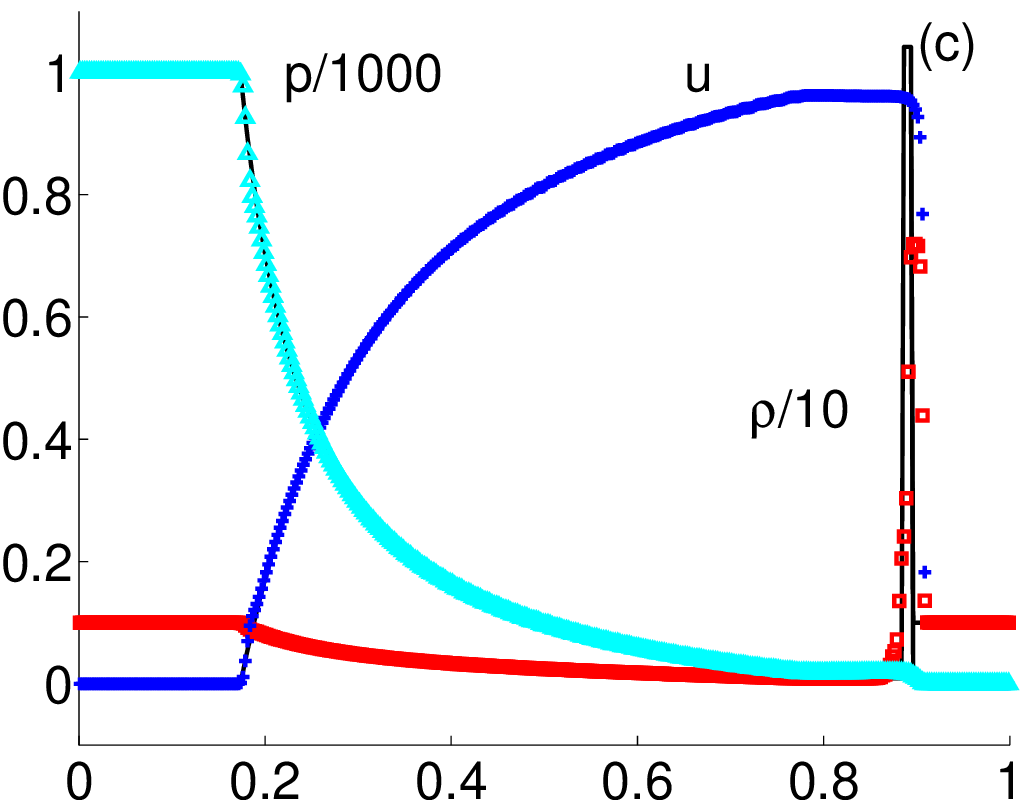}
\includegraphics[scale=0.4]{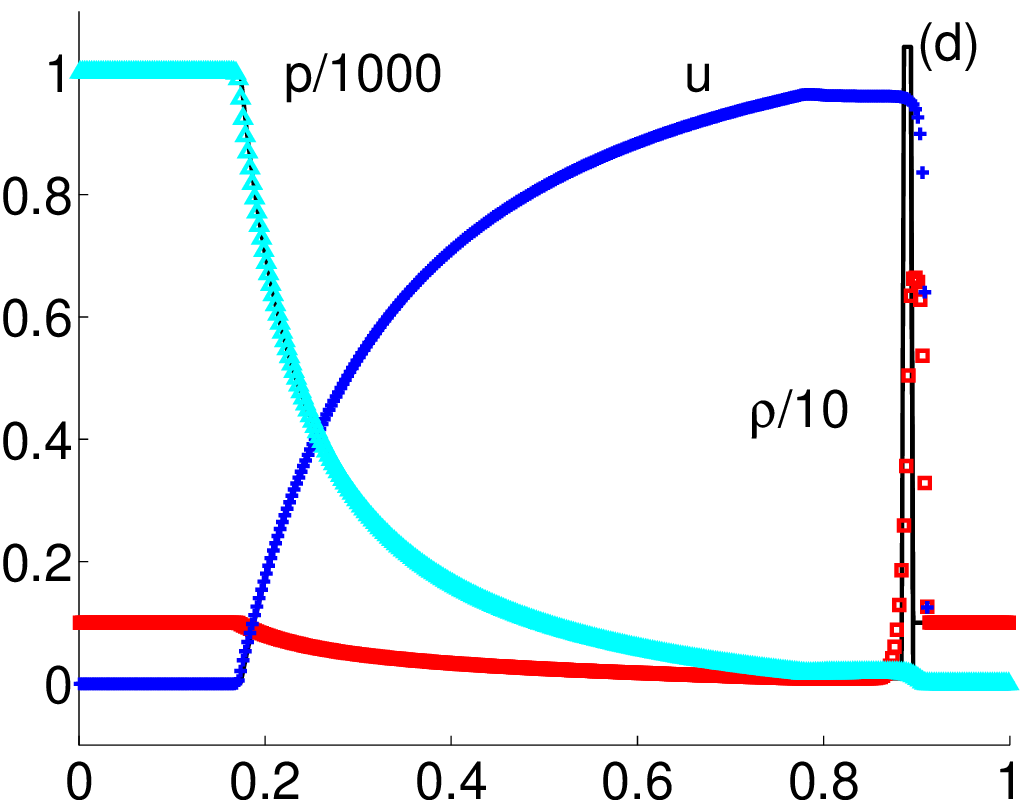}\\
\includegraphics[scale=0.4]{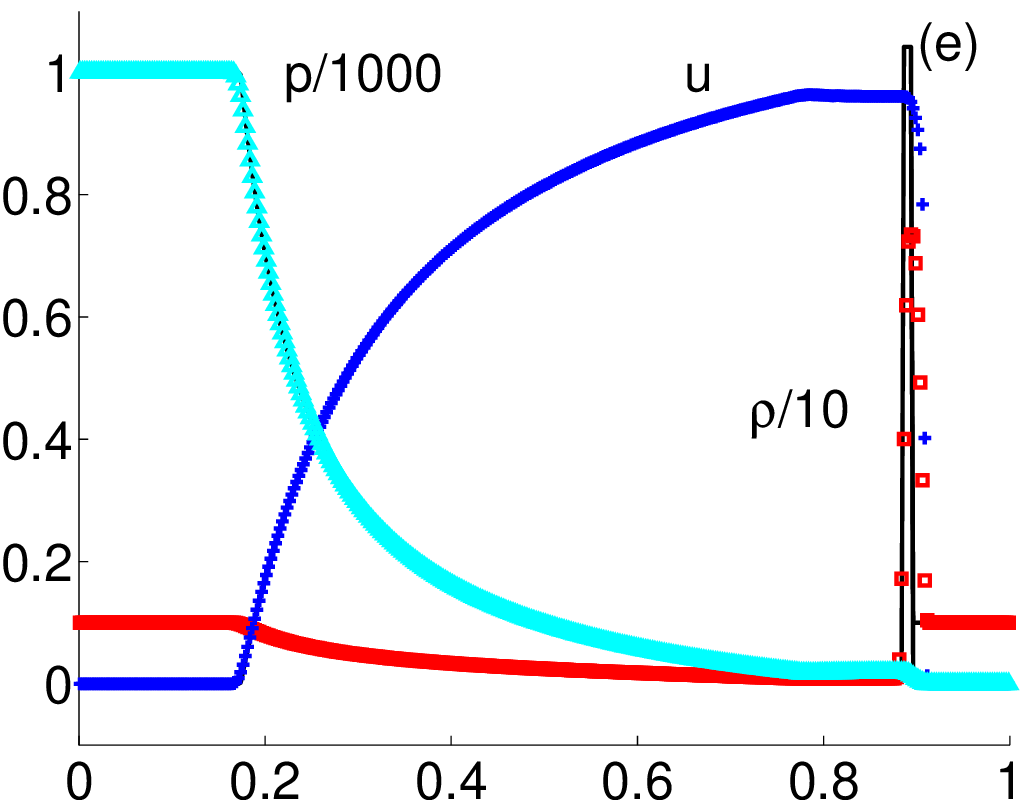}
\includegraphics[scale=0.4]{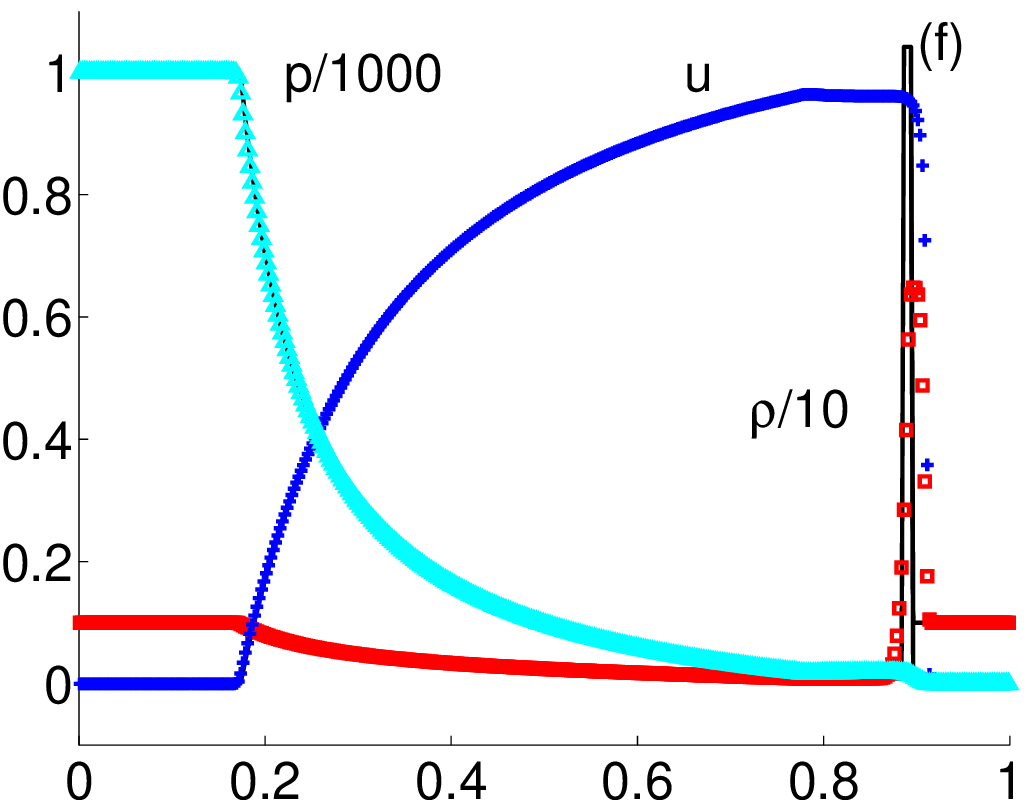}
\caption{One-dimensional Riemann problem 2 at $t=0.4$. Results for
 six schemes: (a) CT-PPM, (b) CT-PLM, (c) RK2-PPM, (d) RK2-PLM, (e) RK3-PPM and (f) RK3-PLM are shown.
 The computational grid consists of 400 zones. Numerical results are
 shown in symbols, whereas the exact solution is shown in solid
 lines. We show proper mass density (square), pressure (triangle)
 and velocity (plus sign)
\protect\label{fig:rie1d2}}
\end{figure*}

\begin{figure*}
\centering
\includegraphics[scale=0.7]{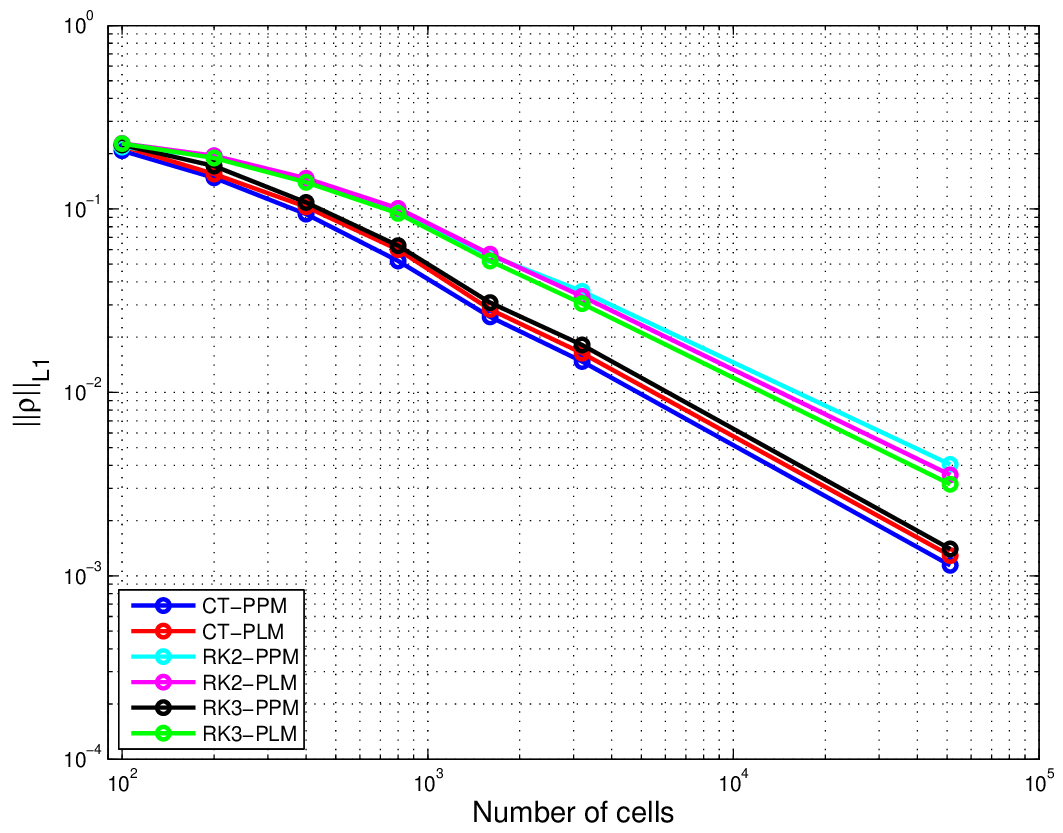}\\
\includegraphics[scale=0.7]{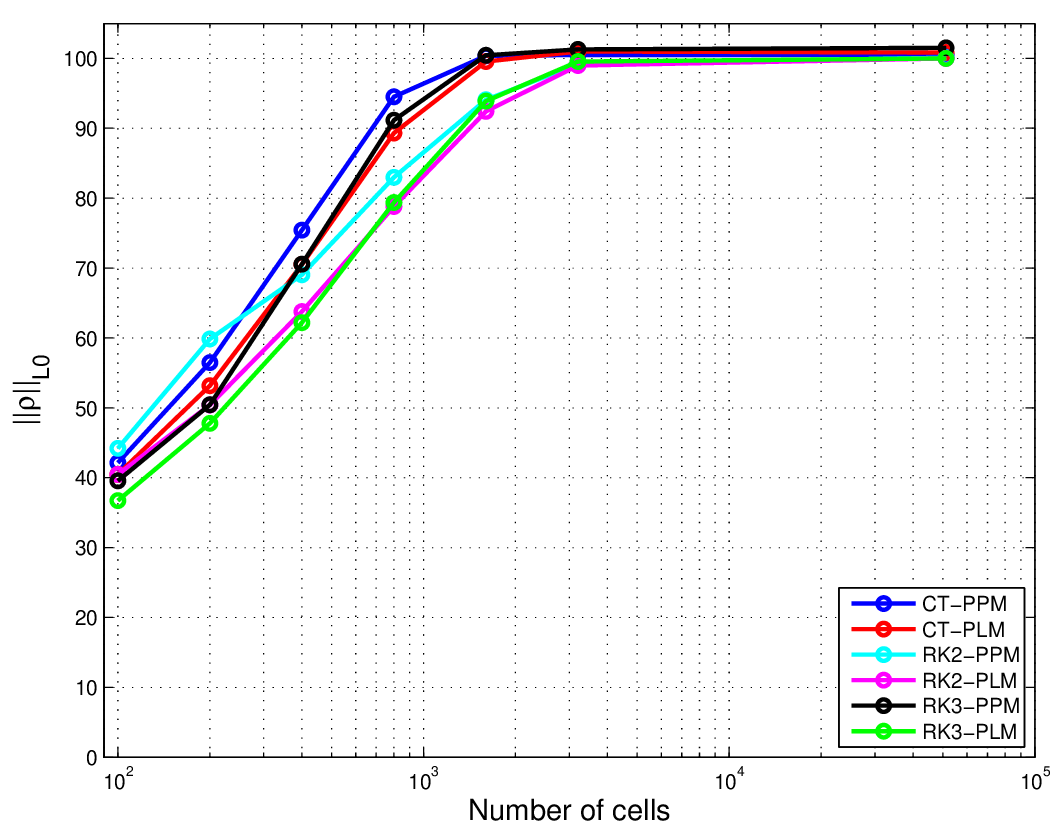}
\caption{$L_1$ (top) and $L_0$ (bottom) errors of the density for
the 1D Riemann Problem
 2. Seven different uniform grid resolutions (100,200,400,800,1600,3200 and 51200 cells) are
 shown at $t = 0.4$. Results for
 six schemes: (blue) CT-PPM, (red) CT-PLM, (cyan) RK2-PPM, (magenta) RK2-PLM, (black) RK3-PPM and (green) RK3-PLM are shown.
\protect\label{fig:conv-rie1d2}}
\end{figure*}

\begin{figure*}
\centering
\includegraphics[scale=0.4]{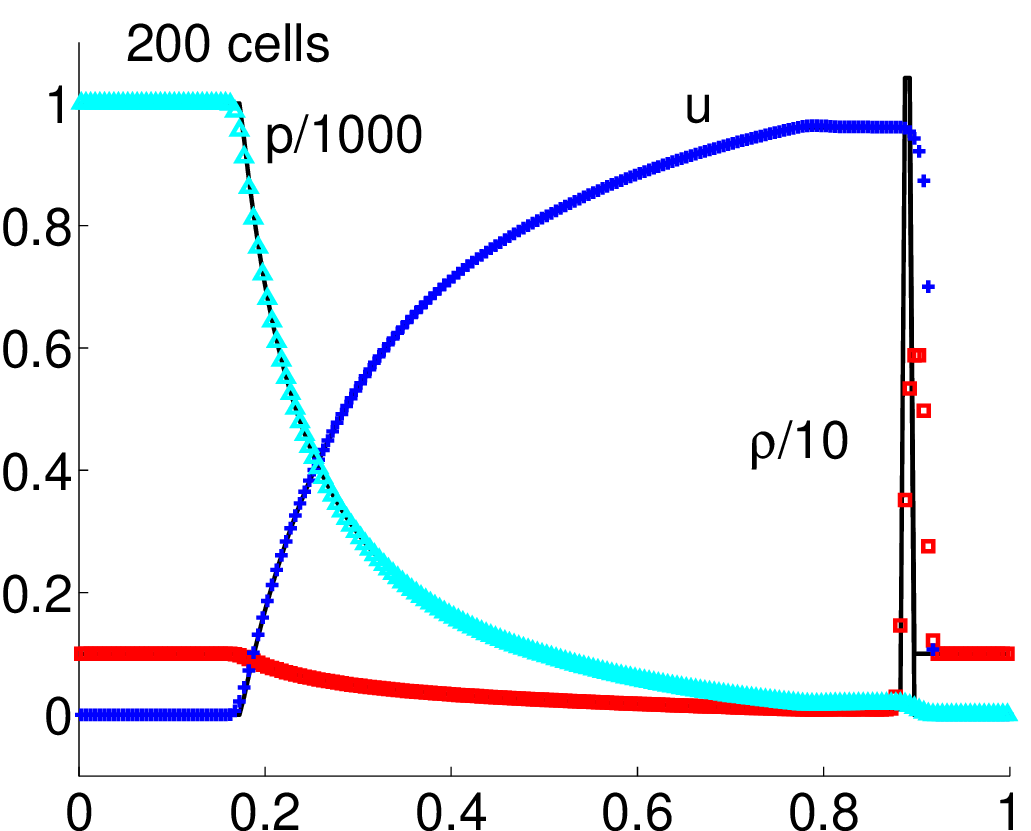}
\includegraphics[scale=0.4]{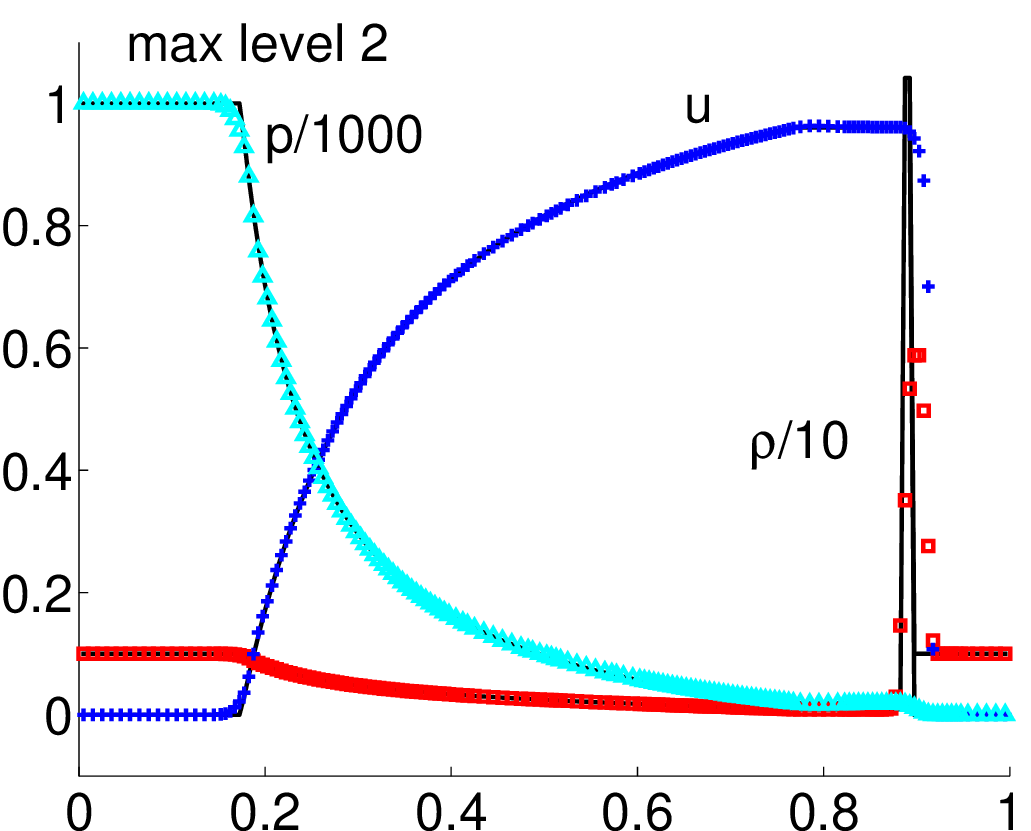}\\
\includegraphics[scale=0.4]{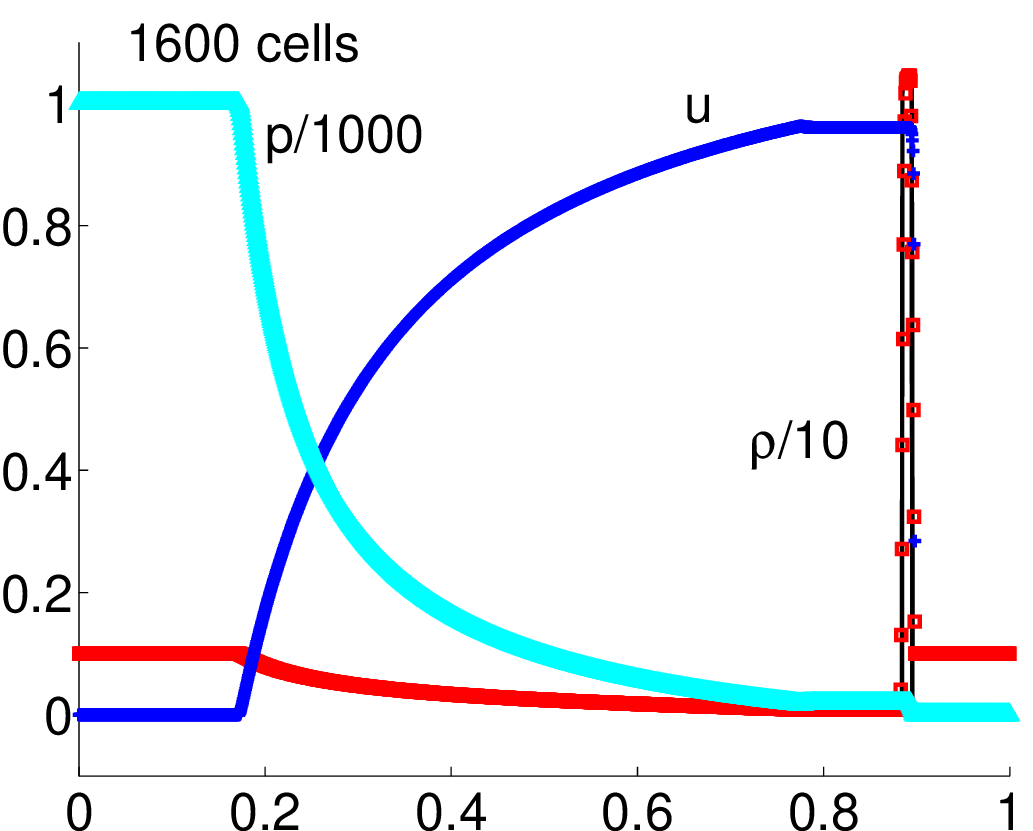}
\includegraphics[scale=0.4]{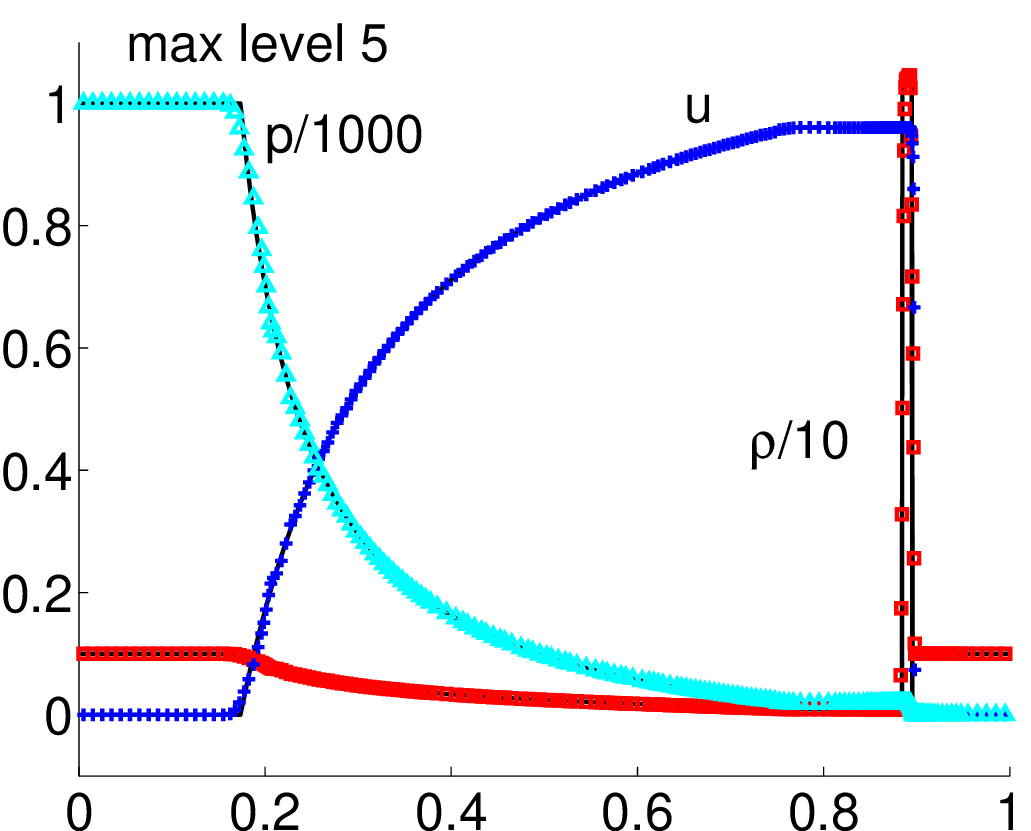}\\
\includegraphics[scale=0.4]{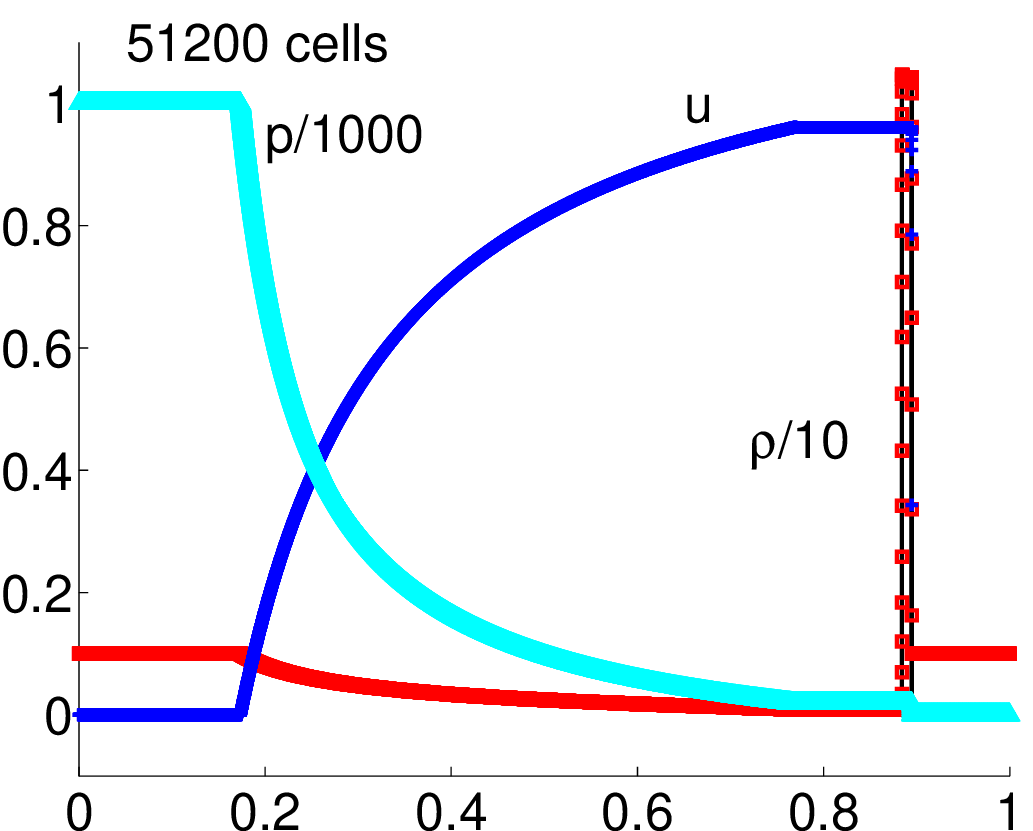}
\includegraphics[scale=0.4]{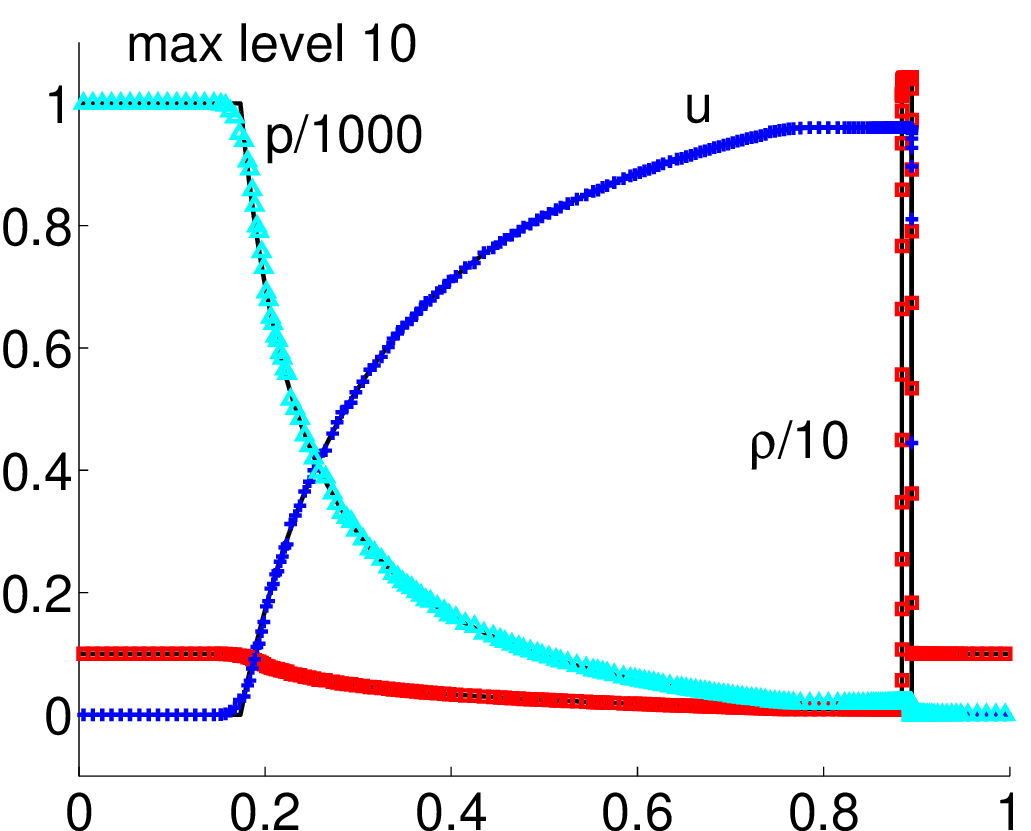}
\caption{One-dimensional Riemann problem 2 at $t=0.4$. Results for
AMR calculations using CT-PPM are shown.
 At the right hand side are constant grid calculations,
 and on the left hand side are equivalent resolution AMR calculations,
 using a base mesh resolution of 100 zones. Numerical results are
 shown in symbols, whereas the exact solution is shown in solid
 lines. We show proper mass density (square), pressure (triangle)
 and velocity (plus sign)
\protect\label{fig:rie1d2-amr}}
\end{figure*}

The results of the six schemes of RELDAFNA for this problem at $t=0.4$
are shown in Fig.~\ref{fig:rie1d2}. In this test problem, the
evolution of the initial discontinuity gives rise to a right-moving
shock, a left-moving rarefaction wave, and a contact discontinuity
in between. Behind the shock, there is an extremely thin dense
shell. The width of the shell is only $0.01056$ at $t=0.4$. The
$L_1$ and $L_0$ norm errors of density at $t=0.4$, for six schemes
with various grid resolutions, are shown in
Fig~\ref{fig:conv-rie1d2}. The results are consistent with other
published results, e.g. \cite{zhang06,del-zanna02,wang08,tchekhovskoy07} (see Table~\ref{tab:rie1d2}). We also
made a calculation with CT-PPM using AMR. In a calculation with
maximum level of $10$, which is equivalent to a constant grid of
$51200$ cells, the maximum number of cells used (during one cycle
only) was $706$ cells, which is only $1.38\%$. The results of a
calculation using AMR with CT-PPM scheme at $t=0.4$, is shown in
Fig.~\ref{fig:rie1d2-amr}.

Fig.~\ref{fig:rie1d2} shows that the thin structure behind the shock
is very difficult to capture using $400$ zones in all six schemes. Its
location is calculated quite well but the internal structure is not
formed correctly, in particular the very high maximum density
(factor $\sim 10$, see section~\ref{features}).
%Still in this problem, since the
%locations of the jumps are calculated right, it is only the density
%profile which differs between the schemes.
%We see that it is more critical to use a higher order spatial reconstruction than
%using one of the two time integration methods, since given a time
%integration method, one gets an apparently higher density at the
%thin shell using PPM. This problem is a good example for the use of
%$L_0$ norm. In Fig.~\ref{fig:conv-rie1d2} (bottom) we can see that
%RK-PPM finds the maximum density considerably better than RK-PLM,
%but since $L_1$ is critically sensitive to the location of
%structures, and since in this problem the shell of maximum density
%is very thin, we see in Fig.~\ref{fig:conv-rie1d2} (top) that in the
%$L_1$ norm they are equivalent. Again we find, that in the two
%norms, using a selected spatial reconstruction method (either PPM or
%PLM), one would find that the CT time integration method is a better
%choice for this problem. In Fig.~\ref{fig:rie1d2-amr} we can see
%that the use of AMR makes the low base mesh calculation very
%accurate. We can capture the thin structures both in location and in
%internal structure including the very high maximum density. Again,
%the high tail of the rarefaction wave could be sharpened by tuning
%the refinement criterion.

% tab 2
\begin{table}
\scriptsize
 \centering
 \begin{tabular}{ccccc}
$Scheme$ & $Number \; of \; cells$ & $L_1 \; Error$ & $Convergence \; Rate$ & $L_0 \; Error \, [\%]$\\
\hline
CT-PPM & 100 & 2.07e-1 & & 42.14 \\
 & 200 & 1.14e-1 & 0.49 & 56.48 \\
 & 400 & 9.37e-2 & 0.66 & 75.43 \\
 & 800 & 5.21e-2 & 0.85 & 94.49 \\
 & 1600 & 2.58e-2 & 1.01 & 100.36 \\
 & 3200 & 1.47e-2 & 0.81 & 100.5 \\
 & 51200 & 1.14e-3 & & 100.5 \\
\hline
RK2-PPM & 100 & 2.17e-1 & & 44.2 \\
 & 200 & 1.95e-1 & 0.16 & 59.85 \\
 & 400 & 1.44e-1 & 0.43 & 62.04 \\
 & 800 & 9.68e-2 & 0.58 & 82.98 \\
 & 1600 & 5.58e-2 & 0.79 & 94.09 \\
 & 3200 & 3.55e-2 & 0.65 & 99.18 \\
 & 51200 & 4.04e-3 & & 100.08 \\
\hline
RK3-PPM & 100 & 2.23e-1 & & 39.56 \\
 & 200 & 1.71e-1 & 0.38 & 50.44 \\
 & 400 & 1.08e-1 & 0.66 & 70.55 \\
 & 800 & 6.30e-2 & 0.78 & 91.16 \\
 & 1600 & 3.08e-2 & 1.03 & 100.44 \\
 & 3200 & 1.81e-2 & 0.77 & 101.29 \\
 & 51200 & 1.4e-3 & & 101.5 \\
\hline
CT-PLM & 100 & 2.2e-1 & & 40.47 \\
 & 200 & 1.55e-1 & 0.5 & 53.17 \\
 & 400 & 1.03e-1 & 0.6 & 70.5 \\
 & 800 & 6e-2 & 0.78 & 89.3 \\
 & 1600 & 2.83e-2 & 1.08 & 99.54 \\
 & 3200 & 1.63e-2 & 0.79 & 100.98 \\
 & 51200 & 1.28e-3 & & 100.84 \\
\hline
RK2-PLM & 100 & 2.28e-1 & & 40.48 \\
 & 200 & 1.95e-1 & 0.22 & 50.4 \\
 & 400 & 1.47e-1 & 0.41 & 63.76 \\
 & 800 & 1.00e-1 & 0.54 & 78.81 \\
 & 1600 & 5.66e-2 & 0.83 & 92.42 \\
 & 3200 & 3.33e-2 & 0.76 & 98.97 \\
 & 51200 & 3.55e-3 & & 100 \\
\hline
RK3-PLM & 100 & 2.27e-1 & & 36.73 \\
 & 200 & 1.89e-1 & 0.26 & 47.78 \\
 & 400 & 1.40e-1 & 0.44 & 62.18 \\
 & 800 & 9.46e-2 & 0.56 & 79.38 \\
 & 1600 & 5.20e-2 & 0.86 & 93.89 \\
 & 3200 & 3.04e-2 & 0.77 & 99.55 \\
 & 51200 & 3.16e-3 & & 100 \\
\hline
 F-WENO \cite{zhang06} & 100  & 2.10e-1 &   &   \\
       & 200  & 1.42e-1 & 0.56 & \\
       & 400  & 9.29e-2 & 0.61 & \\
       & 800  & 5.54e-2 & 0.75 & \\
       & 1600 & 2.54e-2 & 1.1 & \\
       & 3200 & 1.51e-2 & 0.75 & \\
\hline
F-PLM \cite{zhang06} & 100  & 1.96e-1 &      & \\
       & 200  & 1.42e-1 & 0.46 & \\
       & 400  & 1.06e-1 & 0.42 &\\
       & 800  & 7.21e-2 & 0.56 & \\
       & 1600 & 3.92e-2 & 0.88 & \\
       & 3200 & 2.44e-2 & 0.68 & \\
\hline
U-PPM \cite{zhang06} & 100  & 2.18e-1 &    &  \\
       & 200  & 1.52e-1 & 0.52 &  \\
       & 400  & 9.52e-2 & 0.68 & \\
       & 800  & 5.42e-2 & 0.81 & \\
       & 1600 & 2.67e-2 & 1.0  & \\
       & 3200 & 1.67e-2 & 0.68 & \\
\hline
U-PLM \cite{zhang06} & 100  & 2.13e-1 &      & \\
       & 200  & 1.65e-1 & 0.37 & \\
       & 400  & 1.25e-1 & 0.41 & \\
       & 800  & 8.68e-2 & 0.53 & \\
       & 1600 & 4.49e-2 & 0.95 & \\
       & 3200 & 2.71e-2 & 0.73 &
 \end{tabular}
\caption{Error norms of the density for the 1D Riemann Problem
 2. Six schemes of RELDAFNA and four schemes of RAM \cite{zhang06} with various resolutions with uniform spacing are
 shown at $t = 0.4$.
\protect\label{tab:rie1d2}}
\end{table}

\newpage
\subsection{One-Dimensional Riemann Problem 3}
\label{rie1d3}

\begin{figure*}
\centering
\includegraphics[scale=0.4]{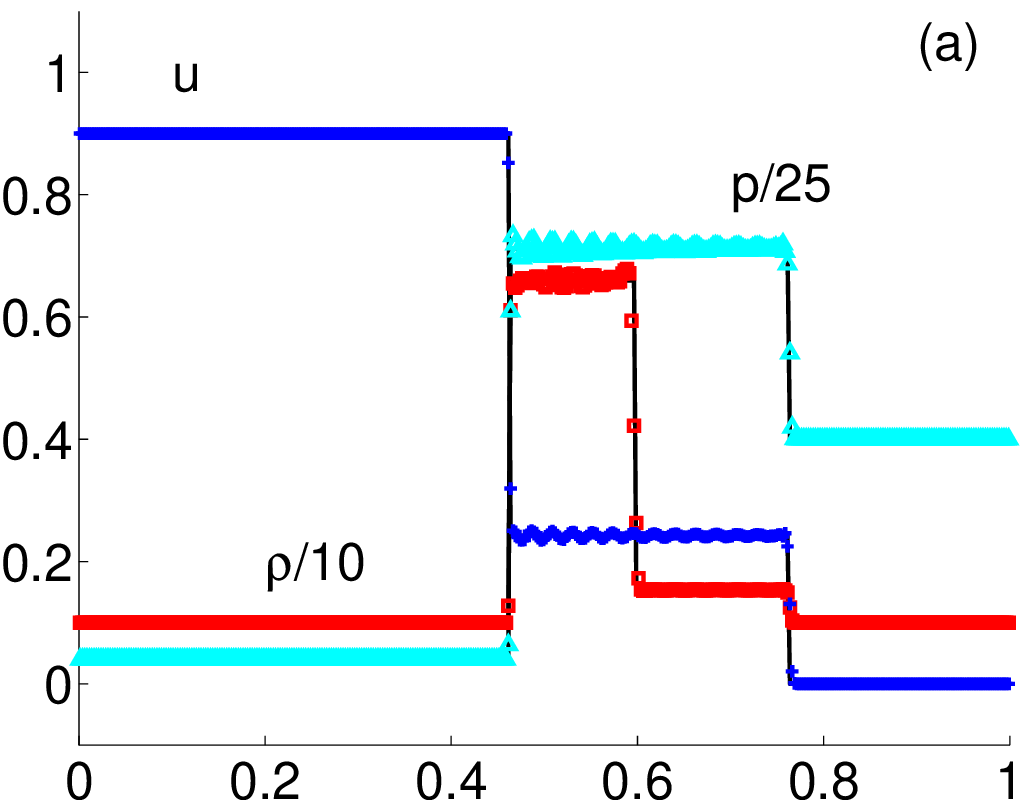}
\includegraphics[scale=0.4]{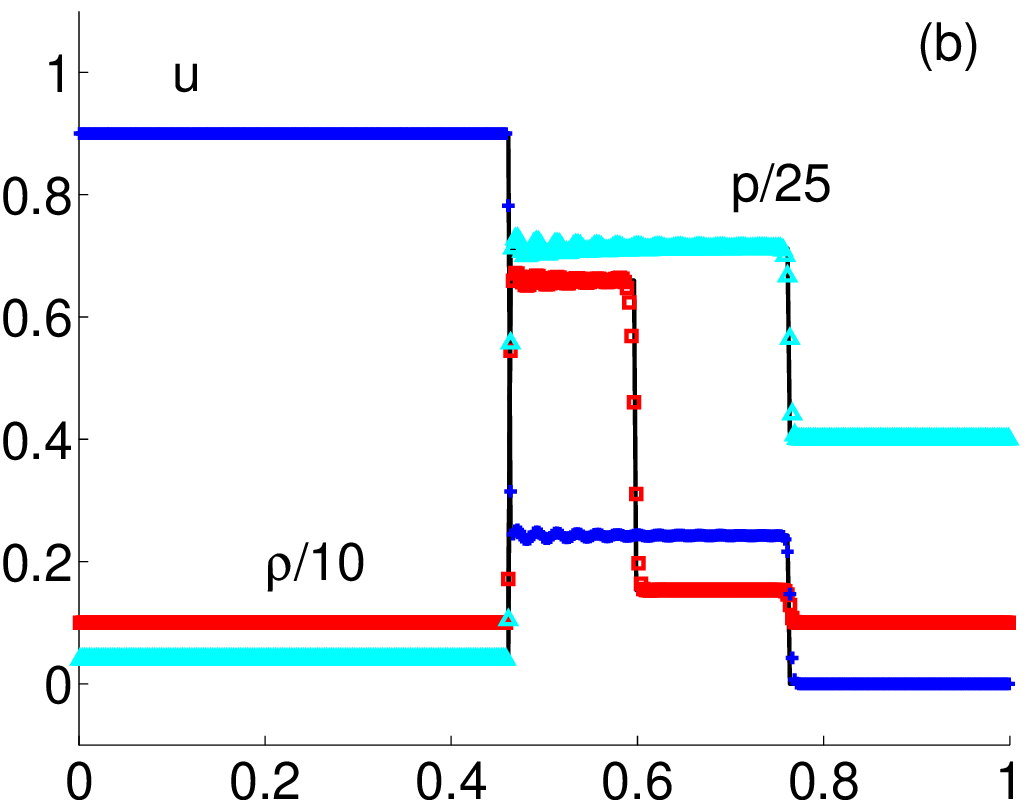}\\
\includegraphics[scale=0.4]{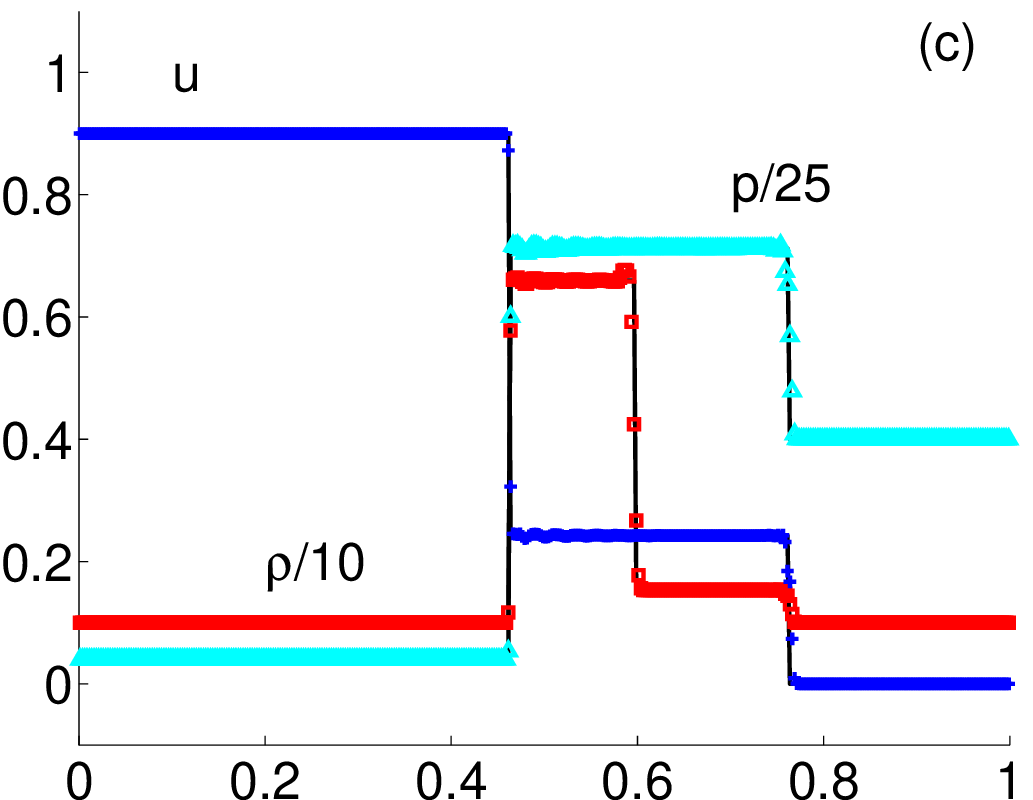}
\includegraphics[scale=0.4]{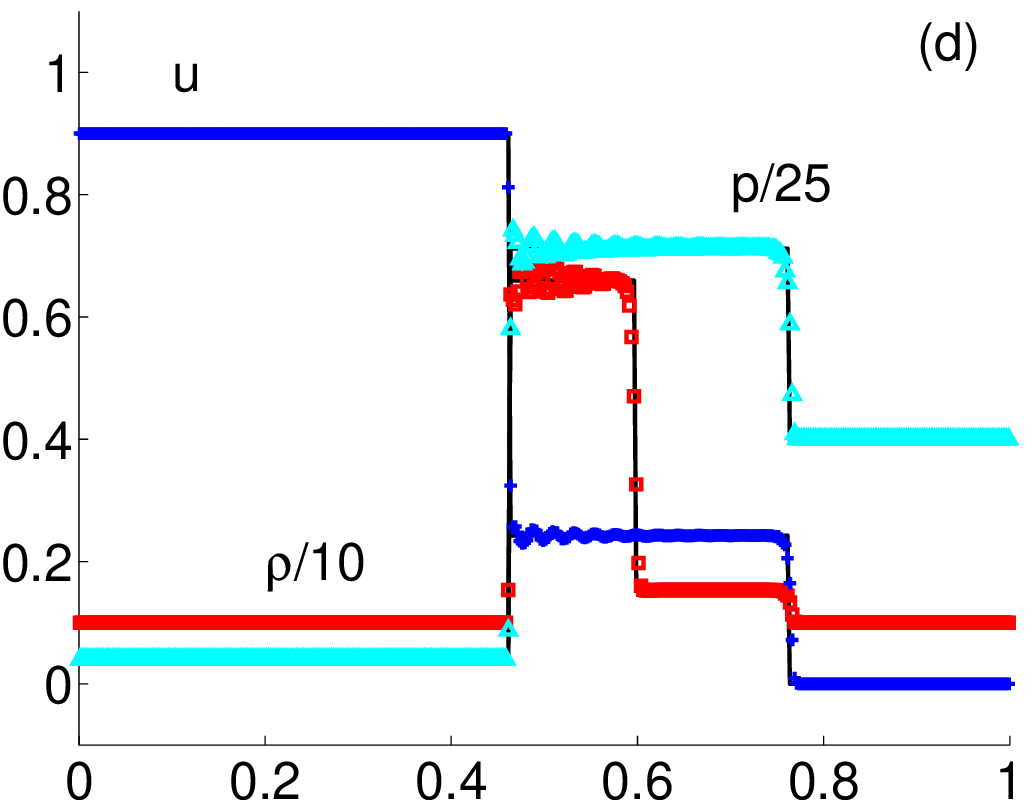}\\
\includegraphics[scale=0.4]{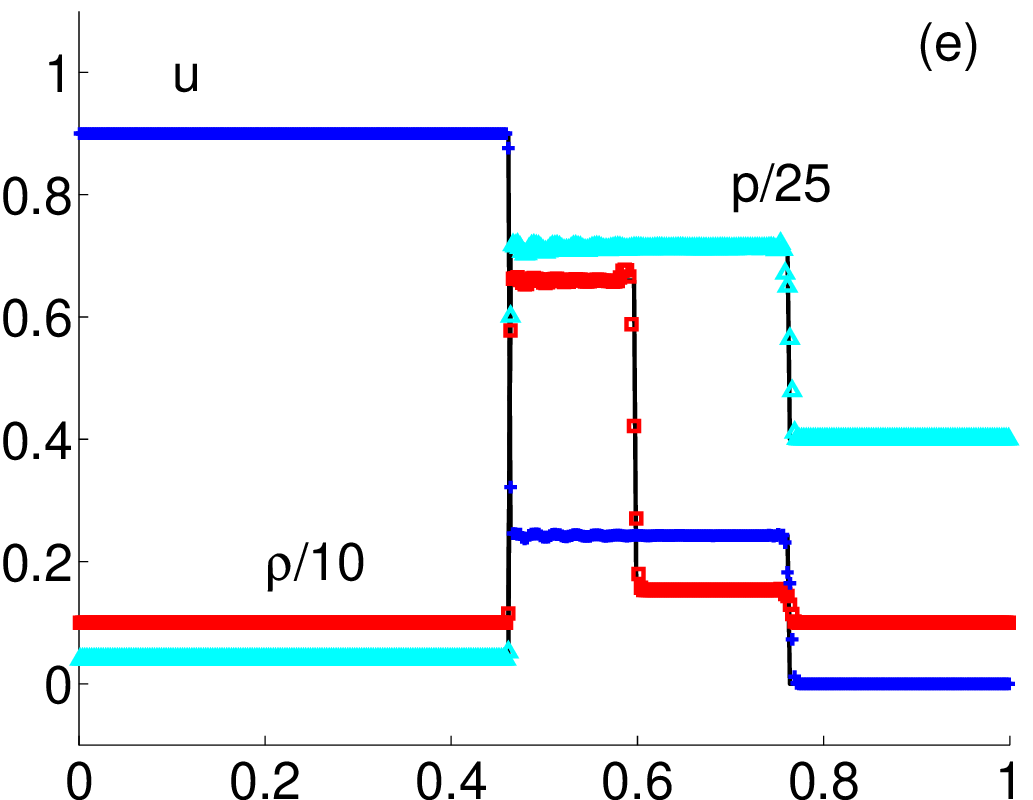}
\includegraphics[scale=0.4]{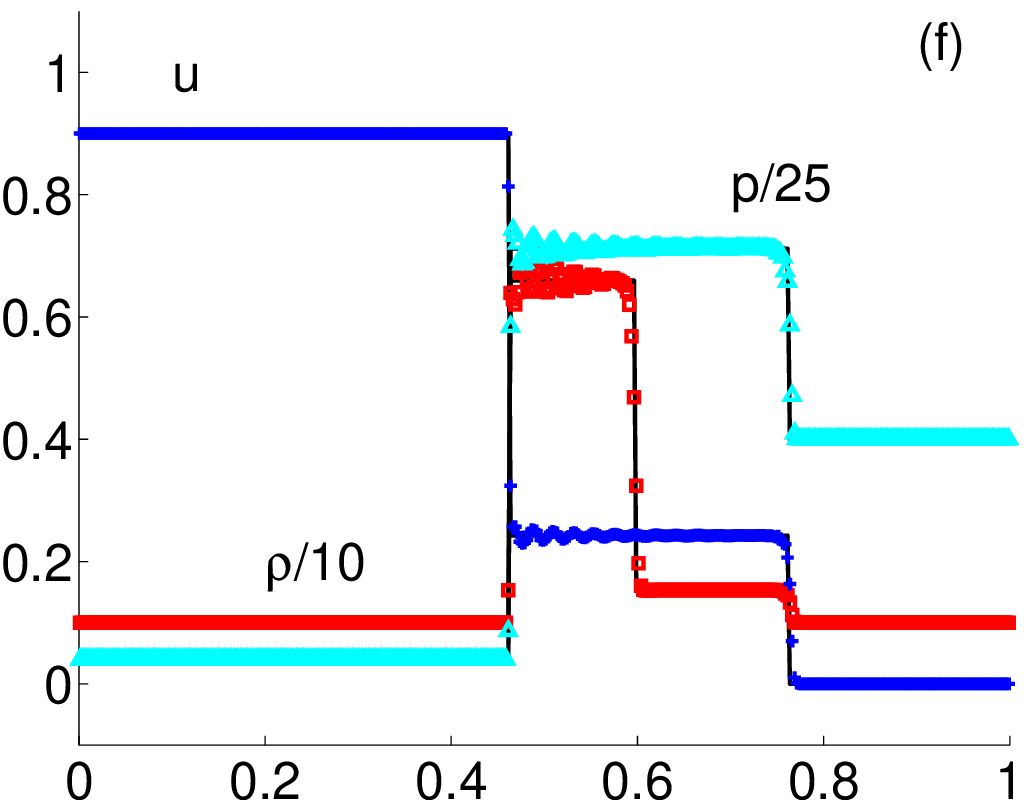}
\caption{One-dimensional Riemann problem 3 at $t=0.4$. Results for
 six schemes: (a) CT-PPM, (b) CT-PLM, (c) RK2-PPM, (d) RK2-PLM, (e) RK3-PPM and (f) RK3-PLM are shown.
 The computational grid consists of 400 zones. Numerical results are
 shown in symbols, whereas the exact solution is shown in solid
 lines. We show proper mass density (square), pressure (triangle)
 and velocity (plus sign)
\protect\label{fig:rie1d3}}
\end{figure*}

\begin{figure*}
\centering
\includegraphics[scale=0.7]{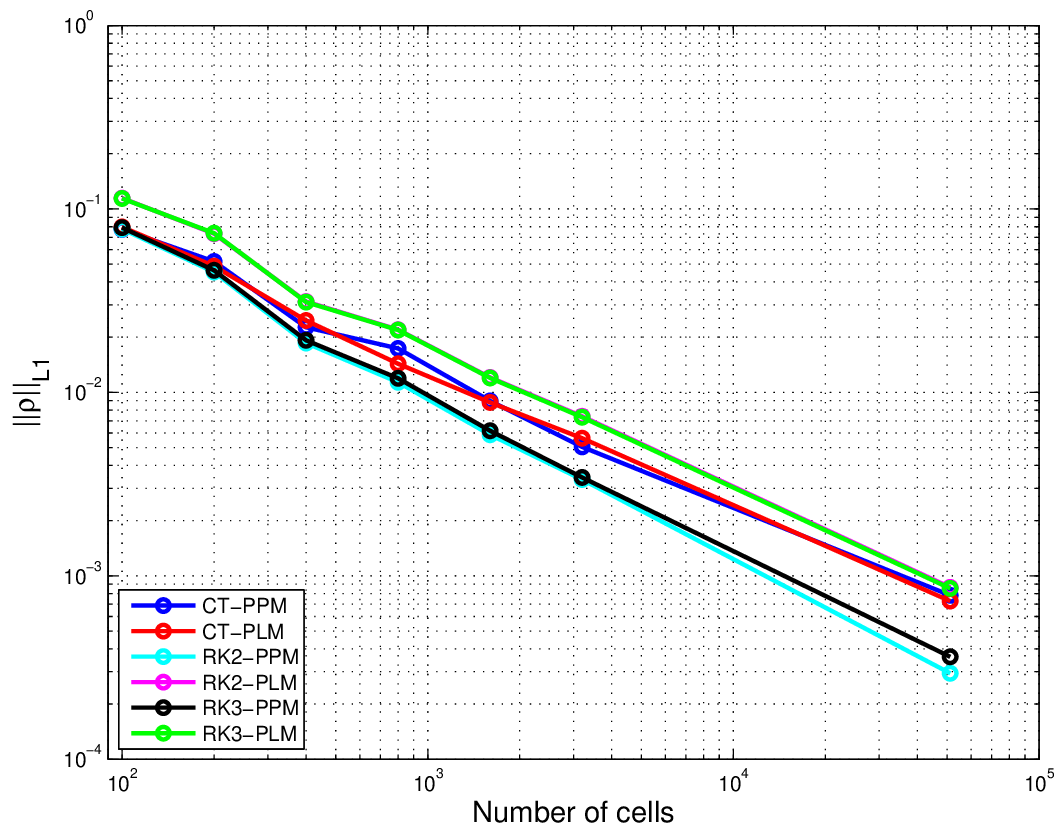}
\caption{$L_1$ errors of the density for the 1D Riemann Problem
 3. Seven different uniform grid resolutions (100,200,400,800,1600,3200 and 51200 cells) are
 shown at $t = 0.4$. Results for
 six schemes: (blue) CT-PPM, (red) CT-PLM, (cyan) RK2-PPM, (magenta) RK2-PLM, (black) RK3-PPM and (green) RK3-PLM are shown.
\protect\label{fig:conv-rie1d3}}
\end{figure*}

\begin{figure*}
\centering
\includegraphics[scale=0.4]{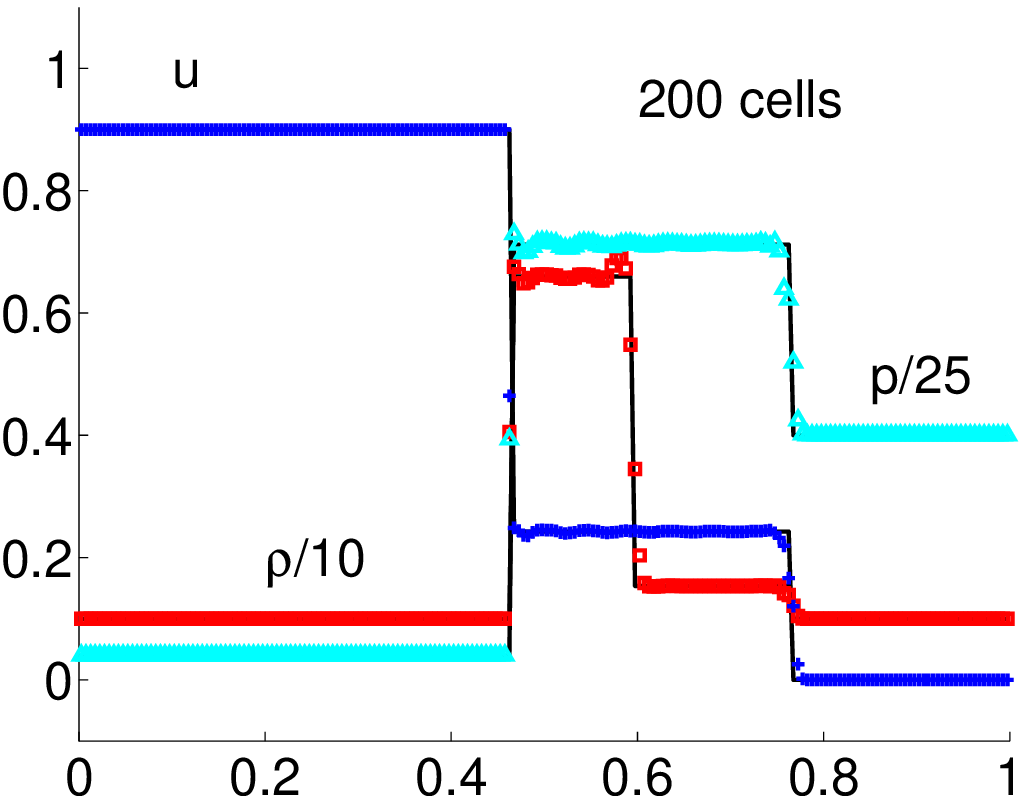}
\includegraphics[scale=0.4]{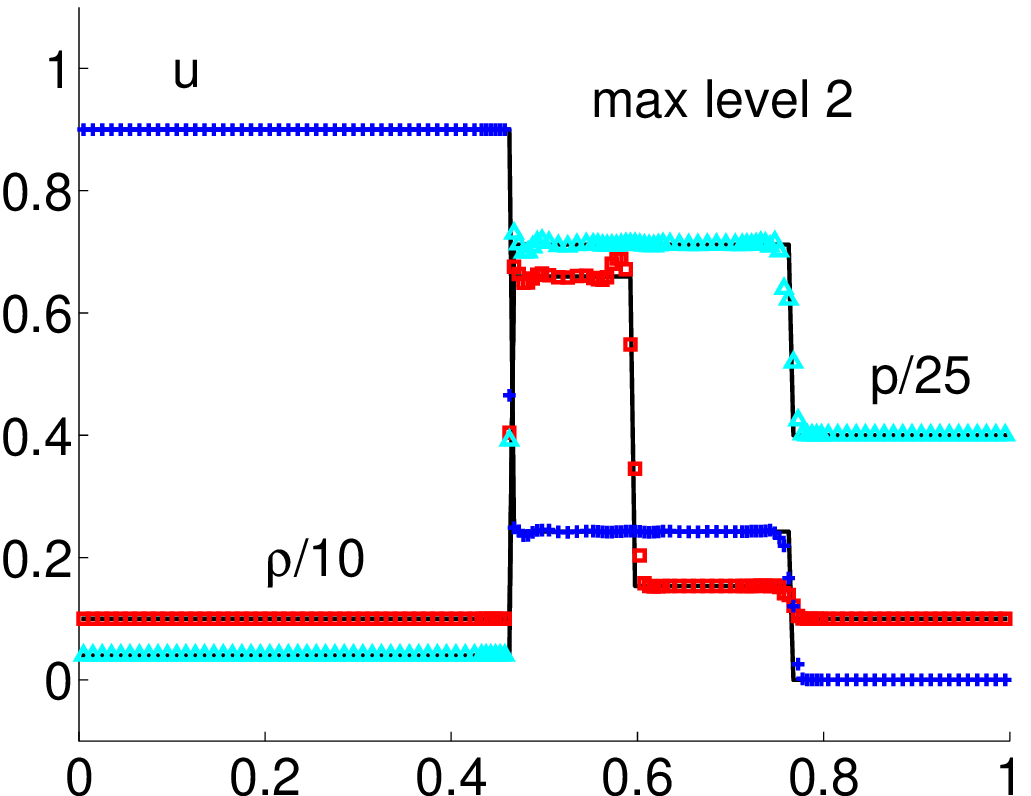}\\
\includegraphics[scale=0.4]{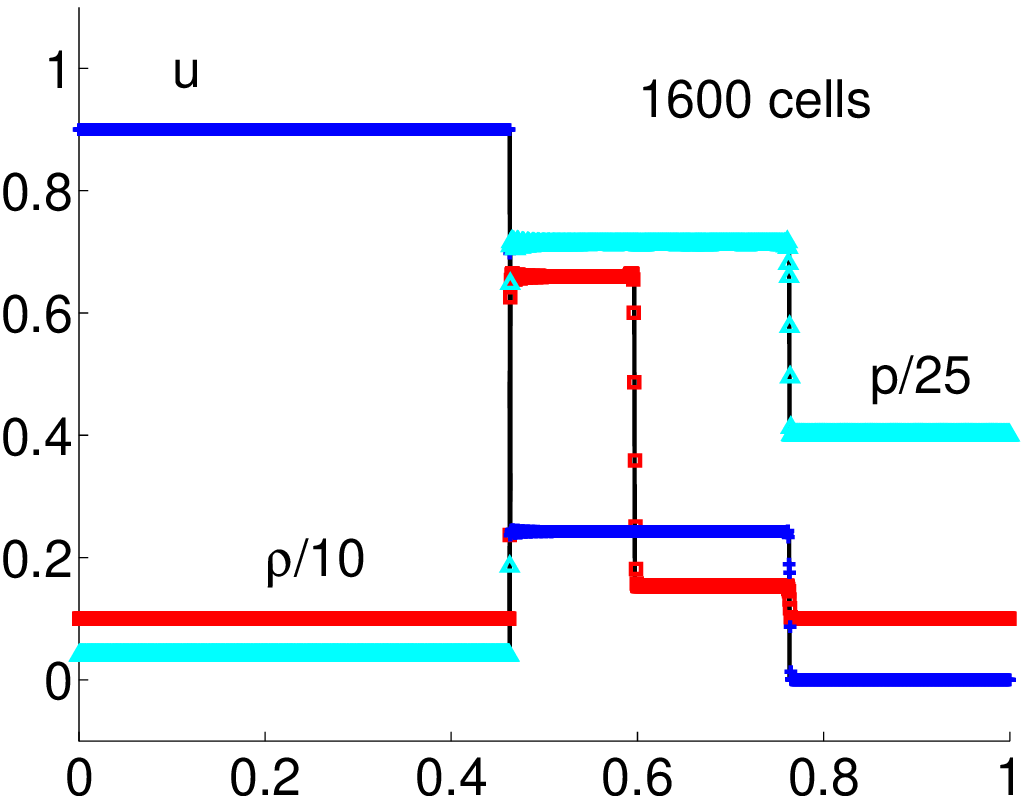}
\includegraphics[scale=0.4]{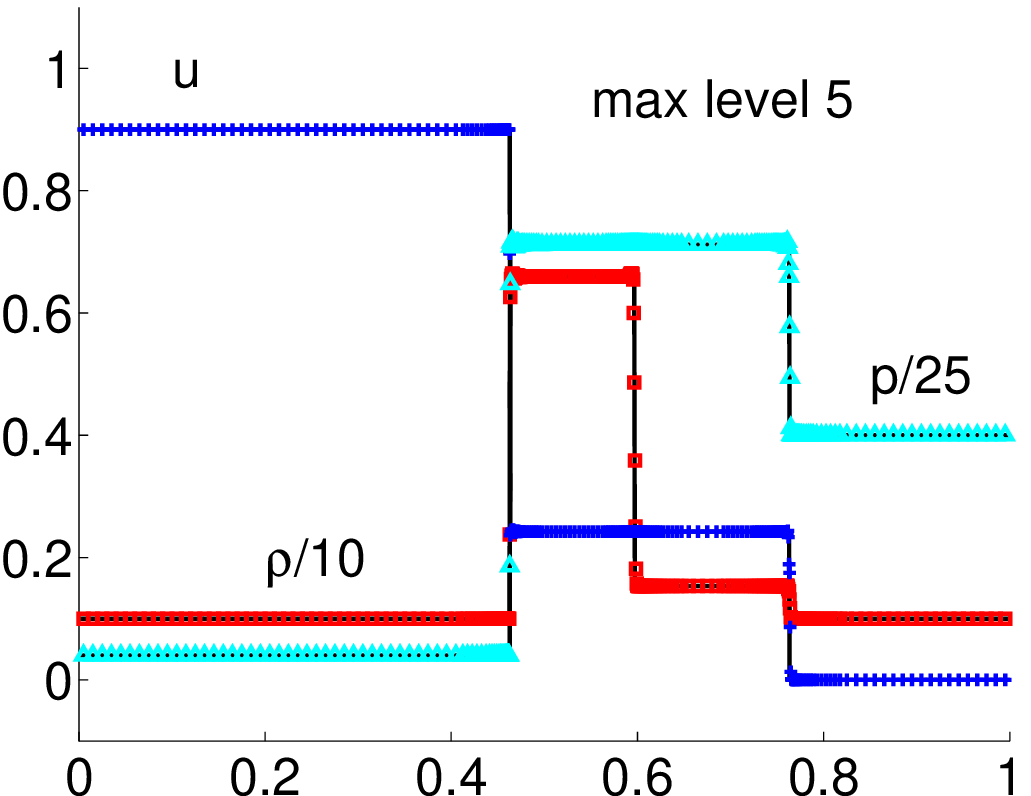}\\
\includegraphics[scale=0.4]{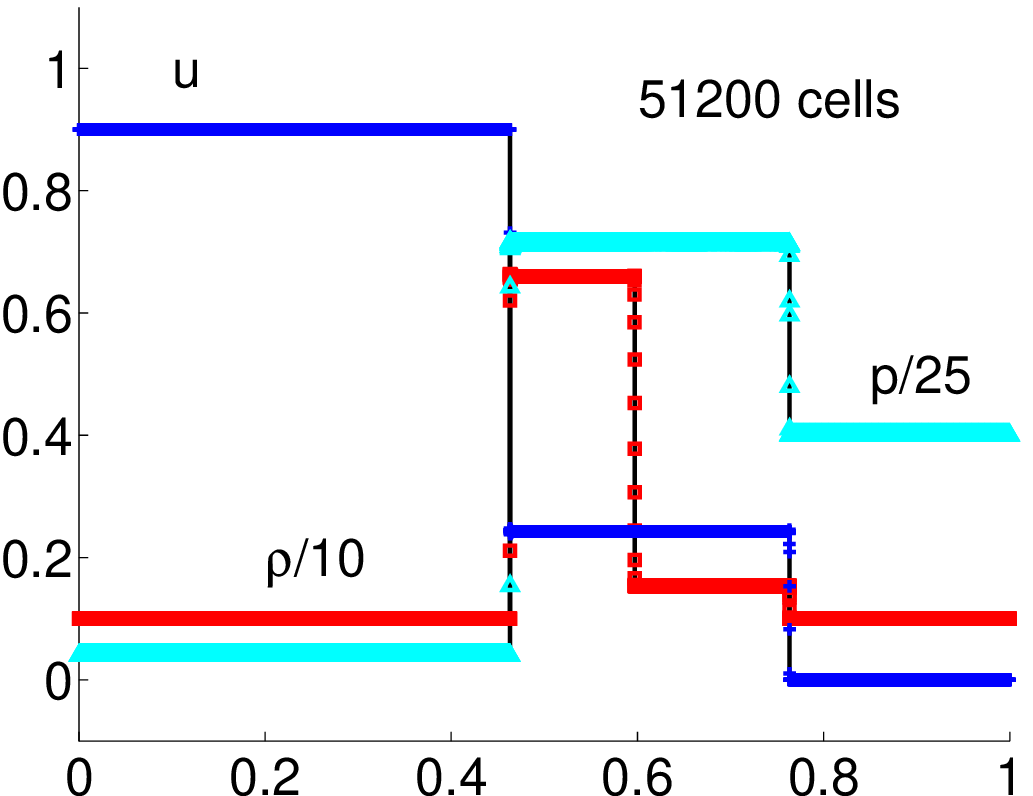}
\includegraphics[scale=0.4]{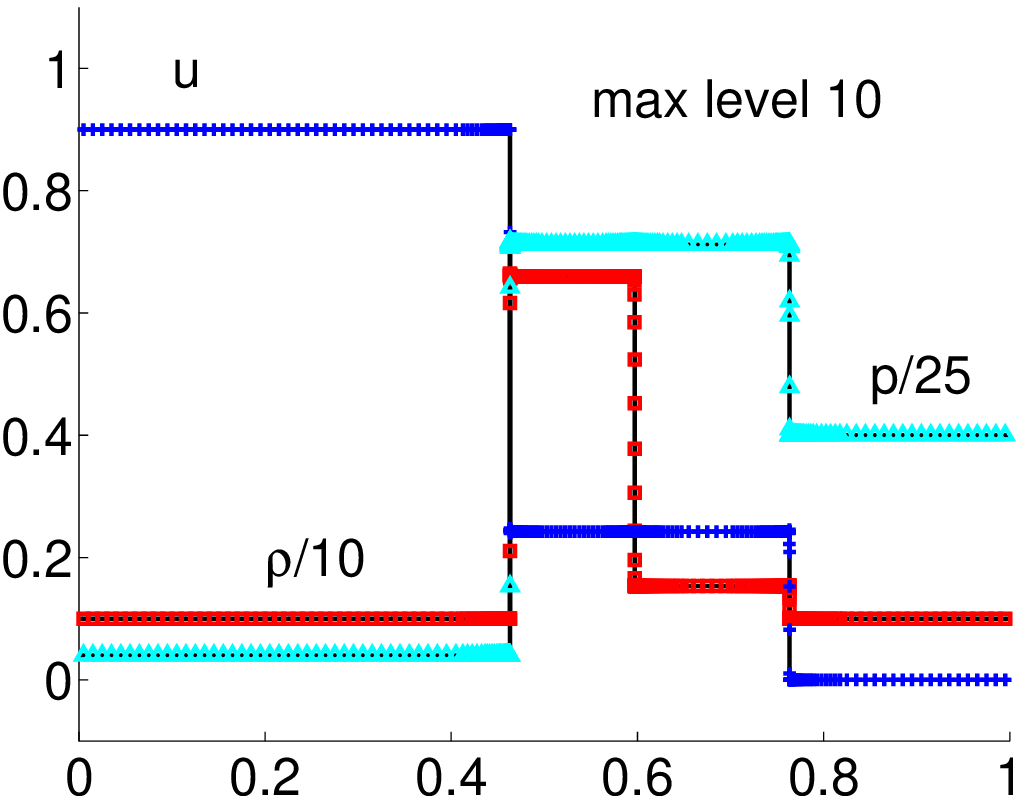}
\caption{One-dimensional Riemann problem 3 at $t=0.4$. Results for
AMR calculations using RK2-PPM. are shown.
 At the right hand side are constant grid calculations, and on the left hand side are equivalent resolution AMR calculations using a base mesh resolution of 100 zones. Numerical results are
 shown in symbols, whereas the exact solution is shown in solid
 lines. We show proper mass density (square), pressure (triangle)
 and velocity (plus sign)
\protect\label{fig:rie1d3-amr}}
\end{figure*}

The results of the six schemes of RELDAFNA for this problem at $t=0.4$ are shown in Fig.~\ref{fig:rie1d3}. In this problem a strong reverse shock forms in which post-shock oscillations are visible. The $L_1$ norm error of density at $t=0.4$, for six schemes with various grid resolutions, are shown in Fig~\ref{fig:conv-rie1d3}. The results are similar to other published results, e.g. \cite{zhang06,tchekhovskoy07} (see Table~\ref{tab:rie1d3}), and are slightly better than the ones shown in \cite{wang08}. We see that the smearing of shock in our CT-PPM calculation is better than given in the mentioned references. We also made a calculation with RK2-PPM using AMR. The maximum number of cells, at a given time, used in a calculation with maximum level of $10$, which is equivalent to a constant grid of $51200$ cells, was $736$ cells which is only $1.44\%$. The results of a calculation using AMR with RK2-PPM scheme at $t=0.4$, is shown in Fig.~\ref{fig:rie1d3-amr}.

\newpage

This problem includes the development of a strong reverse shock and
no rarefaction wave. As can be seen in Fig.~\ref{fig:rie1d3}, all
the schemes develop oscillations in that region. The oscillations
behind the reverse shock appear in all published results and in all
schemes.
%Apart from the oscillations, Fig.~\ref{fig:rie1d3} shows
%that the location of all the jumps is solved quite well using quite
%a low resolution ($400$ zones).
%In this problem, the oscillations are the test to differ between the schemes. When one
%uses PLM spatial reconstruction, it is best to use CT time
%integration, to get the minimal amplitude of oscillations in the
%reverse shock. In the RK-PLM and RK3-PLM the oscillations of the
%density profile in the middle zone ($0.45\lesssim x \lesssim 0.6$),
%affect the contact discontinuity jump which is smeared over a
%considerably higher number of cells than CT-PLM ($\sim 6$ cells). In
%the PPM schemes one finds that RK produces less oscillations, and
%therefore as seen in Fig.~\ref{fig:conv-rie1d3} it gets the lowest
%$L_1$ norm. In Fig.~\ref{fig:rie1d3-amr} we can see that using a
%higher resolution helps reducing the oscillations considerably. We
%can also see, that the dynamics of the AMR can still reduce the
%oscillations, and again we see its strength in overcoming a
%numerical problem using just a slight increase in the number of grid
%zones, to describe the hydrodynamic states through the calculated
%domain.

%% tab 3
\begin{table}
\scriptsize
 \centering
 \begin{tabular}{cccc}
$Scheme$ & $Number \; of \; cells$ & $L_1 \; Error$ & $Convergence \; Rate$\\
\hline
CT-PPM & 100 & 7.74e-2 & \\
 & 200 & 5.16e-2 & 0.58 \\
 & 400 & 2.26e-2 & 1.19 \\
 & 800 & 1.73e-2 & 0.38 \\
 & 1600 & 8.98e-3 & 0.95 \\
 & 3200 & 5.04e-3 & 0.83 \\
 & 51200 & 7.83e-4 & \\
\hline
RK2-PPM & 100 & 7.78e-2 & \\
 & 200 & 4.53e-2 & 0.78 \\
 & 400 & 1.85e-2 & 1.29 \\
 & 800 & 1.13e-2 & 0.71 \\
 & 1600 & 5.87e-3 & 0.95 \\
 & 3200 & 3.35e-3 & 0.81 \\
 & 51200 & 2.94e-4 & \\
\hline
RK3-PPM & 100 & 7.90e-2 & \\
 & 200 & 4.62e-2 & 0.77 \\
 & 400 & 1.92e-2 & 1.26 \\
 & 800 & 1.19e-2 & 0.69 \\
 & 1600 & 6.17e-2 & 0.95 \\
 & 3200 & 3.44e-3 & 0.84 \\
 & 51200 & 3.61e-3 & \\
\hline
CT-PLM & 100 & 7.98e-2 & \\
 & 200 & 4.87e-2 & 0.71 \\
 & 400 & 2.46e-2 & 0.98 \\
 & 800 & 1.43e-2 & 0.78 \\
 & 1600 & 8.8e-3 & 0.7 \\
 & 3200 & 5.62e-3 & 0.65 \\
 & 51200 & 7.28e-4 & \\
\hline
RK2-PLM & 100 & 1.14e-1 & \\
 & 200 & 7.33e-2 & 0.64 \\
 & 400 & 3.13e-2 & 1.23 \\
 & 800 & 2.19e-2 & 0.51 \\
 & 1600 & 1.20e-2 & 0.86 \\
 & 3200 & 7.40e-3 & 0.70 \\
 & 51200 & 8.66e-4 & \\
\hline
RK3-PLM & 100 & 1.14e-1 & \\
 & 200 & 7.38e-2 & 0.63 \\
 & 400 & 3.11e-2 & 1.25 \\
 & 800 & 2.19e-2 & 0.51 \\
 & 1600 & 1.20e-2 & 0.87 \\
 & 3200 & 7.33e-3 & 0.71 \\
 & 51200 & 8.53e-4 & \\
\hline
F-WENO \cite{zhang06} & 100  & 9.97e-2 &      \\
       & 200  & 6.29e-2 & 0.67 \\
       & 400  & 3.01e-2 & 1.1  \\
       & 800  & 1.69e-2 & 0.83 \\
       & 1600 & 9.48e-3 & 0.83 \\
       & 3200 & 5.24e-3 & 0.86 \\
\hline
F-PLM \cite{zhang06} & 100  & 1.12e-1 &      \\
       & 200  & 6.98e-2 & 0.68  \\
       & 400  & 3.45e-2 & 1.0  \\
       & 800  & 1.94e-2 & 0.83 \\
       & 1600 & 1.13e-2 & 0.78 \\
       & 3200 & 6.54e-3 & 0.79 \\
\hline
U-PPM \cite{zhang06} & 100  & 9.72e-2 &      \\
       & 200  & 5.60e-2 & 0.80 \\
       & 400  & 2.49e-2 & 1.2  \\
       & 800  & 1.30e-2 & 0.94 \\
       & 1600 & 6.06e-3 & 1.1  \\
       & 3200 & 3.11e-3 & 0.96 \\
\hline
U-PLM \cite{zhang06} & 100  & 9.53e-2 &      \\
       & 200  & 6.32e-2 & 0.59 \\
       & 400  & 2.99e-2 & 1.1  \\
       & 800  & 1.78e-2 & 0.75 \\
       & 1600 & 1.04e-2 & 0.78 \\
       & 3200 & 6.10e-3 & 0.77
\end{tabular}
\caption{$L_1$ errors of the density for the 1D Riemann
Problem 3. Six schemes of RELDAFNA and four schemes of RAM \cite{zhang06} with various resolutions are shown at
$t=0.4$.} \label{tab:rie1d3}
\end{table}

\newpage
\subsection{One-Dimensional Riemann Problem With Non-Zero Transverse
 Velocity: Problem 1}
\label{sec:rievy1}

\begin{figure*}
\centering
\includegraphics[scale=0.4]{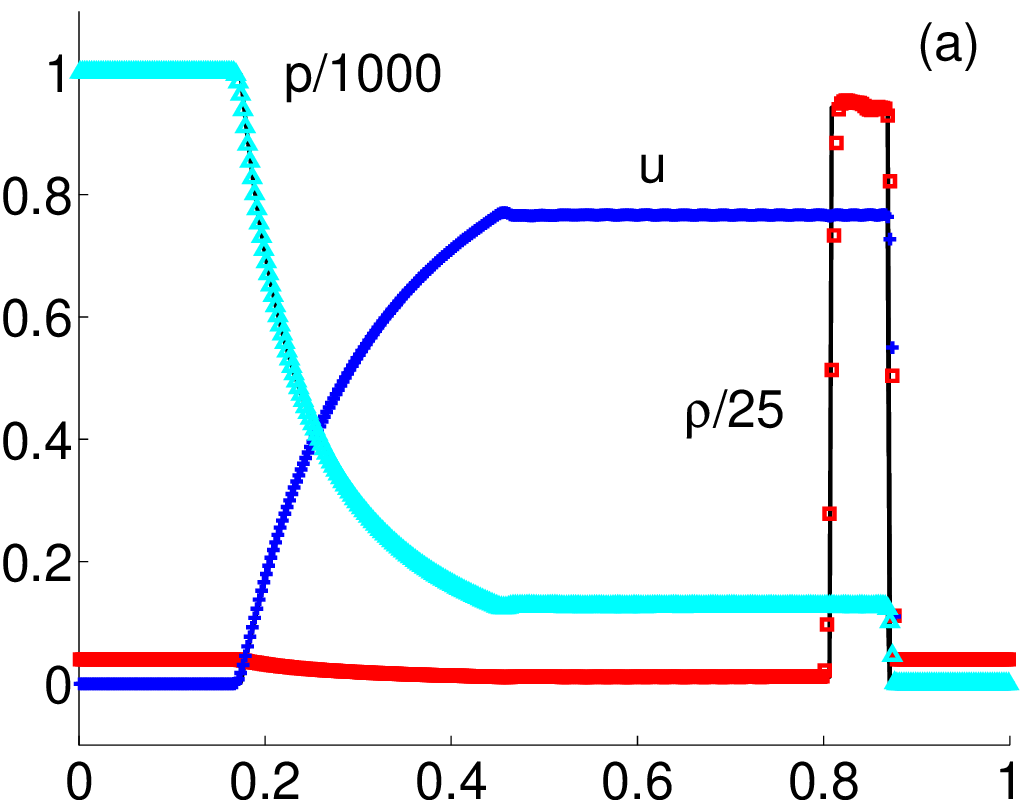}
\includegraphics[scale=0.4]{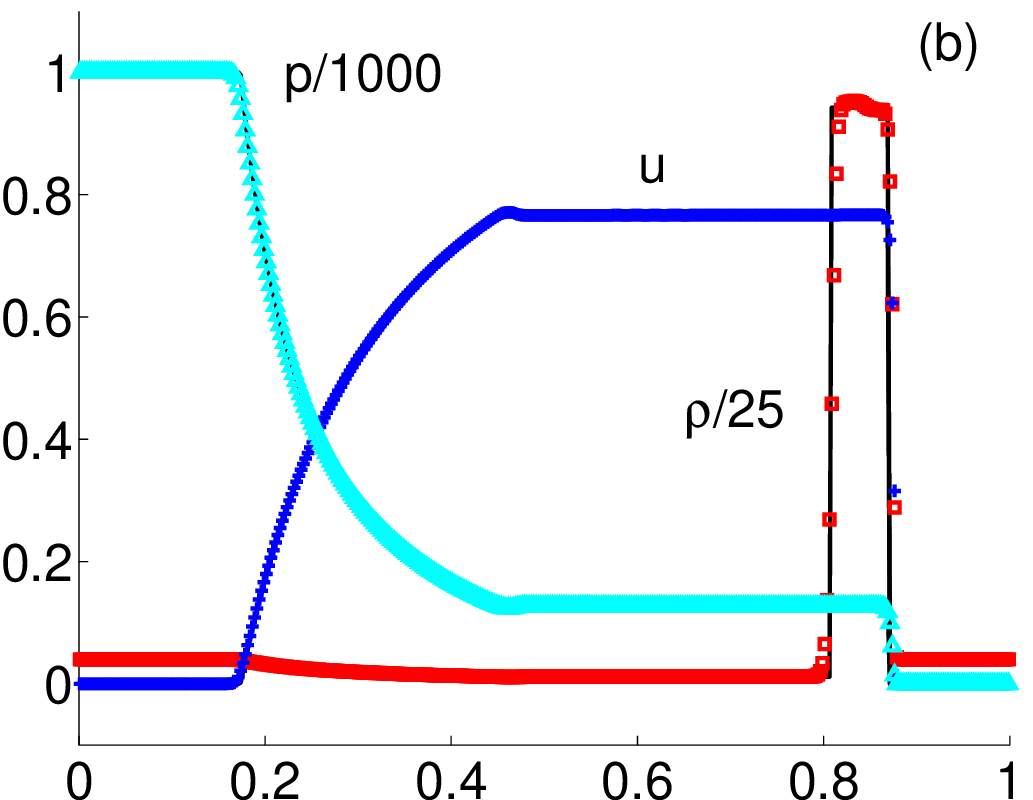}\\
\includegraphics[scale=0.4]{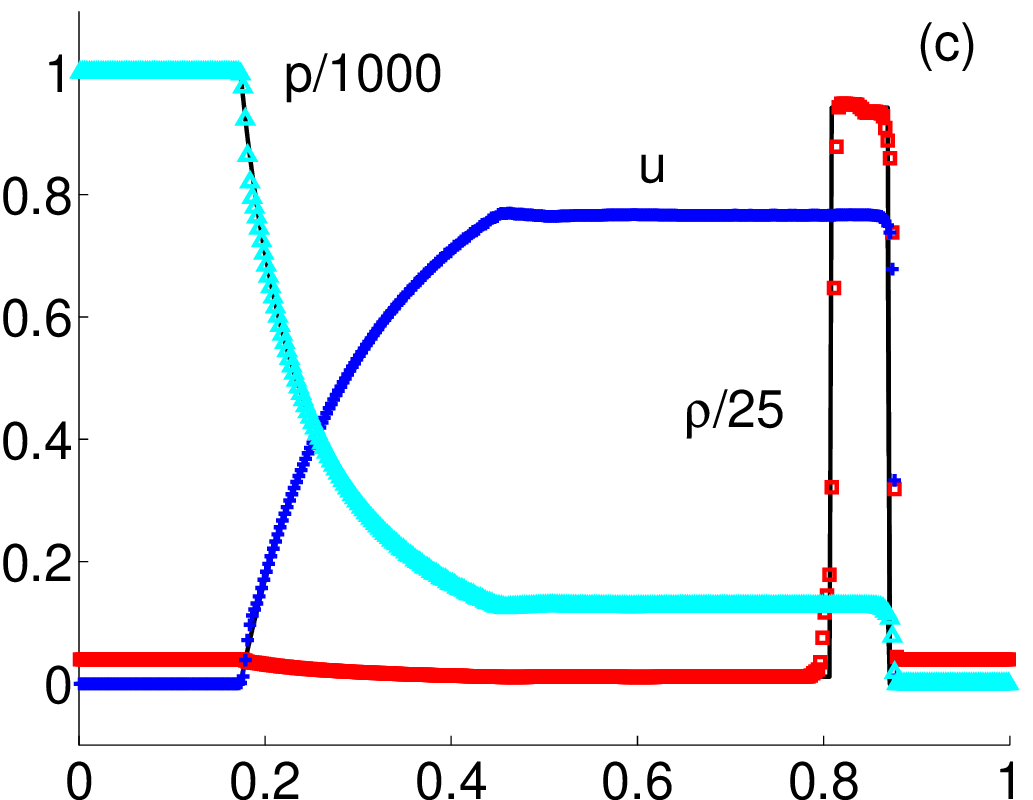}
\includegraphics[scale=0.4]{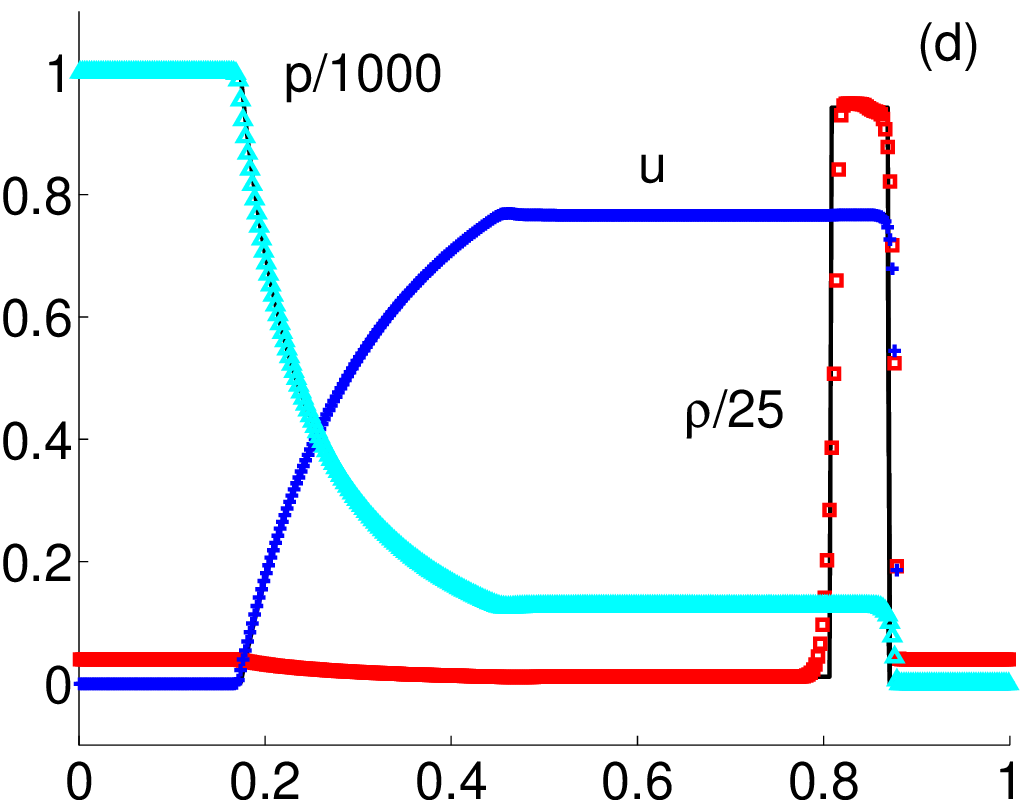}\\
\includegraphics[scale=0.4]{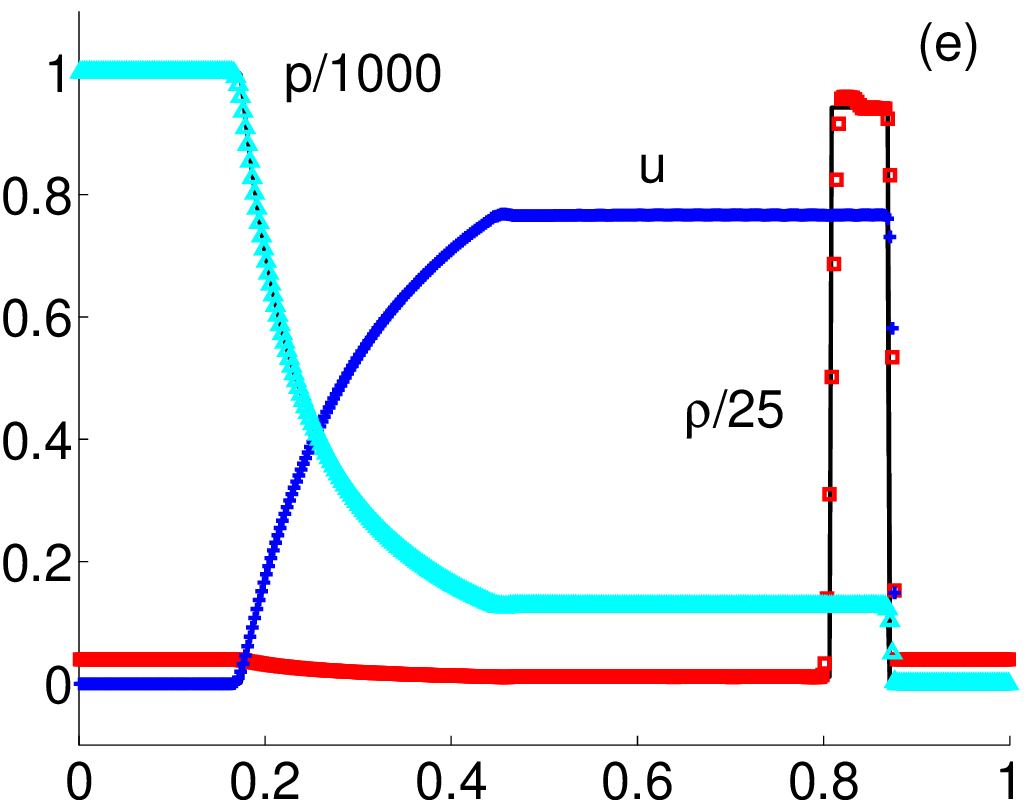}
\includegraphics[scale=0.4]{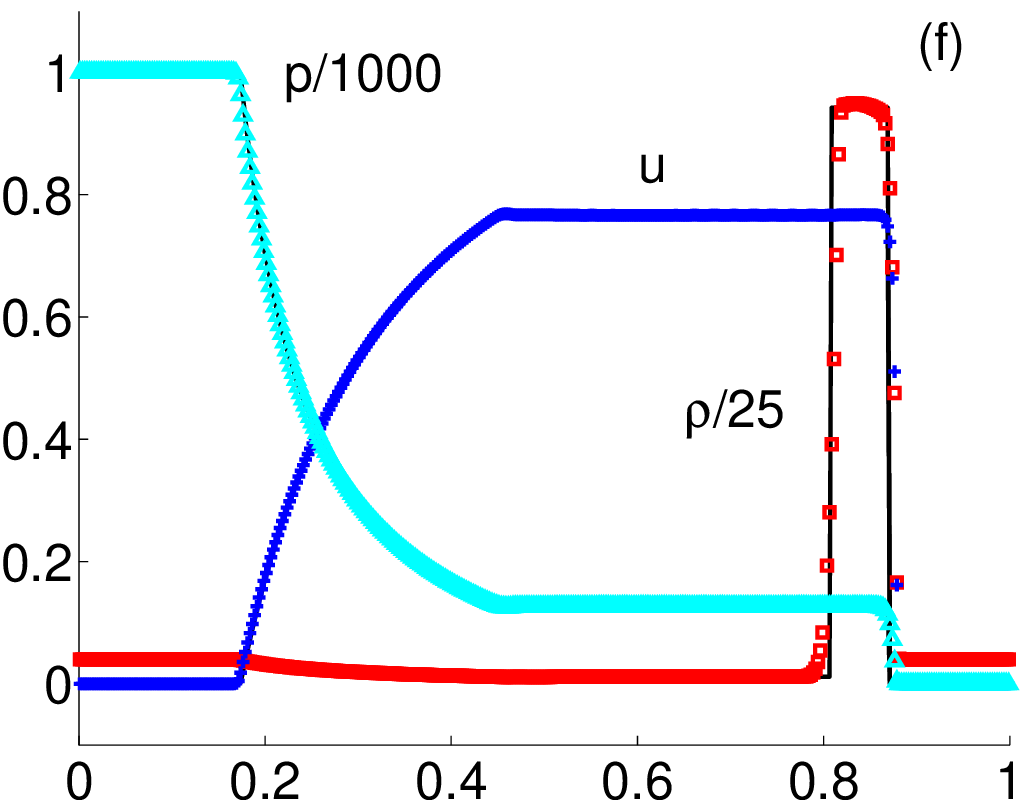}
\caption{One-dimensional Riemann problem with transverse velocity 1
at $t=0.4$. Results for
 six schemes: (a) CT-PPM, (b) CT-PLM, (c) RK2-PPM, (d) RK2-PLM, (e) RK3-PPM and (f) RK3-PLM are shown.
 The computational grid consists of 400 zones. Numerical results are
 shown in symbols, whereas the exact solution is shown in solid
 lines. We show proper mass density (square), pressure (triangle)
 and velocity (plus sign)
\protect\label{fig:rievy1}}
\end{figure*}

\begin{figure*}
\centering
\includegraphics[scale=0.7]{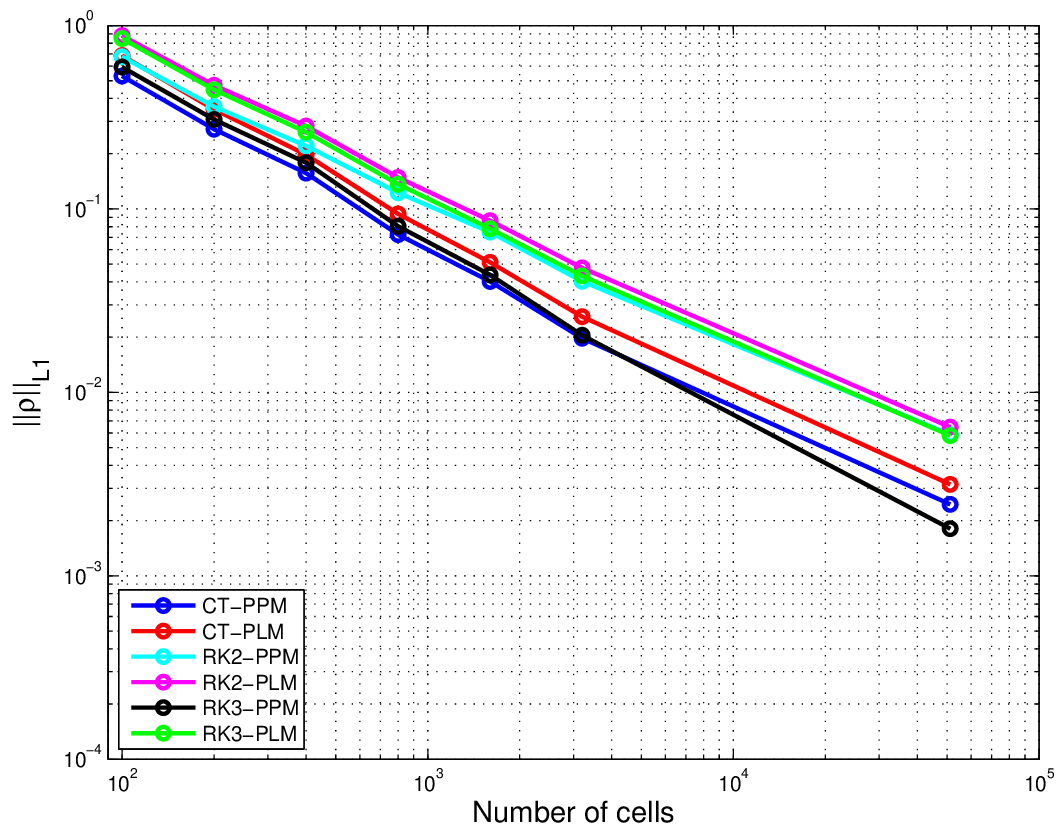}
\caption{$L_1$ errors of the density for the One-dimensional Riemann problem with transverse velocity 1. Seven different uniform grid resolutions (100,200,400,800,1600,3200 and 51200 cells) are
 shown at $t = 0.4$. Results for
 six schemes: (blue) CT-PPM, (red) CT-PLM, (cyan) RK2-PPM, (magenta) RK2-PLM, (black) RK3-PPM and (green) RK3-PLM are shown.
\protect\label{fig:conv-rievy1}}
\end{figure*}

\begin{figure*}
\centering
\includegraphics[scale=0.4]{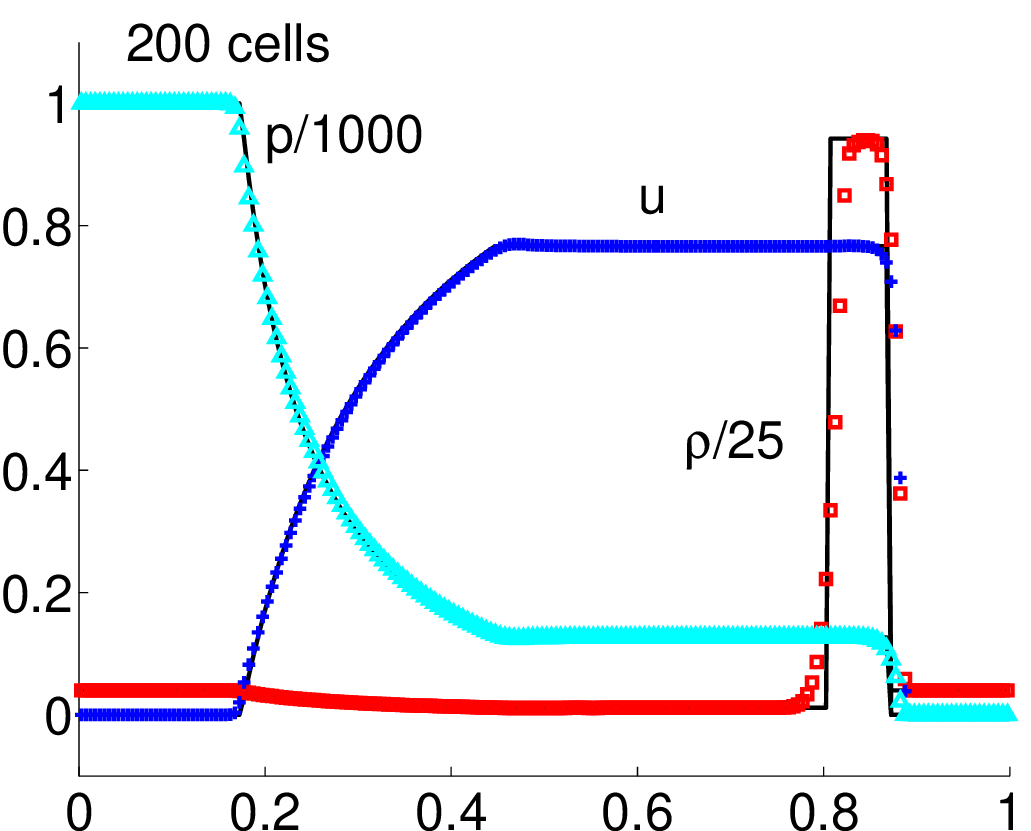}
\includegraphics[scale=0.4]{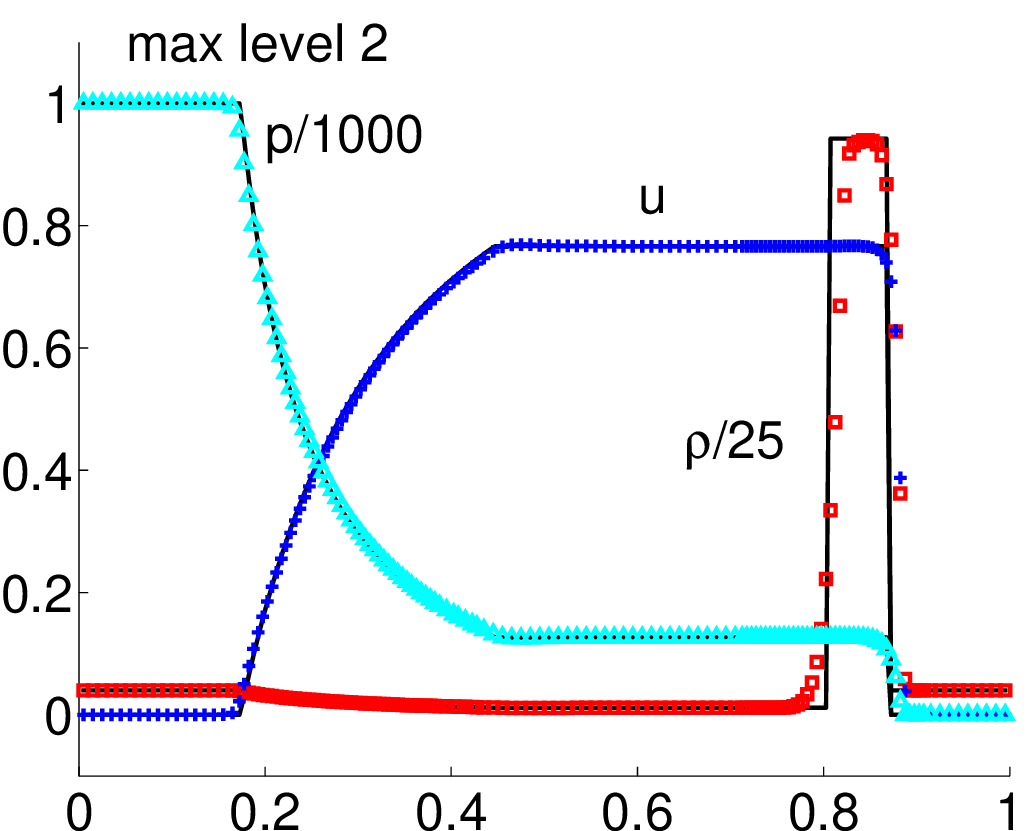}\\
\includegraphics[scale=0.4]{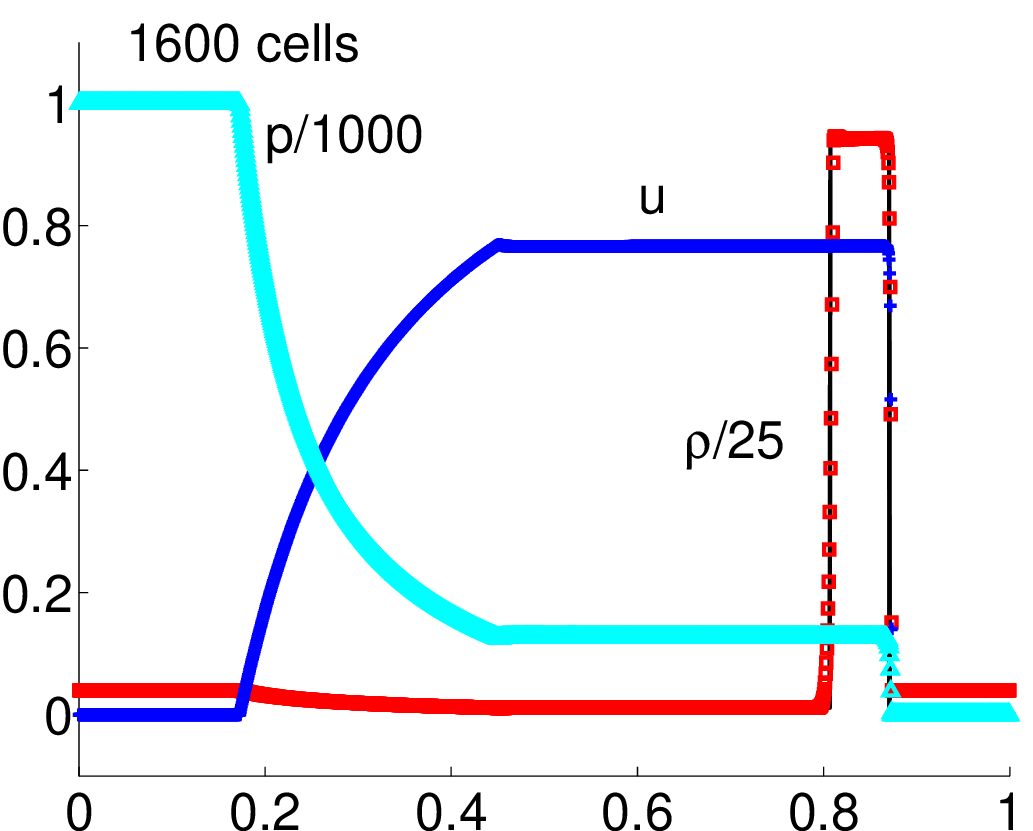}
\includegraphics[scale=0.4]{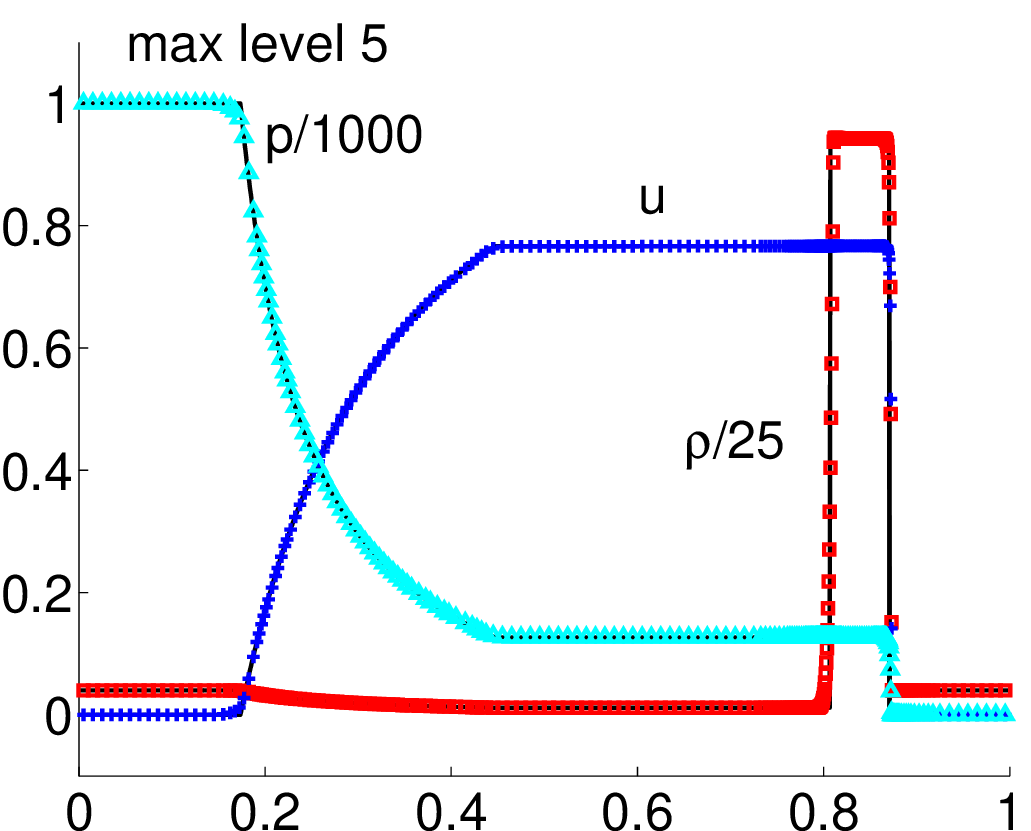}\\
\includegraphics[scale=0.4]{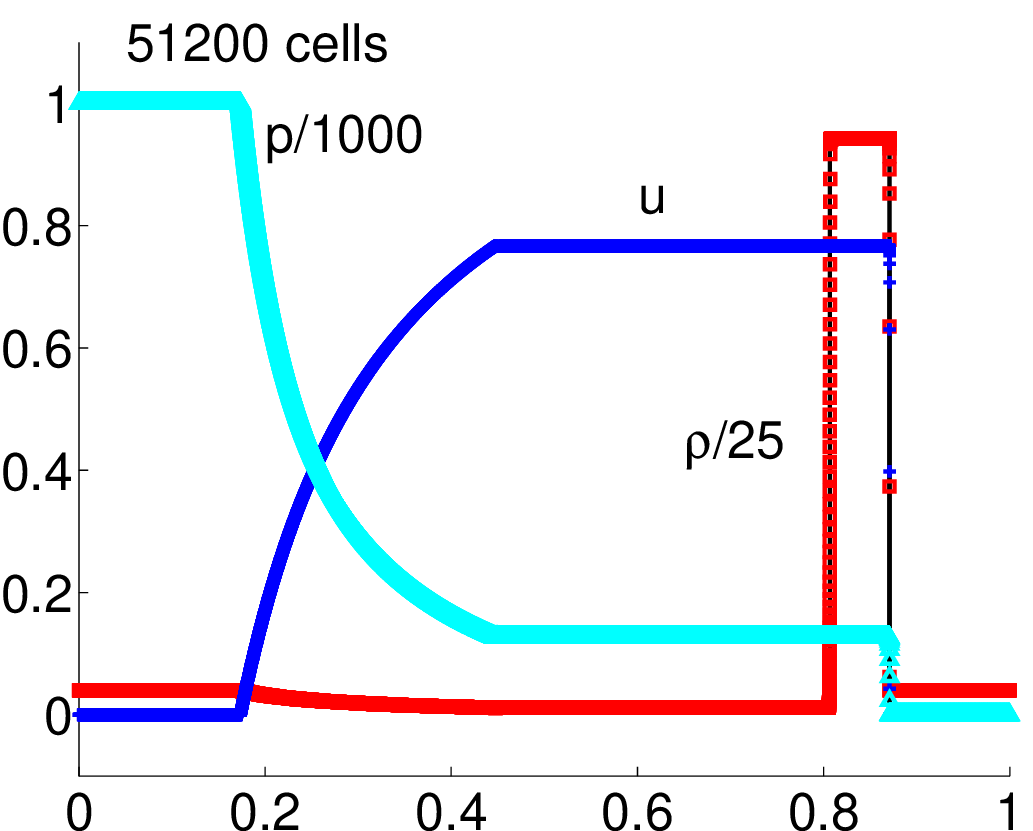}
\includegraphics[scale=0.4]{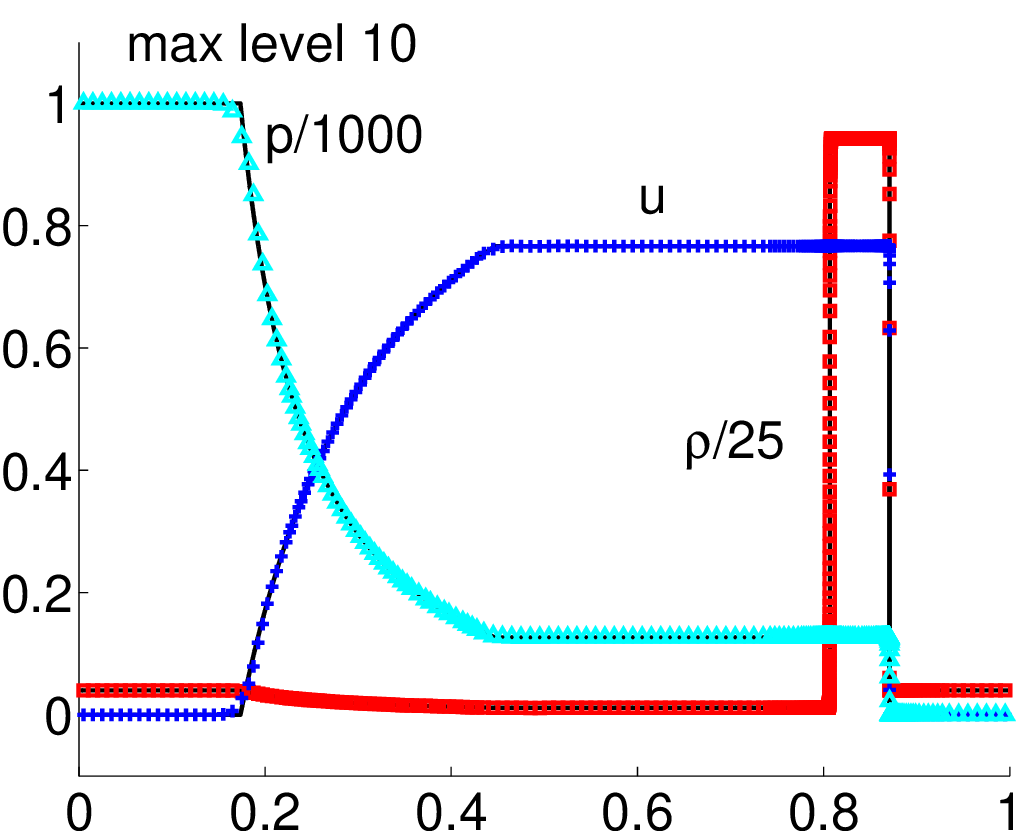}
\caption{One-dimensional Riemann problem with transverse velocity 1 at $t=0.4$. Results for
AMR calculations using RK2-PLM. are shown.
 At the right hand side are constant grid calculations, and on the left hand side are equivalent resolution AMR calculations using a base mesh resolution of 100 zones. Numerical results are
 shown in symbols, whereas the exact solution is shown in solid
 lines. We show proper mass density (square), pressure (triangle)
 and velocity (plus sign)
\protect\label{fig:rievy1-amr}}
\end{figure*}

In this test, the structures formed to section~\ref{sec:rie1d2}, but the thin
density jump is widened along with an increase in the density jump
over the shock (factor $\sim 25$). The results at $t=0.4$ are shown
in Fig.~\ref{fig:rievy1}. Again, our results are similar to the ones
published in \cite{zhang06,tchekhovskoy07} (see Table~\ref{tab:rievy1}). The shocks in our calculations are more
sharp than both mentioned references. Fig~\ref{fig:conv-rievy1}
presents the $L_1$ norm errors for the density profile at $t=0.4$ in
this problem. We also made a calculation with RK2-PLM using AMR. The
maximum number of cells, at a given time, used in a calculation with
maximum level of $10$, which is equivalent to a constant grid of
$51200$ cells, was $742$ cells which is only $1.45\%$. The results
of a calculation using AMR with RK2-PLM scheme at $t=0.4$, is shown
in Fig.~\ref{fig:rievy1-amr}.

Fig.~\ref{fig:rievy1} shows that the thin structure behind the shock
which is created in section~\ref{sec:rie1d2}, is widened by the
presence of transversal velocity component in the flow, and also
heightened by a factor of $\sim 2.5$.
%This widening makes the
%capturing of the high density region much easier, and we see that
%all six schemes find the structure and location of all states quite
%well using only $400$ cells, despite the high energy jump across the
%shock, $6$ times higher than the Newtonian limit.
%Similar to section~\ref{rie1d1}, we conclude that PPM schemes make an apparent
%difference in the sharp structure of the contact discontinuity,
%compared to PLM schemes. We also see, that the use of CT time
%integration method smears the shock on $\sim 3$ cells, which is much
%lower than the RK schemes. In Fig.~\ref{fig:conv-rievy1} we see
%again that CT-PLM is better than RK schemes when considering $L_1$
%norm.
In Fig.~\ref{fig:rievy1-amr} we can see that the widening of
the shell between the shock ad the contact discontinuity makes it
possible to find the maximum density even with very low resolution
and RK-PLM scheme (the scheme with the least number of mathematical
operations, and therefore the fastest one).
%We also see, that still
%in very high resolutions ($\gtrsim 50000$ zones), one needs a lot of
%cells ($\sim 25$) to describe the jump in density over the contact
%discontinuity. The AMR calculation does not affect that, not
%reducing the number of cells needed for that jump on one hand, and
%on the other hand resolving the structures similarly to the
%equivalent uniform grids.

%% tab 4
\begin{table}
\scriptsize
 \centering
 \begin{tabular}{cccc}
$Scheme$ & $Number \; of \; cells$ & $L_1 \; Error$ & $Convergence \; Rate$\\
\hline
CT-PPM & 100 & 5.3e-1 & \\
 & 200 & 2.72e-1 & 0.96 \\
 & 400 & 1.57e-1 & 0.8 \\
 & 800 & 7.25e-2 & 1.11 \\
 & 1600 & 4.03e-2 & 0.84 \\
 & 3200 & 1.97e-2 & 1.02 \\
 & 51200 & 2.45e-3 & \\
\hline
RK2-PPM & 100 & 6.75e-1 & \\
 & 200 & 3.61e-1 & 0.9 \\
 & 400 & 2.20e-1 & 0.71 \\
 & 800 & 1.22e-1 & 0.85 \\
 & 1600 & 7.47e-2 & 0.71 \\
 & 3200 & 4.04e-2 & 0.89 \\
 & 51200 & 5.9e-3 & \\
\hline
RK3-PPM & 100 & 5.94e-1 & \\
 & 200 & 3.07e-1 & 0.95 \\
 & 400 & 1.78e-1 & 0.78 \\
 & 800 & 8.04e-2 & 1.15 \\
 & 1600 & 4.37e-2 & 0.88 \\
 & 3200 & 2.05e-2 & 1.09 \\
 & 51200 & 1.8e-3 & \\
\hline
CT-PLM & 100 & 6.84e-1 & \\
 & 200 & 3.48e-1 & 0.97 \\
 & 400 & 1.97e-1 & 0.82 \\
 & 800 & 9.4e-2 & 1.07 \\
 & 1600 & 5.11e-2 & 0.87 \\
 & 3200 & 2.56e-2 & 0.98 \\
 & 51200 & 3.15e-3 & \\
\hline
RK2-PLM & 100 & 8.8e-1 & \\
 & 200 & 4.72e-1 & 0.9 \\
 & 400 & 2.83e-1 & 0.74 \\
 & 800 & 1.48e-1 & 0.93 \\
 & 1600 & 8.62e-2 & 0.78 \\
 & 3200 & 4.77e-2 & 0.85 \\
 & 51200 & 6.46e-3 & \\
\hline
RK3-PLM & 100 & 8.47e-1 & \\
 & 200 & 4.46e-1 & 0.92 \\
 & 400 & 2.62e-1 & 0.77 \\
 & 800 & 1.36e-1 & 0.94 \\
 & 1600 & 7.81e-2 & 0.8 \\
 & 3200 & 4.32e-2 & 0.85 \\
 & 51200 & 5.82e-3 & \\
\hline
F-WENO & 100  & 7.58e-1 &      \\
       & 200  & 3.92e-1 & 0.95 \\
       & 400  & 2.31e-1 & 0.76 \\
       & 800  & 1.18e-1 & 0.97 \\
       & 1600 & 6.58e-2 & 0.84 \\
       & 3200 & 3.44e-2 & 0.94 \\
\hline
F-PLM \cite{zhang06} & 100  & 8.26e-1 &      \\
       & 200  & 4.59e-1 & 0.85  \\
       & 400  & 2.77e-1 & 0.73  \\
       & 800  & 1.49e-1 & 0.89 \\
       & 1600 & 8.00e-2 & 0.90 \\
       & 3200 & 4.63e-2 & 0.79 \\
\hline
U-PPM \cite{zhang06} & 100  & 8.48e-1 &      \\
       & 200  & 4.25e-1 & 1.0  \\
       & 400  & 2.41e-1 & 0.82  \\
       & 800  & 1.27e-1 & 0.92 \\
       & 1600 & 6.43e-2 & 0.99  \\
       & 3200 & 3.34e-2 & 0.95 \\
\hline
U-PLM \cite{zhang06} & 100  & 9.00e-1 &      \\
       & 200  & 4.72e-1 & 0.93 \\
       & 400  & 2.88e-1 & 0.71  \\
       & 800  & 1.52e-1 & 0.92 \\
       & 1600 & 8.86e-2 & 0.78 \\
       & 3200 & 4.95e-2 & 0.84
\end{tabular}
\caption{$L_1$ errors of the density for the 1D Riemann problem with
non-zero transverse velocity: problem 1 at $t=0.4$. Six schemes of RELDAFNA and four schemes of RAM \cite{zhang06} with
various resolutions using a uniform grid are shown.}
\label{tab:rievy1}
\end{table}

\newpage
\subsection{One-Dimensional Riemann Problem With Transverse Velocity:
 Problem 2}
\label{sec:rievy2}

\begin{figure*}
\centering
\includegraphics[scale=0.4]{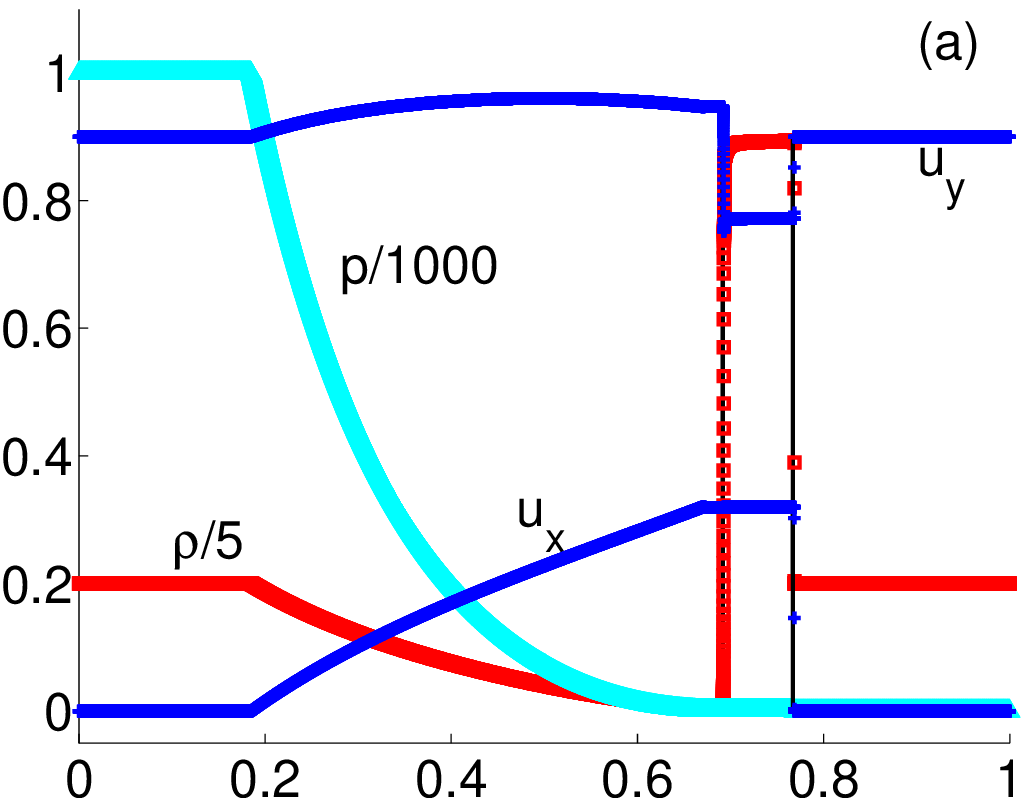}
\includegraphics[scale=0.4]{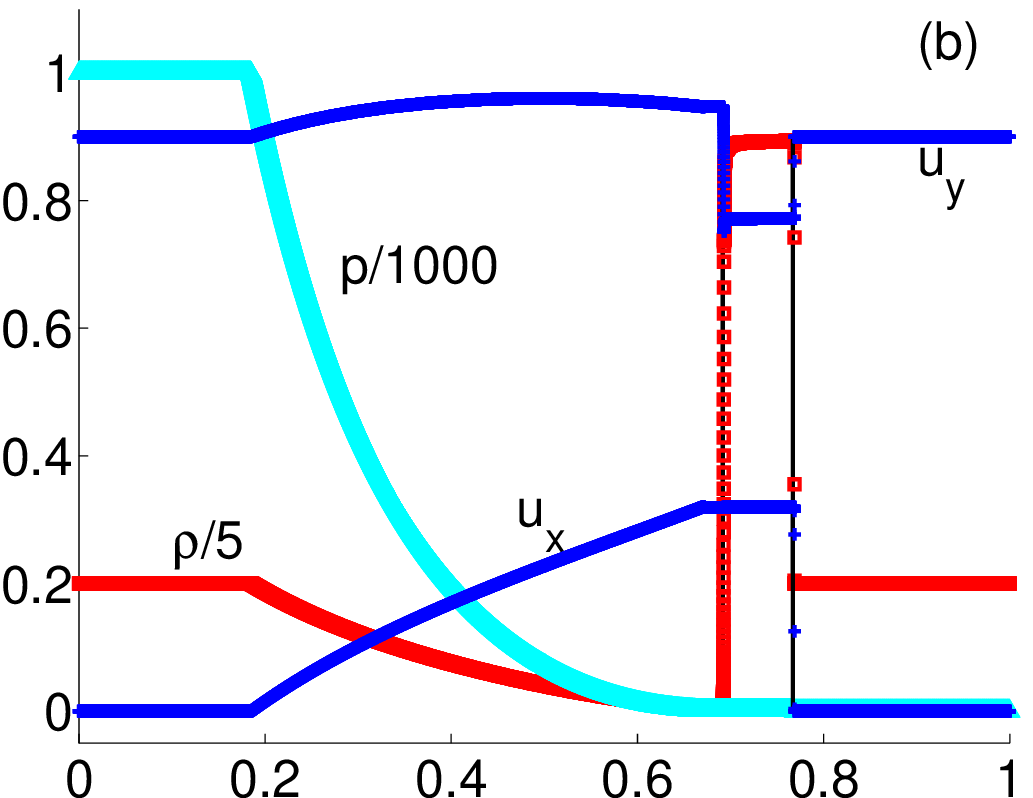}\\
\includegraphics[scale=0.4]{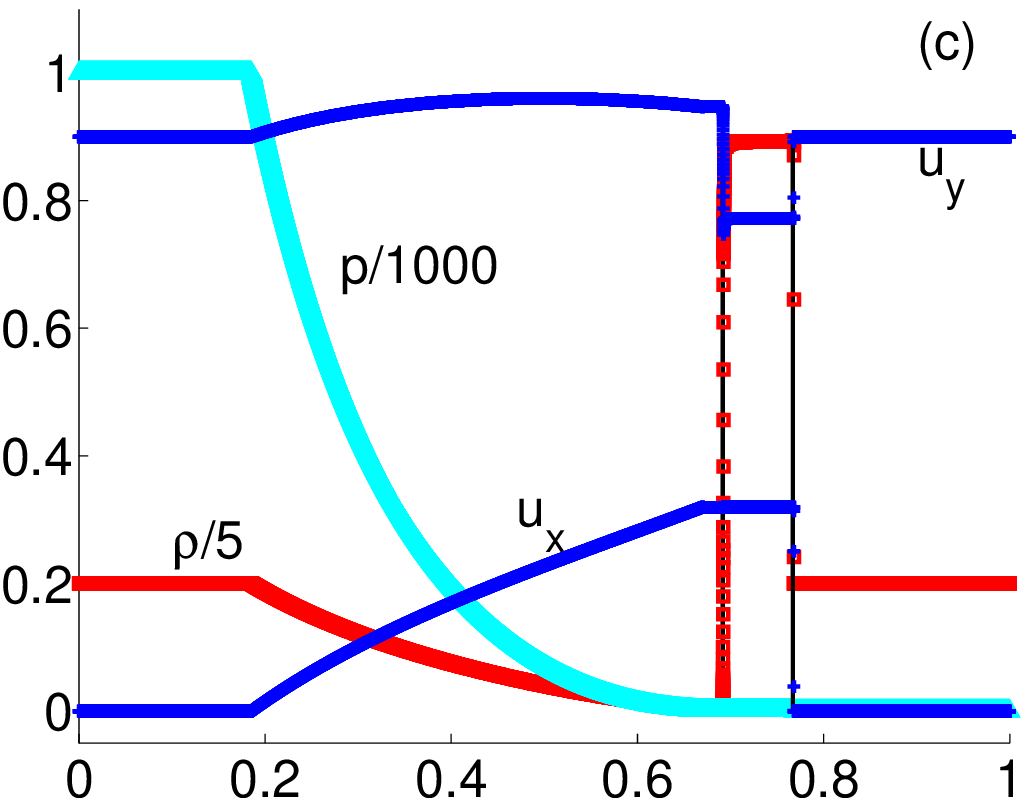}
\includegraphics[scale=0.4]{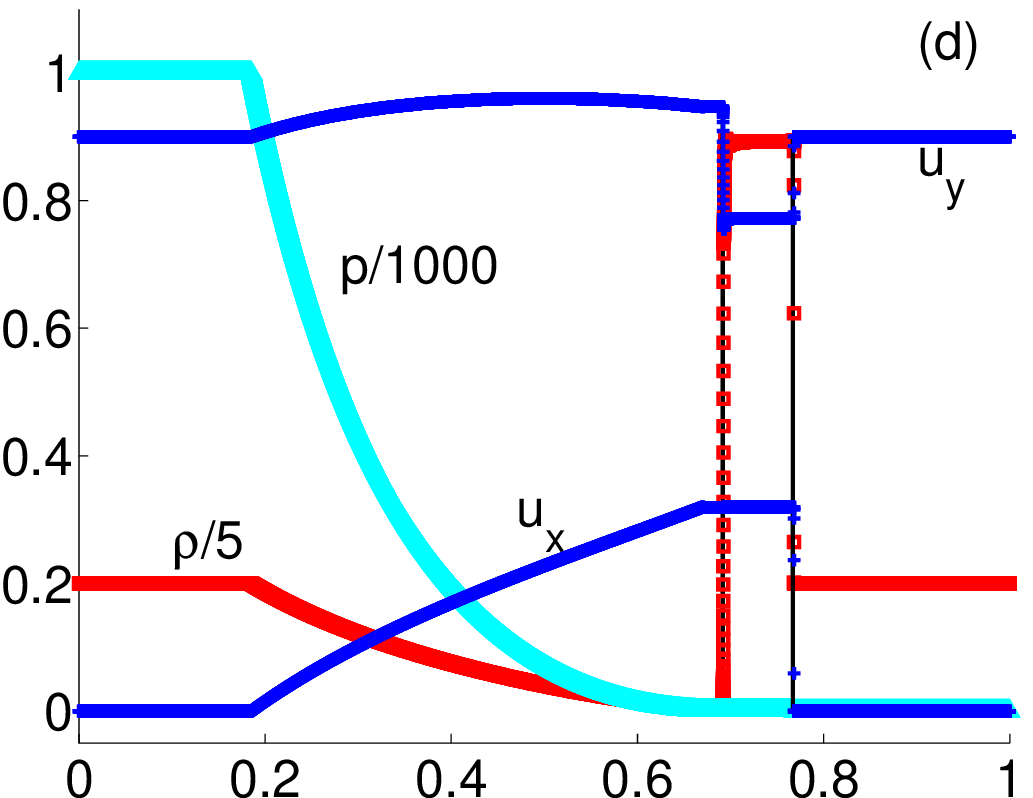}\\
\includegraphics[scale=0.4]{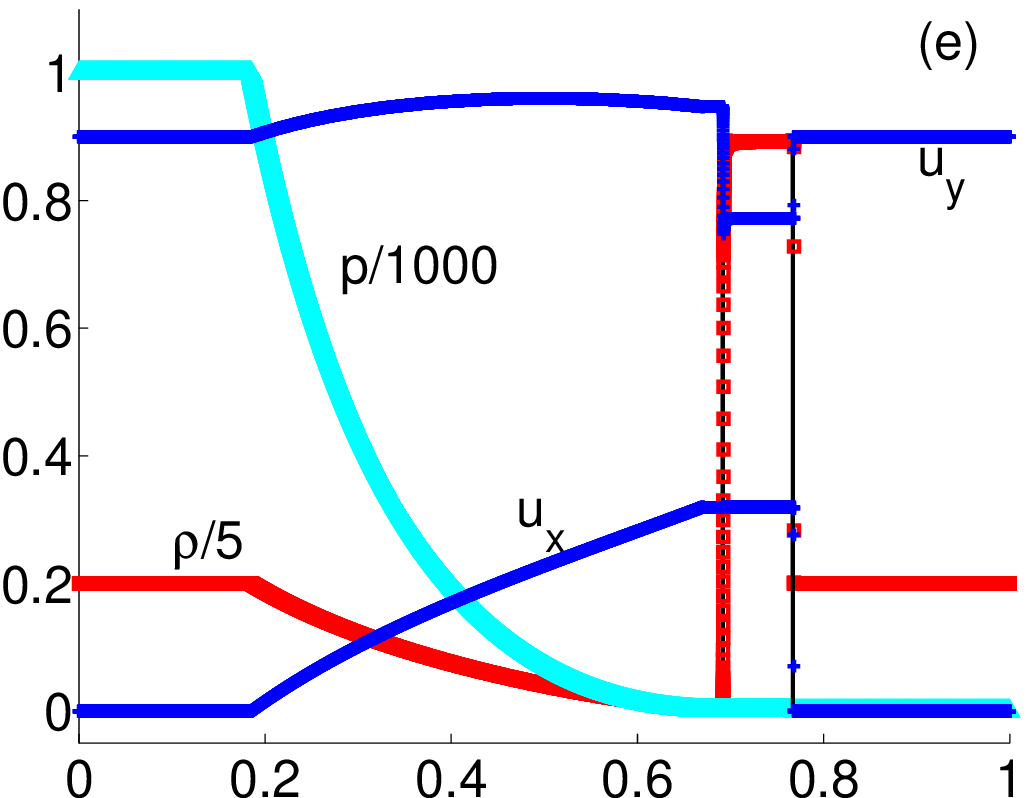}
\includegraphics[scale=0.4]{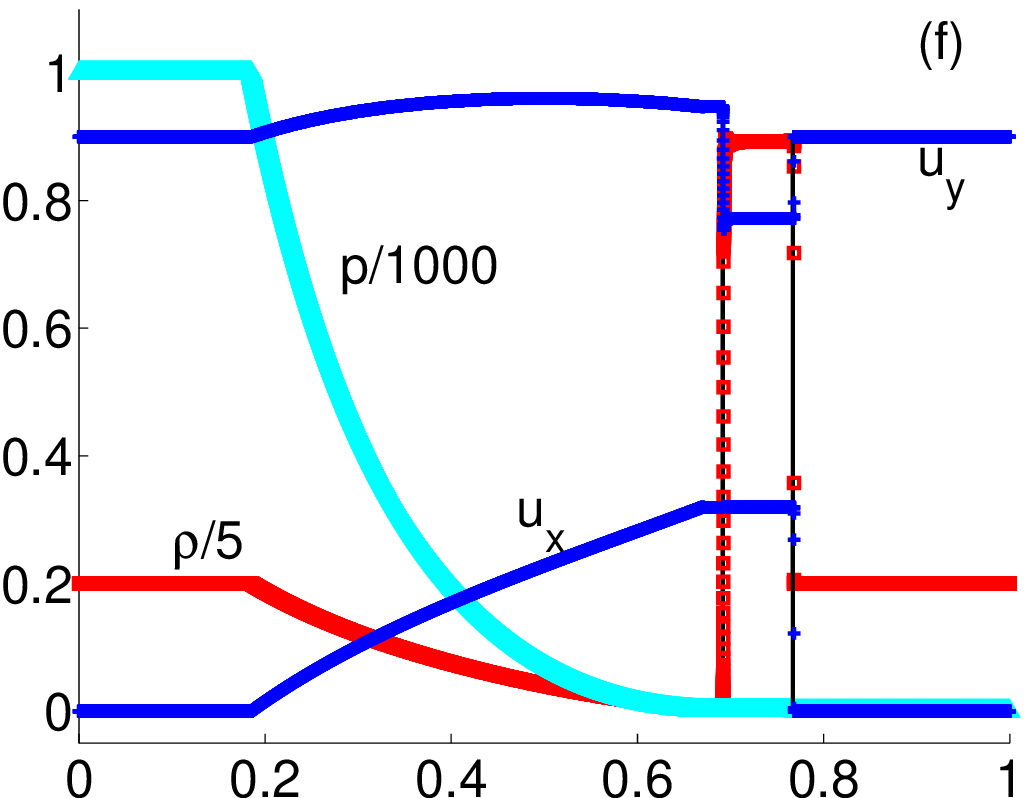}
\caption{One-dimensional Riemann problem with transverse velocity 2
at $t=0.6$. Results for
 six schemes: (a) CT-PPM, (b) CT-PLM, (c) RK2-PPM, (d) RK2-PLM, (e) RK3-PPM and (f) RK3-PLM are shown.
 The computational grid consists of 51200 zones. Numerical results are
 shown in symbols, whereas the exact solution is shown in solid
 lines. We show proper mass density (square), pressure (triangle)
 and velocity (plus sign)
\protect\label{fig:rievy2}}
\end{figure*}

\begin{figure*}
\centering
\includegraphics[scale=0.7]{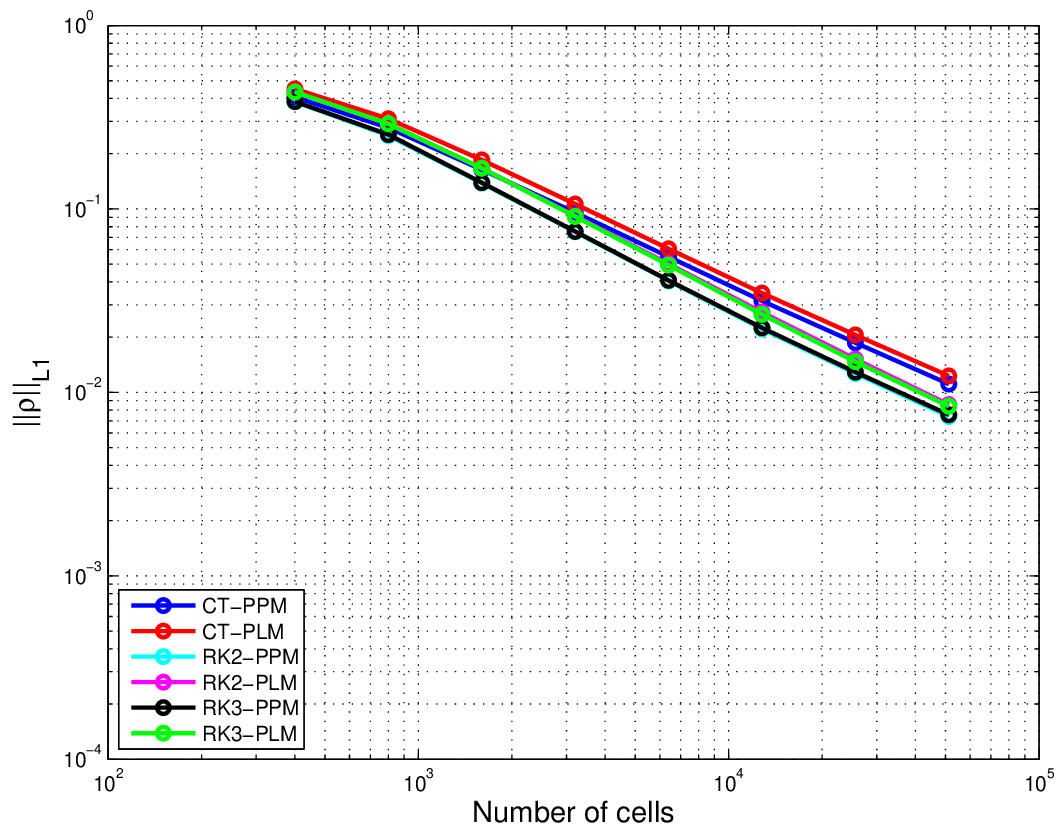}
\caption{$L_1$ errors of the density for the One-dimensional Riemann problem with transverse velocity 2. Eight different uniform grid resolutions (400,800,1600,3200,6400,12800,25600 and 51200 cells) are
 shown at $t = 0.4$. Results for
 six schemes: (blue) CT-PPM, (red) CT-PLM, (cyan) RK2-PPM, (magenta) RK2-PLM, (black) RK3-PPM and (green) RK3-PLM are shown.
\protect\label{fig:conv-rievy2}}
\end{figure*}

\begin{figure*}
\centering
\includegraphics[scale=0.4]{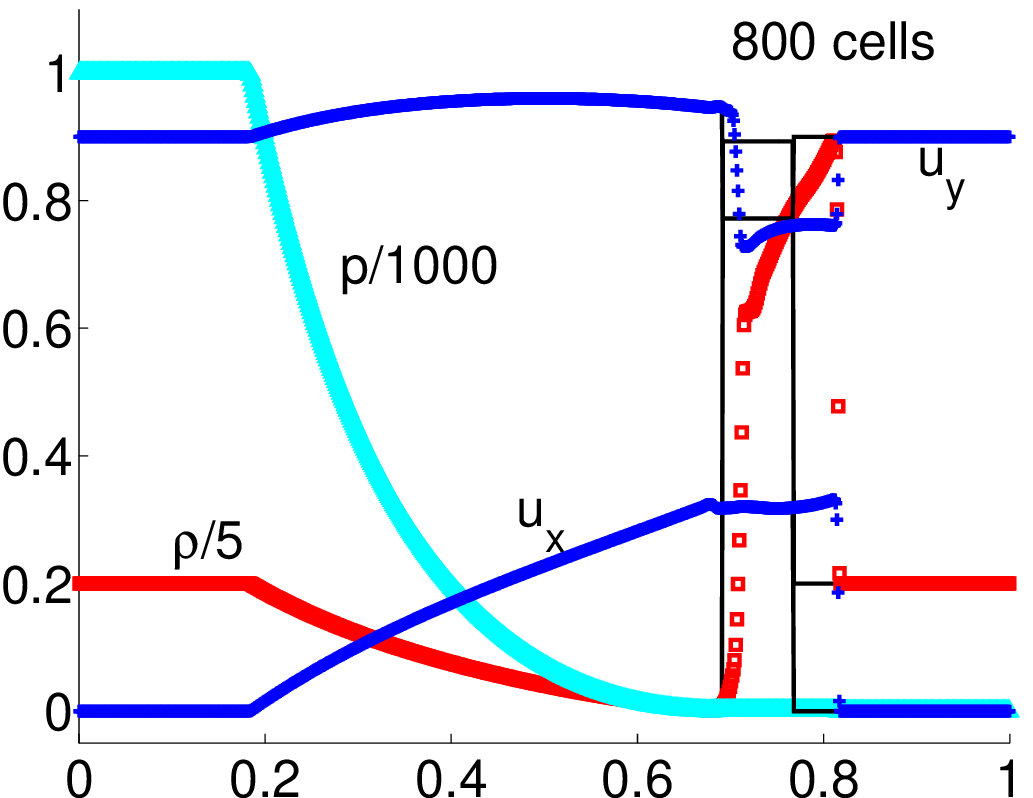}
\includegraphics[scale=0.4]{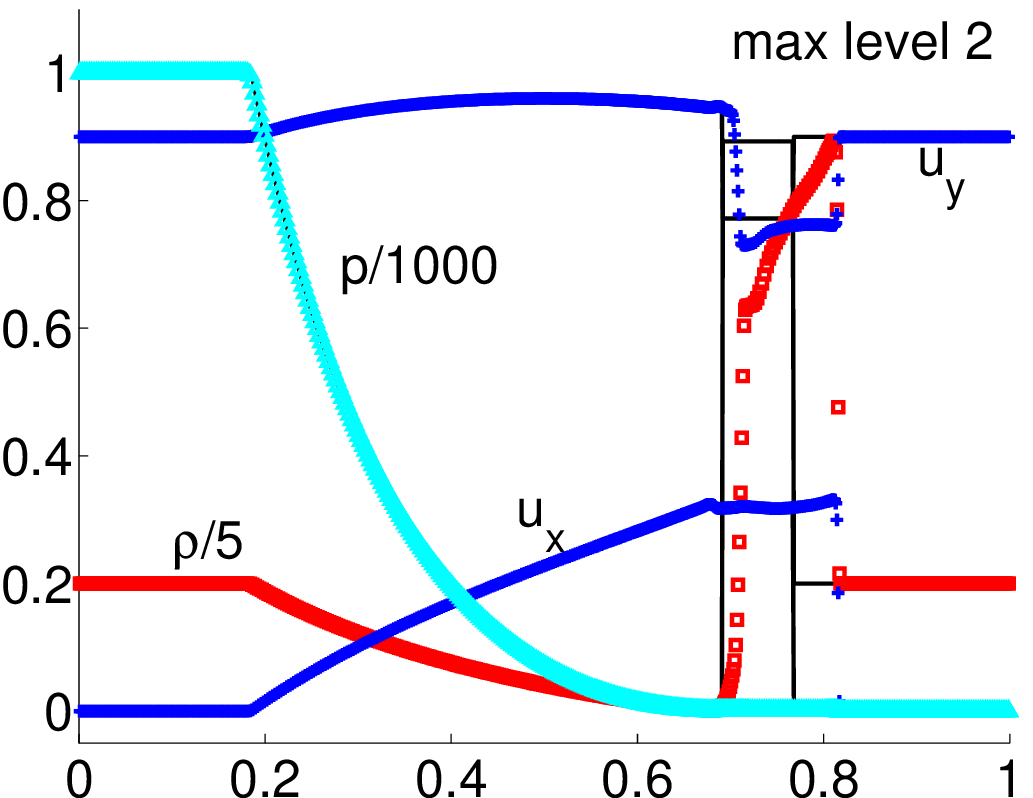}\\
\includegraphics[scale=0.4]{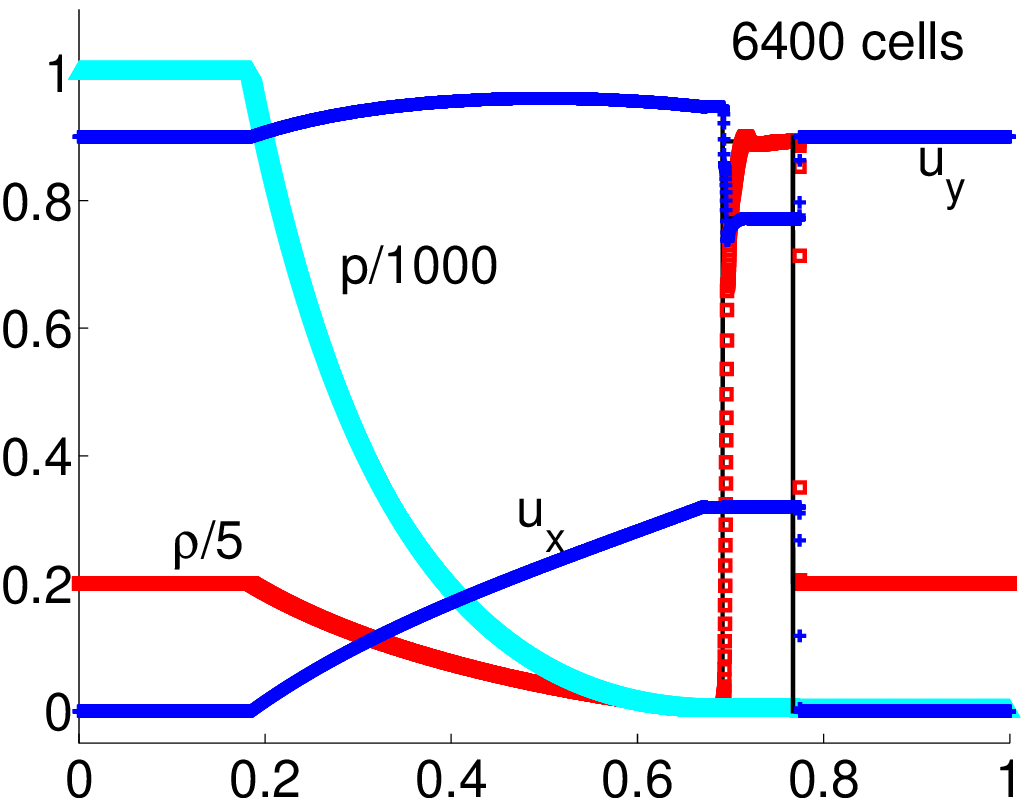}
\includegraphics[scale=0.4]{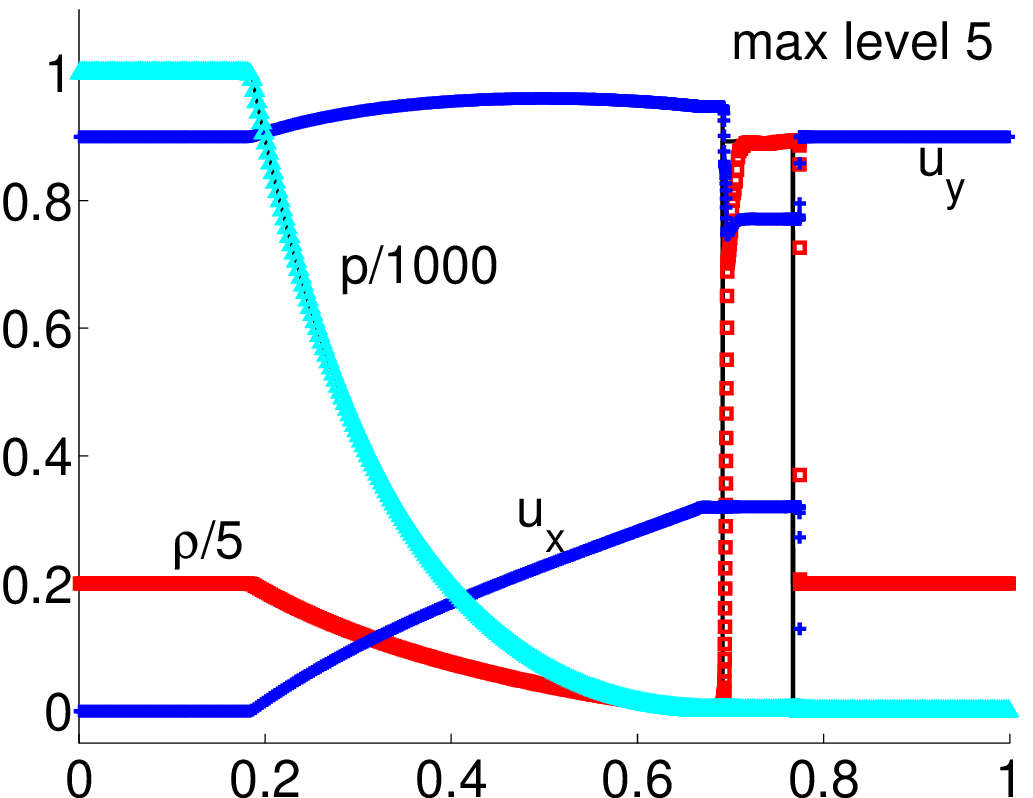}\\
\includegraphics[scale=0.4]{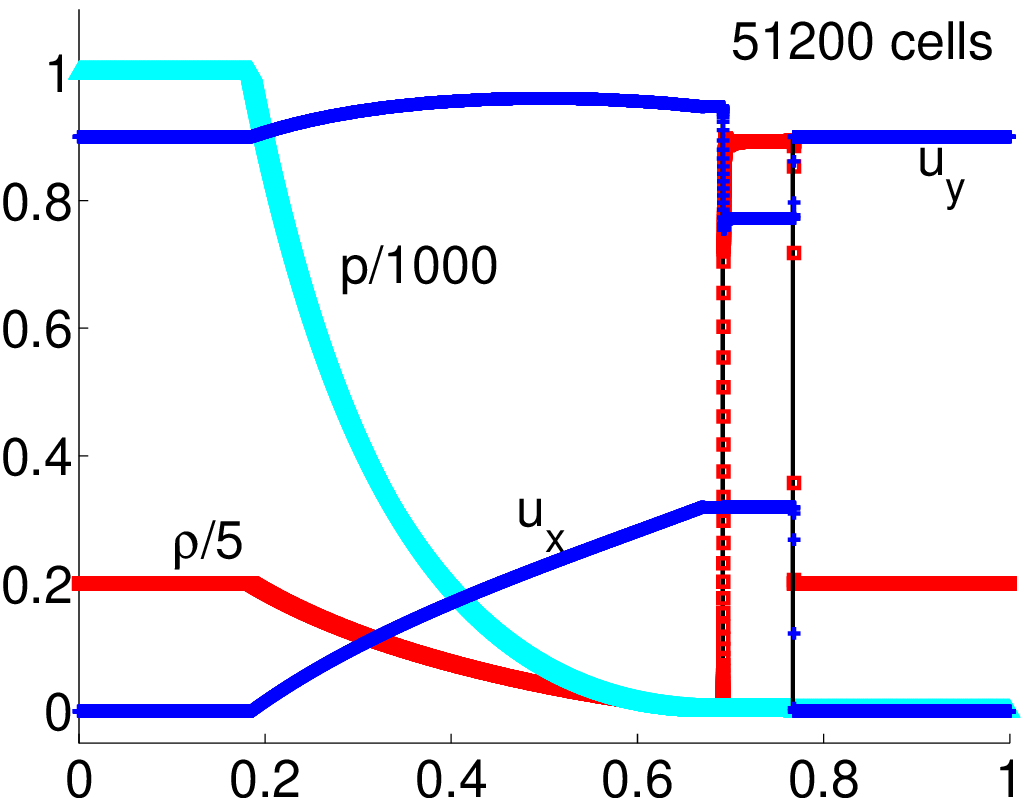}
\includegraphics[scale=0.4]{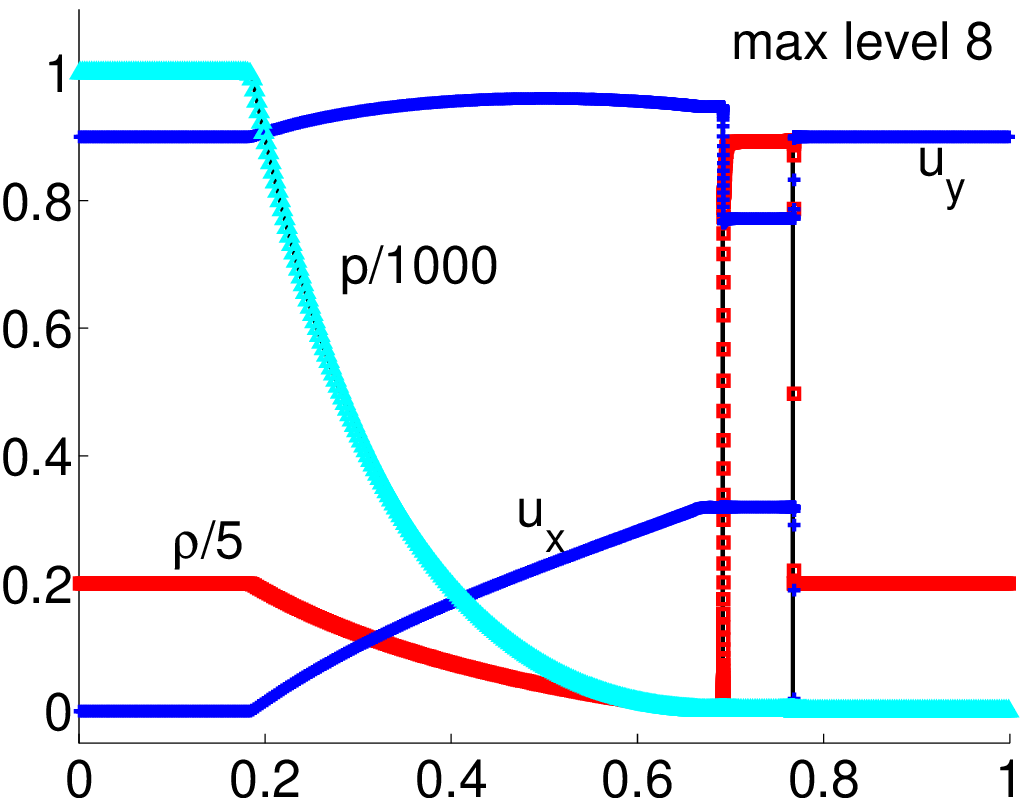}
\caption{One-dimensional Riemann problem with transverse velocity 2 at $t=0.6$. Results for
AMR calculations using RK3-PLM are shown. At the right hand side are
constant grid calculations, and on the left hand side are equivalent
resolution AMR calculations using a base mesh resolution of 400
zones. Numerical results are
 shown in symbols, whereas the exact solution is shown in solid
 lines. We show proper mass density (square), pressure (triangle)
 and velocity (plus sign)
\protect\label{fig:rievy2-amr}}
\end{figure*}

In section~\ref{sec:rievy1}, we tested out schemes on a one-dimensional
Riemann problem with non-zero transverse velocity. As was shown,
$400$ zones were enough to solve that test problem in a
satisfactory manner. As the structure of the transverse velocity
becomes complicated, one needs a much higher resolution in order to
calculate all the different parts of the flow, as will be shown in this section.\\
The results of the six schemes of RELDAFNA for this problem at $t=0.6$ are shown in Fig.~\ref{fig:rievy2}.
Fig.~\ref{fig:conv-rievy2} presents the $L_1$ norm errors for this
problem. For a comparison with one scheme of RAM \cite{zhang06} with published $L_1$ norms see Table~\ref{tab:rievy2}. We also made a calculation with RK3-PLM using AMR. The
maximum number of cells, at a given time, used in a calculation with
maximum level of $8$, which is equivalent to a constant grid of
$51200$ cells, was $1110$ cells which is only $2.17\%$. The results
of a calculation using AMR with RK3-PLM scheme is shown in
Fig.~\ref{fig:rievy2-amr}.

This problem is similar in its initial settings to the problems
introduced in section~\ref{sec:rie1d2} and ~\ref{sec:rievy1}. The
difference is the presence of tangential velocity component in the
hot side. Just like the previous two problems, the breakup of the
initial discontinuity creates a shock moving towards the right, a
rarefaction wave moving towards the left, and a contact
discontinuity in between. Fig.~\ref{fig:rievy2} shows that the thin
structure behind the shock which is created in
section~\ref{sec:rie1d2}, and widened in section~\ref{sec:rievy1},
is widened even more in this problem. Also, the density jump is
lower in this problem. However, other difficulties arise in the
numerical solution of this problem. The state between the
rarefaction wave and the contact discontinuity is very thin in this
problem. In order to capture that thin shell correctly, we need to
use extremely high resolutions ($\sim 50000$ cells), a phenomena
reported also in \cite{zhang06,wang08}. A successful attempt was made to
resolve such problems with a much lower number of cells using the
random-choice method \cite{cannizzo08}, but these attempts did not
manage to develop a multi dimensional scheme using that method.\\
%The need to use this extremely high resolution with all six schemes, as
%shown in Fig.~\ref{fig:rievy2}, says that all of them can solve this
%problem quite well, and we need to use Fig.~\ref{fig:conv-rievy2} to
%study the difference between the schemes. We see again that for such
%problems, with a tangential velocity component, the CT time
%integration method gives the highest $L_1$ norm, and there is no
%apparent advantage for the usage of high order spatial
%reconstruction (PPM over PLM).
%Comparing the values of the $L_1$
%norm of all the previous problems presented, to the one shown in
%Fig.~\ref{fig:conv-rievy2} shows that this problem is much more
%difficult to converge in the $L_1$ norm. Fig.~\ref{fig:rievy2-amr}
%shows that similar to section~\ref{sec:rievy1} it is possible to
%find the maximum density even with very low resolution and PLM
%scheme. We can also see, that since one needs a lot of cells to
%accurately solve this problem on a uniform grid, the use of AMR for
%this problem is crucial. We can see that the AMR calculations
%reproduce the same results as their equivalent uniform grids, even
%if these solutions are not accurate, i.e. if a uniform grid
%solution suffers from inaccuracy in structures location and internal
%profile, so will the AMR calculation with the same fine grid
%resolution but with a considerably less number of cells ($\sim 1-2
%\%$).

%% tab 5
\begin{table}
\scriptsize
 \centering
 \begin{tabular}{cccc}
$Scheme$ & $Number \; of \; cells$ & $L_1 \; Error$ & $Convergence \; Rate$\\
\hline
CT-PPM & 400 & 4.1e-1 & \\
 & 800 & 2.8e-1 & 0.56 \\
 & 1600 & 1.64e-1 & 0.76 \\
 & 3200 & 9.56e-2 & 0.78 \\
 & 6400 & 5.49e-2 & 0.8 \\
 & 12800 & 3.16e-2 & 0.8 \\
 & 25600 & 1.87e-2 & 0.76 \\
 & 51200 & 1.11e-2 & 0.75 \\
\hline
RK2-PPM & 400 & 3.84e-1 & \\
 & 800 & 2.52e-1 & 0.61 \\
 & 1600 & 1.38e-1 & 0.86 \\
 & 3200 & 7.5e-2 & 0.88 \\
 & 6400 & 4.05e-2 & 0.89 \\
 & 12800 & 2.21e-2 & 0.87 \\
 & 25600 & 1.27e-2 & 0.8 \\
 & 51200 & 7.4e-3 & 0.78 \\
\hline
RK3-PPM & 400 & 3.85e-1 & \\
 & 800 & 2.54e-1 & 0.6 \\
 & 1600 & 1.4e-1 & 0.87 \\
 & 3200 & 7.53e-2 & 0.88 \\
 & 6400 & 4.08e-2 & 0.88 \\
 & 12800 & 2.25e-2 & 0.86 \\
 & 25600 & 1.3e-2 & 0.8 \\
 & 51200 & 7.55e-2 & 0.77 \\
\hline
CT-PLM & 400 & 4.51e-1 & \\
 & 800 & 3.1e-1 & 0.54 \\
 & 1600 & 1.85e-1 & 0.74 \\
 & 3200 & 1.06e-1 & 0.8 \\
 & 6400 & 6.06e-2 & 0.81 \\
 & 12800 & 3.48e-2 & 0.8 \\
 & 25600 & 2.06e-2 & 0.76 \\
 & 51200 & 1.23e-2 & 0.74 \\
\hline
RK2-PLM & 400 & 4.33e-1 & \\
 & 800 & 2.9e-1 & 0.58 \\
 & 1600 & 1.67e-1 & 0.8 \\
 & 3200 & 9.16e-2 & 0.86 \\
 & 6400 & 5.02e-2 & 0.87 \\
 & 12800 & 2.74e-2 & 0.87 \\
 & 25600 & 1.52e-2 & 0.85 \\
 & 51200 & 8.58e-3 & 0.82 \\
\hline
RK3-PLM & 400 & 4.34e-1 & \\
 & 800 & 2.91e-1 & 0.57 \\
 & 1600 & 1.67e-1 & 0.8 \\
 & 3200 & 9.1e-2 & 0.87 \\
 & 6400 & 4.94e-2 & 0.88 \\
 & 12800 & 2.66e-2 & 0.89 \\
 & 25600 & 1.47e-2 & 0.86 \\
 & 51200 & 8.39e-2 & 0.81 \\
\hline
F-WENO-A \cite{zhang06}  &     400  &  5.21e-1  &       \\
           &    800  &  3.63e-1  &  0.52 \\
           &   1600  &  2.33e-1  &  0.64 \\
           &   3200  &  1.26e-1  &  0.89 \\
           &   6400  &  6.49e-2  &  0.96 \\
           &  12800  &  3.38e-2  &  0.94 \\
           &  25600  &  1.80e-2  &  0.91 \\
           &  51200  &  9.95e-3  &  0.86 \\

\end{tabular}
\caption{$L_1$ errors of the density for the 1D Riemann problem with
non-zero transverse velocity: problem 2 at $t=0.6$. Six schemes of RELDAFNA and one scheme of RAM \cite{zhang06} with various resolutions are shown.}
\label{tab:rievy2}
\end{table}

\newpage
We shortly summarize here the different conclusions from the one dimensional tests with analytical solution.
\begin{itemize}

\item In all the problems tested, the quantity which differs between the schemes is the density, since it is discontinuous over the contact discontinuity as well as the shock wave. It is apparent that PPM schemes used less cells to capture the jump in the density (CT-PPM only $\sim 5$ cells in the zero tangential velocity problems).

\item Using a selected spatial reconstruction method (either PPM or PLM), one would find that the CT time integration method is a better choice, since its flat density between the contact discontinuity and the shock wave seems uniform, and is closest to the analytic solution. Another feature seen to differ CT time integration method is the smearing of shock fronts. Using CT in problems with zero tangential velocity smears shock fronts on $\sim3$~cells, lower than using RK methods.

\item We see that it is more critical to use a higher order spatial reconstruction (PPM rather than PLM) than
using one of the two time integration methods, since given a time integration method, one gets an apparently higher density at the thin shells using PPM. \textbf{However, comparison to spatially higher order schemes, does not show any justifying improvements for the additional computational work. The norms of errors of analytically solved problems are similar to those obtained by other groups using higher order reconstruction methods.}

\item The use of AMR makes the low base mesh calculation very accurate. We can capture the thin structures both in location and in internal structure including the very high maximum density. However, a user must tune his/her refinement and derefinement criterions in order to sharpen structures even more (especially rarefaction fans). A wise choice of refinement criterions can solve a given problem very accurately without use of a very high number of cells.

\item Problems involving slowly moving structures produce oscillations. Such a phenomena is visible in all RELDAFNA schemes although RK$\#$-PPM methods produce smaller oscillations. This feature is common to all other published results. As expected, use of higher resolution helps reducing the oscillations considerably. Of course, the use AMR reduces the oscillations without a notable increase in the number cells. The source of such oscillations and the way to overcome them is one of the challenges future research of numerical methods to solve the SRHD Equations would encounter.

\item Naturally, the location of thin structures is demanding a higher resolution than solving their initial structure, as can be seen by comparing problems no.~$2$ and $4$ where a dense shell with a high density jump, when widened in problem $4$ was solved quite well with a moderate resolution.

\item When non-zero tangential velocity components are present, we observe a smearing of the sharp interfaces, using all schemes of RELDAFNA. This feature becomes very demanding in problem no.~$5$, where the use of very high resolution is needed. This problem is well known in the literature \cite{zhang06,wang08} and is due to the fact the Lorenz factor in these cases is not continuous across contact discontinuities. These problems are solved accurately and with a moderate number of cells only by using AMR. The AMR calculations reproduce the same results as their  equivalent uniform grids, even if these solutions are not accurate, i.e. if a uniform grid solution suffers from inaccuracy in structures location and internal profile, so will the AMR calculation with the same fine grid resolution but with a considerably less number of cells ($\sim 1-2\%$). Since much of our multidimensional problems from astrophysics involve flows which are one dimensionally similar to problem no.~$5$, such as jets heating the ISM, we learn that use of AMR is crucial, and one dimensional study of the effective resolution needed can be made before turning to a multidimensional study. Different ways to overcome this problem without using a very high resolution are being tested, and since they are critical for multidimensional calculations, where high resolution becomes computationally hard, this subject is in the focus of future numerical work on the SRHD equations.
\end{itemize}

\newpage

\section{Two-Dimensional Riemann problem} \label{sec:rie2d}

Unlike one dimensional Riemann problems that have an analytic
solution \cite{marti94,pons00}, in two dimensions the breakout of 4
separate touching constant states can not be analytically solved.
Therefore, the best measures of a numerical solution to such a
problem, would be comparison to other published numerical results,
and conservation of symmetry of the solution with
time.

\begin{figure*}
\centering
\includegraphics[scale=0.35]{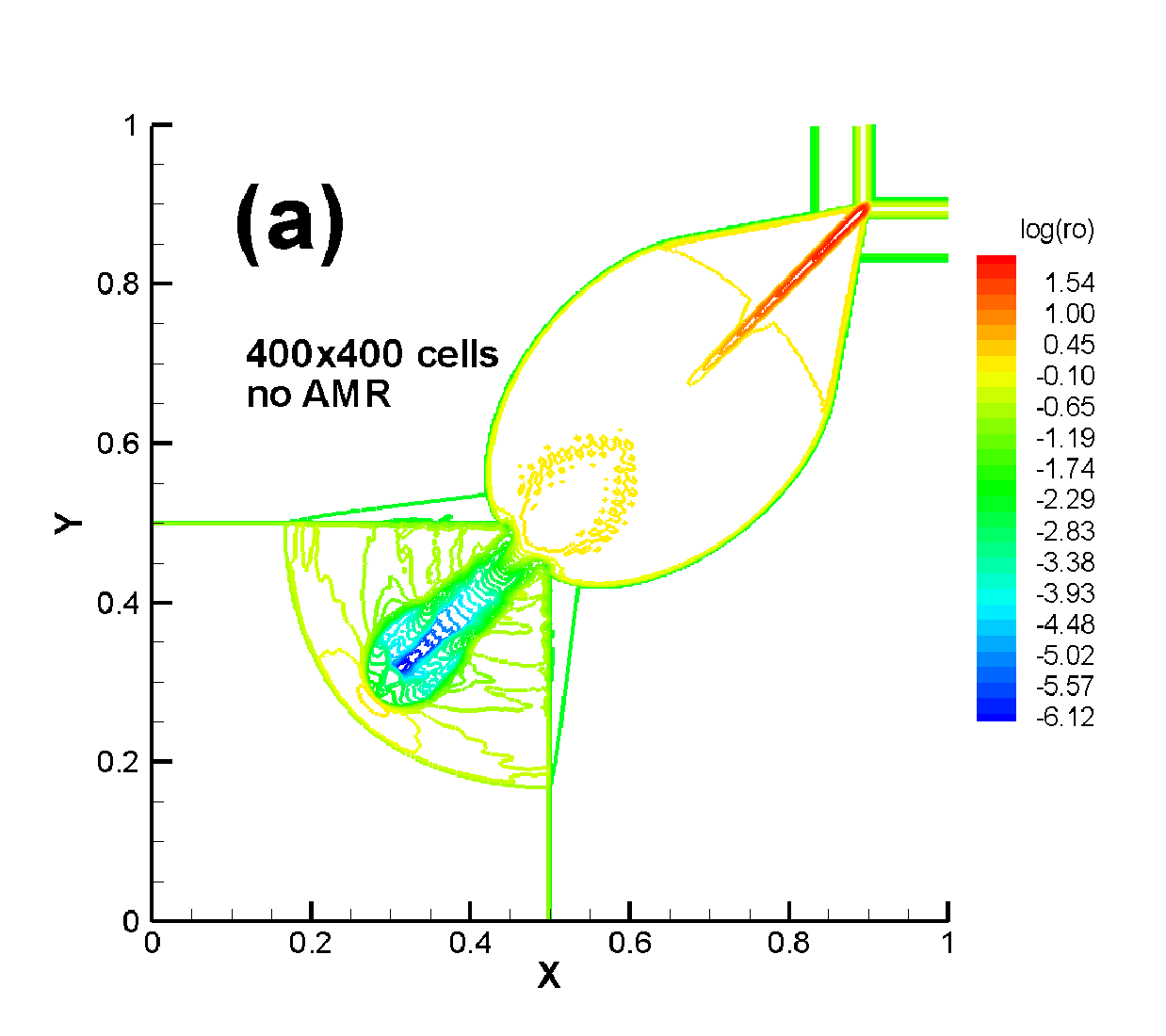}
\includegraphics[scale=0.35]{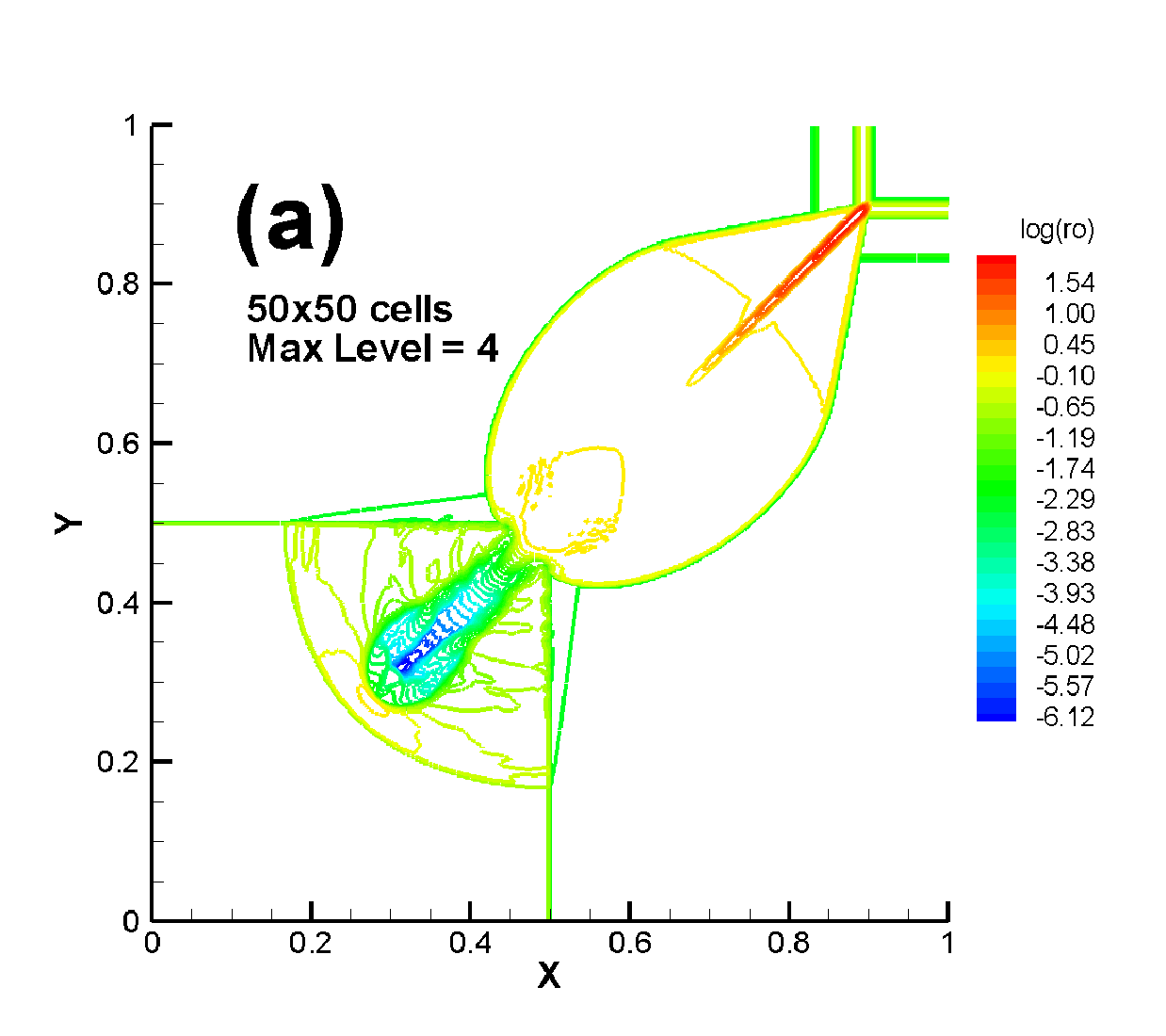}\\
\includegraphics[scale=0.35]{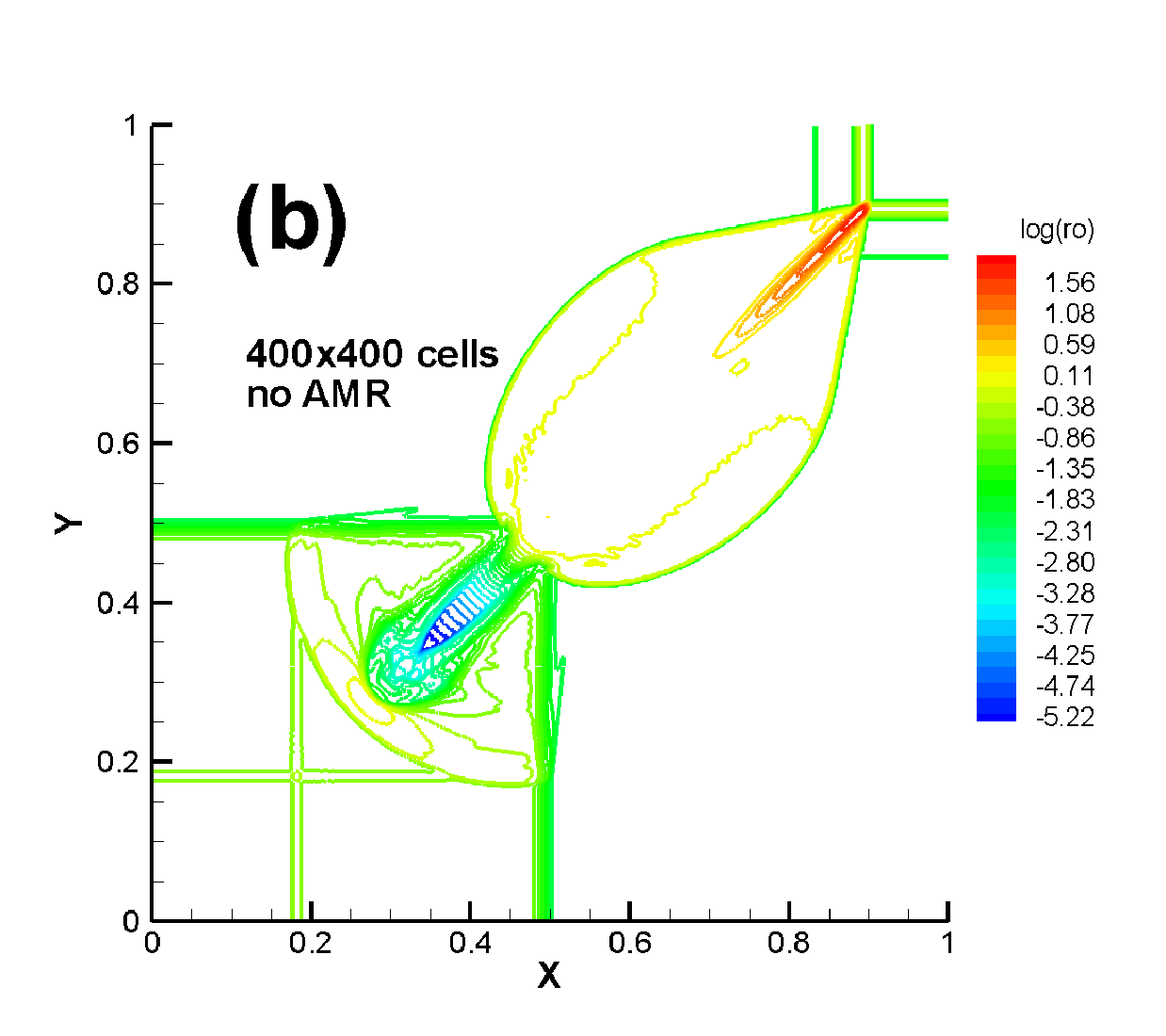}
\includegraphics[scale=0.35]{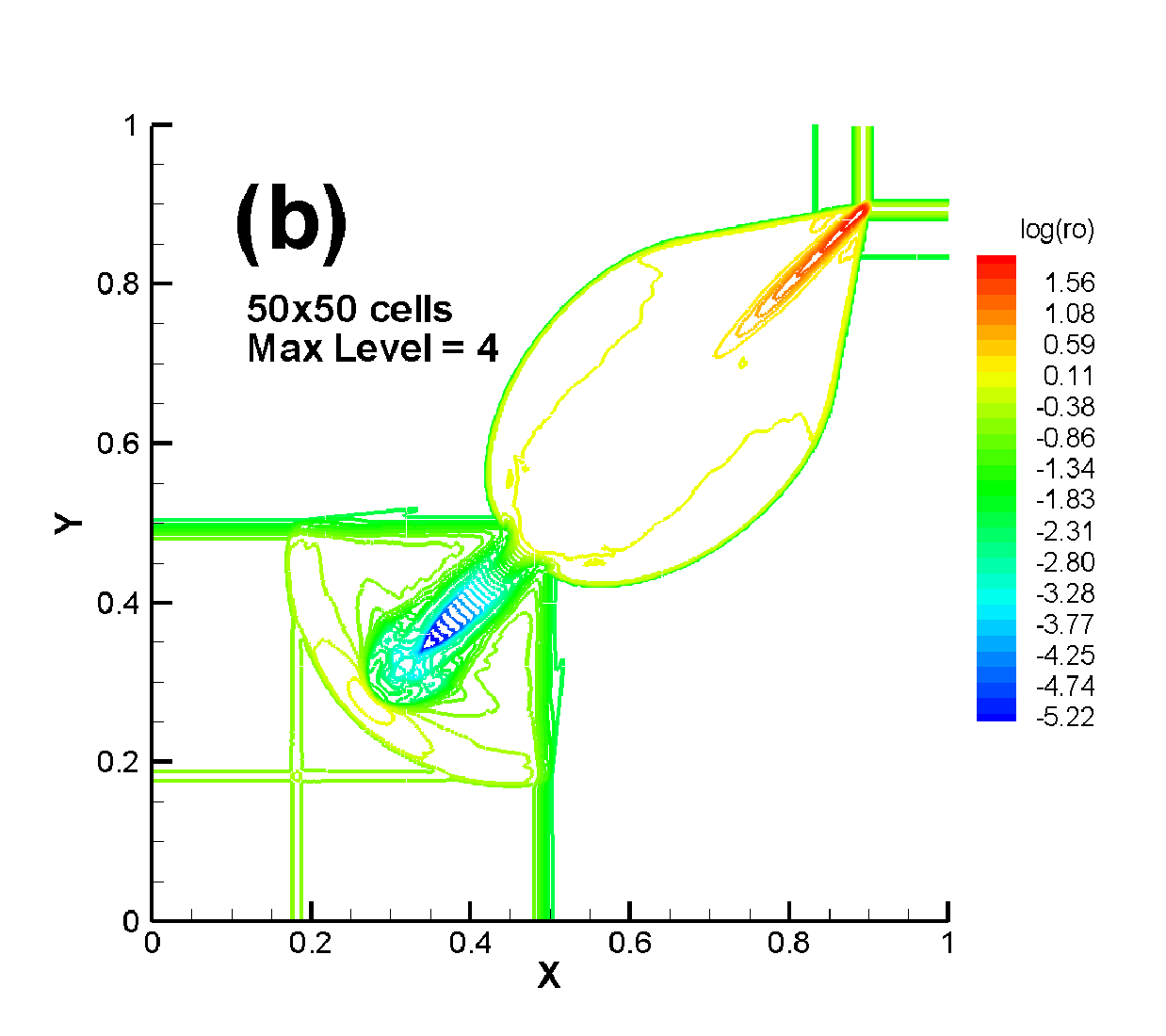}\\
\includegraphics[scale=0.35]{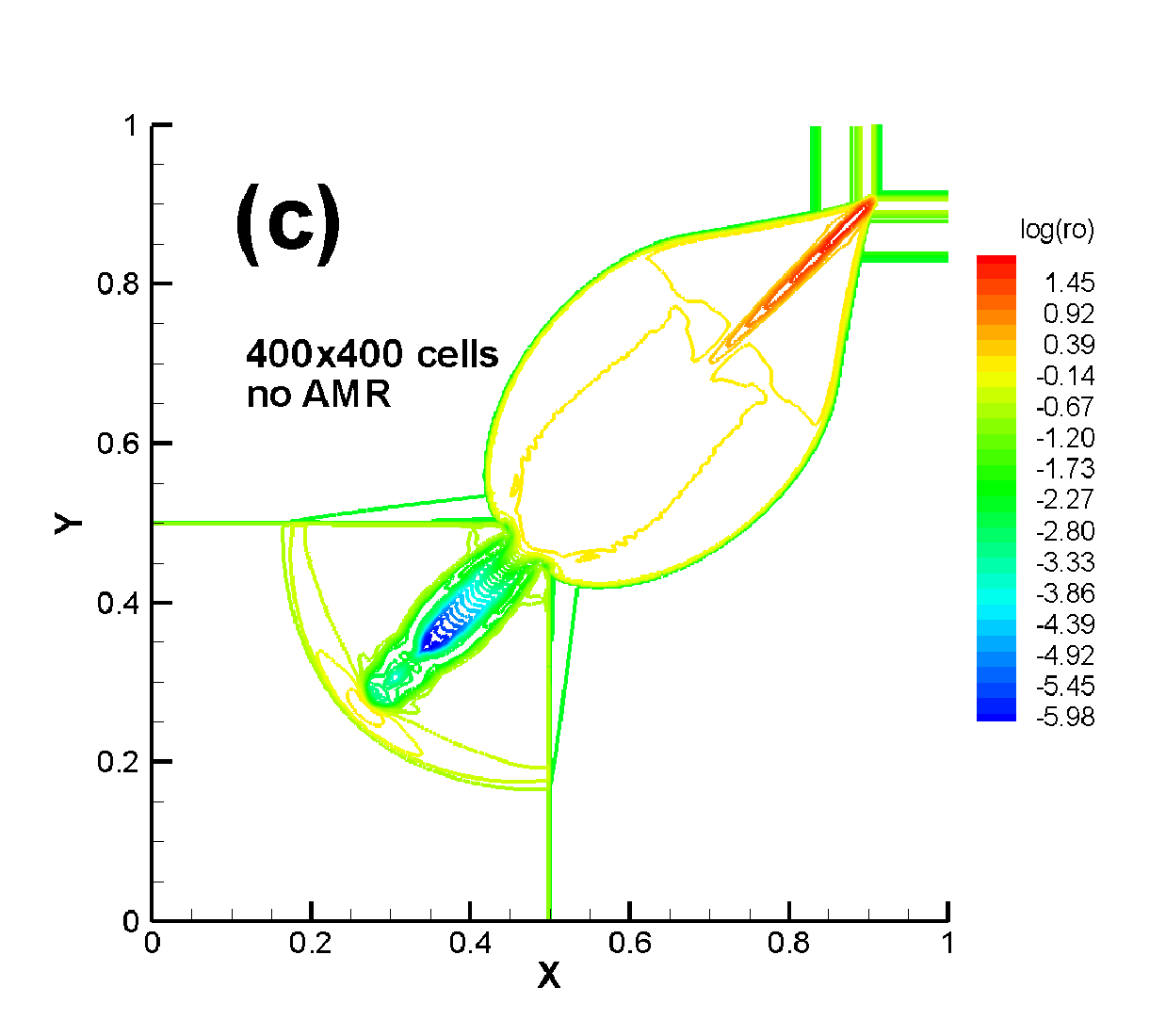}
\includegraphics[scale=0.35]{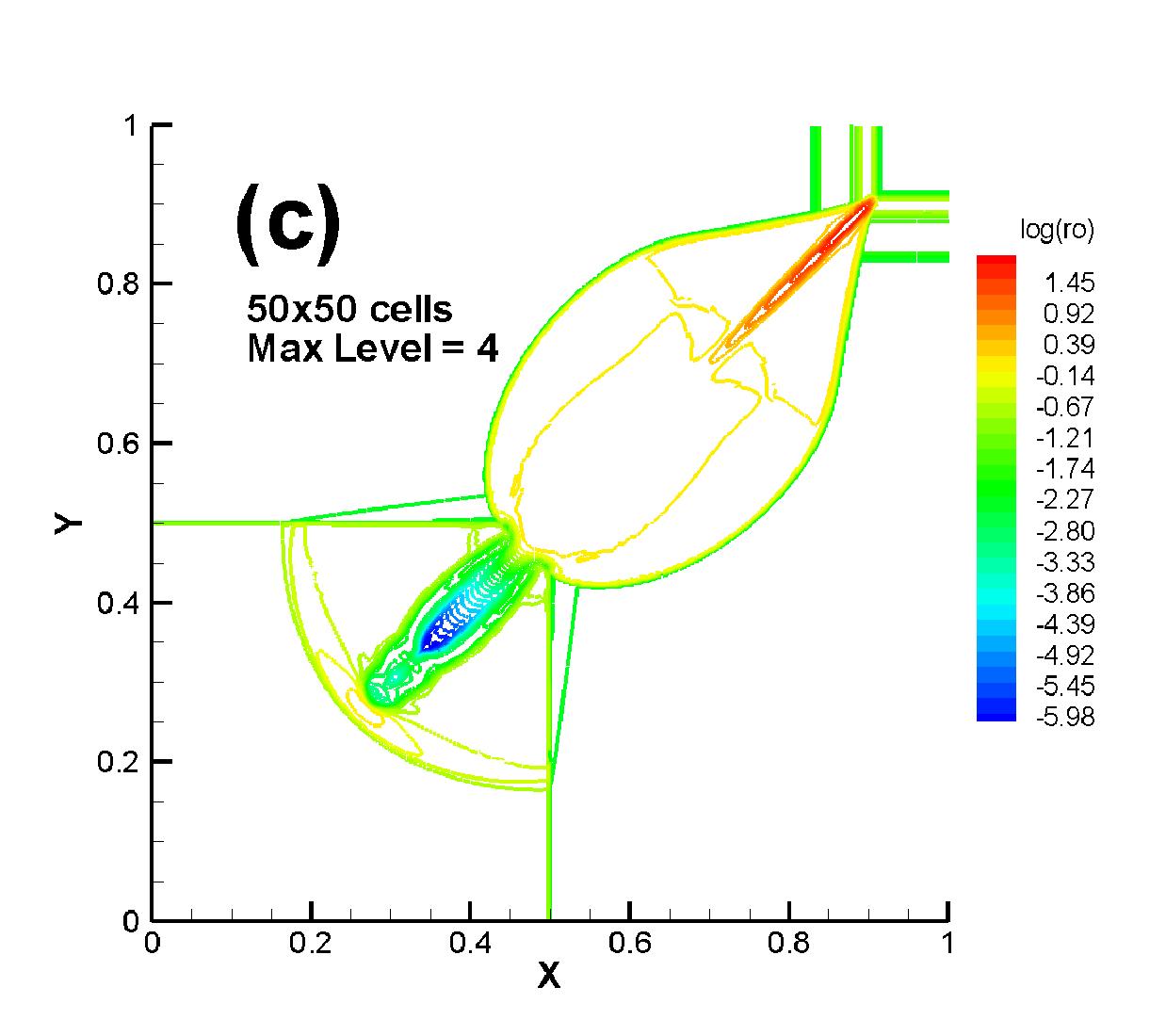}\\
\caption{ Two dimensional Riemann problem at $t = 0.4$. Results of
calculations using CT-PPM-HLLC, CT-PPM-HLL, and RK3-PLM-HLLC are
shown in panels (a), (b), and (c), respectively. Thirty equally
spaced contours of the logarithm of proper density are plotted. On
the left hand side are constant grid calculations, and on the right
hand side are equivalent resolution AMR calculations.
\protect\label{fig:riemann2d}}
\end{figure*}

In the following, the spatial domain is constructed of the
square $(x,y) \in [0,1]\times[0,1]$ filled with an ideal gas with an
adiabatic index of $\gamma=\frac{5}{3}$. Putting a superscript $NE$ refers to
the square $(x,y) \in [0.5,1]\times[0.5,1]$. Putting a superscript
$NW$ refers to the square $(x,y) \in [0,0.5]\times[0.5,1]$. Putting
a superscript $SE$ refers to the square $(x,y) \in
[0.5,1]\times[0,0.5]$. Putting a superscript $SW$ refers to the square
$(x,y) \in [0,0.5]\times[0,0.5]$. The four different zones defined
are filled with gas at the following thermodynamic states:
$$ \begin{array}{ll}
(\rho,v_x,v_y,p)^{NE} = & (0.1,0,0,0.01)\,, \\
(\rho,v_x,v_y,p)^{NW} = & (0.1,0.99,0,1)\,, \\
(\rho,v_x,v_y,p)^{SW} = & (0.5,0,0,1)\,, \\
(\rho,v_x,v_y,p)^{SE} = & (0.1,0,0.99,1)\,.
\end{array} $$
This test is also a standard test for numerical codes. It was suggested by \cite{del-zanna02} and repeated by \cite{lucas04,zhang06,wang08,meliani07,mignone05a,tchekhovskoy07}. We use constant zoning of $400 \times 400$
zones, and also AMR zoning of a base mesh with $50 \times 50$ and a
max level of refinement of 4 which is equivalent to the $400 \times
400$ zones mesh. The cycle with the maximum number of cells had
$~56,000$ cells which is $35\%$ of the total number of cells in the
constant zoning calculation, $160,000$. We have outflow boundary
conditions in all directions. Our results of
three schemes are shown in Fig.~\ref{fig:riemann2d}. Comparing our
result with other groups' result \cite{zhang06,wang08} shows good
agreement. The cross in the lower left corners of (b) CT-PPM-HLL, is
a numerical artifact due to the inability to maintain a contact
discontinuity perfectly, which is absent in the results using the
HLLC solver (a) and (c). This agrees with the result of
\cite{mignone05a}, that the HLLC solver behaves better in this
problem than other Riemann solvers, because of its ability to
resolve contact discontinuities. The appearance of two curved
shocks and the elongated diagonal spike of density in between in
panel NE is in agreement with the results of \cite{lucas04}, whereas
the diagonal feature is much less prominent in the results of
\cite{del-zanna02}. It is also seen that RELDAFNA could resolve this
2D problem with AMR with no distinguished and prominent differences
compared to the equivalent constant zoning mesh solution.

\addcontentsline{toc}{chapter}{Bibliography}
\bibliographystyle{ws-procs975x65}
\bibliography{thesis}

\end{document}